\pdfoutput=0
\documentclass[a4paper]{JHEP3}
\usepackage{amsmath}
\usepackage{amsfonts,amscd,dsfont}
\usepackage{feynmp}
\usepackage{amsthm}
\usepackage{amssymb}
\usepackage{graphicx}

\usepackage[toc,page]{appendix}

\usepackage{etoolbox}

\usepackage{cite}

\usepackage[caption=false]{subfig}

\newcommand{\eps}{\epsilon}
\newcommand{\ord}{\begin{cal}O\end{cal}}

\def\be{\begin{equation}}
\def\ee{\end{equation}}
\def\beq{\begin{equation}}
\def\eeq{\end{equation}}
\def\bsp#1\esp{\begin{split}#1\end{split}}

\newcommand{\cA}{\begin{cal}A\end{cal}}

\newcommand{\cE}{\begin{cal}E\end{cal}}
\newcommand{\cF}{\begin{cal}F\end{cal}}
\newcommand{\cG}{\begin{cal}G\end{cal}}
\newcommand{\cH}{\begin{cal}H\end{cal}}

\newcommand{\cL}{\begin{cal}L\end{cal}}
\newcommand{\cM}{\begin{cal}M\end{cal}}
\newcommand{\cN}{\begin{cal}N\end{cal}}

\newcommand{\cR}{\begin{cal}R\end{cal}}

\newcommand{\cX}{\begin{cal}X\end{cal}}

\newcommand{\cZ}{\begin{cal}Z\end{cal}}

\newcommand{\fa}{\mathfrak{a}}
\newcommand{\fb}{\mathfrak{b}}
\newcommand{\fc}{\mathfrak{c}}
\newcommand{\fd}{\mathfrak{d}}

\newcommand{\zbar}{{\bar z}}

\newcommand{\zb}{{\bar z}}

\newcommand{\bx}{{\bf x}}

\newcommand{\wb}{\bar{w}}

\newcommand{\bfs}{{\bf s}}

\newtheorem{claim}{Claim}

\def \sha{{\,\amalg\hskip -3.6pt\amalg\,}}
\appto\appendix{\addtocontents{toc}{\protect\setcounter{tocdepth}{1}}}

\appto\listoffigures{\addtocontents{lof}{\protect\setcounter{tocdepth}{1}}}
\appto\listoftables{\addtocontents{lot}{\protect\setcounter{tocdepth}{1}}}

\title{
Multi-Regge kinematics and the moduli space of Riemann spheres with marked points
}

\author{Vittorio Del Duca$^{a}$\footnote{On leave from Istituto Nazionale di Fisica Nucleare,
Laboratori Nazionali di Frascati, Italy.},
Stefan Druc$^b$,
James Drummond$^b$,
Claude Duhr$^{c,d}$\footnote{On leave from the ``Fonds National de la Recherche Scientifique'' (FNRS), Belgium.},
Falko Dulat$^e$,
Robin Marzucca$^d$,
Georgios Papathanasiou$^e$,
Bram Verbeek$^d$\\
${}^a$ Institute for Theoretical Physics, ETH Z\"urich, 8093 Z\"urich, Switzerland.\\
${}^b$ School of Physics \& Astronomy, University of Southampton, \\
\phantom{${}^b$} Highfield, Southampton, SO17 1BJ, United Kingdom.\\
${}^c$ Theoretical Physics Department, CERN, CH-1211 Geneva 23, Switzerland.\\
${}^d$ Center for Cosmology, Particle Physics and Phenomenology (CP3),\\
\phantom{{}$^d$} Universit\'e catholique de Louvain,\\
\phantom{{}$^d$} Chemin du Cyclotron 2, 1348 Louvain-La-Neuve, Belgium.\\
${}^e$ SLAC National Accelerator Laboratory, Stanford University, \\
\phantom{${}^e$} Stanford, CA 94309, USA.
}

\preprint{CP3-16-32, CERN-TH-2016-143, SLAC-PUB-16659}

\abstract{
We show that scattering amplitudes in planar $\cN=4$ Super Yang-Mills in multi-Regge kinematics
can naturally be expressed in terms of single-valued iterated integrals on the moduli space of
Riemann spheres with marked points. As a consequence, scattering amplitudes in this limit can be expressed
as convolutions that can easily be computed using Stokes' theorem. We apply this framework to MHV amplitudes
to leading-logarithmic accuracy (LLA),
and we prove that at $L$ loops all MHV amplitudes are determined by amplitudes with up to $L+4$ external legs.
We also investigate non-MHV amplitudes, and we show that they can be obtained by convoluting the MHV results with a certain helicity flip kernel. We classify all leading singularities that appear at LLA in the Regge limit for arbitrary helicity configurations and any number of external legs. Finally, we use our new framework to obtain explicit analytic results at LLA for all MHV amplitudes up to five loops and all non-MHV amplitudes with up to eight external legs and four loops.
}


\def\skeleton{
                  \fmfstraight
                     \fmfleft{p1,pdn1,pdn2,pd5,pd4,p2}
                     \fmfright{x1,x2,x3,x5,x6,x7}
                     \fmf{phantom}{p1,u1,v1n,u2,pn,u3,x1}
                     \fmf{phantom}{p2,o1,v23,o2,p3,o3,x7}
                     \fmffreeze
                     \fmf{phantom}{pn,pn1,pn2,pax6,pax5,pax4,pax3,pax2,pax1,p5,p4,p3}
                     \fmffreeze
                     \fmf{plain}{p2,v23}
                     \fmf{plain}{v23,p3}
                     \fmf{phantom}{v23,v4,v5,ax1,ax2,ax3,ax4,ax5,ax6,vn2,vn1,v1n}
                     \fmfv{decoration.shape=circle,decoration.size=2,label=$\scriptstyle x_2$,l.a=90}{v23}
                     \fmffreeze
                     \fmf{zigzag}{v23,v4}
                     \fmf{zigzag}{v4,v5}
                     \fmf{phantom}{v5,v67,ax1}
                     \fmfv{label=$\vdots$,l.d=-1mm}{v67}
                     \fmf{zigzag}{ax1,ax2}
                     \fmf{zigzag}{ax2,ax3}
                    \fmf{zigzag}{ax3,ax4} 
                    \fmf{zigzag}{ax4,ax5}
                    \fmf{zigzag}{ax5,ax6}
                     \fmfv{label=$\vdots$,l.d=-1mm}{ax6}
                     \fmf{zigzag}{vn2,vn1}
                     \fmf{zigzag}{vn1,v1n}
                     \fmf{plain}{p1,v1n}
                     \fmf{plain}{v1n,pn}
                     \fmfv{decoration.shape=circle,decoration.size=2,label=$\scriptstyle x_N$,l.a=-90}{v1n}
                     \fmffreeze
                     \fmf{plain}{v4,p4}
                     \fmf{plain}{ax2,pax2}
                      \fmf{plain}{ax3,pax3}
                     \fmf{plain}{ax4,pax4}
                     \fmf{plain}{ax5,pax5}
                     \fmf{plain}{vn1,pn1}
                     \fmfv{label=$\scriptstyle{k}_1$,l.a=0}{p4}
                     \fmfv{label=$\scriptstyle{k}_{i-2}$,l.a=0}{pax2}
                     \fmfv{label=$\scriptstyle{k}_{i-1}$,l.a=0}{pax3}
                     \fmfv{label=$\scriptstyle{k}_{i}$,l.a=0}{pax4}
                     \fmfv{label=$\scriptstyle{k}_{i+1}$,l.a=0}{pax5}
                     \fmfv{label=$\scriptstyle{k}_{N-4}$,l.a=0}{pn1}
                     \fmffreeze
                                          \fmf{phantom}{v5,v67,vn2}
                     \fmfv{label=$\vdots$,l.d=-1mm}{v67}
                     \fmffreeze
                     \fmf{phantom}{p2,xfu,v23}
                     \fmf{phantom}{p1,xfd,v1n}
                     \fmf{phantom}{v23,x1u,p3}
                     \fmf{phantom}{v4,xp4,p4}
                     \fmf{phantom}{v5,xp5,p5}
                     \fmf{phantom}{ax1,xpax1,pax1}
                     \fmf{phantom}{ax2,xpax2,pax2}
                     \fmf{phantom}{ax3,xpax3,pax3}
                     \fmf{phantom}{ax4,xpax4,pax4}
                     \fmf{phantom}{ax5,xpax5,pax5}
                     \fmf{phantom}{ax6,xpax6,pax6}
                     \fmf{phantom}{vn2,xpn2,pn2}
                     \fmf{phantom}{vn1,xpn1,pn1}
                     \fmf{phantom}{v1n,x1d,pn}
                     \fmffreeze
                     \fmf{phantom}{xfu,xf,xfd}
                     \fmfv{decoration.shape=circle,decoration.size=2,label=$\scriptstyle{x}_{1}$,l.a=180}{xf}
                     \fmf{phantom}{x1u,xx1,xp4,xx2,xp5,xpax1,xxi1,xpax2,xxi2,xpax3,xxi3,xpax4,xxi4,xpax5,xxi5,xpax6,xpn2,xxn4,xpn1,xxn3,x1d}
                     \fmfv{decoration.shape=circle,decoration.size=2,label=$\scriptstyle{x}_{3}$,l.a=0}{xx1}
                     \fmfv{decoration.shape=circle,decoration.size=2,label=$\scriptstyle{x}_{4}$,l.a=0}{xx2}
                     \fmfv{decoration.shape=circle,decoration.size=2,label=$\scriptstyle{x}_{i+1}$,l.a=0}{xxi1}
                     \fmfv{decoration.shape=circle,decoration.size=2,label=$\scriptstyle{x}_{i+2}$,l.a=0}{xxi2}
                     \fmfv{decoration.shape=circle,decoration.size=2,label=$\scriptstyle{x}_{i+3}$,l.a=0}{xxi3}
                     \fmfv{decoration.shape=circle,decoration.size=2,label=$\scriptstyle{x}_{i+4}$,l.a=0}{xxi4}
                     \fmfv{decoration.shape=circle,decoration.size=2,label=$\scriptstyle{x}_{i+5}$,l.a=0}{xxi5}
                     \fmfv{decoration.shape=circle,decoration.size=2,label=$\scriptstyle{x}_{N-2}$,l.a=0}{xxn4}
                     \fmfv{decoration.shape=circle,decoration.size=2,label=$\scriptstyle{x}_{N-1}$,l.a=0}{xxn3}
                     \fmffreeze
     }

     \def\skeletonQ{
                  \fmfstraight
                     \fmfleft{p1,pdn1,pdn2,pd5,pd4,p2}
                     \fmfright{x1,x2,x3,x5,x6,x7}
                     \fmf{phantom}{p1,u1,v1n,u2,pn,u3,x1}
                     \fmf{phantom}{p2,o1,v23,o2,p3,o3,x7}
                     \fmffreeze
                     \fmf{phantom}{pn,pn1,pn2,pax6,pax5,pax4,pax3,pax2,pax1,p5,p4,p3}
                     \fmffreeze
                     \fmf{plain}{p2,v23}
                     \fmf{plain}{v23,p3}
                     \fmf{phantom}{v23,v4,v5,ax1,ax2,ax3,ax4,ax5,ax6,vn2,vn1,v1n}
                     \fmfv{decoration.shape=circle,decoration.size=2,label=$\scriptstyle x_2$,l.a=90}{v23}
                     \fmffreeze
                     \fmf{zigzag,label=$\scriptstyle q_0$}{v23,v4}
                     \fmf{zigzag,label=$\scriptstyle q_1$}{v4,v5}
                     \fmf{phantom}{v5,v67,ax1}
                     \fmfv{label=$\vdots$,l.d=-1mm}{v67}
                     \fmf{zigzag}{ax1,ax2}
                     \fmf{zigzag}{ax2,ax3}
                    \fmf{zigzag,label=$\scriptstyle q_i$}{ax3,ax4} 
                    \fmf{zigzag}{ax4,ax5}
                    \fmf{zigzag}{ax5,ax6}
                     \fmfv{label=$\vdots$,l.d=-1mm}{ax6}
                     \fmf{zigzag,label=$\scriptstyle q_{N-5}$}{vn2,vn1}
                     \fmf{zigzag,label=$\scriptstyle q_{N-4}$}{vn1,v1n}
                     \fmf{plain}{p1,v1n}
                     \fmf{plain}{v1n,pn}
                     \fmfv{decoration.shape=circle,decoration.size=2,label=$\scriptstyle x_N$,l.a=-90}{v1n}
                     \fmffreeze
                     \fmf{plain}{v4,p4}
                     \fmf{plain}{ax2,pax2}
                      \fmf{plain}{ax3,pax3}
                     \fmf{plain}{ax4,pax4}
                     \fmf{plain}{ax5,pax5}
                     \fmf{plain}{vn1,pn1}
                     \fmfv{label=$\scriptstyle{k}_1$,l.a=0}{p4}
                     \fmfv{label=$\scriptstyle{k}_{i-2}$,l.a=0}{pax2}
                     \fmfv{label=$\scriptstyle{k}_{i-1}$,l.a=0}{pax3}
                     \fmfv{label=$\scriptstyle{k}_{i}$,l.a=0}{pax4}
                     \fmfv{label=$\scriptstyle{k}_{i+1}$,l.a=0}{pax5}
                     \fmfv{label=$\scriptstyle{k}_{N-4}$,l.a=0}{pn1}
                     \fmffreeze
                                          \fmf{phantom}{v5,v67,vn2}
                     \fmfv{label=$\vdots$,l.d=-1mm}{v67}
                     \fmffreeze
                     \fmf{phantom}{p2,xfu,v23}
                     \fmf{phantom}{p1,xfd,v1n}
                     \fmf{phantom}{v23,x1u,p3}
                     \fmf{phantom}{v4,xp4,p4}
                     \fmf{phantom}{v5,xp5,p5}
                     \fmf{phantom}{ax1,xpax1,pax1}
                     \fmf{phantom}{ax2,xpax2,pax2}
                     \fmf{phantom}{ax3,xpax3,pax3}
                     \fmf{phantom}{ax4,xpax4,pax4}
                     \fmf{phantom}{ax5,xpax5,pax5}
                     \fmf{phantom}{ax6,xpax6,pax6}
                     \fmf{phantom}{vn2,xpn2,pn2}
                     \fmf{phantom}{vn1,xpn1,pn1}
                     \fmf{phantom}{v1n,x1d,pn}
                     \fmffreeze
                     \fmf{phantom}{xfu,xf,xfd}
                     \fmfv{decoration.shape=circle,decoration.size=2,label=$\scriptstyle{x}_{1}$,l.a=180}{xf}
                     \fmf{phantom}{x1u,xx1,xp4,xx2,xp5,xpax1,xxi1,xpax2,xxi2,xpax3,xxi3,xpax4,xxi4,xpax5,xxi5,xpax6,xpn2,xxn4,xpn1,xxn3,x1d}
                     \fmfv{decoration.shape=circle,decoration.size=2,label=$\scriptstyle{x}_{3}$,l.a=0}{xx1}
                     \fmfv{decoration.shape=circle,decoration.size=2,label=$\scriptstyle{x}_{4}$,l.a=0}{xx2}
                     \fmfv{decoration.shape=circle,decoration.size=2,label=$\scriptstyle{x}_{i+1}$,l.a=0}{xxi1}
                     \fmfv{decoration.shape=circle,decoration.size=2,label=$\scriptstyle{x}_{i+2}$,l.a=0}{xxi2}
                     \fmfv{decoration.shape=circle,decoration.size=2,label=$\scriptstyle{x}_{i+3}$,l.a=0}{xxi3}
                     \fmfv{decoration.shape=circle,decoration.size=2,label=$\scriptstyle{x}_{i+4}$,l.a=0}{xxi4}
                     \fmfv{decoration.shape=circle,decoration.size=2,label=$\scriptstyle{x}_{i+5}$,l.a=0}{xxi5}
                     \fmfv{decoration.shape=circle,decoration.size=2,label=$\scriptstyle{x}_{N-2}$,l.a=0}{xxn4}
                     \fmfv{decoration.shape=circle,decoration.size=2,label=$\scriptstyle{x}_{N-1}$,l.a=0}{xxn3}
                     \fmffreeze
     }

 \def\transverseskeleton{
   \fmfstraight
                     \fmfleft{p1,pdn1,pdn2,pd5,pd4,p2}
                     \fmfright{x1,x2,x3,x5,x6,x7}
                     \fmf{phantom}{p1,u1,v1n,u2,pn,u3,x1}
                     \fmf{phantom}{p2,o1,v23,o2,p3,o3,x7}
                     \fmffreeze
                     \fmf{phantom}{pn,pn1,pn2,p5,p4,p3}
                     \fmffreeze
                     \fmf{dashes}{p2,v23}
                     \fmf{dashes}{v23,p3}
                     \fmf{phantom}{v23,v4,v5,vn2,vn1,v1n}
                     \fmffreeze
                     \fmf{plain,label=$\scriptstyle{\bf q}_{0}$,label.side=right,l.d=0.055w}{v23,v4}
                     \fmf{plain,label=$\scriptstyle{\bf q}_{1}$,label.side=right,l.d=0.055w}{v4,v5}
                     \fmf{plain,label=$\scriptstyle{\bf q}_{N-5}$,label.side=right,l.d=0.055w}{vn2,vn1}
                     \fmf{plain,label=$\scriptstyle{\bf q}_{N-4}$,label.side=right,l.d=0.055w}{vn1,v1n}
                     \fmf{dashes}{p1,v1n}
                     \fmf{dashes}{v1n,pn}
                     \fmf{plain}{v4,p4}
                     \fmf{plain}{v5,p5}
                     \fmf{plain}{vn2,pn2}
                     \fmf{plain}{vn1,pn1}
                     \fmfv{label=$\scriptstyle{\bf k}_1$,l.a=0}{p4}
                     \fmfv{label=$\scriptstyle{\bf k}_2$,l.a=0}{p5}
                     \fmfv{label=$\scriptstyle{\bf k}_{N-5}$,l.a=0}{pn2}
                     \fmfv{label=$\scriptstyle{\bf k}_{N-4}$,l.a=0}{pn1}
                     \fmffreeze
                                          \fmf{phantom}{v5,v67,vn2}
                     \fmfv{label=$\vdots$,l.d=-1mm}{v67}
                     \fmffreeze
                     \fmf{phantom}{p2,xfu,v23}
                     \fmf{phantom}{p1,xfd,v1n}
                     \fmf{phantom}{v23,x1u,p3}
                     \fmf{phantom}{v4,xp4,p4}
                     \fmf{phantom}{v5,xp5,p5}
                     \fmf{phantom}{vn2,xpn2,pn2}
                     \fmf{phantom}{vn1,xpn1,pn1}
                     \fmf{phantom}{v1n,x1d,pn}
                     \fmffreeze
                     \fmf{phantom}{xfu,xf,xfd}
                     \fmfv{decoration.shape=circle,decoration.size=2,label=$\scriptstyle{\bf x}_{1}$,l.a=180}{xf}
                     \fmf{phantom}{x1u,xx1,xp4,xx2,xp5,xpn2,xxn4,xpn1,xxn3,x1d}
                     \fmfv{decoration.shape=circle,decoration.size=2,label=$\scriptstyle{\bf x}_{2}$,l.a=0}{xx1}
                     \fmfv{decoration.shape=circle,decoration.size=2,label=$\scriptstyle{\bf x}_{3}$,l.a=0}{xx2}
                     \fmfv{decoration.shape=circle,decoration.size=2,label=$\scriptstyle{\bf x}_{N-3}$,l.a=0}{xxn4}
                     \fmfv{decoration.shape=circle,decoration.size=2,label=$\scriptstyle{\bf x}_{N-2}$,l.a=0}{xxn3}
}
     
 \def\notation{
   \fmfstraight
                     \fmfleft{p1,pdn1,pdn2,pd5,pd4,p2}
                     \fmfright{x1,x2,x3,x5,x6,x7}
                     \fmf{phantom}{p1,u1,v1n,u2,pn,u3,x1}
                     \fmf{phantom}{p2,o1,v23,o2,p3,o3,x7}
                     \fmffreeze
                     \fmf{phantom}{pn,pn1,pn2,p5,p4,p3}
                     \fmffreeze
                     \fmf{dashes}{p2,v23}
                     \fmf{dashes}{v23,p3}
                     \fmf{phantom}{v23,v4,v5,vn2,vn1,v1n}
                     \fmffreeze
                     \fmf{plain}{v23,v4}
                     \fmf{plain,label=\tiny $\quad\quad\rho_1\ i_1$,label.side=left}{v4,v5}
                     \fmf{plain,label=\tiny $\ \rho_{N-5}\ i_{N-5}$,l.side=left}{vn2,vn1}
                     \fmf{plain}{vn1,v1n}
                     \fmf{dashes}{p1,v1n}
                     \fmf{dashes}{v1n,pn}
                     \fmf{plain}{v4,p4}
                     \fmf{plain}{v5,p5}
                     \fmf{plain}{vn2,pn2}
                     \fmf{plain}{vn1,pn1}
                     \fmffreeze
                                          \fmf{dots}{v5,vn2}
                    \fmffreeze
                    \fmf{phantom}{p2,xfu,v23}
                     \fmf{phantom}{p1,xfd,v1n}
                     \fmf{phantom}{v23,x1u,p3}
                     \fmf{phantom}{v4,xp4,p4}
                     \fmf{phantom}{v5,xp5,p5}
                     \fmf{phantom}{vn2,xpn2,pn2}
                     \fmf{phantom}{vn1,xpn1,pn1}
                     \fmf{phantom}{v1n,x1d,pn}
                     \fmffreeze
                     \fmf{phantom}{xfu,xf,xfd}
                     \fmfv{label=\tiny $\!\!\!1$,l.a=0}{xf}
                     \fmf{phantom}{x1u,xx1,xp4,xx2,xp5,xpn2,xxn4,xpn1,xxn3,x1d}
                     \fmfv{label=\tiny $\!\!\!0$,l.a=0}{xx1}
                     \fmfv{label=\tiny $\!\!\!\infty$,l.a=0}{xxn3}

\fmffreeze
                    \fmfv{label=\tiny $\!\!\!h_1$,l.a=0}{p4}
                    \fmfv{label=\tiny $\!\!\!h_2$,l.a=0}{p5}
                     \fmfv{label=\tiny $\!\!\!h_{N-5}$,l.a=0}{pn2}
                     \fmfv{label=\tiny $\!\!\!h_{N-4}$,l.a=0}{pn1}
}

 \def\thmleft{
   \fmfstraight
  \fmfright{p2,k1,kim1,ki,kip1,kip2,kn4,pn1}
  \fmfleft{p1,a1,a2,a3,a4,a5,a6,pn}
  \fmf{dashes}{p1,v1,p2}
  \fmf{dashes}{pn,v8,pn1}
  \fmffreeze
  \fmf{phantom}{v1,v2,v3,v4,v5,v6,v7,v8}
  \fmf{plain}{v1,v2}
  \fmf{dots}{v2,v3}
  \fmf{plain,label=\tiny $\rho_c\ i_c$,l.side=right}{v3,v4}
  \fmf{plain,label=\tiny $\rho_b\ 0$,l.side=right}{v4,v5}
  \fmf{plain,label=\tiny $\rho_a\ i_a$,l.side=right}{v5,v6}
  \fmf{dots}{v6,v7}
  \fmf{plain}{v7,v8}
  \fmffreeze
  \fmf{plain}{v2,k1}
  \fmf{plain}{v3,kim1}
  \fmf{plain}{v4,ki}
  \fmf{plain}{v5,kip1}
  \fmf{plain}{v6,kip2}
  \fmf{plain}{v7,kn4}
  \fmffreeze
  \fmfv{label=\tiny $\!\!h$,l.a=0}{ki}
  \fmfv{label=\tiny $\!\!h$,l.a=0}{kip1}
 }

 \def\thmright{
   \fmfstraight
  \fmfright{p2,k1,kim1,ki,kip1,kn4,pn1}
  \fmfleft{p1,a1,a2,a3,a4,a5,pn}
  \fmf{dashes}{p1,v1,p2}
  \fmf{dashes}{pn,v7,pn1}
  \fmffreeze
  \fmf{phantom}{v1,v2,v3,v4,v5,v6,v7}
  \fmf{plain}{v1,v2}
  \fmf{dots}{v2,v3}
  \fmf{plain,label=\tiny $\rho_c\ i_c$,l.side=right}{v3,v4}
  \fmf{plain,label=\tiny $\rho_a\ i_a$,l.side=right}{v4,v5}
  \fmf{dots}{v5,v6}
  \fmf{plain}{v6,v7}
  \fmffreeze
  \fmf{plain}{v2,k1}
  \fmf{plain}{v3,kim1}
  \fmf{plain}{v4,ki}
  \fmf{plain}{v5,kip1}
  \fmf{plain}{v6,kn4}
  \fmffreeze
  \fmfv{label=\tiny $\!\!h$,l.a=0}{ki}
 }

\def\reggecut    {
   \fmfstraight
   \fmftop{p1,p2}
   \fmfbottom{pn,p3}
   \fmf{plain}{p1,x1,y1,pn}
   \fmf{plain}{p2,x2,y2,p3}
   \fmf{phantom}{pn,kn4,kn3,e11,e12,kj,kjm1,e21,e22,ki1,ki,e31,e32,k2,k1,p3}
   \fmffreeze
   \fmf{phantom}{y1,ykn4,ykn3,ye11,ye12,ykj,ykjm1,ye21,ye22,yki1,yki,ye31,ye32,yk2,yk1,y2}
   \fmf{zigzag}{x1,xkn4,xkn3,xe11}
   \fmf{dots}{xe11,xe12}
   \fmf{zigzag}{xe12,xkj,xkjm1,xe21}
   \fmf{dots}{xe21,xe22}
   \fmf{zigzag}{xe22,xki1,xki,xe31}
   \fmf{dots}{xe31,xe32}
   \fmf{zigzag}{xe32,xk2,xk1,x2}
   \fmffreeze
   \fmf{zigzag}{ykj,ykjm1,ye21}
   \fmf{dots}{ye21,ye22}
   \fmf{zigzag}{ye22,yki1,yki}
   \fmffreeze
   \fmf{plain}{ykn4,xkn4,kn4}
   \fmf{plain}{ykn3,xkn3,kn3}
   \fmf{plain}{ykj,xkj,kj}
   \fmf{plain}{ykjm1,xkjm1,kjm1}
   \fmf{plain}{yki1,xki1,ki1}
   \fmf{plain}{yki,xki,ki}
   \fmf{plain}{yk2,xk2,k2}
   \fmf{plain}{yk1,xk1,k1}
   \fmffreeze
   \fmf{phantom}{ye12,s1a,xe12}
   \fmf{phantom}{ykj,s1b,xkj}
   \fmf{phantom}{ye31,s2a,xe31}
   \fmf{phantom}{yki,s2b,xki}
   \fmffreeze
   \fmf{phantom}{s1a,ss1,s1b}
   \fmf{phantom}{s2a,ss2,s2b}
   \fmffreeze
   \fmf{dashes}{ss1,ss2}
   \fmfv{label=$\scriptstyle k_{N-4}$,l.a=-90}{kn4}
   \fmfv{label=$\scriptstyle k_{N-3}$,l.a=-90}{kn3}
   \fmfv{label=$\scriptstyle k_{q}$,l.a=-90}{kj}
   \fmfv{label=$\scriptstyle k_{q-1}$,l.a=-90}{kjm1}
   \fmfv{label=$\scriptstyle k_{p+1}$,l.a=-90}{ki1}
   \fmfv{label=$\scriptstyle k_{p}$,l.a=-90}{ki}
   \fmfv{label=$\scriptstyle k_{2}$,l.a=-90}{k2}
   \fmfv{label=$\scriptstyle k_{1}$,l.a=-90}{k1}
   \fmfv{label=$\scriptstyle p_{N}$,l.a=-90}{pn}
   \fmfv{label=$\scriptstyle p_{3}$,l.a=-90}{p3}
   \fmfv{label=$\scriptstyle p_{1}$,l.a=90}{p1}
   \fmfv{label=$\scriptstyle p_{2}$,l.a=90}{p2}
   }

\begin{document}

\maketitle

\section{Introduction}
\label{sec:intro}

Over the last years there has been tremendous progress in understanding the structure of the $S$-matrix of the ${\cal N}=4$ Super Yang-Mills (SYM) theory. 
In the planar limit the $S$-matrix exhibits a dual conformal symmetry~\cite{Drummond:2006rz,Bern:2006ew,Bern:2007ct,Alday:2007hr,Drummond:2007aua,Brandhuber:2007yx}, and the dual conformal algebra closes together with the ordinary superconformal 
algebra in configuration space into an infinite dimensional Yangian algebra~\cite{Drummond:2009fd}.
The dual conformal invariance is broken by infrared divergences~\cite{Drummond:2007cf,Drummond:2007au}, but it is possible to construct finite, dual conformally invariant ratios in which all infrared divergences cancel. As a consequence, the analytic structure of scattering amplitudes in planar $\cN=4$ SYM is highly constrained. In particular, the four and five-point amplitudes are fixed to all loop orders by symmetries in terms of the one-loop amplitudes and the cusp anomalous dimension~\cite{Bern:2005iz,Drummond:2007au}, which is known exactly from integrability methods~\cite{Beisert:2006ez}. The first time non-trivial dual conformally invariant corrections appear is then for amplitudes with at least six external legs~\cite{Drummond:2007au,Bern:2008ap,Drummond:2008aq}. The ordinary and dual conformal symmetries are also at the heart of a duality between (colour-ordered) scattering amplitudes and Wilson loops computed along a light-like polygonal contour~\cite{Alday:2007hr,Alday:2007he,Berkovits:2008ic,Drummond:2007aua,Brandhuber:2007yx,Drummond:2007au,Drummond:2007cf,Drummond:2008aq,CaronHuot:2010ek,Mason:2010yk}. The Wilson loops can be described using an operator product expansion (OPE) near the collinear limit~\cite{Alday:2010ku,Gaiotto:2010fk,Gaiotto:2011dt,Sever:2011da}. The OPE approach to Wilson loops was recently extended by interpreting it as an exchange of excitations of a flux tube sourced by the sides of the light-like polygon. The spectrum of excitations of this flux tube, as well as their $S$-matrix, can in turn be determined at all values of the coupling by integrability methods~~\cite{Basso:2010in,Basso:2013vsa,Basso:2013aha,Basso:2014koa,Basso:2014jfa,Basso:2014nra,Basso:2014hfa,Basso:2015rta,Basso:2015uxa}.

The progress in understanding the structure of the $S$-matrix in planar ${\cal N}=4$ SYM is not only due to symmetries and integrability, but also due to an improved understanding of the mathematical structures underlying perturbative scattering amplitudes. Indeed, Yangian invariance is intimately related to the appearance of certain geometric and algebraic structures whose relevance for scattering amplitudes was not appreciated before. For example, the kinematics of an amplitude can be encoded in terms of momentum twistors~\cite{Hodges:2009hk}, elements of a three-dimensional projective space on which the dual conformal group acts linearly. In terms of momentum twistors a kinematic configuration is equivalent to a configuration of points in three-dimensional projective space $\mathbb{CP}^3$~\cite{Golden:2013xva}. Scattering amplitudes with $N$ external legs are then expected to be iterated integrals of certain differential one-forms~\cite{symbolsC} defined on the space of configurations of points $\textrm{Conf}_N(\mathbb{CP}^3)$~\cite{Golden:2014xqa}. The singularities of the iterated integrals should be described by a certain cluster algebra that is naturally associated with the space $\textrm{Conf}_N(\mathbb{CP}^3)$~\cite{cluster1,cluster2,scott,gekhtman,keller,Golden:2013xva}. The simplest instance of iterated integrals that one encounters when computing scattering amplitudes are the so-called multiple polylogarithms~\cite{Goncharov:2001,Brown:2009qja} which correspond to iterated integrals over rational functions. It is believed that all maximally helicity violating (MHV) and and next-to-MHV (NMHV) amplitudes in $\cN=4$ SYM can be expressed in terms of multiple polylogarithms of uniform transcendental weight~\cite{ArkaniHamed:2012nw}. In particular, this implies that all six and seven-point amplitudes are polylogarithmic functions.

The collinear OPE combined with the improved understanding of the geometry underlying planar scattering amplitudes has led to tremendous progress in determining the perturbative $S$-matrix in planar $\cN=4$ SYM, at least in the cases where the functions are expressible in terms of multiple polylogarithms. In particular, the six-point MHV and NMHV amplitudes are known explicitly up to five loops~\cite{DelDuca:2009au,DelDuca:2010zg,Goncharov:2010jf,Dixon:2011pw,Dixon:2011nj,Dixon:2013eka,Dixon:2014iba,Dixon:2014voa,Dixon:2015iva} while the seven-point MHV amplitude is known analytically at two loops~\cite{Golden:2014xqf}. Beyond seven points and two loops complete analytic results are currently unavailable, although some analytic results for more loops and legs are known in the situation where the kinematics is restricted to lie in a two-dimensional plane~\cite{DelDuca:2010zp,Heslop:2010kq,Caron-Huot:2013vda}. In addition, we know the symbols~\cite{symbolsC,symbols,Brown:2009qja,Goncharov:2010jf,Duhr:2011zq} of all two-loop MHV amplitudes~\cite{CaronHuot:2011ky} and of the three-loop seven-point MHV amplitude~\cite{Drummond:2014ffa}. The reasons for this lack of explicit analytic results for amplitudes with more loops and legs are, among others, that the corresponding cluster algebra is infinite starting from eight points and that it is expected that for higher-point amplitudes new classes of functions may appear that can no longer be expressed in terms of multiple polylogarithms~\cite{CaronHuot:2012ab}. As a consequence, it is no longer possible to classify all the singularities and classes of functions that may appear in the analytic results for these functions.

The aim of this paper is to study a kinematic limit, the multi-Regge limit, where we can completely describe the geometry underlying the scattering for any number external particles, and hence we can completely classify all the iterated integrals that appear in the final result. 
The study of this limit has its origins not in $\cN=4$ SYM, but it has been known since the early days of QCD that
in the Regge limit $s\gg |t|$ scattering amplitudes exhibit a rich analytic structure.
The paradigm example is the BFKL equation in QCD, which resums the radiative corrections in $\log(s/|t|)$
to parton-parton scattering at leading logarithmic accuracy (LLA)~\cite{Kuraev:1976ge,Kuraev:1977fs,Balitsky:1978ic} and
next-to-LLA (NLLA)~\cite{Fadin:1998py,Camici:1997ij,Ciafaloni:1998gs}.
The building blocks of the BFKL resummation at LLA are the multi-gluon amplitudes, which are evaluated in
multi-Regge kinematics (MRK), {i.e.,} in the approximation of a strong rapidity ordering of the outgoing gluons.
The multi-Regge limit is thus the kinematic cornerstone of the BFKL resummation at LLA.
In establishing the BFKL equation, the gluon rapidities are then integrated out,
and the BFKL equation is reduced to a two-dimensional problem in terms of purely transverse degrees of freedom: 
{i.e.,} the evolution of a $t$-channel gluon ladder in transverse momentum space and Mellin moment 
space.

In planar $\cN=4$ SYM in the Euclidean region where all Mandelstam invariants are negative, scattering amplitudes in MRK factorise to all orders in perturbation theory into certain building blocks describing the resummed effective propagators in the $t$-channel and the emission of gluons along the $t$-channel ladder formed by the effective propagators. These building are determined to all orders by the four and five-point amplitudes, and hence scattering amplitudes in MRK are trivial in the Euclidean region~\cite{Bartels:2008ce,Bartels:2008sc,Brower:2008nm,Brower:2008ia,DelDuca:2008jg}. Starting from six external particles scattering amplitudes have Regge cuts that are not captured correctly by the Regge-factorised form. As a consequence, amplitudes in MRK are no longer trivial if the multi-Regge limit is taken after analytic continuation to a Mandelstam region~\cite{Bartels:2008ce,Bartels:2008sc}. The discontinuity across the cut is described to all orders by a dispersion relation closely related to the BFKL evolution equation. The integrand of the dispersion integral factorises in Fourier-Mellin space, and the building blocks describing the factorisation are closely related to the energy spectrum and the $S$-matrix of the flux-tube excitations in the collinear OPE approach~\cite{Basso:2014pla,Drummond:2015jea}.

A lot of effort has recently gone into determining six-point scattering amplitudes in planar $\cN=4$ SYM in MRK, both at strong~\cite{Bartels:2010ej,Bartels:2013dja} and at weak coupling~\cite{Lipatov:2010ad,Lipatov:2010qg,Lipatov:2012gk,Dixon:2011pw,Dixon:2012yy,Dixon:2014voa,Pennington:2012zj,Basso:2014pla,Drummond:2015jea,Broedel:2015nfp}. In particular it was observed in ref.~\cite{Dixon:2012yy} that the six-point amplitude in MRK can be expressed perturbatively in terms of single-valued harmonic polylogarithms~\cite{BrownSVHPLs}. Moreover, one can write down a generating functional for all six-point amplitudes in MRK at LLA~\cite{Pennington:2012zj,Broedel:2015nfp}. Beyond six points only the two-loop MHV amplitudes are known, fully analytically to LLA~\cite{Prygarin:2011gd,Bartels:2011ge} and up to terms proportional to multiple-zeta values at NLLA~\cite{Bargheer:2015djt}. The seven-point amplitude in MRK has also been considered at strong coupling~\cite{Bartels:2012gq,Bartels:2014ppa,Bartels:2014mka}. One of the issues to push computations to more loops and legs is that evaluating the dispersion integrals leads to very complicated multiple sums, and the number of such sums increases with the number of external legs.

In this paper we study scattering amplitudes in planar $\cN=4$ SYM in MRK for any number $N$ of external legs and arbitrary helicity configurations. Since the central emission block describing the emission of a gluon along the $t$-channel ladder is currently only known to leading order, we restrict ourselves to LLA in this work. We foresee, however, that the methods and the tools introduced in this paper are generic and can be applied beyond LLA as well, and even to cross sections outside of $\cN=4$ SYM, which were shown to exhibit the same features as scattering amplitudes in MRK in $\cN=4$ SYM at least at LLA~\cite{DelDuca:2013lma}. Due to the strong ordering in rapidities, the only non-trivial kinematical dependence in the Regge limit can be through the transverse components of the external momenta. The cornerstone of our method is the realisation that the geometry in the transverse space can be described by a configuration of points in the complex plane. In other words, MRK is described by the geometry of the moduli space $\mathfrak{M}_{0,n}$ of Riemann spheres with $n$ marked points. The geometry of $\mathfrak{M}_{0,n}$ is well understood. In particular, the cluster algebra associated to $\mathfrak{M}_{0,n}$ is always of finite type and corresponds to the Dynkin diagram $A_{n-3}$. The algebra of iterated integrals on this space can also be completely described: they are iterated integrals of $d\log$-forms with singularities when some of the marked points coincide. It can be shown that the algebra of iterated integrals on $\mathfrak{M}_{0,n}$ factors through certain hyperlogarithm algebras~\cite{Brown:2009qja}. In other words, iterated integrals on $\mathfrak{M}_{0,n}$ can always be expressed in terms of multiple polylogarithms and rational functions with singularities when two marked points coincide.
The analytic structure of scattering amplitudes is further constrained by unitarity. We show that this requirement constrains the iterated integrals that can appear in MRK to single-valued functions, i.e., linear combinations of products of iterated integrals on $\mathfrak{M}_{0,n}$ (and their complex conjugates) such that all branch cuts cancel. This generalises the findings of ref.~\cite{Dixon:2012yy} that the six-point amplitude in MRK can be expressed in terms of single-valued harmonic polylogarithms for any number of external legs.

The framework of single-valued iterated integrals on $\mathfrak{M}_{0,n}$ allows us to compute scattering amplitudes in MRK for many loops and external legs. 
Starting from a (conjectural) representation of the amplitude in a Mandelstam region as a multiple Fourier-Mellin transform, we increase the number of loops by convoluting with the Fourier-Mellin transform of the BFKL eigenvalue. The single-valued nature of the functions reduces the computation of the convolution integral to a simple application of Stokes' theorem. In addition, we prove a certain factorisation theorem for amplitudes in MRK that generalises to any number of loops and legs a factorisation property observed for two-loop MHV amplitudes in MRK~\cite{Prygarin:2011gd,Bartels:2011ge}. The factorisation theorem implies in particular that MHV amplitudes at $L$ loops are completely determined by MHV amplitudes with up to $(L+4)$ external legs. We use this property to present for the first time analytic results for MHV amplitudes in MRK up to five loops with an arbitrary number of external legs. We also show that non-MHV amplitudes can be constructed via convolutions with a certain helicity flip kernel. Convolutions with this kernel introduce rational prefactors, and we perform a classification of all the leading singularities that appear in MRK at LLA. We also present for the first time explicit analytic results for non-MHV amplitudes with up to eight external legs and up to four loops.

This paper is organised as follows. In Section~\ref{sec:sec_mrk} we present a conjectural representation of scattering amplitudes in MRK at  LLA, valid for any number of loops or external legs. In Section~\ref{sec:svmpls} we discuss the connection between MRK and the moduli space of Riemann spheres with marked points, and we discuss in particular three different ways to construct single-valued iterated integrals on $\mathfrak{M}_{0,n}$. In Section~\ref{sec:mhv} we apply our technology to the case of MHV amplitudes. We prove the factorisation theorem for MHV amplitudes and obtain explicit analytic results for all MHV amplitudes with up to five loops with an arbitrary number of external legs. In Section~\ref{sec:nmhv} we introduce the helicity flip operator and we extend the results of Section~\ref{sec:mhv} to the non-MHV case. We also present a complete classification of leading singularities in MRK to LLA and obtain explicit result for all non-MHV amplitudes up to four loops and with up eight external legs. In Section~\ref{sec:MPL_proof} we discuss the implications of our work on the analytic structure of scattering amplitudes in MRK, and we prove that amplitudes in MRK can always be expressed in terms of single-valued polylogarithms, independently of the helicity configuration and the number of loops and legs. In Section~\ref{sec:conclusion} we draw our conclusions and discuss how our results can be extended beyond LLA. We include several appendices with technical details on the construction of single-valued functions and explicit results for amplitudes with up to eight external legs and three loops. All the results obtained in this paper are provided in computer-readable form as ancillary material with the arXiv submission of this paper.


\section{Scattering amplitudes in MRK to LLA}
\label{sec:sec_mrk}

\subsection{Scattering amplitudes and cluster algebras}
\label{sec:cluster}
\begin{figure}[!t]
    \centering
                \includegraphics[scale=0.42]{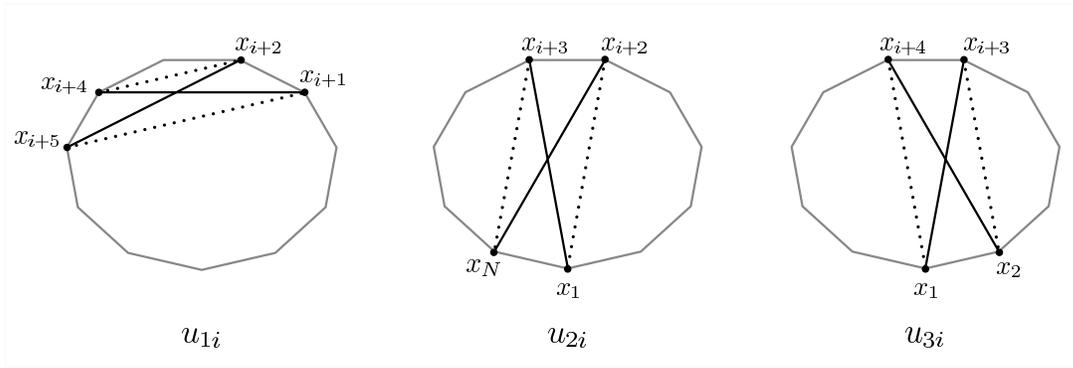}
    \caption{The algebraically independent set of $3N-15$ cross-ratios.}
    \label{fig:cross-ratios}
\end{figure}

In this section we review some background material on planar scattering amplitudes in the $\cN=4$ Super Yang-Mills (SYM) theory.
We begin by recalling some basic facts about the kinematical dependence of gauge theory amplitudes. An $N$-point colour-ordered gluonic scattering amplitude is a function of $N$ massless momenta $p_i$, $1\le i\le N$, with a specific cyclic ordering and subject to momentum conservation
\be
\sum_{i=1}^N p_i =0\,.
\ee
Null momenta $p_i$ in four dimensions may be described by choosing a pair of commuting spinorial variables so that
\be
p_i^{\alpha \dot{\alpha}} = \lambda_i^{\alpha} \tilde{\lambda}_i^{\dot \alpha}\,.
\label{momspinors}
\ee
An ordered sequence of null momenta $p_i$ obeying the momentum conservation condition
may also be described by a lightlike polygon in a dual space by choosing dual coordinates $x_i$ such that
\be
x_{i}-x_{i-1} = p_i\,.
\ee
Momentum conservation implies the closure of the polygon. In planar $\mathcal{N}=4$ SYM the amplitudes exhibit dual conformal symmetry, meaning that the essential kinematical dependence is captured by the conformal cross-ratios
\be\label{eq:Uijkl_def}
U_{ijkl} = \frac{x_{ij}^2 x_{kl}^2}{x_{ik}^2 x_{jl}^2}\,,
\ee
with $x_{ij}=x_i-x_j$. In four dimensions, only $3N-15$ of the cross ratios one can form out of $N$ points are independent.  Following ref.~\cite{Bartels:2012gq,Bartels:2014mka}, from the set of all the $U_{ijkl}$ one can pick a particular algebraically independent set of $3N-15$ cross ratios as (see Fig.~\ref{fig:cross-ratios}),
\be\label{eq:ui_def}
u_{1i} = \frac{x_{i+1,i+5}^2 x_{i+2,i+4}^2}{x_{i+1,i+4}^2x_{i+2,i+5}^2}\,,\qquad u_{2i} = \frac{x_{N,i+3}^2 x_{1,i+2}^2}{x_{N, i+2}^2 x_{1,i+3}^2}\,, \qquad u_{3i} = \frac{x_{1\,i+4}^2 x_{2, i+3}^2}{x_{1,i+3}^2 x_{2,i+4}^2}\,.
\ee
All other cross ratios can be expressed as algebraic functions in these $3N-15$ independent cross ratios.

The sequence of null lines describing the polygon may in turn be described by twistor variables\footnote{The unusual choice of the lower index in the definition of the twistor variable $Z_i$ is made in order to render common but competing conventions consistent with each other.}
\be
Z_i^A = (\lambda_{i+1}^\alpha , \mu_{i+1}^{\dot \alpha})\,.
\ee
The variables $Z_i$ are called \emph{momentum twistors}~\cite{Hodges:2009hk} due to their relation with the momenta. The twistors obey the incidence relation
\be
\mu_i^{\dot \alpha} = x_i^{\alpha \dot{\alpha}} \lambda_{i \alpha}\,.
\label{incidence}
\ee
The momentum twistors $Z_i$ are free variables\footnote{For real Minkowski signature momentum variables we should impose a reality condition upon the $Z_i$. For now we keep them complex.}, and every point $x_i$ is represented in momentum twistor space by the line passing through the twistors $Z_{i-1}$ and $Z_i$.
Since the relation \eqref{momspinors} allows for a rescaling
\be
\lambda_i \rightarrow \kappa_i \lambda_i\,, \qquad \tilde{\lambda}_i \rightarrow \kappa_i^{-1} \tilde{\lambda}_i\,,
\ee
and the incidence relation~\eqref{incidence} is homogeneous, we find that the $Z_i$ are actually elements of $\mathbb{CP}^3$. In terms of the twistor variables the kinematical variables are given by
\be
x_{ij}^2= (p_{i+1} + \ldots +p_{j})^2 = \frac{\langle i-1\,i\,j-1\,j \rangle}{\langle i-1\, i\,I \rangle \langle j-1\, j\, I\rangle}\,,
\label{twistorinvariants}
\ee
where the twistor four-bracket is defined as the determinant of the column vectors, $\langle ijkl \rangle = \det(Z_i\,Z_j\,Z_k\,Z_l)$. $I$ is the so-called \emph{infinity twistor} and corresponds to a choice of null cone at infinity in the coordinate patch given by the $x_i$. The dependence on the infinity twistor must drop out from all dual conformally invariant quantities. In particular, the dual conformally invariant cross ratios of eq.~\eqref{eq:Uijkl_def} can be written in the form
\be
U_{ijkl} = \frac{\langle i-1\,i\,j-1\,j\rangle\langle k-1\,k\,l-1\,l\rangle}{\langle i-1\,i\,k-1\,k\rangle\langle j-1\,j\,l-1\,l\rangle}\,.
\ee

\begin{figure}[!t]
    \centering
        \includegraphics[scale=0.42]{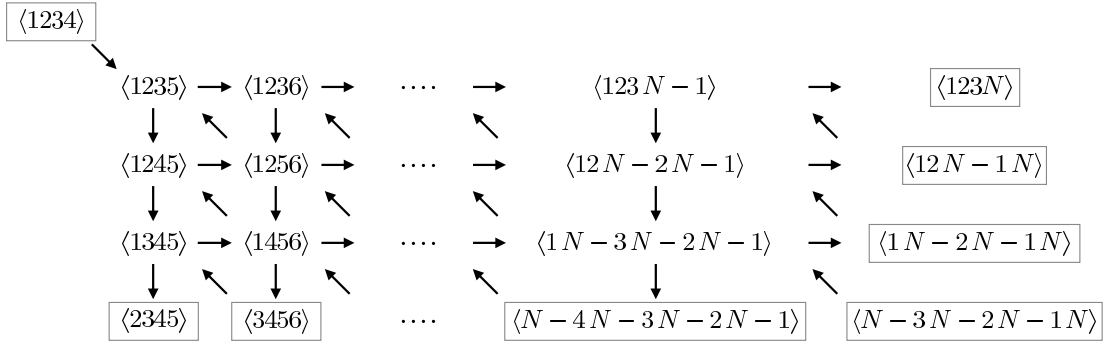}
    \caption{The $\mathcal{A}$-coordinates for the initial quiver for ${\rm Gr}(4,N)$ with frozen nodes in boxes.}
    \label{fig:clustercoords}
\end{figure}

\begin{figure}[!t]
    \centering
    \vspace{0.3cm}
                \includegraphics[scale=0.4]{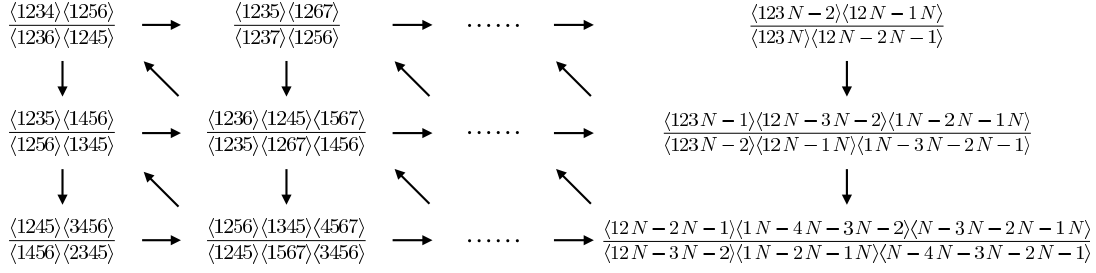}
    \caption{The $\mathcal{X}$-coordinates for the initial quiver for ${\rm Gr}(4,N)$.}
    \label{fig:clusterXcoords}
\end{figure}

As momentum twistors are free variables, we can describe the kinematics of colour-ordered partial amplitudes by a configuration of $N$ points in $\mathbb{CP}^3$~\cite{Golden:2013xva}. We denote the set of all such configurations by
\be
{\rm Conf}_N(\mathbb{CP}^3) \simeq {\rm Gr}(4,N)/(\mathbb{C}^*)^{N-1}\,.
\ee
Naturally associated to the spaces ${\rm Conf}_N(\mathbb{CP}^3)$ are cluster algebra structures~\cite{cluster1,cluster2,scott,gekhtman,keller}, which play a role in describing the singularity structure of scattering amplitudes or light-like Wilson loops in planar $\mathcal{N}=4$ SYM theory~\cite{Golden:2013xva}.
The $\mathcal{A}$-coordinates of the cluster algebras are homogeneous polynomials in the Pl\"ucker coordinates $\langle ijkl \rangle$. For the cluster algebras associated to ${\rm Gr}(4,N)$ one defines an initial cluster given by the quiver diagram in Fig. \ref{fig:clustercoords}. Other clusters are obtained by a repeated process called mutation. The $\mathcal{A}$-coordinates in the initial cluster are given by certain Pl\"ucker coordinates. The nodes in boxes are called \emph{frozen nodes} and the others are called \emph{unfrozen}. For each unfrozen node one can form  $\mathcal{X}$-coordinates by taking the product of all $\mathcal{A}$-coordinates connected by incoming arrows and dividing by the product of all $\mathcal{A}$-coordinates connected by outgoing ones. We label the $\mathcal{X}$-coordinates as $\mathcal{X}_{ij}$ for $i=1,2,3$ and $j=1,\ldots,N-5$ following the obvious structure of Fig. \ref{fig:clusterXcoords}. Explicitly, they are given by
\be\bsp\label{eq:X_coords}
\cX_{1j} &\,= \frac{\langle1\,2\,3\,j+3\rangle\langle1\,2\,j+4\,j+5\rangle}{\langle1\,2\,3\,j+5\rangle\langle1\,2\,j+3\,j+4\rangle}\,,\\
\cX_{2j} &\,= \frac{\langle1\,2\,3\,j+4\rangle\langle1\,2\,j+2\,j+3\rangle\langle1\,j+3\,j+4\,j+5\rangle}{\langle1\,2\,3\,j+3\rangle\langle1\,2\,j+4\,j+5\rangle\langle1\,j+2\,j+3\,j+4\rangle}\,,\\
\cX_{3j} &\,= \frac{\langle1\,2\,j+3\,j+4\rangle\langle1\,j+1\,j+2\,j+3\rangle\langle j+2\,j+3\,j+4\,j+5\rangle}{\langle1\,2\,j+2\,j+3\rangle\langle1\,j+3\,j+4\,j+5\rangle\langle j+1\,j+2\,j+3\,j+4\rangle}\,.
\esp\ee
 The $\mathcal{X}$-coordinates of any given cluster, in particular the initial one outlined above, form a complete set of coordinates for the kinematical dependence of the scattering amplitude or Wilson loop.

\subsection{Multi-Regge kinematics} \label{sec:mrk}

\begin{figure}[!t]
\begin{center}
\begin{minipage}{0.3\linewidth}
     \begin{fmffile}{u1i}
     \begin{fmfgraph*}(120,150)
           \skeletonQ
           \fmf{plain,left=.2}{xxi2,xxi4}
           \fmf{plain,right=.3}{xxi1,xxi5}
           \fmf{dashes,right=.2}{xxi1,xxi4}
           \fmf{dashes,left=.3}{xxi2,xxi5}
     \end{fmfgraph*}
     \end{fmffile}
\end{minipage}
\begin{minipage}{0.3\linewidth}
     \begin{fmffile}{u2i}
     \begin{fmfgraph*}(120,150)
           \skeleton
           \fmf{plain}{xxi3,v1n}
           \fmf{plain}{xxi2,xf}
           \fmffreeze
           \fmf{dashes}{xxi3,xf}
           \fmf{dashes}{xxi2,v1n}
     \end{fmfgraph*}
     \end{fmffile}
\end{minipage}
\begin{minipage}{0.3\linewidth}
     \begin{fmffile}{u3i}
     \begin{fmfgraph*}(120,150)
           \skeleton
           \fmf{plain}{xxi3,v23}
           \fmf{plain}{xxi4,xf}
           \fmffreeze
           \fmf{dashes}{xxi3,xf}
           \fmf{dashes}{xxi4,v23}
     \end{fmfgraph*}
     \end{fmffile}
\end{minipage}
\end{center}
\caption{\label{fig:u_small}The three cross ratios associated to the reggeized propagator $|{\bf q}_{i}|^2$: $u_{1i}$ (left), $u_{2i}$ (center) and $u_{3i}$ (right). Solid lines denote square distances in the numerator, and dashed lines in the denominator, respectively.}
\end{figure}
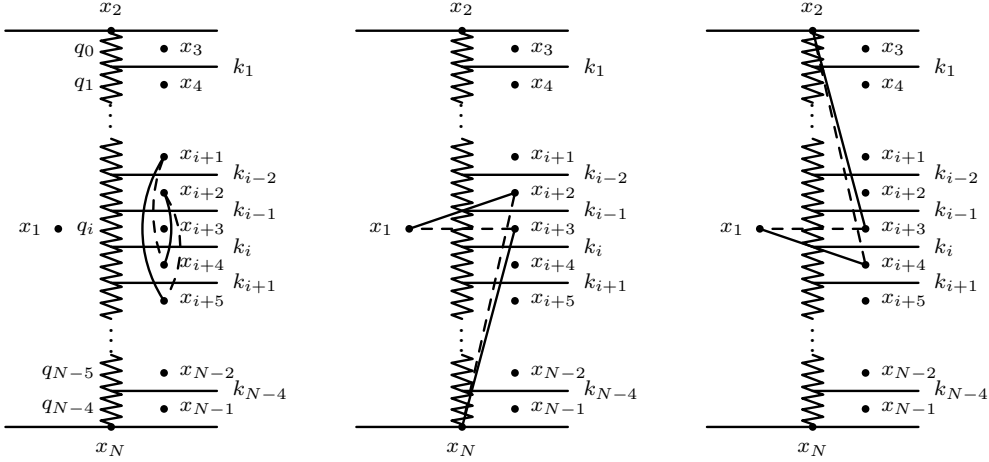

The focus of this paper are planar colour-ordered scattering amplitudes in $\cN=4$ SYM in a special kinematic limit of 2-to-$(N-2)$-gluon scattering, the so-called multi-Regge kinematics (MRK) \cite{Fadin:2011we}. 
In order to define this limit, it is convenient to work in conventions where all momenta are taken as outgoing. We define lightcone and (complex) transverse momenta
\be
p^\pm\equiv p^0\pm p^z\,,\quad {\bf p}_k \equiv p_{k\bot} = p_k^x + i p_k^y\,.
\ee
Using this decomposition, the scalar product between two four vectors $p$ and $q$ is given by
\be
2p\cdot q=p^+ q^-+p^- q^+- {\bf p}{\bf \bar q}- {\bf \bar p}{\bf  q}\,.
\ee
Without loss of generality we may choose a reference frame such that the momenta of the initial state gluons $p_1,p_2$ lie on the $z$-axis with $p_2^z=p_2^0$, which implies $p_1^+=p_2^-={\bf p}_1={\bf p}_2=0$. Then the multi-Regge limit is defined as the limit where the outgoing gluons with momenta $p_i$, $i\ge3$, are strongly ordered in rapidity (or equivalently in the lightcone +-coordinates) while having comparable transverse momenta,
\beq\label{eq:MRK_def}
p_3^+\gg p_4^+\gg \ldots p_{N-1}^+\gg p_N^+\,, \qquad |{\bf p}_3|\simeq\ldots \simeq|{\bf p}_N|\,.
\eeq
The mass-shell condition $p_i^2=p_i^+ p_i^- -|{\bf p}_i|^2=0$ implies that
\beq\label{eq:pminus_ordering}
p_N^-\gg p_{N-1}^-\gg \ldots p_{4}^-\gg p_3^-\,.
\eeq
The ordering between the lightcone coordinates in eq.~\eqref{eq:MRK_def} implies the following hierarchy between the Lorentz invariants,
\be\bsp\label{eq:MRK_def2}
s_{12}\gg s_{3\cdots N-1},s_{4\cdots N}&\,\gg s_{3\cdots N-2},s_{4\cdots N-1}, s_{5\cdots N}\gg\cdots\\
&\ldots\gg s_{34},\ldots, s_{N-1N}\gg-t_1,\cdots, -t_{N-3}\,,
\esp\ee
with $t_i$ held fixed, where
\begin{align}
s_{i(i+1)\ldots j}&\equiv(p_i+p_{i+1}+\ldots+p_{j})^2=x^2_{(i-1)j}\,,\\
t_{i+1}&\equiv q_{i}^2\,, \quad q_{i}\equiv -p_2-\ldots -p_{i+3}=x_{(i+3)1}\,.\label{eq:t_def}
\end{align}
Let us briefly sketch how the hierarchy in eq.~\eqref{eq:MRK_def2} follows from the strong ordering in lightcone coordinates, eq.~\eqref{eq:MRK_def}. In MRK momentum conservation can be written in the form
\be\label{eq:p_conservation}
\begin{aligned}
p_1^-&=-\sum_{i=3}^N p_i^-\simeq -p_N^-\,,\quad
p_2^+=-\sum_{i=3}^N p_i^+\simeq -p_3^+\,,
0=\sum_{i=3}^N {\bf p}_i\,,
\end{aligned}
\ee
and the two-particle invariants in MRK become
\be
\begin{aligned}\label{eq:Mandelstam_MRK}
 s_{12}&=2p_1\cdot p_2\simeq p_3^+ p_N^-\\
s_{1i}&=2p_1\cdot p_i\simeq - p_i^+ p_N^-\\
s_{2i}&=2p_2\cdot p_i\simeq -p_3^+ p_i^-\\
s_{ij}&=2p_i\cdot p_j\simeq p_i^+ p_j^-\,,\quad 3\le i<j\le N\,.
\end{aligned}
\ee
From the last line of eq.~\eqref{eq:Mandelstam_MRK}, it is evident that all Mandelstam invariants made of $k$ consecutive final state momenta $s_{ii+1\ldots i+k}\simeq s_{ii+k}$ will be comparable in size, and much larger than invariants made of $k-1$ consecutive momenta. This proves the hierarchy between $s$-channel invariants in eq.~\eqref{eq:MRK_def2}.
For the scale separation between $s$-like and $t$-like variables, we start by noting that the transverse component of the momentum transfer $q_{i}$ defined in eq.~(\ref{eq:t_def}) will be a sum of the final state transverse momenta, and thus also comparable with them in size. In addition, (\ref{eq:MRK_def})-(\ref{eq:t_def}) and (\ref{eq:p_conservation}) imply that $q_{i}^+\simeq p_{i+4}^+$, $q_{i}^-\simeq -p_{i+3}^-$ and therefore $-q_{i}^+ q_{i}^-\ll p_{i+3}^+p_{i+3}^-\simeq |{\bf q}_{i}|^2$. In other words, the $q_{i}$ are dominated by their transverse components, $q_{i}^2\simeq -|{\bf q}_{i}|^2$, and will thus be much smaller than the $s_{j(j+1)}$, $j=3,\ldots,N-1$, which are dominated by their lightcone components (\ref{eq:Mandelstam_MRK}). This then completes the proof that eq.~(\ref{eq:MRK_def2}) follows from eq.~(\ref{eq:MRK_def}).

The analysis of MRK thus far only relied on Lorentz symmetry. Let us now specialize to planar $\mathcal{N}=4$ SYM, which in addition exhibits dual conformal invariance. The three conformally invariant cross ratios $(u_{1i}, u_{2i}, u_{3i})$ of eq.(\ref{eq:ui_def}) can be associated to the $t$-channel invariants (\ref{eq:t_def}), which have transverse momentum $|{\bf q}_{i}|^2$ (see Fig.~\ref{fig:u_small})~\cite{Bartels:2012gq,Bartels:2014mka}. In MRK these cross ratios take the form
\beq\bsp\label{eq:udef}
u_{1i} &\,=  1 - \delta_i\,\frac{|{\bf k}_{i} + {\bf k}_{i+1}|^2}{|{\bf k}_{i+1}|^2}+\ord(\delta_i^2)\,,\\
u_{2i}  &\,= \delta_i\,\frac{|{\bf q}_{i-1}|^2}{|{\bf q}_{i}|^2}+\ord(\delta_i^2)\,,\\
u_{3i} &\,=  \delta_i\,\frac{|{\bf q}_{i+1}|^2\,|{\bf k}_i|^2}{|{\bf q}_{i}|^2\,|{\bf k}_{i+1}|^2}+\ord(\delta_i^2)\,,
\esp\eeq
where ${k}_i \equiv {p}_{i+3}$, $1\le i\le N-4$, denote the momenta of the gluons emitted along the $t$-channel ladder, and we define the ratio $\delta_i\equiv k_{i+1}^+/k_i^+$. From eq.~(\ref{eq:MRK_def}) it is evident that in MRK we have $\delta_i\to0$, and so we see that all the $u_{1i}$ tend to 1 at the same speed as the $u_{2i}$ and $u_{3i}$ vanish. It is convenient to define
the reduced cross ratios~\cite{Bartels:2012gq,Bartels:2014mka},
\beq\bsp\label{eq:reduced_cross_ratios}
\tilde{u}_{2i} = \frac{u_{2i}}{1-u_{1i}} &\,=\frac{|{\bf q}_{i-1}|^2\,|{\bf k}_{i+1}|^2}{|{\bf q}_{i}|^2\,|{\bf k}_{i} + {\bf k}_{i+1}|^2}+\ord(\delta_i)\,,\\
\tilde{u}_{3i} = \frac{u_{3i}}{1-u_{1i}} &\,= \frac{|{\bf q}_{i+1}|^2\,|{\bf k}_{i}|^2}{|{\bf q}_{i}|^2\,|{\bf k}_{i} + {\bf k}_{i+1}|^2}+\ord(\delta_i)\,.
\esp\eeq

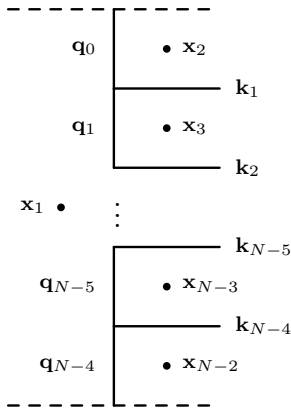
\begin{figure}[!t]
\begin{center}
  \begin{fmffile}{dual_graph}
                  \begin{fmfgraph*}(120,150)
                \transverseskeleton
                           \end{fmfgraph*}
                             \end{fmffile}
                             \end{center}
\caption{\label{fig:dual_coordinates}The dual coordinates in the transverse space. Dashed lines indicate the forward momenta with zero transverse momentum, which are strictly speaking absent in the transverse momentum space because they are orthogonal to it.}
\end{figure}

We now introduce dual coordinates in the transverse space $\mathbb{CP}^1$ by (see Fig.~\ref{fig:dual_coordinates})
\beq
{\bf q}_i = {\bf x}_{i+2} - {\bf x}_1 {\rm ~~and~~} {\bf k}_i = {\bf x}_{i+2} - {\bf x}_{i+1}\,.
\eeq
The reduced cross ratios $\tilde{u}_{2i}$ and $\tilde{u}_{3i}$ can then we written as (squares of) cross ratios in $\mathbb{CP}^1$,
\beq\label{eq:ui_xi}
\tilde{u}_{2i} \simeq |\xi_{2i}|^2 {\rm~~and~~}\tilde{u}_{3i} \simeq |\xi_{3i}|^2\,,
\eeq
with
\beq
\xi_{2i} = \frac{({\bf x}_1 -{\bf x}_{i+1})\,({\bf x}_{i+3} -{\bf x}_{i+2})}{({\bf x}_1 -{\bf x}_{i+2})\,({\bf x}_{i+3} -{\bf x}_{i+1})} {\rm~~and~~}
\xi_{3i} = \frac{({\bf x}_1 -{\bf x}_{i+3})\,({\bf x}_{i+2} -{\bf x}_{i+1})}{({\bf x}_1 -{\bf x}_{i+2})\,({\bf x}_{i+3} -{\bf x}_{i+1})} \,.
\eeq
It is easy to check that
\beq
\xi_i \equiv \xi_{2i} = 1-\xi_{3i}\,.
\eeq
We also introduce the transverse cross ratios
\beq\label{eq:z_i_def}
z_i\equiv 1-\frac{1}{\xi_i} = \frac{({\bf x}_1 -{\bf x}_{i+3})\,({\bf x}_{i+2} -{\bf x}_{i+1})}{({\bf x}_1 -{\bf x}_{i+1})\,({\bf x}_{i+2} -{\bf x}_{i+3})}  = -\frac{{\bf q}_{i+1}\,{\bf k}_{i}}{{\bf q}_{i-1}\,{\bf k}_{i+1}}\,.
\eeq
In the literature it is customary to use the variables $w_i\equiv -z_i$.

It is easy to see from Fig.~\ref{fig:dual_coordinates} that the MRK setup has a natural $\mathbb{Z}_2$ symmetry, called \emph{target-projectile symmetry}~{\cite{Bartels:2012gq,Bartels:2014mka}}, which acts by reflecting all the points along the horizontal symmetry axis. This symmetry acts on the points ${\bf x}_i$ via
\beq
{\bf x}_i \mapsto \left\{\begin{array}{ll}
{\bf x}_1\,, & \textrm{ if }i=1\,,\\
{\bf x}_{N-i}\,, &\,\textrm{ if } 2\le i\le N-2\,.
\end{array}\right.
\eeq
On the cross ratios $z_i$ target-projectile symmetry acts by
\beq
z_i \mapsto 1/z_{N-4-i}\,.
\eeq

In the previous section we have seen that the kinematics of scattering amplitudes in planar $\cN=4$ SYM are naturally encoded through a configuration of $N$ momentum twistors in three-dimensional projective space $\mathbb{CP}^3$. In the remainder of this section we show that there is a very natural geometrical interpretation of MRK in terms of momentum twistors. More precisely, we will show that the dual conformal invariance of planar $\cN=4$ SYM implies that the multi-Regge limit defined in eq.~\eqref{eq:MRK_def} is conformally equivalent to the strongly-ordered multi-soft limit where the momenta $p_i$, $3\le i\le N-3$, are soft, with $p_i$ softer than $p_{i+1}$.

Before proving the connection between the multi-Regge and soft limits, let us discuss in more detail how to take a single soft limit in momentum twistor space. In terms of dual coordinates, the momentum $p_{i+1}$ is soft if the points $x_{i}$ and $x_{i+1}$ coincide. As the points $x_i$ correspond to lines in momentum twistor space, the soft limit corresponds to the limit where the momentum twistors $Z_{i-1}$, $Z_{i}$ and $Z_{i+1}$ are aligned.
In other words, to set the momentum $p_{i+1}$ to zero all we have to do is to take the twistor $Z_i$ to lie on the line between $Z_{i-1}$ and $Z_{i+1}$. The limit leaves two degrees of freedom from the three associated to $Z_i$. The remaining degrees of freedom can be thought of (using real twistor space as an analogy) as the distance along the line between $Z_{i-1}$ and $Z_{i+1}$ and the angle of approach to the line in which the limit is taken.
More generally, let us consider a limit where the twistor $Z_i$ approaches the line between $Z_j$ and $Z_k$. We parametrise this situation as follows
\be\label{eq:soft_parametrisation}
Z_i \rightarrow \hat{Z}_i = Z_{j} + \alpha_i \tfrac{\langle j\,j-1\,k+1\,k+2 \rangle}{\langle k \,j-1\,k+1\,k+2\rangle} Z_{k} + \epsilon_i \tfrac{\langle j\,k\,k+1\,k+2 \rangle}{\langle j-1\,k\,k+1\,k+2 \rangle} Z_{j-1} - \epsilon_i \beta_i \tfrac{\langle j\,j-1\,k\,k+2 \rangle}{\langle k+1\,j-1\,k\,k+2\rangle} Z_{k+1}\,,
\ee
and the limit where $Z_i$, $Z_j$ and $Z_k$ are aligned corresponds to the limit $\epsilon_i \rightarrow 0$. The existence of the last two terms in eq.~(\ref{eq:soft_parametrisation}) ensures that $x_{i-1 i+1}^2, x_{ii+2}^2\sim \epsilon_i$ as we approach the limit, as can be shown from eq.~(\ref{twistorinvariants}) and the fact that we can choose the infinity twistor such that $\langle i\, j\, I\rangle=\epsilon_{\alpha \beta} \lambda_{i+1}^\alpha \lambda_{j+1}^\beta$.

The multi-soft limit we wish to consider is one where we sequentially take the momenta $p_i$, $3\le i\le N-3$, to be soft. This corresponds to taking twistor $Z_2$ to the line $(Z_1Z_3)$, then $Z_3$ to the line $(Z_1Z_4)$ and so on. In this limit the cross ratios~\eqref{eq:ui_def} behave like
\be
u_{1i} \rightarrow 1\,, \qquad u_{2i} \rightarrow 0\,, \qquad u_{3i} \rightarrow 0\,,
\ee
i.e., the cross ratios behave in the same way as in MRK, cf. eq.~\eqref{eq:MRK_def}. This is, however, still insufficient to conclude that this multi-soft limit is equivalent to MRK, and we still need to show that the cross ratios approach their limiting values at the same speed. Equivalently, we need to show that the reduced cross ratios ~\eqref{eq:reduced_cross_ratios} are finite in the limit. This is indeed the case, and we find
\be\bsp\label{eq:reduced_soft_ratios}
\tilde{u}_{2i} &\,= \frac{u_{2i}}{1-u_{1i}} \rightarrow \frac{\alpha_{i+1} \beta_{i+1}}{(1+\alpha_{i+1})(1+\beta_{i+1})}\,,\\
\tilde{u}_{3i} &\,= \frac{u_{3i}}{1-u_{1i}} \rightarrow \frac{1}{(1+\alpha_{i+1})(1+\beta_{i+1})}\,.
\esp\ee
Hence, we conclude that this particular multi-soft limit is conformally equivalent to the multi-Regge limit.
Comparing eq.~\eqref{eq:reduced_soft_ratios} to eq.~\eqref{eq:ui_xi} and eq.~\eqref{eq:z_i_def}, we see that we can identify the parameters $\alpha_{i+1}$ and $\beta_{i+1}$ that describe the reduced cross ratios in the multi-soft limit with the $\mathbb{CP}^1$ cross ratio that appear in MRK,
\beq
\alpha_{i+1} = -1/z_i {\rm~~and~~} \beta_{i+1} = -1/\bar{z}_i\,.
\eeq

\subsection{Planar SYM amplitudes in multi-Regge kinematics}

\begin{figure}[!t]
\begin{center}
  \begin{fmffile}{regge_cut}
                  \begin{fmfgraph*}(400,70)
                \reggecut
                           \end{fmfgraph*}
                             \end{fmffile}
                             \end{center}
\caption{\label{cutter}Diagrammatic representation of the Mandelstam region $[p,q]$. The discontinuity in the $(k_p+\ldots+k_q)^2$ channel is indicated by the dashed line.}
\end{figure}
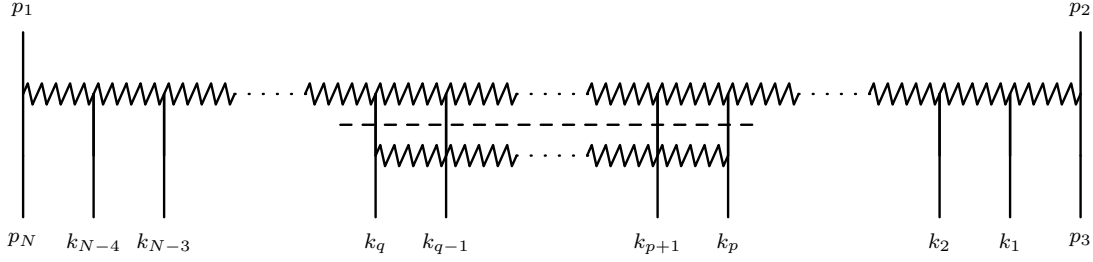

So far all the considerations were purely kinematical.
In this section we present the (conjectural) representation of an amplitude in MRK to leading logarithmic accuracy (LLA). Helicity must be conserved by the gluons going very forward, so that the different helicity configurations are distinguished only by the helicities of the gluons emitted along the ladder. Denoting these helicities by $h_1,\ldots,h_{N-4}$, we define the ratio
\beq\label{eq:R_definition}
e^{i\Phi_{h_1,\ldots,h_{N-4}}}\,\cR_{h_1,\ldots,h_{N-4}} \equiv \left[\frac{A_N(-,+,h_1,\ldots,h_{N-4},+,-)}{A_N^{\textrm{BDS}}(-,+,\ldots,+,-)}\right]_{|\textrm{MRK, LLA}}\,,
\eeq
where $A_N(-,+,h_1,\ldots,h_{N-4},+,-)$ is the (colour-ordered) amplitude for the production of $N-4$ gluons emitted along the ladder, and $A_N^{\textrm{BDS}}(-,+,\ldots,+,-)$ is the corresponding BDS amplitude. The function $\cR_{h_1,\ldots,h_{N-4}}$ is finite, and thus dual conformally invariant. It can easily be related to the well-known remainder and ratio functions.
Since Regge factorisation holds in the Euclidean region, the ratio in the left-hand side of eq.~\eqref{eq:R_definition} tends to a phase in this region. The exact form of this phase is immaterial in the following, because it is simply obtained as the ratio of the corresponding tree amplitudes~\cite{DelDuca:1995zy}. We normalise the left-hand side of eq.~\eqref{eq:R_definition} such that $\cR_{h_1,\ldots,h_{N-4}}=1$ in the Euclidean region.

If we take a discontinuity corresponding to a consecutive subset of final-state momenta $k_l$, $l\in [p,q]\subseteq\{1,\ldots,N-4\}$, i.e., a discontinuity with respect to the invariant $(k_p+\ldots+k_q)^2$, then $\cR_{h_1,\ldots,h_{N-4}}$ is no longer trivial due to the presence of a Regge cut (see Fig.~\ref{cutter})~\cite{Bartels:2014mka,Bartels:2008ce,Bartels:2008sc,Lipatov:2010ad,Lipatov:2010qg,Bartels:2011ge,Lipatov:2012gk,Bartels:2013jna,Bartels:2014jya}. In terms of the dual conformal cross ratios,
\beq\label{eq:Uij_def}
U_{ij}\equiv U_{i,j+1,j,i+1}=\frac{x_{ij+1}^2x_{i+1j}^2}{x_{ij}^2x_{i+1j+1}^2}\,,
\eeq
taking this discontinuity corresponds to analytically continuing $U_{pq+2}$ around the origin while all other cross ratios $U_{ij}$ are held fixed. In the following we denote the value of the ratio $\cR_{h_1,\ldots,h_{N-4}}$ in this so-called \emph{Mandelstam region} $[p,q]$ by $\cR_{h_1,\ldots,h_{N-4}}^{[p,q]}$.
We conjecture that $\cR_{h_1,\ldots,h_{N-4}}^{[p,q]}$ in MRK to LLA can be cast in the form of a multiple Fourier-Mellin integral
\begin{align}\label{eq:MRK_conjecture}
\cR&_{h_1\ldots h_{N-4}}^{[p,q]}(\{\tau_k,z_k\}_{p\le k\le q-1}) = 1 + a\,i\pi\,r^{[p,q],(1)}_{h_1\ldots h_{N-4}}\\
\nonumber
&\, + a\,i\pi\,(-1)^{q-p} \,\left[\prod_{k=p}^{q-1}\sum_{n_k=-\infty}^{+\infty}\left(\frac{z_k}{\zbar_k}\right)^{n_k/2}\int_{-\infty}^{+\infty}\frac{d\nu_k}{2\pi}|z_k|^{2i\nu_k}\right]\\
\nonumber&\times\!\!
\left[-1+\prod_{k=p}^{q-1}\tau_{k}^{aE_{\nu_kn_k}}\right]\,
\chi^{h_p}({\nu_p,n_p})\,
\left[\prod_{k=p}^{q-2}C^{h_{k+1}}(\nu_k,n_k,\nu_{k+1},n_{k+1})\,\right]\,\chi^{-h_{q}}({\nu_{q-1},n_{q-1}})\,.
\end{align}
In this expression, we defined $\tau_k \equiv \sqrt{u_{2k}u_{3k}}$, and $a$ is the 't Hooft coupling. To LLA, the value of $\tau_k$ is independent of $k$, but we prefer to keep the $\tau_k$ different for reasons that will become clear in subsequent sections.  The one-loop coefficients receive contributions from both the Regge pole and cut. They are sums of logarithms whose functional form is irrelevant for the remainder of this paper. $E_{\nu n}$ is the leading-order (LO) BFKL eigenvalue,
\beq
E_{\nu n} = -\frac{1}{2}\frac{|n|}{\nu^2+\frac{n^2}{2}}+\psi\left(1+i\nu+\frac{|n|}{2}\right)+\psi\left(1-i\nu+\frac{|n|}{2}\right)-2\psi(1)\,,
\eeq
and $\chi^h({\nu,n})$ is the LO impact factor~\cite{Bartels:2008ce,Bartels:2008sc}
\beq\label{eq:chi_def}
\chi^{\pm}({\nu,n}) = \frac{1}{i\nu\pm\frac{n}{2}} = \left[\chi^{\mp}({-\nu,-n}) \right]^* \,.
\eeq
The central emission block for the emission of a positive-helicity gluon is\cite{Bartels:2011ge}
\beq\bsp\label{eq:C_def}
C^+(\nu,n,\mu,m)
 &\,= \frac{\Gamma(1-i\nu-\frac{n}{2})\,\Gamma(i\mu+\frac{m}{2})\,\Gamma(i(\nu-\mu)+\frac{m-n}{2})}{\Gamma(1+i\nu-\frac{n}{2})\,\Gamma(-i\mu+\frac{m}{2})\,\Gamma(1-i(\nu-\mu)+\frac{m-n}{2})}\,.
\esp\eeq
The central emission block for the emission of a negative-helicity gluon takes the form
\beq
C^-(\nu,n,\mu,m)=[C^+(-\nu,-n,-\mu,-m)]^* =C^+(\nu,-n,\mu,-m)\,.
\eeq
Equation~\eqref{eq:MRK_conjecture} reproduces the known Fourier-Mellin representation of the six-point MHV and NMHV amplitudes in MRK to LLA~\cite{Bartels:2008ce,Bartels:2008sc,Bartels:2011ge}, and also of the seven-point MHV amplitude to LLA~\cite{Bartels:2011ge}
In the remainder of this section we give further support to our conjecture by showing that it is consistent with target-projectile symmetry and with the factorisation of the amplitude in infrared limits. The function $\cR_{h_1,\ldots,h_{N-4}}^{[p,q]}$ is identical to the ratio where all the gluons  that are not taken into account in the discontinuity have been removed. In other words, if we know the results for the Mandelstam regions $[1,N-4]$, then we can reconstruct all other cases. Hence, in the following we only discuss this particular case, and we simply write $\cR_{h_1,\ldots,h_{N-4}}$ for $\cR_{h_1,\ldots,h_{N-4}}^{[1,N-4]}$.

In order to fully define the expression for $\cR_{h_1,\ldots,h_{N-4}}^{[p,q]}$ in eq. (\ref{eq:MRK_conjecture}) we must specify the contours of integration. The integrals over the $\nu_k$ are taken along the real $\nu_k$-axes, however the quantities $\chi$ and $C$ have poles on the real axes for certain values of the $n_k$. Our contour prescription for avoiding these poles is as follows. For $n_p=0$ we replace
\beq
\chi^{h_p}(\nu_p,0) \rightarrow \frac{1}{i \nu_p - \epsilon}\,.
\label{rep1}
\eeq
For $n_{k-1} = n_k$ we find that $C(\nu_{k-1},n_{k-1},\nu_k,n_k)$ exhibits a pole at $\nu_{k-1} = \nu_k$, as can be seen by inspecting the third factor in the numerator of eq. (\ref{eq:C_def}). We avoid this pole by replacing it as follows,
\beq
\frac{1}{i(\nu_{k-1} - \nu_k)} \rightarrow \frac{1}{i(\nu_{k-1} - \nu_k) + \epsilon}\,.
\label{rep2}
\eeq
For $n_{q-1}=0$ we replace
\beq
\chi^{-h_{q-1}}(\nu_{q-1},0) \rightarrow \frac{1}{i \nu_{q-1} + \epsilon}\,.
\label{rep3}
\eeq
In all cases we take $\epsilon$ to be an infinitesimal positive number.

The effect of the replacement (\ref{rep1}) is to shift the pole from $\chi^{h_p}(\nu_p,0)$ slightly into the lower half $\nu_p$ plane. The shift (\ref{rep2}) means the pole is slightly shifted into the upper half $\nu_{k-1}$ plane (or the lower half $\nu_k$ plane). Finally the shift (\ref{rep3}) takes the pole slightly into the upper half $\nu_{q-1}$ plane.

Equation~\eqref{eq:MRK_conjecture} can be written as an inverse multiple Fourier-Mellin transform. The (inverse) Fourier-Mellin transform of a function $F(\nu,n)$ is defined as
\beq\label{eq:fourier_mellin_def}
f(z) = \cF[F(\nu,n)] = \sum_{n=-\infty}^{+\infty}\left(\frac{z}{\zb}\right)^{n/2}\,\int_{-\infty}^{+\infty}\frac{d\nu}{2\pi}\,|z|^{2i\nu}\,F(\nu,n)\,,
\eeq
where $z\in\mathbb{C}$. This integral transform is invertible, and its inverse is given by
\beq\label{eq:inv_FM}
\cF^{-1}[f(z)] = \int\frac{d^2z}{\pi}\, z^{-1-i\nu-n/2}\, \zb^{-1-i\nu+n/2}\,f(z)\,,
\eeq
with the usual metric on the complex plane
\beq
d^2z = -\frac{dz\wedge d\bar{z}}{2i} = dx\wedge dy = r\,dr\wedge d\varphi\,,\quad \textrm{for } z=x+iy=re^{i\varphi}\,.
\eeq
The Fourier-Mellin transform has the property that it maps ordinary products into convolutions. More precisely,  if $\cF[F]=f$ and $\cF[G]=g$, then
\beq\label{eq:conv_thm}
\cF[F\cdot G] = \cF[F]\ast\cF[G] = f\ast g\,,
\eeq
where the convolution is defined by
\beq\bsp\label{eq:conv_def}
(f\ast g)(z)
&\,= \frac{1}{\pi}\int \frac{d^2w}{|w|^2}\,f(w)\,\,g\left(\frac{z}{w}\right)\,.
\esp\eeq
A proof of the convolution theorem for the Fourier-Mellin transform is given in Appendix~\ref{app:fourier_mellin}. It is easy to see that the convolution product is associative and commutative, and the distribution $\pi\,\delta^{(2)}(1-z)$ is a neutral element.

We conclude this section by quoting some properties of the Fourier-Mellin space functions that enter eq.~\eqref{eq:MRK_conjecture}.
For $n_k=0$, the BFKL eigenvalue and the central emission block have the following properties~\cite{Fadin:2011we,Lipatov:1976zz,Kuraev:1976ge,Balitsky:1978ic,Fadin:2006bj,Caron-Huot:2013fea,Basso:2014pla},
\begin{align}\label{eq:E_bootstrap}
\lim_{\nu\to0}E_{\nu 0} &\,= 0\,,\\
\label{eq:C1_bootstrap}
\lim_{\nu\to0}C^{\pm}(\nu,0,\mu,m) &\,= \phantom{\pm}\chi^{\pm}({\mu,m})\,,\\
\label{eq:C2_bootstrap}
\lim_{\mu\to0}C^{\pm}(\nu,n,\mu,0) &\,= -\chi^{\mp}({\nu,n})\,,\\
\label{eq:C12_bootstrap}
\textrm{Res}_{\nu=\mu}C^{\pm}(\nu,n,\mu,n) &\,= (-1)^{n}\,i\,.
\end{align}
Note that $E_{\nu0}$ vanishes quadratically as $\nu \rightarrow 0$ due to its symmetry under $\nu \leftrightarrow -\nu$.
As we will see shortly, the above relations guarantee that eq.~\eqref{eq:MRK_conjecture} has the correct soft behaviour.
In order to prove the last relation (\ref{eq:C12_bootstrap}), we need the following identity,
\beq
\frac{\sin\pi(\frac{n}{2}+i\nu)}{\sin\pi(\frac{n}{2}-i\nu)} = (-1)^{n+1}\,,\quad n\in\mathbb{Z}\,.
\eeq
In order to show this identity, let us define
\beq
S_n = \frac{\sin\pi(\frac{n}{2}+i\nu)}{\sin\pi(\frac{n}{2}-i\nu)}\,.
\eeq
Obviously, $S_0=-1$ and $S_1=1$. Moreover, $S_n$ satisfies a recursion of order two,
\beq
S_{n+2} =  \frac{\sin\left[\pi+\pi(\frac{n}{2}+i\nu)\right]}{\sin\left[\pi+\pi(\frac{n}{2}-i\nu)\right]} =  \frac{\sin\pi(\frac{n}{2}+i\nu)}{\sin\pi(\frac{n}{2}-i\nu)} = S_n\,.
\eeq
Hence, $S_n=(-1)^{n+1}$.
Finally, we note the following relation between the central emission block and the impact factor,
\beq\label{eq:helflip_C}
\frac{C^-(\nu,n,\mu,m)}{C^+(\nu,n,\mu,m)} = \frac{\chi^+(\nu,n)\,\chi^-(\mu,m)}{\chi^-(\nu,n)\,\chi^+(\mu,m)}\,.
\eeq

\subsection{Soft limits}
\label{sec:soft_limits}
In this section we show that the function $\cR_{h_1,\ldots,h_{N-4}}$ has the correct behavior in all infrared limits. Due to the strong ordering in the rapidities (or equivalently, in the +-lightcone coordinates), there are no collinear singularities. All the singularities of an amplitude in MRK can therefore be associated to some final-state partons being soft. In addition, there are no soft singularities associated with the two final-state particles at the end of the ladder. Hence, an amplitude in MRK has soft singularities only in the limits where one of the momenta $k_i$, $1\le i\le N-4$, vanishes. The vanishing of $k_i$ implies that in particular its transverse component ${\bf k}_i$ goes to zero. Using eq.~\eqref{eq:z_i_def}, we see that the limit ${\bf k}_i\to 0$ corresponds to the limit where some of the cross ratios $z_k$ take a special value.
There are three distinct cases to consider:
\begin{enumerate}
\item If ${\bf k}_1\to0$, $z_1\to0$, and all other cross ratios remain finite.
\item If ${\bf k}_{N-4}\to0$, $z_{N-5}\to\infty$, and all other cross ratios remain finite.
\item If ${\bf k}_i\to0$, for $2\le i\le N-5$, $z_{i}\to0$ and $z_{i-1}\to \infty$, but the product $z_{i-1}z_i$ remains finite. All other cross ratios remain finite.
\end{enumerate}

In the remainder of this section we show that eq.~\eqref{eq:MRK_conjecture} has the correct behaviour in the soft limit, where the function $\cR_{h_1,\ldots,h_{N-4}}$ behaves like
\beq
\lim_{k_j\to 0}\cR_{h_1,\ldots,h_{N-4}}(\{\tau_k,z_k\}_{1\le k\le N-5}) = \cR_{h_1,\ldots,\hat{h}_j,\ldots,h_{N-4}}(\{\tau'_k,z'_k\}_{1\le k\le N-6})\,,
\eeq
where the $\tau'_k$ and $z'_k$ are combinations of the $\tau_k$ and $z_k$, and the hat indicates that the corresponding element is absent.

Let us start by analysing the limit ${\bf k}_1\to 0$, which corresponds to $z_1\to 0$. We want to obtain the leading behaviour of the Fourier-Mellin integral~\eqref{eq:MRK_conjecture} in this limit.
We therefore close the contour in the lower half $\nu_1$-plane, giving a series of terms with non-negative powers of $|z_1|$.
The leading contribution comes from the pole from $\chi^{h_1}(\nu_1,n_1)$ which is shifted slightly away from the real axis into the lower half $\nu_1$ plane by the prescription (\ref{rep1}). This pole is only present for $n_1=0$.
In other words we have a pole at $\nu_1=-i\epsilon$ for $n_1=0$. If we take the residue at this pole we find that, due to eq.(\ref{eq:C1_bootstrap}), the factor $C^{h_2}(\nu_1,n_1,\nu_2,n_2)$ contributes
\be
C^{h_2}(-i\epsilon,0,\nu_2,n_2) \rightarrow \chi^{h_2}(\nu_2  + i \epsilon,n_2)\,.
\label{Ctochilim1}
\ee
This is precisely the factor present in $\cR_{h_2,\ldots,h_{N-4}}$ with the correct contour prescription for the pole on the real axis at $n_2=0$. Note that the prescriptions (\ref{rep1}) and (\ref{rep2}) are consistent in this regard, since they both dictate that for $n_2=0$ the remaining pole on the real axis is shifted into the lower half $\nu_2$ plane. Looking at the other factors in the integrand we see that all signs and factors of $2 \pi$ are as needed and the factor $|z_1|^{i \nu_1}$ becomes $1$. The factor $\tau_1^{E_{\nu_10}}$ in the product term in the square brackets in the final line of (\ref{eq:MRK_conjecture}) also becomes $1$ due to the relation (\ref{eq:E_bootstrap}). Thus we reproduce precisely the Fourier-Mellin integral expression for $\cR_{h_2,\ldots,h_{N-4}}$, as required,
\beq
\lim_{{\bf k}_1\to 0}\cR_{h_1,\ldots,h_{N-4}}(\tau_1,z_1,\ldots,\tau_{N-5},z_{N-5}) = \cR_{h_2,\ldots,h_{N-4}}(\tau_2,z_2,\ldots,\tau_{N-5},z_{N-5})\,.
\eeq

Next, let us look at the limit ${\bf k}_{N-4} \to 0$, which corresponds to $z_{N-5}\to\infty$. In this case we should close the contour in the upper half $\nu_{N-5}$ plane to obtain a series in non-negative powers of $|z_{N-5}|^{-1}$. By an argument similar to the one in the previous case, the leading behaviour comes from the pole in $\chi^{-h_{N-4}}(\nu_{N-5},n_{N-5})$ which is shifted slightly into the upper half plane by the prescription (\ref{rep3}). Thus we have a pole at $\nu_{N-5}=i\epsilon$ and this pole is present only for $n_{N-5}=0$. The residue from this pole contributes a factor
\be
C^{h_{N-5}}(\nu_{N-6},n_{N-6},i\epsilon,0) \rightarrow - \chi^{h_{N-5}}(\nu_{N-6} - i \epsilon,n_{N-6})\,.
\ee
The sign difference, compared to eq.~\eqref{Ctochilim1}, is compensated by the fact the contour is taken in the opposite orientation compared to the $\nu_1$ contour in the ${\bf k}_1\to 0$ limit. The rest of the analysis is similar to the ${\bf k}_1\to 0$ case. Hence, we again obtain the expected soft behaviour,
\beq
\lim_{{\bf k}_{N-4}\to 0}\cR_{h_1,\ldots,h_{N-4}}(\tau_1,z_1,\ldots,\tau_{N-5},z_{N-5}) = \cR_{h_1,\ldots,h_{N-3}}(\tau_1,z_1,\ldots,\tau_{N-6},z_{N-6})\,.
\eeq

The remaining soft limits are slightly more subtle, because they involve two cross ratios at the same time. Consider ${\bf k}_j\to0$, which corresponds to $z_{j-1}\to\infty$ and $z_j\to0$, with their product held fixed. We define $\nu = \nu_{j-1} - \nu_j$ and we find that the powers of the moduli can be reorganised as
\beq
|z_{j-1}|^{2i\nu_{j-1}}|z_j|^{2i\nu_j} = |z_{j-1}|^{2i\nu}|z_{j-1} z_j|^{2 i \nu_j}\,.
\eeq
Since $|z_{j-1}| \rightarrow \infty$ with $|z_{j-1} z_j|$ fixed we need to perform an integration over $\nu$, closing it in the upper half plane to obtain a series in non-negative powers of $|z_{j-1}|^{-1}$. The leading contribution comes from the pole at $\nu = i \epsilon$ from the factor $C(\nu_{j-1},n_{j-1},\nu_j,n_j)$, which is shifted slightly into the upper half $\nu$-plane by the prescription (\ref{rep2}). This pole is only present for $n_{j-1}=n_j$. The integral over $\nu$ then contributes the residue given in eq. (\ref{eq:C12_bootstrap}). The factor $(-1)^{n_j}$ thus obtained should be absorbed into the phase factor,
\beq
(-1)^{n_j} \biggl(\frac{z_{j-1} z_j}{\bar{z}_{j-1} \bar{z}_j}\biggr)^{n_j /2} =  \biggl(\frac{z}{\bar{z}}\biggr)^{n_{j} /2} \,,
\eeq
where $z = - z_{j-1} z_j$. Similarly the remaining modulus factor may be written as $|z_{j-1} z_j |^{2 i \nu_j} = |z|^{2 i \nu_j}$. Moreover, from eq.~\eqref{eq:udef} we see that
\beq
\tau_{j-1}\tau_j = \sqrt{u_{2j-1}u_{3j-1}u_{2j}u_{3j}} = \left|\frac{{\bf q}_{j-2}{\bf q}_{j+1}{\bf k}_{j-1}}{{\bf q}_{j-1}{\bf q}_{j}{\bf k}_{j+1}}\right|\,,
\eeq
which is precisely the argument of the large logarithm in the soft limit.  Finally, we note that since we evaluate $\nu_{j-1}$ at the value $\nu_j + i \epsilon$, the remaining $C$-factor dependent on $\nu_{j-1}$ becomes
\beq
C^{h_{j-1}}(\nu_{j-2},n_{j-2},\nu_{j-1},n_{j-1}) \rightarrow C^{h_{j-1}}(\nu_{j-2},n_{j-2},\nu_{j}+ i \epsilon,n_{j})
\eeq
and we see that the remaining pole on the real axis at $\nu_{j-2} = \nu_j$ is correctly shifted slightly into the upper half $\nu_{j-2}$ plane (or lower half $\nu_j$ plane), consistent with the prescription~\eqref{rep2}. In the case $j=2$ we have instead a factor $\chi^{h_1}(\nu_{j-1},n_{j-1})$ which becomes (again correctly) $\chi^{h_1}(\nu_{j} + i \epsilon,n_j)$.
We thus get the expected behaviour, with a combined cross ratio $z = - z_{j-1}z_j$,
\beq\bsp
\lim_{{\bf k}_{j}\to 0}&\cR_{h_1,\ldots,h_{N-4}}(\tau_1,z_1,\ldots,\tau_{N-5},z_{N-5})\\
&\, = \cR_{h_1,\ldots,\hat{h}_j,\ldots,h_{N-3}}(\tau_1,z_1,\ldots,\tau_{j-1}\tau_j,-z_{j-1}z_j,\ldots,\tau_{N-6},z_{N-6})\,,
\esp\eeq
where the hat on $h_j$ means this gluon is eliminated.

We see that the nature of the contour prescription, described in eqs.~(\ref{rep1} - \ref{rep3}) is intimately tied to the correct behaviour under soft limits. In order to have the correct behaviour in the soft limits, we have no choice but to implement the contours we have outlined. The reader may wonder if our discussion remains valid when many poles overlap. Indeed, it is possible that our contour can become pinched between poles which are separated only as a result of our $\epsilon$ prescription. This troublesome behaviour can been seen already in the six-point case where for $n_1=0$ we have poles at $\pm i \epsilon$ from $\chi^{h_1}(\nu_1,0)$ and $\chi^{-h_2}(\nu_1,0)$ respectively, with the contour running between. Evaluating the $\nu_1$ integral in, say, the lower half plane we obtain a factor of $1/\epsilon$ which looks dangerous. However the factor in the square brackets in the last line of eq.~\eqref{eq:MRK_conjecture} comes to the rescue since we find
\be
[-1+\tau_1^{a E_{\nu_10}}] = a E_{\nu_10} \log \tau_1 + \ldots  = O(\epsilon^2)\,.
\ee
Thus the singularity from the pinched pole is killed by the fact that we have written $[-1 + \prod \tau_k^{a E_{\nu_kn_k}}]$, eliminating any one-loop contribution from the Fourier-Mellin integral which we have instead made explicit in the term $r^{(1)}$ in the first line of eq.~\eqref{eq:MRK_conjecture}. Similar behaviour occurs for higher $N$ when many poles on the real axis coincide. Thus our formula (\ref{eq:MRK_conjecture}) does make sense with the $\epsilon$ prescription given and indeed the above argument verifies the correct (vanishing) soft behaviour in the six-point case. However, the existence of potential pinched poles is still troubling and we expect that it should be regulated by shifts of the pole positions in the real $\nu_k$ directions by amounts which vanish as the coupling $a$ tends to zero, similarly to the discussion of ref.~\cite{Caron-Huot:2013fea,Basso:2014pla} in the six-point case.

\subsection{Symmetries}

In this section we show how the ratio $\cR_{h_1,\ldots,h_{N-4}}$ transforms under symmetry transformations on the cross ratios $z_i$, in particular target-projectile symmetry and complex conjugation.

The exchange $z_i\leftrightarrow \bar{z}_i$ acts on the ratio by reversing the helicities,
\beq\label{eq:sym_CC}
\cR_{h_1,\ldots,h_{N-4}}(\tau_1,\bar{z}_1,\ldots,\tau_{N-5},\bar{z}_{N-5}) =
\cR_{-h_1,\ldots,-h_{N-4}}(\tau_1,{z}_1,\ldots,\tau_{N-5},{z}_{N-5}) \,.
\eeq
Indeed, we can do the replacement $n_k\to-n_k$ in the Fourier sum, and we observe that the eigenvalue is an even function of $n$, and the impact factor and the central emission block have the property
\beq
\chi^{\pm}(\nu,-n) = \chi^{\mp}(\nu,n) {\rm~~and~~} C^{\pm}(\nu,-n,\mu,-m) = C^{\mp}(\nu,n,\mu,m)\,,
\eeq
and so eq.~\eqref{eq:sym_CC} follows.

Next, let us analyse how target-projectile symmetry acts on the ratio. Target-projectile symmetry acts on both the $z_k$ and the $\tau_k$ variables,
\begin{align}
z_k &\mapsto 1/z_{N-4-k} {\rm~~and~~}\tau_k \mapsto \tau_{N-4-k} \,.
\end{align}
It corresponds to the transformation $(\nu_k,n_k)\mapsto (-\nu_{N-4-k},-n_{N-4-k})$ in Fourier-Mellin space. The factor $[-1 + \prod \tau_k^{a E_{\nu_k,n_k}}]$ involving the BFKL eigenvalue is invariant, and the product of impact factors transforms as
\beq\bsp
\chi^{h_1}(\nu_1,n_1)\,\chi^{-h_{N-5}}(\nu_{N-5},n_{N-5}) &\,\mapsto \chi^{h_1}(-\nu_{N-5},-n_{N-5})\,\chi^{h_{N-5}}(-\nu_1,-n_1)\\
&\, = \chi^{-h_{N-5}}(\nu_1,n_1)\,\chi^{-(-h_{1})}(\nu_{N-5},n_{N-5})\,.
\label{TPimpfactors}
\esp\eeq
Similarly, we have for the central emission blocks,
\beq
C^{\pm}(-\nu_{j},-n_j,-\nu_{j-1},-n_{j-1}) = C^{\mp}(\nu_{j-1},n_{j-1},\nu_{j},n_j)\,.
\label{TPCblock}
\eeq
Indeed, we have
\beq\bsp
C^+&(-\mu,-m,-\nu,-n) = \frac{\Gamma(1+i\mu+\frac{m}{2})\,\Gamma(-i\nu-\frac{n}{2})\,\Gamma(i(\nu-\mu)+\frac{m-n}{2})}{\Gamma(1-i\mu+\frac{m}{2})\,\Gamma(i\nu-\frac{n}{2})\,\Gamma(1-i(\nu-\mu)+\frac{m-n}{2})}\\
&\,= \frac{\left(i\nu-\frac{n}{2}\right)\,\left(i\mu+\frac{n}{2}\right)}{\left(i\nu+\frac{n}{2}\right)\,\left(i\mu-\frac{n}{2}\right)}\,\frac{\Gamma(i\mu+\frac{m}{2})\,\Gamma(1-i\nu-\frac{n}{2})\,\Gamma(i(\nu-\mu)+\frac{m-n}{2})}{\Gamma(-i\mu+\frac{m}{2})\,\Gamma(1+i\nu-\frac{n}{2})\,\Gamma(1-i(\nu-\mu)+\frac{m-n}{2})}\\
&\,=\frac{\chi^+(\nu,n)\,\chi^-(\mu,m)}{\chi^-(\nu,n)\,\chi^+(\mu,m)}\,C^+(\nu,n,\mu,m)\\
&\,=C^-(\nu,n,\mu,m)\,.
\esp\eeq
Note that both transformations (\ref{TPimpfactors}) and (\ref{TPCblock}) also respect our $\epsilon$ prescription for the poles on the real axis.
Hence, target-projectile symmetry acts on the ratio by
\beq
\cR_{h_1,\ldots,h_{N-4}}\left(\tau_{1},\frac{1}{z_{1}},\ldots,\tau_{N-5},\frac{1}{z_{N-5}}\right) =
\cR_{-h_{N-4},\ldots,-h_1}(\tau_{N-5},{z}_{N-5},\ldots,\tau_{1},{z}_{1}) \,.
\eeq

\subsection{Perturbative expansion of the ratio $\mathcal{R}_{h_1,\ldots,h_{N-4}}$}
So far all the considerations were made before the perturbative expansion of the function $\mathcal{R}_{h_1,\ldots,h_{N-4}}$. If we expand the integrand in eq.~\eqref{eq:MRK_conjecture} perturbatively, then at each order we obtain logarithms of $\tau_k$. The coefficients of these logarithms are the main objects of interest in the rest of this paper. We write the perturbative expansion of the function $\mathcal{R}_{h_1,\ldots,h_{N-4}}$ as
\beq\bsp
\cR&_{h_1,\ldots,h_{N-4}}\left(\tau_{1},{z_{1}},\ldots,\tau_{N-5},{z_{N-5}}\right) =1+a\,i\pi\,r_{h_1,\ldots,h_{N-4}}^{(1)}\\
&\,+ 2\pi i \sum_{i=2}^{\infty}\sum_{i_1+\ldots+i_{N-5}=i-1}a^i\,\left(\prod_{k=1}^{N-5}\frac{1}{i_k!}\log^{i_k}\tau_k\right)\,{g}_{h_1,\ldots,h_{N-4}}^{(i_1,\ldots,i_{N-5})}(z_1,\ldots,z_{N-5})\,.
\esp\eeq
The perturbative coefficients are completely known for $N=6$ for both MHV and NMHV~\cite{Lipatov:2010qg,Lipatov:2010ad,Lipatov:2012gk,Dixon:2012yy,Pennington:2012zj,Broedel:2015nfp}, and for all MHV amplitudes at two loops~\cite{Bartels:2011ge,Prygarin:2011gd,Bargheer:2015djt}. Comparing the perturbative expansion to eq.~\eqref{eq:MRK_conjecture}, we see that the perturbative coefficients admit a representation as a Fourier-Mellin transform,
\begin{align}\nonumber
{g}&_{h_1,\ldots,h_{N-4}}^{(i_1,\ldots,i_{N-5})}(z_1,\ldots,z_{N-5}) =
\frac{(-1)^{N+1}}{2} \,\left[\prod_{k=1}^{N-5}\sum_{n_k=-\infty}^{+\infty}\left(\frac{z_k}{\zbar_k}\right)^{n_k/2}\int_{-\infty}^{+\infty}\frac{d\nu_k}{2\pi}|z_k|^{2i\nu_k}E_{\nu_kn_k}^{i_k}\right]\\
\label{eq:g_n_def}&\times
\chi^{h_1}({\nu_1,n_1})\,
\left[\prod_{j=1}^{N-6}C^{h_j}(\nu_j,n_j,\nu_{j+1},n_{j+1})\,\right]\,\chi^{-h_{N-5}}({\nu_{N-5},n_{N-5}})\,.
\end{align}
The poles on the real axis are dealt with by the prescription already outlined in (\ref{rep1}) - (\ref{rep3}).

The symmetries of the ratio $\mathcal{R}_{h_1,\ldots,h_{N-4}}$ discussed in the previous section induce similar symmetries on the perturbative coefficients,
\beq\bsp
{g}_{h_1,\ldots,h_{N-4}}^{(i_1,\ldots,i_{N-5})}(z_1,\ldots,z_{N-5}) &\,= {g}_{-h_{1},\ldots,-h_{N-4}}^{(i_1,\ldots,i_{N-5})}(\bar{z}_{1},\ldots,\bar{z}_{N-5})\\
&\,={g}_{-h_{N-4},\ldots,-h_{1}}^{(i_{N-5},\ldots,i_{1})}\!\!\left(\frac{1}{z_{N-5}},\ldots,\frac{1}{z_1}\right)\,.
\esp\eeq

In the soft limits, the perturbative coefficients must reduce to lower-point functions. The limits where either ${\bf k}_1$ or ${\bf k}_{N-4}$ vanish are easy to describe: the perturbative coefficients reduce to the corresponding coefficients with the soft momentum removed, except if the corresponding large logarithm is present, in which case the perturbative coefficient vanishes in the limit. More precisely,
\beq\bsp
\lim_{z_1\to0} {g}_{h_1,\ldots,h_{N-4}}^{(i_1,\ldots,i_{N-5})}(z_1,\ldots,z_{N-5}) &\,= \delta_{i_10}\,{g}_{h_2,\ldots,h_{N-4}}^{(i_2,\ldots,i_{N-5})}(z_2,\ldots,z_{N-5})\,,\\
\lim_{z_{N-5}\to\infty} {g}_{h_1,\ldots,h_{N-4}}^{(i_1,\ldots,i_{N-5})}(z_1,\ldots,z_{N-5}) &\,= \delta_{i_{N-5}0}\,{g}_{h_1,\ldots,h_{N-5}}^{(i_1,\ldots,i_{N-6})}(z_1,\ldots,z_{N-6})\,.
\esp\eeq
If ${\bf k}_j$, with $j\notin\{1,N-4\}$ is soft, then the perturbative coefficients behave like,
\beq\bsp
\lim_{\substack{(z_{j-1},z_j)\to(\infty,0)\\ z_{j-1}z_j\textrm{ fixed}}}&{g}_{h_1,\ldots,h_{N-4}}^{(i_1,\ldots,i_{N-5})}(z_1,\ldots,z_{N-5})\\
&\, = {g}_{h_1,\ldots,\hat{h}_{j},\ldots,h_{N-4}}^{(i_1,\ldots,i_{j-1}+i_j,\ldots,i_{N-5})}(z_1,\ldots,-z_{j-1}z_j,\ldots,z_{N-5})\,.
\esp\eeq
Indeed, we have
\begin{align}
&\lim_{{\bf k}_j\to 0}\cR_{h_1,\ldots,h_{N-4}}\left(\tau_{1},{z_{1}},\ldots,\tau_{N-5},{z_{N-5}}\right)\\
\nonumber&\,=2\pi i \sum_{i=2}^{\infty}\sum_{i_1+\ldots+i_{N-5}=i-1}a^i\,\left(\prod_{k=1}^{N-5}\frac{1}{i_k!}\log^{i_k}\tau_k\right)\,{g}_{h_1,\ldots,\hat{h}_{j},\ldots,h_{N-4}}^{(i_1,\ldots,i_{j-1}+i_j,\ldots,i_{N-5})}\\
\nonumber&\,=2\pi i \sum_{i=2}^{\infty}\sum_{i_1+\ldots+i'+\ldots+i_{N-5}=i-1}\sum_{i_{j-1}+i_j=i'}a^i\,\left(\prod_{k=1}^{N-5}\frac{1}{i_k!}\log^{i_k}\tau_k\right)\,{g}_{h_1,\ldots,\hat{h}_{j},\ldots,h_{N-4}}^{(i_1,\ldots,i',\ldots,i_{N-5})}\\
\nonumber&\,=2\pi i \sum_{i=2}^{\infty}\sum_{i_1+\ldots+i'+\ldots+i_{N-5}=i-1}a^i\,\frac{1}{i'!}\log^{i'}(\tau_{j-1}\tau_j)\,\left(\prod_{\substack{k=1\\k\notin\{j-1,j\}}}^{N-5}\frac{1}{i_k!}\log^{i_k}\tau_k\right)\,{g}_{h_1,\ldots,\hat{h}_{j},\ldots,h_{N-4}}^{(i_1,\ldots,i',\ldots,i_{N-5})}\,,
\end{align}
where the last step follows from the binomial theorem,
\beq
\sum_{i_{j-1}+i_j=i'}\frac{1}{i_{j-1}!i_j!}\,\log^{i_{j-1}}\tau_{j-1}\,\log^{i_{j}}\tau_{j} = \frac{1}{i'!}\log^{i'}(\tau_{j-1}\tau_j)\,.
\eeq


\section{MRK and the moduli space of genus zero curves with marked points}
\label{sec:svmpls}
\subsection{MRK and the moduli space $\mathfrak{M}_{0,N-2}$}

In this section we argue that it is possible to describe the space of functions of scattering amplitudes in planar $\cN=4$ SYM in MRK. We start by noting that in MRK the only non-trivial functional dependence is through the transverse momenta. 
In the previous section we have seen that the kinematics in the transverse space is described by $n\equiv N-2$ dual coordinates $\bx _i$. 
Hence, in the multi-Regge limit the kinematics is described by a configuration of $(N-2)$ points in $\mathbb{CP}^1$. The space of such configurations is equivalent to the moduli space of genus zero curves with $(N-2)$ marked points,
\beq 
\textrm{Conf}_{N-2}(\mathbb{CP}^1) \simeq \mathfrak{M}_{0,N-2}\,.
\eeq
In Section~\ref{sec:cluster} we have seen that the cluster algebra attached to the configuration space describing the kinematics of an amplitude is related to the singularities of the amplitude. 
From the previous discussion it is thus natural to expect that amplitudes in planar $\cN=4$ SYM in MRK can be expressed in terms of iterated integrals on $\mathfrak{M}_{0,N-2}$. We now show that this is indeed the case. More precisely, we show that the cluster algebra associated to $\textrm{Conf}_{N}(\mathbb{CP}^3)$ in full kinematics reduces to the cluster algebra of $\mathfrak{M}_{0,N-2}$.

We start from the duality between MRK and multi-soft limits discussed in Section~\ref{sec:mrk}. We insert the  parametrisation of eq.~\eqref{eq:soft_parametrisation} into the cluster $\cX$-coordinates of eq.~\eqref{eq:X_coords} and we take the limit $\eps_i\to0$. We see that all $\cX$-coordinates of the form $\cX_{2j}$ vanish in the limit, while the others reduce to either holomorphic or anti-holomorphic cross ratios in $\mathbb{CP}^1$,
\beq\label{eq:Xcoordinates_MrK}
\cX_{1j} = \frac{(\overline{\bx}_2-\overline{\bx}_{j+2})(\overline{\bx}_{j+3}-\overline{\bx}_{j+4})}{(\overline{\bx}_2-\overline{\bx}_{j+4})(\overline{\bx}_{j+2}-\overline{\bx}_{j+3})}\,,\quad \cX_{2j}=0\,,\quad
\cX_{3j} = \frac{({\bx}_1-{\bx}_{j+1})({\bx}_{j+2}-{\bx}_{j+3})}{({\bx}_1-{\bx}_{j+3})({\bx}_{j+1}-{\bx}_{j+2})}\,.
\eeq
We see that the $\cX$-coordinates are singular when two points $\bx_i$ coincide, which is precisely the singularity structure of the moduli  space $\mathfrak{M}_{0,N-2}$. However, we have obtained two copies of points, a holomorphic and an anti-holomorphic one. This can be understood from the cluster algebra in Fig.~\ref{fig:clusterXcoords}. Indeed, in the multi-Regge limit the middle line in the quiver vanishes, and so the cluster algebra splits into two disconnected parts, one which only depends on holomorphic variables and the other one only on anti-holomorphic variables. Each of these two parts is isomorphic to the cluster algebra $A_{N-5}$, which is the cluster algebra that describes the singularity structure of $\textrm{Conf}_{N-2}(\mathbb{CP}^1)\simeq \mathfrak{M}_{0,N-2}$. Hence, we conclude that in MRK the cluster algebra of $\textrm{Conf}_N(\mathbb{CP}^3)$ reduces to the cluster algebra $A_{N-5}\times A_{N-5}$, and the two copies of $A_{N-5}$ are complex conjugate to each other in the case of real $2$-to-$(N-2)$ scattering. As a consequence, we expect that planar scattering amplitudes in $\cN=4$ SYM in MRK can be expressed through iterated integral with singularities precisely when the $\cX$-coordinates in eq.~\eqref{eq:Xcoordinates_MrK} are singular, i.e., iterated integrals over integrable words made out of the one-forms $d\log(\bx_i-\bx_j)$ (and their complex conjugates).
Note that scattering amplitudes in MRK are singular whenever one of the final-state gluons is soft, ${\bf k}_i\to0$, (see Section~\ref{sec:soft_limits}) which happens precisely when $\bx_i=\bx_{i+1}$, $2\le i\le N-4$. It is remarkable that the cluster algebra in MRK is of finite type, independently of the number $N$ of external particles. Indeed, it is known that a cluster algebra is of finite type precisely if one of the quivers that represent its seeds is a Dynkin diagram~\cite{cluster2}. The cluster algebras associated to the six and seven-point amplitudes are of finite type (the corresponding Dynkin diagrams are $A_3$ and $E_6$), but starting from $N=8$ the cluster algebra is infinite~\cite{scott,Golden:2013xva}. Remarkably, the cluster algebra in general kinematics always reduces to a cluster algebra of finite type in MRK.

Scattering amplitudes, however, cannot be arbitrary combinations of iterated integrals built on $A_{N-5}\times A_{N-5}$, but the branch cuts of the amplitudes are constrained by physics arguments. 
In particular, massless scattering amplitudes can have branch points at most when a Mandelstam invariant vanishes or becomes infinite, which puts strong constraints on the first letter in the word defining the iterated integral\footnote{We note that this condition is independent of whether the iterated integral can be evaluated in terms of multiple polylogarithms.}~\cite{Gaiotto:2011dt}. Dual conformal invariance implies that the first letter of the word must be a cross ratio $d\log U_{ijkl}$. In the Mandelstam region $[p,q]$, however, integrability combined with the first entry condition implies that on this Riemann sheet the branch points are determined by products of cross ratios that become equal to 0, 1 or $\infty$. In other words, in a Mandelstam region the first letter is either a cross ratio $d\log U_{ijkl}$ or $d\log(1-\prod_{ijkl}U_{ijkl}^{n_{ijkl}})$. In the following we show that this implies that in MRK the first entries are necessarily absolute values squared of cross ratios in $\mathbb{CP}^1$.

To start, we note that there are $N(N-5)/2$ multiplicatively independent cross ratios, which we may choose as
\begin{align}\nonumber
u_{1i}\,, u_{2i}\,,u_{3i}\,,&\quad 1\le i\le N-5\,,\\
\label{eq:indep_cr}
{U}_{ij}\,, &\quad 2 \leq i \leq j-4 \leq N-5\,,
\end{align}
where these cross ratios have been defined in eq.~\eqref{eq:ui_def} and~\eqref{eq:Uij_def}. The multi-Regge limit of $(u_{1i}, u_{2i},u_{3i})$ was already analysed in Section~\ref{sec:mrk}. Using the duality between MRK and the multi-soft limit, it is easy to show that all the $U_{ij}$ tend to 1 in MRK. We introduce new reduced cross ratios which have a finite multi-Regge limit,
\beq\label{eq:reduced_2}
\tilde{U}_{ij} \equiv \frac{1-U_{ij}}{\prod_{k=i-1}^{j-4} (1-u_{1k})} \to \left|\frac{\bx_i-\bx_{j-1}}{\bx_i-\bx_{i+2}}\,\prod_{k=i+1}^{j-3}\frac{\bx_k-\bx_{k+1}}{\bx_k-\bx_{k+2}}\right|^2\,.
\eeq
From eq.~\eqref{eq:reduced_2} we see that all the $U_{ij}$ approach 1 at different speeds in the multi-soft limit. Indeed, the multi-soft limit is approached sequentially according to $\eps_2\ll\eps_3\ll\ldots\ll\eps_{N-4}$, where $\eps_i$ are the small parameters introduced in eq.~\eqref{eq:soft_parametrisation}. Since $u_{1i}=1+\ord(\eps_{i+1})$, we see that $U_{ij} =1+ \ord(\eps_{i}\ldots\eps_{j-4})$, and so all the $U_{ij}$ approach 1 at a different speed.

We now show that the first entries of the perturbative coefficients reduce to absolute values squared of cross ratios in $\mathbb{CP}^1$ (up to logarithmically divergent terms that are absorbed into the definition of the $\tau_k$). Let us first look at the case where the first letter is $d\log U_{ijkl}$. It is sufficient to analyse the multiplicatively independent cross ratios in eq.~\eqref{eq:indep_cr}. They all tend to 1, except for $u_{2i}$ and $u_{3i}$, which we may exchange for the corresponding reduced cross ratios $\tilde{u}_{2i}$ and $\tilde{u}_{3i}$. The latter reduce to absolute values squared of cross ratios in $\mathbb{CP}^1$, see eq.~\eqref{eq:ui_xi}.

Next, let us analyse the case of a letter of the type $d\log(1-\prod_{ijkl}U_{ijkl}^{n_{ijkl}})$. It is sufficient to assume that the factors in the product are taken from eq.~\eqref{eq:indep_cr}. If one of the factors goes to zero in MRK, then the claim is true, because we have for example,
\beq
d\log(1-u_{2i}^{n}\,U) \to \left\{\begin{array}{ll}
n\,d\log u_{2i} + d\log U\,,& \textrm{ if } n<0\,,\\
0\,,&\textrm{ if } n>0\,,
\end{array}\right.
\eeq
where $U$ is any product of cross ratios that tend to 1 in MRK.
If all the factors in the product $\prod_{ijkl}U_{ijkl}^{n_{ijkl}}$ tend to 1, then we know that one of the factors tends to one much slower than the others. Hence, up to terms that are power-suppressed in MRK, we only need to keep this factor. The claim then follows from eq.~\eqref{eq:reduced_2}.

The previous discussion implies that the coefficients appearing in the perturbative expansion of scattering amplitudes  in planar $\cN=4$ SYM are iterated integrals with singularities described by the cluster algebra $A_{N-5}\times A_{N-5}$ and whose first letters are absolute values squared of cross ratios. As the first entries describe the branch points of the function, we conclude that the perturbative coefficients have no branch cuts when seen as functions of the complex points $\bx_i$. 
In other words, these iterated integrals must define single-valued functions on the moduli space of Riemann spheres with $N-2$ marked points.
In the remainder of this section we review the theory of single-valued iterated integrals on $\mathfrak{M}_{0,N-2}$. We first discuss ordinary, not necessarily single-valued, iterated integrals on $\mathfrak{M}_{0,N-2}$, and we turn to the construction of their single-valued analogues at the end of the section.

\subsection{Coordinate systems on $\mathfrak{M}_{0,n}$}
\label{sec:coordinates}
In this section we review various coordinate systems on $\mathfrak{M}_{0,n}$ which are useful to study iterated integrals  and the multi-Regge limit. 
As a geometric space, we can describe $\mathfrak{M}_{0,n}$ by configurations of $n$ distinct points on the Riemann sphere. We identify configurations that are related by conformal transformations. As $SL(2,\mathbb{C})$ has complex dimension 3, we immediately see that
\beq
\dim_{\mathbb{C}}\mathfrak{M}_{0,n} = n-3\,.
\eeq

Roughly speaking, since $\mathfrak{M}_{0,n}$ is $SL(2,\mathbb{C})$-invariant, a system of coordinates is given by a set of cross ratios formed out of the points ${\bf x}_i$. There is no global coordinate system on $\mathfrak{M}_{0,n}$. 
One such set of cross ratios is given by the cross ratios $z_i$ defined in eq.~\eqref{eq:z_i_def}. We will refer to these coordinates as \emph{Fourier-Mellin coordinates}. These coordinates are well suited to write down the Fourier-Mellin transforms that describe amplitudes in MRK. These coordinates, however, are not ideal to describe the iterated integrals on $\mathfrak{M}_{0,n}$. 

In ref.~\cite{Brown:2009qja} various local systems of coordinates are discussed that are well suited to study iterated integrals on $\mathfrak{M}_{0,n}$. A particularly simple set of local coordinates are the \emph{simplicial coordinates}, obtained by using the $SL(2,\mathbb{C})$ invariance to fix three of the $n$ points to $0$, $1$ and $\infty$, e.g.,
\beq
(\bx_1,\ldots,\bx_{n}) \to (0,1,\infty,t_1,\ldots,t_{n-3})\,, \textrm{ with } t_{i-3} = \frac{(\bx_{i}-\bx_1)(\bx_{2}-\bx_3)}{(\bx_{i}-\bx_3)(\bx_{2}-\bx_1)}\,, \, 4\le i\le n\,.
\eeq
Note that there are $6\,\binom{n}{3}=n(n-1)(n-2)$ different choices for simplicial coordinates, depending on which three points we fix to $(0,1,\infty)$.
Using simplicial coordinates we can describe $\mathfrak{M}_{0,n}$ (roughly speaking) as the space
\beq
\{(t_1,\ldots,t_{n-3})\in \mathbb{C}^{n-3}| t_i\neq 0,1 \textrm{ and } t_i\neq t_j\}\,.
\eeq

While there is in principle no reason to prefer one particular choice of simplicial coordinates over the other, some choices are more suited to MRK than others. In particular, it is useful to choose the coordinates so that they transform nicely under the symmetries of the problem. In our case, we prefer to choose simplicial coordinates on which target-projectile symmetry  acts in a simple way.
It is easy to check that the simplicial coordinate systems with this property are defined by fixing the points $(\bx_1,\bx_k,\bx_{N-k})$, $2\le k\le \left\lceil\frac{N-1}{2}\right\rceil$. In addition, for $N$ even the set of simplicial coordinates defined by fixing $(\bx_{N/2},\bx_k,\bx_{N-k})$ also has this property.

There is one particular choice of simplicial coordinates with the nice property that in these coordinates the two-loop MHV amplitudes factorise into sums of six-point amplitudes~\cite{Bartels:2011ge,Prygarin:2011gd,Bargheer:2015djt}. They are defined by 
\beq
(\bx_1,\ldots,\bx_{N-2}) \to (1,0,\rho_1,\ldots,\rho_{N-5},\infty)\,.
\eeq
We refer to these coordinates as \emph{simplicial MRK coordinates}. 
From the previous discussion it follows that simplicial MRK coordinates transform nicely under target projectile symmetry,
\beq
(\rho_1,\ldots,\rho_{N-5}) \mapsto \left(1/\rho_{N-5},\ldots,1/\rho_{1}\right)\,.
\eeq
Simplicial MRK coordinates are related to the Fourier-Mellin coordinates by
\beq\label{eq:z_to_rho}
z_i = \frac{(\rho_i-\rho_{i-1})(\rho_{i+1}-1)}{(\rho_i-\rho_{i+1})(\rho_{i-1}-1)}\,,
\eeq 
with $\rho_{0}=0$ and $\rho_{N-4}=\infty$. In these coordinates the two-loop MHV amplitude takes a particularly simple form~\cite{Bartels:2011ge,Prygarin:2011gd}
\beq\label{eq:2-loop_MHV}
{g}_{+\ldots+}^{(0,\ldots,0,1,0,\ldots,0)}(\rho_1,\ldots,\rho_{N-5}) = \frac{1}{4}\,\log\left|1-\rho_k\right|^2\,\,\log\left|1-\frac{1}{\rho_k}\right|^2\,,
\eeq
where $k$ denotes the position of the 1 in $(0,\ldots,0,1,0,\ldots,0)$.
Finally, we point out that soft limits are very easy to describe in simplicial MRK coordinates. In the limit where ${\bf k}_i$ is soft we have $\rho_{i-1}=\rho_i$ (with $\rho_{0}=0$ and $\rho_{N-4}=\infty$).

There is another class of simplicial coordinates which will be important in the remainder of this paper. Let us start from the Fourier-Mellin coordinates, and let us single out one of them, say $z_i$. Then there is always a (non unique) set of simplicial coordinates $(t_1^{(i)},\ldots, t_{N-5}^{(i)})$ such that $t^{(i)}_i=z_i$. Indeed, from eq.~\eqref{eq:z_i_def} we see that we can define these coordinates by
\beq
(\bx_1,\ldots,\bx_{N-2}) \to (\infty,t_1^{(i)},\ldots,t_i^{(i)},0,1,\ldots,t_{N-5}^{(i)})\,.
\eeq
We will refer to these simplicial coordinates as \emph{simplicial coordinates based at $z_i$}. They do in general not possess any simple transformation properties under target-projectile symmetry, but they will be essential in order to carry out all the Fourier-Mellin integrations, because they `interpolate' between the Fourier-Mellin and simplicial MRK coordinates.

Sometimes it is helpful to describe the moduli space $\mathfrak{M}_{0,n}$ in projective terms. To do so we can introduce $n$ elements $r_i \in \mathbb{CP}^1$, that is $n$ two-component complex vectors modulo non-zero complex scalings. We may return to the $\bx_i$ coordinates by making use of the scalings so that $r_i = (1,\bx_i)$. In the projective language $SL(2,\mathbb{C})$ invariance means that all quantities should be expressed in terms of the $SL(2,\mathbb{C})$ invariant two-brackets
\beq
(ij) = \epsilon_{ab}r_i^a r_j^b\,,
\label{2bracket}
\eeq
where $\epsilon_{ab}$ is the two-index antisymmetric tensor with $\epsilon_{12} = 1$. Moreover, since we must maintain the projective nature of the $r_i$ we must form only quantities which are homogeneous of degree zero. Such quantities are given by cross-ratios. 

If we choose an ordering of our points (corresponding to the one induced by the colour ordering of the scattering amplitude) we may introduce a particular set of cross-ratios, the {\sl dihedral coordinates},
\be
v_{ij} = 
\frac{(i\,j+1)(i+1\, j)}{(ij)(i+1\, j+1)} = 
\frac{(\bx_i - \bx_{j+1})(\bx_{i+1} - \bx_j)}{(\bx_i - \bx_j)(\bx_{i+1} - \bx_{j+1})} \,,
\label{dihedrals}
\ee
where indices are treated modulo $n$ and we have given both projective and coordinate-fixed forms. Note that only $(n-3)$ of the $v_{ij}$ are algebraically independent, since this is the dimension of the moduli space $\mathfrak{M}_{0,n}$. To continue, we pick a dihedral structure $\eta$ on $\mathfrak{M}_{0,n}$, i.e. a cyclic ordering of the $n$ points $r_i$ modulo reflections
. 
In our case the points ${\bf x}_i$, and hence also the $r_i$, come with a natural dihedral structure induced by the colour ordering and target projectile symmetry. We therefore assume from now on that $\mathfrak{M}_{0,n}$ is equipped with this particular dihedral structure, and we will often omit the dependence on the choice of $\eta$ explicitly.

Dihedral coordinates will play an important role in the next section when defining iterated integrals on $\mathfrak{M}_{0,n}$. Moreover, they allow one to give a nice geometric interpretation of real moduli space $\mathfrak{M}_{0,n}(\mathbb{R})$, which we describe in the remainder of this section.
In the real moduli space, the region of $\mathfrak{M}_{0,n}$ defined by $0 < v_{ij} < 1$ describes the interior of a Stasheff polytope or associahedron. The full real moduli space is tiled by $n!/(2n)$ such regions, each one corresponding to a different choice of dihedral structure $\eta$. The codimension one faces of the polytope are each obtained by taking one of the $v_{ij}$ to zero (while maintaining $0 < v_{ij} < 1$ for the others). One can then continue to codimension two boundaries of the boundary face etc. This process can be continued all the way until one reaches the codimension $(n-3)$ (i.e. dimension zero) vertices. 

The combinatorics describing the various boundaries are such that each vertex $V$ of the Stasheff polytope is labelled by a triangulation $T_V$ of an $n$-sided polygon  (which in our case corresponds to the polygon formed by the dual coordinates ${\bf x}_i$ in the natural order induced by the color ordering, see Fig.~\ref{fig:dual_coordinates}), with the chords $\{i,j\} \in T_V$ defining the triangulation given by the set of $v_{ij}$ that are zero at the vertex $V$. The other $v_{ij}$ are equal to one at this vertex. This structure is described in detail in ref.~\cite{Brown:2009qja} and we refer the reader there for more details. Let us note however that two vertices $V$ and $V'$ which are separated by a single edge correspond to two triangulations which differ by a single chord. In other words, to obtain $T_{V'}$ from $T_V$, one removes some chord $\{i,j\}$ from $T_V$ and replaces it with a crossing chord $\{k,l\}$ such that the result is still a triangulation.
The projective and dihedral coordinates will be useful in the discussion of the Knizhnik-Zamolodchikov equation on $\mathfrak{M}_{0,n}$ which follows.

\subsection{Iterated integrals on $\mathfrak{M}_{0,n}$}
\label{sec:iterated_integrals}
In this section we summarise the theory of iterated integrals on $\mathfrak{M}_{0,n}$, before describing their single-valued analogues in the next section.
A very helpful way to think about iterated integrals on $\mathfrak{M}_{0,n}$ is to think of them as being described in terms of generating functions which obey the Knizhnik-Zamolodchikov (KZ) equation~\cite{Brown:2009qja}. The KZ equation on $\mathfrak{M}_{0,n}$ can be written 
in terms of the projective variables $r_i$ introduced above eq. (\ref{2bracket}) 
as follows,
\be
dL = \Omega L\,, \qquad \Omega = \sum_{i<j} \Omega_{ij}\,, \qquad \Omega_{ij} = X_{ij} d \log (ij)\,.
\label{KZ}
\ee
Here the $X_{ij}$ are a collection of formal generators obeying
\be
X_{ij} = X_{ji}\,, \qquad X_{ii}=0\,, \qquad \sum_i X_{ij} = 0\,, \qquad [X_{ij},X_{kl}] = 0\,, \quad \{i,j,k,l\} \text{ distinct.}
\label{Xbraidrels}
\ee
The first two relations in eq.~(\ref{Xbraidrels}) are conventional, ensuring that there are as many generators as there are one-forms $d \log(ij)$. The third relation ensures that the connection $\Omega$ is homogeneous under rescalings of the $r_i$, so that it is indeed a connection on the moduli space of points in $\mathbb{CP}^1$. 
The final relation in eq.~(\ref{Xbraidrels}) completes a centre-free presentation of the infinitesimal pure braid relations on the $X_{ij}$ and it ensures that the connection $\Omega$ obeys
\be
\Omega \wedge \Omega = 0\,.
\label{flatness}
\ee
Since $\Omega$ also trivially obeys $d\Omega=0$, the condition (\ref{flatness}) implies that the connection is flat. We can consider solutions of eq.~(\ref{KZ}) which take the form,
\be
L = 1 + \text{higher-order terms in the $X_{ij}$}\,.
\ee
Such solutions are formal series in the generators $X_{ij}$, i.e., they are a sum over all words in the $X_{ij}$ of any length, modulo the relations (\ref{Xbraidrels}). The coefficients of the independent words are given by iterated integrals on $\mathfrak{M}_{0,n}$, and hence the solutions $L$ can be viewed as generating functions of the class of $A_{n-3}$ cluster polylogarithms. Iterated integrals form a shuffle algebra, and in the following we denote by $B_n$ the shuffle algebra over $\mathbb{Q}$ of all iterated integrals on $\mathfrak{M}_{0,n}$. As a vector space, $B_n$ is generated by the coefficients of the independent words in $L$.

The description of the KZ equation given in eq.~(\ref{KZ}) and (\ref{Xbraidrels}) is manifestly invariant under all permutations of the $r_i$. In other words it did not depend on our initial choice of ordering $r_1,\ldots,r_n$. It will be useful however to present another description, presented in detail in ref.~\cite{Brown:2009qja}, which manifests only a dihedral symmetry. The construction depends on the choice of dihedral structure, and as before we choose the one induced by the colour ordering. In terms of the dihedral coordinates $v_{ij}$ the KZ equation takes the form,
\be
d L = \Omega L\,, \qquad \Omega = \sum_{\{i,j\}} \delta_{ij}\, d \log v_{ij}\,.
\label{KZdihedral}
\ee
The sum is over all pairs $\{i,j\}$ where the indices $i$ and $j$ are separated by at least two, with all indices treated modulo $n$. 
We can identify a pair $\{i,j\}$ with the corresponding chord of the polygon built on the points $r_i$, or equivalently $\bx_i$ (see Section~\ref{sec:coordinates}).
The generators $\delta_{ij}$ are related to the $X_{ij}$ via
\be
X_{ij} = \delta_{i\,j+1} + \delta_{i+1\,j} - \delta_{ij} - \delta_{i+1\,j+1}\,,
\ee
and consequently obey
\be
[\delta_{i\,j+1} + \delta_{i+1\,j} - \delta_{ij} - \delta_{i+1\,j+1},\delta_{k\,l+1} + \delta_{k+1\,l} - \delta_{kl} - \delta_{k+1\,l+1}] = 0\,, \quad \{i,j,k,l\} \text{  distinct.}
\label{dihedralbraidrels}
\ee
We also define $\delta_{ii}= \delta_{i\,i+1} =0$. Note that the above relations imply that two generators $\delta_{ij}$ and $\delta_{kl}$ commute if the chords $\{i,j\}$ and $\{k,l\}$ of the polygon do not intersect. This implies in particular that all the $\delta_{ij}$ associated to a triangulation, and hence to a vertex $V$ of the Stasheff polytope, commute,
\be
[\delta_{ij},\delta_{kl}]=0\, \qquad \{i,j\} , \{k,l\} \in T_V\,.
\ee

We may now define canonically normalised solutions $L_V$ to the KZ equation (\ref{KZdihedral}) associated to each vertex $V$ on the boundary of the polytope defining the positive region such that $L_V$ is real-valued in the interior of the Stasheff polytope, i.e., where all $v_{ij}$ obey $0 < v_{ij} < 1$. The solution $L_V$ that we want is chosen to have the following behaviour in a neighbourhood of $V$
\be
L_V = L_{V,\rm an} \Biggl( \prod_{\{i,j\}\in T_V} v_{ij}^{\delta_{ij}} \Biggr)  \,,
\label{asymptotics}
\ee
where $L_{V, \rm an}$ is analytic in a neighbourhood of $V$. To linear order we have
\be
L_V = 1 + \sum_{\{i,j\}} \delta_{ij} \log v_{ij} + \ldots \,.
\label{Lvleadingterms}
\ee
The behaviour (\ref{Lvleadingterms}) is in fact independent of the choice of $V$, with the dependence on $V$ arising at quadratic and higher order. We may regard $L_V$ as a shuffle regularised path-ordered exponential in the connection $\Omega$. The coefficients of the independent words in $L_V$ are again iterated integrals on $\mathfrak{M}_{0,n}$. In fact, these coefficients simply provide an alternative set of generators for the shuffle algebra $B_n$. Note that, although the set of generators depends on the choice of the vertex $V$ used to define the generating function $L_V$, the shuffle algebra $B_n$ is independent of the vertex $V$.

Let us discuss how the generators obtained from different choices of $V$ are related.
The KZ equation being homogeneous, different solutions, associated with different vertices $V$ and $V'$, are related by a parallel transport by a constant series $\Phi_{V,V'}$,
\be\label{eq:transport}
L_{V'} = L_V\, \Phi_{V,V'}\,.
\ee
By considering the case where two vertices are connected by a single edge on the boundary of the polytope $\mathfrak{M}_{0,n}(\mathbb{R})$, we find that the constant series is given by the canonical Drinfeld associator, given by a sum over shuffle regularised multiple-zeta values,
\be\label{eq:Drinfeld}
\Phi(e_0,e_1) = \sum_w w (-1)^{d(w)} \zeta_\sha(w)\,,
\ee
where the sum is over all words $w$ in two non-commuting generators $e_0$ and $e_1$ and $\zeta_\sha(w)$ is the shuffle regularised multiple zeta value labelled by the word $w$. The quantity $d(w)$ is the number of $e_1$ generators in the word $w$ and is present in order to be coherent with the usual definition of multiple zeta values. To complete the relation between $L_V$ and $L_{V'}$ we still need to determine the values of $e_0$ and $e_1$ that enter eq.~\eqref{eq:Drinfeld}. In Section~\ref{sec:coordinates} we have seen that to every vertex $V$ of the Stasheff polytope we can associate a triangulation $T_V$ of the polygon formed by the points ${\bf x}_i$, and the triangulation associated to two vertices connected by a single edge differ by exactly one chord. Since to every chord $\{i,j\}$ we can associate a letter $\delta_{ij}$, we can determine the $e_0$ and $e_1$ from the two chords by which the triangulations differ.
More precisely, to move between two adjacent vertices of the polytope we apply the associator $\Phi(\delta,\delta')$ where the arguments $\delta$ and $\delta'$ correspond to the generators $\delta_{ij}$ associated to the codimension one face being left behind and $\delta_{kl}$ associated to the one being moved to respectively. Note that since these two faces do not intersect on the boundary of the Stasheff polytope, the two generators $\delta$ and $\delta'$ will never commute. This corresponds precisely with the fact that one obtains the triangulation $T_{V'}$ from $T_V$ by removing the chord $\{i,j\}$ and replacing it with a crossing chord $\{k,l\}$.

Iterated integrals are in general not single-valued.
The monodromies of $L_V$ around the singularities defined by $v_{ij} = 0$ for $\{i,j\} \in T_V$ immediately follow from the asymptotic behaviour of eq.~(\ref{asymptotics}). If we denote the monodromy operator associated with the singularity $v_{ij}=0$ by $\cM_{ij}$, we have
\be
\mathcal{M}_{ij} L_V = L_V\, e^{2 \pi i \delta_{ij}}\,, \qquad \{i,j\} \in T_V\,.
\label{simplemonods}
\ee
To compute the monodromies around another singularity, one first applies a parallel transport from the vertex $V$ to the vertex $V'$ which sits on that singularity via eq.~\eqref{eq:transport}, then performs the monodromy canonically according to the prescription~\eqref{simplemonods}, and then parallel transports back again,
\be\label{eq:Mono_1}
\mathcal{M}_{ij} L_V = L_V\, \Phi_{V,V'}\, e^{2 \pi i \delta_{ij}}\, \Phi_{V',V}\,.
\ee
This formula can be taken for all $\{i,j\}$. It reduces to (\ref{simplemonods}) in the case where the vertex already sits on the singularity labelled by $\{i,j\}$ since in that case $\delta_{ij}$ commutes with $\Phi_{V,V'}$ and 
\be
\Phi_{V,V'}\, \Phi_{V',V} = 1\,.
\ee


In practice it is often useful to work with an explicit basis for the iterated integrals generated by solutions of the KZ equation. The basis we will use is given in terms of hyperlogarithms. We can simply relate this to the previous description of the KZ equation and its solutions as follows. We work in simplicial coordinates of the form
\be
\{\bx_1 , \bx_2 , \ldots , \bx_n\} = \{\infty,0,1,t_1,\ldots,t_{n-3}\}\,.
\ee
The KZ connection on $\mathfrak{M}_{0,n}$ becomes
\be
\Omega^{(n)} = \sum_{4\leq i \leq n} [X_{2i}\, d \log t_{i-3} + X_{3i}\, d\log(1-t_{i-3})] + \sum_{4\leq i<j\leq n} X_{ij}\, d\log (t_{i-3} - t_{j-3}) \,,
\ee
where we have indicated the number $n$ of marked points.
We iteratively factorise solutions of KZ in the form
\be
L_n = F_n L_{n-1}\, ,
\ee
where $L_{n-1}$ is a solution of KZ on $\mathfrak{M}_{0,n-1}$,
\be
d L_{n-1} = \Omega^{(n-1)} L_{n-1}\,
\ee
and $L_3 \equiv 1$. In other words we have a solution of the form
\be
L_n = F_n F_{n-1} \ldots F_4\,.
\ee
Since $F_{n} = L_n (L_{n-1})^{-1}$ we find that 
\begin{align}
d F_n & = d L_n (L_{n-1})^{-1} + L_n d (L_{n-1})^{-1} \, \notag\\
&= \Omega^{(n)} F_n - L_n (L_{n-1})^{-1} \Omega^{(n-1)}\,.
\end{align}
From this it follows that $F_n$ obeys a Picard-Fuchs type equation,
\be
\frac{d F_n}{d t_{n-3}}   = \left(\frac{X_{2n}}{t_{n-3}} + \frac{X_{3n}}{t_{n-3}-1}+  \sum_{i=4}^{n-1} \frac{X_{in}}{t_{n-3} - t_{i-3}}\right)\,F_n\,.
\ee
We are interested in the solution of the above equation given by
\be
F_n = \sum_w w \, G(\sigma_1,\ldots,\sigma_{|w|} ; t_{n-3})\,.
\label{FnseriesforKZ}
\ee
Here the sum is over all words $w \in \langle\langle X_{2n},\ldots,X_{n-1\,,n}\rangle \rangle$ and we denote the `weight' or the length of the word $w$ by $|w|$. The variables $\sigma_1,\ldots,\sigma_w$ are obtained from the word $w$ by the translation of generators $X_{in}$ into letters defined by
\be
X_{2n} \mapsto 0\,, \quad X_{3n} \mapsto 1\,, \quad X_{in} \mapsto t_{i-3} \text{ for } i\geq4\,.
\ee
Finally the functions $G$ are given by hyperlogarithms or iterated integrals of the form
\beq\bsp\label{eq:MPL_def}
G(\sigma_1,\ldots,\sigma_{|w|};t_{n-3})  &\,= \int_{0}^{t_{n-3}}\frac{dt}{t-\sigma_1}\,G(\sigma_2,\ldots,\sigma_{|w|};t) \\
&\,= \int_0^{t_{n-3}} d\log(t_{n-3}-\sigma_{|w|})\ldots d\log(t_{n-3}-\sigma_1)\,.
\esp\eeq

From the above discussion it is clear that the shuffle algebra $B_{n}$ has a recursive structure. In particular, if we work in simplicial coordinates, this recursive structure reads
\beq\label{eq:B_n_rec}
B_{n} \simeq B_{n-1}\otimes_{\mathbb{Q}} L_{\{0,1,t_1,\ldots,t_{n-4}\}}\,,
\eeq
where $L_{\Sigma}$ denotes the shuffle algebra of hyperlogarithms with singularities at $\sigma_i\in \Sigma$ (the $\sigma_i$ are complex constants), i.e., the linear span of all iterated integrals of the form (\ref{eq:MPL_def}).

The recursion starts with $B_3 \equiv \mathbb{Q}$ (because we cannot form a cross ratio with three points), and $B_4$ is  the algebra of harmonic polylogarithms with singularities at most at $0$ and $1$. In other words, if we fix an order on the simplicial coordinates $t_i$, we can describe $B_{n}$ explicitly as linear combinations of hyperlogarithms with singularities at most at $t_{n-3}\in\{0,1,t_1,\ldots,t_{n-4}\}$, and the coefficients in the linear combination are iterated integrals on the moduli space $\mathfrak{M}_{0,n-1}$. A vector-space basis for $L_{\Sigma}$ is simply given by all hyperlogarithms, and so we can easily obtain a basis for $B_{n}$.

We end this discussion by noting that there is an alternative way to construct a basis for $B_{n}$. Since $\mathfrak{M}_{0,n} \simeq G(2,n)$, we can equally well describe $B_{n}$ as the algebra of all $A_{n-3}$ cluster polylogarithms~\cite{Golden:2014xqa}, and a basis for all $A_{n-3}$ cluster polylogarithms was given in ref.~\cite{Parker:2015cia}.

\subsection{Single-valued iterated integrals on $\mathfrak{M}_{0,n}$}
\label{sec:SV_MPLs}

We have seen that scattering amplitudes in MRK can be expressed through single-valued iterated integrals on $\mathfrak{M}_{0,n}$. In this section we present different ways to construct these functions. 

{In the previous section we have seen that the algebra $B_n$ of ordinary, not necessarily single-valued, iterated integrals on $\mathfrak{M}_{0,n}$ factors through hyperlogarithm algebras. For this reason, it is instructive to first understand the construction of single-valued hyperlogarithms in detail before generalising to iterated integrals on $\mathfrak{M}_{0,n}$. We therefore start by reviewing the construction of ref.~\cite{BrownSVHPLs,BrownSVMPLs}, where single-valued solutions to a certain Picard-Fuchs equation in one variable are constructed. The resulting functions are single-valued analogues of the hyperlogarithms. The strategy to construct single-valued iterated integrals on $\mathfrak{M}_{0,n}$ is then to generalise the results of ref.~\cite{BrownSVHPLs,BrownSVMPLs} from the Picard-Fuchs equation in the hyperlogarithm case to the KZ equation~\eqref{KZ} on $\mathfrak{M}_{0,n}$. In both cases the construction of single-valued functions preserves the algebra structure. Hence, since iterated integrals on $\mathfrak{M}_{0,n}$ can always be written in terms of hyperlogarithms, as a byproduct we find that both constructions give consistent results, and every single-valued iterated integral on $\mathfrak{M}_{0,n}$ can be written in terms of single-valued hyperlogarithms.
Finally, inspired by ref.~\cite{Brown:2013gia,Brown_Notes}, we present a purely algebraic way to define single-valued analogues of hyperlogarithms.

\subsubsection{Single-valued hyperlogarithms from Picard-Fuchs equations}
\label{eq:SV_hyperlogs_KZ}
In this section we discuss the construction of single-valued hyperlogarithms, following ref.~\cite{BrownSVHPLs,BrownSVMPLs}. We review the construction in detail, because the techniques introduced in the hyperlogarithms case can be extended to the KZ equation on $\mathfrak{M}_{0,n}$. This will be done in the next section. 

Consider a set of complex constants, $\Sigma=\{\sigma_1,\ldots,\sigma_n\}$. We denote the shuffle algebra of all hyperlogarithms with singularities in $\Sigma$ by $L_{\Sigma}$, see eq.~\eqref{eq:MPL_def}.
In the following it will be useful to take a more abstract viewpoint. Let $X=\{x_1,\ldots,x_n\}$, and $\mathbb{C}\langle X\rangle$ is the complex vector space generated by all words with letters from $X$, and the multiplication is the shuffle product.
Following ref.~\cite{BrownSVHPLs,BrownSVMPLs}, we start by defining the \emph{universal algebra of hyperlogarithms} $\mathcal{HL}_{\Sigma}$ as the algebra $\mathbb{C}\langle X\rangle$, but with rational functions (with poles at most at $z\in \Sigma$) as coefficients, and a derivation $\partial$ which acts on rational functions as $\partial/\partial z$ and on words as
\beq
\partial(x_iw) = \frac{1}{z-\sigma_i}\,w\,.
\eeq
$\mathcal{HL}_{\Sigma}$ is an abstract algebra (with a derivation) which has exactly the same properties as the algebra $L_\Sigma$ (shuffle and differentiation). 
A \emph{realisation of }$\mathcal{HL}_{\Sigma}$ is then an algebra morphism $\rho:\mathcal{HL}_{\Sigma}\to A$ that preserves the derivative. In particular, the hyperlogarithms $L_\Sigma$ are a realisation of $\mathcal{HL}_{\Sigma}$. We will in the following refer to this realisation as the \emph{standard realisation},
\beq
\rho_G:\cH\cL_\Sigma\to L_{\Sigma},\,\, w\mapsto G(w;z)\,,
\eeq
where we made a slight abuse of notation: if the word is $w=x_{i_1}\ldots x_{i_{|w|}}$, with $|w|$ the length of the word $w$, then we define $G(w;z)\equiv G(\sigma_{i_1},\ldots,\sigma_{i_{|w|}};z)$.
In the following also the dual of $\mathcal{HL}_{\Sigma}$ will be important. The dual of $\mathbb{C}\langle X\rangle$ is the space $\mathbb{C}\langle\langle X\rangle\rangle$ of formal power series in words. 

Next, consider a realisation $\rho$ of $\mathcal{HL}_{\Sigma}$, and consider the generating series
\beq
F_{\rho} = \sum_{w\in X^*}\rho(w)\,w\,.
\eeq
This generating series is a general series of the form (\ref{FnseriesforKZ}).
In particular, $F_{\rho}$ satisfies a Picard-Fuchs-type equation 
\beq\label{eq:KZ}
\frac{\partial}{\partial z}F_\rho = \sum_{i=1}^n\frac{x_i}{z-\sigma_i}\,F_{\rho}\,.
\eeq
Conversely, every solution to eq.~\eqref{eq:KZ} gives rise to a realisation of $\mathcal{HL}_{\Sigma}$.}
Moreover, one can check that if $F'$ is any other solution to eq.~\eqref{eq:KZ}, then there is a constant series $T\in \mathbb{C}\langle\langle X\rangle\rangle$ such that $F'=F_{\rho}T$. Finally, it is easy to see that we can find $n$ solutions $L^{\sigma_i}$ such that close to the singularity $z\sim \sigma_i$ we have $L^{\sigma_i}\sim (z-\sigma_i)^{x_i}$. Hence, we conclude that there are constant series $Z_{ij}\in \mathbb{C}\langle\langle X\rangle\rangle$ such that $L^{\sigma_j}=L^{\sigma_i}Z_{ij}$. We refer to the $Z_{ij}$ as \emph{associators} {and they play the same role as the Drinfeld associator in eq.~\eqref{eq:transport}. Note that $Z_{ij}Z_{jk}=Z_{ik}$, and 
their inverses are $Z_{ij}^{-1} = Z_{ji}$. The associators can be obtained as the shuffle-regularised values of $F_{\rho}$ at the singular points~\cite{BrownSVMPLs}. 
In particular, if $\Sigma=\{0,1\}$, we have $Z_{01} = \Phi(x_0,x_1)$, where $\Phi$ is the Drinfeld associator of eq.~\eqref{eq:Drinfeld}.}

From now on we will always identify one of the singular points with 0, say $\sigma_0=0$ (this is always possible using $SL(2,\mathbb{C})$ transformations). We define
\beq
L(z) \equiv L^{\sigma_0}(z) = \sum_{w\in X^*} \rho_G(w)\,w = \sum_{w\in X^*} G(w;z)\,w\,,
\eeq
and we write the associators as $Z^{\sigma_i} \equiv Z_{0i}$, so that $L^{\sigma_i}(z)=L(z)\,Z^{\sigma_i}$.

Due to the presence of the singularities in eq.~\eqref{eq:KZ}, the solutions to eq.~\eqref{eq:KZ} will in general have discontinuities with branch points at $z=\sigma_i$. We denote by $\cM_{\sigma_i}F_{\rho}$ the function obtained by analytically continuing $F_{\rho}$ along a small loop encircling $z=\sigma_i$. It is easy to check that $\cM_{\sigma_i}F_{\rho}$ is still a solution to eq.~\eqref{eq:KZ}, and so there must be a constant series $M_{\sigma_i}$ such that $\cM_{\sigma_i}F_{\rho} = F_{\rho}M_{\sigma_i}$. We obtain
\beq
\cM_{\sigma_j}L^{\sigma_j} = L^{\sigma_i}\,e^{2\pi i x_j} {\rm~~and~~} \cM_{\sigma_j}L = L\,(Z^{\sigma_j})^{-1}\,e^{2\pi i x_j}\,Z^{\sigma_j}\,.
\eeq
{Note the similarity of the previous equations with the monodromies of iterated integrals on $\mathfrak{M}_{0,n}$, eq.~\eqref{simplemonods} and~\eqref{eq:Mono_1}.}


One of the main results of ref.~\cite{BrownSVMPLs} is that there is always a solution to eq.~\eqref{eq:KZ} with a prescribed monodromy. More precisely, if we are given $n$ (grouplike) elements $A_1,\ldots,A_n\in \mathbb{C}\langle\langle X\rangle\rangle$, then there is always a realisation $\rho:\cH\cL_{\Sigma}\to L_{\Sigma}\overline{L}_{\Sigma}$, with $\overline{L}_{\Sigma}$ the complex conjugate of $L_{\Sigma}$, such that $\cM_{\sigma_i}F_{\rho}=F_{\rho}A_i$. There are two particular cases of this:
\begin{enumerate}
\item If we choose $A_k=(Z^{\sigma_k})^{-1}\,e^{2\pi i x_k}\,Z^{\sigma_k}$, $\forall1\le k\le n$, we obtain the standard realisation $\rho_G$.
\item We may also choose $A_k=1$, $\forall1\le k\le n$ and we see that there is a realisation of $\cH\cL_{\Sigma}$ that is single-valued. 
\end{enumerate}

It is possible to write down a generating function for the single-valued realisation, similar to the generating series $L(z)$ for the standard realisation.
Consider two alphabets $X=\{x_1,\ldots,x_n\}$ and $Y=\{y_1,\ldots,y_n\}$ and two generating functions
\beq
L_X(z) = \sum_{w\in X^*}G(w;z)\,w {\rm~~and~~} \widetilde{\overline{L}}_Y(\bar{z}) = \sum_{w\in Y^*}G(\bar{w};\bar{z})\,\tilde{w}\,,
\eeq
where $\tilde{}$ denotes the operation of reversal of words, i.e., $\tilde{w}$ is the word $w$ with all its letters in reversed order. We define
\beq
\cL_X(z) \equiv L_X(z)\widetilde{\overline{L}}_Y(\bar{z})\,.
\eeq
$\cL_X(z)$ is a solution of eq.~\eqref{eq:KZ}, because
\beq
\frac{\partial}{\partial z}\cL_X(z) = \frac{\partial}{\partial z}L_X(z)\widetilde{\overline{L}}_Y(\bar{z})
= \sum_{i=1}^n\frac{x_i}{z-\sigma_i}\,L_X(z)\widetilde{\overline{L}}_Y(\bar{z})
=\sum_{i=1}^n\frac{x_i}{z-\sigma_i}\,\cL(z)\,.
\eeq
The monodromies of $\cL_X(z)$ are
\beq
\cM_{\sigma_k}\cL_X(z) = L_X(z)\,M_{\sigma_k}\widetilde{\overline{L}}_Y(\bar{z})\,,
\eeq
with 
\beq
M_{\sigma_k} = Z^{\sigma_k}(X)^{-1}\,e^{2\pi ix_k}\,Z^{\sigma_k}(X)\,\widetilde{\overline{Z}}{}^{\sigma_k}(Y)\,e^{-2\pi iy_k}\,\widetilde{\overline{Z}}{}^{\sigma_k}(Y)^{-1}\,.
\eeq
Obviously $\cL_X(z)$ is single-valued if $M_{\sigma_k}=1$, $\forall 1\le k\le n$, which implies that the letters in $X$ and $Y$ are not independent. Infinitesimally, this condition becomes
\beq\label{eq:SV_condition}
\widetilde{\overline{Z}}{}^{\sigma_k}(Y)\,y_k\,\widetilde{\overline{Z}}{}^{\sigma_k}(Y)^{-1} = Z^{\sigma_k}(X)^{-1}\,x_k\,Z^{\sigma_k}(X)\,.
\eeq
This equation can be solved perturbatively in the length of the words. 
While solving the constraints~\eqref{eq:SV_condition} is conceptually very algorithmic, explicitly constructing the solutions order-by-order in the length of the words quickly becomes very tedious. Below we construct an explicit solution to the constraints~\eqref{eq:SV_condition}. Before doing so, it will be useful to introduce some more notation that will be useful to write down an explicit solution to the constraints~\eqref{eq:SV_condition}.

Let us for now assume that we have obtained the solution to eq.~\eqref{eq:SV_condition} to any desired order. If we substitute this solution into the definition of $\cL_X$, we obtain in this way the single-valued realisation $\rho_{SV}$ of $\cH\cL_{\Sigma}$,
\beq
\cL_X(z) \equiv \sum_{w\in X^*}\rho_{SV}(w)\,w\,.
\eeq
Some comments are in order:
First, in the case where $\Sigma=\{0,1\}$, the single-valued realisation corresponds to the single-valued harmonic polylogarithms of ref.~\cite{BrownSVHPLs}.
Second, the solution for $Y$ in terms of $X$ is unique order-by-order in the length of the words, and so the single-valued realisation is unique. Finally,  $\rho_{SV}$ and $\rho_G$ are really just two different realisations of the same abstract algebra $\cH\cL_{\Sigma}$ (just like an abstract group may have different representations). In other words, the standard and single-valued realisations have \emph{exactly} the same properties. In particular, they form a shuffle algebra and have the same behaviour under holomorphic differentiation. We stress, however, that the behaviour under anti-holomorphic differentiation is less obvious. We will address this issue in Section~\ref{sec:period}.

In the following we write $\cG(w;z)\equiv\cG_w(z)\equiv \rho_{SV}(w)$. Let us denote the algebra generated by the functions $\cG(w;z)$ by $L_{\Sigma}^{SV}$. We can define a linear map
\begin{eqnarray}
\bfs_{\Sigma}  : L_{\Sigma} \to L_{\Sigma}^{SV}\,,\, G(w;z) \mapsto \cG(w;z)\,.
\end{eqnarray}
As $L_{\Sigma}$ and $L_{\Sigma}^{SV}$ are just different realisations of $\cH\cL_{\Sigma}$, $\bfs_{\Sigma}$ preserves the multiplication,
\beq
\bfs_{\Sigma}(a\cdot b) = \bfs_{\Sigma}(a)\cdot \bfs_{\Sigma}(b)\,.
\eeq


Let us now return to the question of how we can explicitly solve the constraints~\eqref{eq:SV_condition}. In the following we denote by $\cZ$ the algebra of multiple zeta values, and by $\cZ^{SV}$ the algebra of their single-valued analogues. It is possible to construct explicitly a homomorphism $\bfs_{\zeta}:\cZ\to \cZ^{SV}$~\cite{Brown:2013gia}.
One can check that if $G(w,z)\in L_{\Sigma}$, then its regularised version at some singularity reduces to a linear combination of hyperlogarithms with one singularity less and with MZVs as coefficients. In other words, we have
\beq
\textrm{Reg}_{z=\sigma_k}G(w;z) \in \cZ\otimes L_{\Sigma_k}\,,
\eeq
with $\Sigma_k={\Sigma/\{\sigma_k\}}$ and where we see elements of $L_{\Sigma_k}$ as functions of $\sigma_k$.
We denote by $\hat{\bfs}_{\Sigma}$ the natural map 
\beq
\hat{\bfs}_{\Sigma}\equiv \bfs_{\zeta}\otimes \bfs_{\Sigma}: \cZ\otimes L_{\Sigma} \to  \cZ^{SV}\otimes L_{\Sigma}^{SV}\,.
\eeq
The single-valued maps preserve the multiplication, and so they commute with shuffle-regularisation,
\beq
\hat{\bfs}_{\Sigma_k}\left[\textrm{Reg}_{z=\sigma_k}G(w;z)\right] = \textrm{Reg}_{z=\sigma_k}\left[\bfs_{\Sigma}(G(w;z))\right] = \textrm{Reg}_{z=\sigma_k}\cG(w;z)\,.
\eeq

Using these definitions, we can explicitly solve the constraints~\eqref{eq:SV_condition}. We claim that the solution for $y_k$ to eq.~\eqref{eq:SV_condition} is obtained by conjugating $x_k$ by the single-valued analogue of the associator $Z^{\sigma_k}(X)$,
\beq\label{eq:y_sol}
y_k = \hat{\bfs}_{\Sigma_{k}}\left(Z^{\sigma_k}(X)\right)^{-1}\,x_k\,\hat{\bfs}_{\Sigma_{_k}}\left(Z^{\sigma_k}(X)\right)\,.
\eeq
Equation~\eqref{eq:y_sol} states that the single-valued analogues of the hyperlogarithms, and thus the map $\bfs_{\Sigma}$, can be constructed recursively in the number of singularities $\sigma_k$. The recursion starts with the single-valued harmonic polylogarithms, in which case the associator involves only MZVs, and so the map $\hat{\bfs}_{\Sigma_k}$ reduces to $\bfs_{\zeta}$.

In order to see why eq.~\eqref{eq:y_sol} holds, let us cast the constraints~\eqref{eq:SV_condition} in the form
\beq\bsp\label{eq:formal_sol}
y_k &\,= \widetilde{\overline{Z}}{}^{\sigma_k}(Y_X)^{-1}\,Z^{\sigma_k}(X)^{-1}\,x_k\,Z^{\sigma_k}(X)\,\widetilde{\overline{Z}}{}^{\sigma_k}(Y_X)\\
&\, = \left(Z^{\sigma_k}(X)\,\widetilde{\overline{Z}}{}^{\sigma_k}(Y_X)\right)^{-1}\,x_k\,\left(Z^{\sigma_k}(X)\,\widetilde{\overline{Z}}{}^{\sigma_k}(Y_X)\right)\,,
\esp\eeq
where we write $Y_X$ instead of $Y$ in order to indicate that this identity holds on the solution to the constraints~\eqref{eq:SV_condition}, i.e., we have inserted the solution to eq.~\eqref{eq:SV_condition} into the right-hand side of eq.~\eqref{eq:formal_sol}. The right-hand-side then only depends on the letters $x_i$, and so eq.~\eqref{eq:formal_sol} is a formal solution to the constraints. Comparing eq.~\eqref{eq:formal_sol} and eq.~\eqref{eq:y_sol}, we need to show that
\beq
\hat{\bfs}_{\Sigma_k}\left(Z^{\sigma_k}(X)\right) = Z^{\sigma_k}(X)\,\widetilde{\overline{Z}}{}^{\sigma_k}(Y_X)\,.
\eeq
This relation is in fact a generalisation of the relation between Deligne's and  Drinfeld's associators in the case where $\Sigma=\{0,1\}$~\cite{Brown:2013gia}.
We start from the fact that the associator can be written as the shuffle regularised version of $L_X(z)$ at the point $z=\sigma_k$,
\beq
Z^{\sigma_k}(X) = \textrm{Reg}_{z=\sigma_k}L_X(z){\rm~~and~~}\overline{Z}^{\sigma_k}(Y) = \textrm{Reg}_{\bar{z}=\bar{\sigma}_k}\overline{L}_Y(\bar{z})\,.
\eeq
We assume that we have constructed all single-valued hyperlogarithms with a certain number of singularities, and we want to add one more singularity, i.e., we assume that we know how to construct all the $\bfs_{\Sigma_k}$, and we want to construct $\bfs_{\Sigma}$. We have
\beq\bsp
\hat{\bfs}_{\Sigma_k}(Z^{\sigma_k}(X))&\, = \hat{\bfs}_{\Sigma_k}\left[\textrm{Reg}_{z=\sigma_k}L_X(z)\right]\\
&\,= \textrm{Reg}_{z=\sigma_k}\left[\bfs_{\Sigma}(L_X(z))\right]\\
 &\,= \textrm{Reg}_{z=\sigma_k}\left[L_X(z)\widetilde{\overline{L}}_{Y_X}(\bar{z})\right]\\
 &\, = \left[\textrm{Reg}_{z=\sigma_k}L_X(z)\right]\,\left[\textrm{Reg}_{\bar{z}=\bar{\sigma}_k}\widetilde{\overline{L}}_{Y_X}(\bar{z})\right]\,.
\esp\eeq
The first factor immediately gives an associator, $\textrm{Reg}_{z=\sigma_k}L_X(z) = Z^{\sigma_k}(X)$. The second factor also gives an associator. Indeed, the solution $Y_X$ is independent of $z$, and so the shuffle regularisation does not act on the letters $y_i$ and it commutes with the reversal of words. Hence, $\textrm{Reg}_{\bar{z}=\bar{\sigma}_k}\widetilde{\overline{L}}_{Y_X}(\bar{z})=\widetilde{\overline{Z}}{}^{\sigma_k}(Y_X)$, which finishes the proof.
Note that at the same time we have proved the identity
\beq
\hat{\bfs}_{\Sigma_k}(Z^{\sigma_k}(X)) = \textrm{Reg}_{z=\sigma_k}\cL_X(z)\,.
\eeq
In practise, it is often easier to use this last relation to construct the single-valued associators than constructing the standard associators and then acting with the single-valued map. 

\subsubsection{Single-valued iterated integrals from a differential equation on $\mathfrak{M}_{0,n}$}
In this section we extend the construction of the previous section to iterated integrals on $\mathfrak{M}_{0,n}$. Our goal will be to find single-valued solutions to the KZ equation~\eqref{KZ} on $\mathfrak{M}_{0,n}$
To construct a generating series of single-valued polylogarithms on $\mathfrak{M}_{0,n}$ we first take two copies of the infinitesimal pure braid generators, $\delta_{ij}$ and $\delta_{ij}'$, obeying the same relations (\ref{dihedralbraidrels}). We then have two copies of the KZ equation, one based on the $\delta_{ij}$ with dihedral coordinates $v_{ij}$ and one based on the $\delta_{ij}'$ with coordinates $\bar{v}_{ij}$ respectively. We then choose a vertex $V$ and pick a solution $L_V$, a formal series in the $\delta_{ij}$, and the corresponding $\bar{L}_V'$, a series in the $\delta_{ij}'$.

Now we consider
\be
\mathcal{L}_V = L_V \tilde{\bar{L}}_V'\,,
\ee
where the tilde operation means reversing all words in the $\delta_{ij}'$ generators. Now if we impose that the $\bar{v}_{ij}$ coordinates are the complex conjugates of the $v_{ij}$ then we obtain the following results for the general monodromy of $\mathcal{L}_V$,
\be
\mathcal{M}_{ij} \mathcal{L}_V = L_V \,\Phi_{V,V'} \,e^{2 \pi i \delta_{ij}}\, \Phi_{V',V}\, \tilde{\Phi}'_{V',V} \,e^{-2 \pi i \delta_{ij}'}\, \tilde{\Phi}'_{V,V'}\, \tilde{\bar{L}}_V'\,,
\ee
where $V'$ is again some vertex which sits on the singularity denoted by the pair $\{i,j\}$.

Single valuedness means imposing that there is no such monodromy and hence we have
\be
\Phi_{V,V'} e^{2 \pi i \delta_{ij}} \Phi_{V',V} \tilde{\Phi}'_{V',V} e^{-2 \pi i \delta_{ij}'} \tilde{\Phi}'_{V,V'} = 1.
\label{delta'rels}
\ee
for all $\{i,j\}$. This provides exactly the right number of conditions to eliminate the $\delta_{ij}'$ in terms of the $\delta_{ij}$. For the $\{i,j\}$ in the triangulation $T_V$ the relation (\ref{delta'rels}) reduces simply to
\be
\delta_{ij}' = \delta_{ij} \,, \qquad \{i,j\} \in T_V\,.
\ee
for the other $\{i,j\}$ it becomes
\be
\delta'_{ij} = \delta_{ij} + \text{ higher order terms involving MZVs,} \qquad \{i,j\} \notin T_V\,.
\ee
The series $\mathcal{L}_V$ then becomes a generating series for all single-valued multiple polylogarithms on $\mathfrak{M}_{0,n}$. Since it is real-valued inside the polytope $\mathfrak{M}_{0,n}(\mathbb{R})$ and it has no monodromy, it is real valued everywhere in $\mathfrak{M}_{0,n}$.
Expanding $\mathcal{L}_V$ over all words in the $\delta_{ij}$ modulo the pure braid relations (\ref{dihedralbraidrels}) gives all the single-valued multiple polylogarithms as coefficients,
\be
\mathcal{L}_V = \sum_w w \mathcal{L}_{V,w}\,.
\ee
The advantage of this construction is that it shows that the construction of single-valued polylogarithms does not rely directly on the decomposition into hyperlogarithms.
However, since both the generating series of single-valued hyperlogarithms and of single-valued iterated intgerals on $\mathfrak{M}_{0,n}$ satisfy the same holomorphic differential equation as their non-single-valued analogues, we can repeat the very same argument given at the end of Section~\ref{sec:iterated_integrals} to conclude that the algebra $B_n^{SV}$ of single-valued iterated integrals on $\mathfrak{M}_{0,n}$ has a recursive structure similar to the recursive structure of $B_n$ (see eq.~\eqref{eq:B_n_rec}). In particular, working with a specific choice of simplicial coordinates, we have
\beq
B_{n}^{SV} \simeq B_{n-1}^{SV}\otimes_{\mathbb{Q}} L_{\{0,1,t_1,\ldots,t_{n-4}\}}^{SV}\,,
\eeq
i.e., for a given choice of simplicial coordinates, every single-valued iterated integral on $\mathfrak{M}_{0,n}$ can be written as a linear combination of products of single-valued hyperlogarithms.

\subsubsection{A purely algebraic approach to single-valued hyperlogarithms}
\label{sec:period}

So far we have seen that it is possible to define single-valued multiple polylogarithms, and thus single-valued iterated integrals on $\mathfrak{M}_{0,n}$, as solutions to a certain Picard-Fuchs equation with trivial monodromy. While the construction of these solutions is algorithmic, it can be desirable to have a purely combinatorial definition of single-valued multiple polylogarithms that does not require any reference to any differential equation. Inspired by ref.~\cite{Brown:2013gia,Brown_Notes} we present in this section such a purely combinatorial definition. We introduce a map $\bfs$ that only relies on the Hopf algebra structure on multiple polylogarithms, and we show that the resulting functions satisfy the Picard-Fuchs equation of Section~\ref{eq:SV_hyperlogs_KZ} and are single-valued. Hence, they must be identical to the single-valued functions of Section~\ref{eq:SV_hyperlogs_KZ}.

The results of this section make heavy use of the Hopf algebra structure underlying hyperlogarithms~\cite{Goncharov:2001}. We therefore start this section with a short review on this topic. 
In this context it is convenient to use an alternative notation for the hyperlogarithms, where we allow for more general lower integration limits. Following ref.~\cite{Goncharov:2001}, we define
\beq
I(a_0;a_n,\ldots,a_1;z) = \int_{a_0}^z\frac{dt}{t-a_n}\,I(a_0;a_{n-1},\ldots,a_1;t)\,.
\eeq
It is easy to see that $G(a_1,\ldots,a_n;z) = I(0;a_n,\ldots,a_1;z)$. 
In ref.~\cite{Goncharov:2001} it was shown that these functions form a Hopf algebra. In order to write down the coproduct it is useful to introduce some more notation.
%
%
Consider the word $\vec a\equiv a_n\ldots a_1$ and write $I(0;\vec a;z)=I(0;a_n,\ldots,a_1;z)$. If $\vec b$ and $\vec c$ are sub-words of $\vec a$,  then we denote by $I_{\vec b}(0;\vec c;z)$ the iterated integral obtained by integrating over a contour which encircles the points in $\vec b$, in that order. We can always express $I_{\vec b}(0;\vec c;z)$ in terms of hyperlogarithms:
\begin{enumerate}
\item If $\vec b=\emptyset$ is the empty word, then $I_{\emptyset}(0;\vec c;z) = I(0;\vec c;z)$ is just a hyperlogarithm.
\item $I_{\vec b}(0;\vec c;z) = 0$, unless $\vec b$ is a sub-word of $\vec c$, because otherwise we take residues at points where there are no singularities in the integrand.
\item $I_{\vec b}(0;\vec b;z) = (2\pi i)^{|\vec b|}$, where $|\vec b|$ denotes the length of the word $\vec b$.
\item If $\vec b = c_{i_1}\ldots c_{i_m}$ is a proper sub-word of $\vec c = c_1\ldots c_k$, we have
\beq\bsp
I&_{\vec b}(0;\vec c;z) = (2\pi i)^{|\vec b|}I(0;c_1,\ldots,c_{i_1-1};c_{i_1})\ldots
I(c_{i_m};c_{i_m+1},\ldots,c_{k};z)\,.
\esp\eeq
\end{enumerate}
Using this notation, the coproduct on hyperlogarithms can be written in the following compact form~\cite{Goncharov:2001},
\beq\label{eq:coproduct_def}
\Delta(I(a_0;\vec a;z)) = \sum_{\emptyset\subseteq \vec b\subseteq \vec a} I(a_0;\vec b;z) \otimes \left[(2\pi i)^{-|\vec b|}\,I_{\vec b}(a_0;\vec a;z)\right]\,,
\eeq
where the sum runs over all subwords of $\vec a$. The resulting bialgebra is graded by the weight (i.e., the length of the words) and elements of weight zero are precisely the rational numbers. It follows that there is a unique way to promote this bialgebra to a Hopf algebra where the antipode is determined recursively in the weight through the condition
\beq\label{eq:antipode_def}
\mu(S\otimes \textrm{id})\Delta(G(\vec a;z))) = \mu(\textrm{id}\otimes S)\Delta(G(\vec a;z))) = 0\,, \textrm {if } |\vec a|\ge 1\,,
\eeq
where $\mu(a\otimes b) = a\cdot b$ denotes the multiplication.
For example, if $|\vec a|=1$, eq.~\eqref{eq:antipode_def} takes the form
\beq
S(G(a;z)) + G(a;z) = 0\,,
\eeq
and so the antipode of hyperlogarithms of weight one is uniquely determined. Similarly, for hyperlogarithms of weight two eq.~\eqref{eq:antipode_def} reduces to
\beq\bsp\label{eq:G2_antipode_def}
0  &\,=S(G(a,b;z))+S(G(a;z))\, G(b;a)+S(G(b;z))\, [G(a;z)- G(a;b)] + G(a,b;z)\\
&\,=S(G(a,b;z))-G(a;z)\, G(b;a)-G(b;z)\, [G(a;z)- G(a;b)] + G(a,b;z)\,,
\esp\eeq
where the last step follows upon inserting the result for the antipode of hyperlogarithms of weight one. We see that $S(G(a,b;z))$ is uniquely determined by eq.~\eqref{eq:G2_antipode_def}. Repeating the same construction for higher weights, we see that the antipode is uniquely determined in a recursive manner through the coproduct. The antipode is an involution, $S^2=\textrm{id}$, and it preserves the product and the coproduct,
\beq
S(a\cdot b) = S(b)\cdot S(a) {\rm~~and~~} \Delta S = (S\otimes S)\tau\Delta\,,
\eeq
with $\tau(a\otimes b)=b\otimes a$.

Let us now show how we can use the coproduct and the antipode to define single-valued hyperlogarithms. We use the notation of Section~\ref{eq:SV_hyperlogs_KZ} and we write $L_{\Sigma}$ for the shuffle algebra of all hyperlogarithms with singularities in $\Sigma$, $\overline{L}_{\Sigma}$ is its complex conjugate and $L_{\Sigma}\overline{L}_{\Sigma}\simeq L_{\Sigma}\otimes\overline{L}_{\Sigma}$. Note that each of these algebras is actually a Hopf algebra for the coproduct in eq.~\eqref{eq:coproduct_def}. Let us define a map
\beq\label{eq:s_tilde_def}
\tilde{S}:L_{\Sigma} \to \overline{L}_{\Sigma}\,;\quad G(\vec a;z) \mapsto (-1)^{|\vec a|}\,\overline{S}(G(\vec a;z))\,,
\eeq
where $\overline{S}$ denotes the complex conjugate of the antipode. It is easy to check that $\tilde{S}$ inherits many properties from $S$. In particular, it is an involution and it satisfies
\beq
\tilde{S}(a\cdot b) = \tilde{S}(b)\cdot \tilde{S}(a) {\rm~~and~~} \Delta \tilde{S} = (\tilde{S}\otimes \tilde{S})\tau\Delta\,.
\eeq
Unlike the antipode, $\tilde{S}$ does not satisfy eq.~\eqref{eq:antipode_def}. Rather, the equivalent equation for $\tilde{S}$ defines the single-valued map (see also ref.~\cite{Brown:2013gia}),
\beq\label{eq:SV_def}
\bfs = \mu(\tilde{S}\otimes\textrm{id})\Delta\,,
\eeq
i.e., we claim that $\cG(\vec a; z) = \bfs(G(\vec a;z))$ is the single-valued analogue of $G(\vec a;z)$. Before proving single-valuedness, let us discuss some of the properties of the single-valued map $\bfs$. Unlike the definition of the map $\bfs_{\Sigma}$ of Section~\ref{eq:SV_hyperlogs_KZ}, the definition~\eqref{eq:SV_def} is purely combinatorial and does not depend on the set of singularities. It is easy to see that $\bfs$ is $\mathbb{Q}$-linear and that it preserves the multiplication (see Appendix~\ref{app:hopf} for a detailed proof), 
\beq
\bfs(a\cdot b) = \bfs(a)\cdot \bfs(b)\,.
\eeq
 We stress at this point that $\bfs$ is only linear with respect to rational numbers. In particular, this means that $\bfs$ may act non-trivially on non-algebraic periods. Indeed, we have~\cite{Brown:2013gia}
\beq\label{eq:sv_zeta}
\bfs(i\pi) = 0 {\rm~~and~~} \bfs(\zeta_{n}) = 2\zeta_n\,, \, \textrm{ for }n\textrm{ odd}\,.
\eeq

Let us denote by $L_{\Sigma}^{SV}\subset L_{\Sigma}\overline{L}_{\Sigma}$ the image of $L_{\Sigma}$ under the map $\bfs$. We use suggestively the same notation as for the shuffle algebra of single-valued hyperlogarithms from Section~\ref{eq:SV_hyperlogs_KZ}.
While $L_{\Sigma}$ and $L_{\Sigma}\overline{L}_{\Sigma}$ are Hopf algebras, the algebra $L_{\Sigma}^{SV}$ is not a sub-Hopf algebra of $L_{\Sigma}\overline{L}_{\Sigma}$, but the Hopf algebra structure on $L_{\Sigma}\overline{L}_{\Sigma}$ turns $L_{\Sigma}^{SV}$ into a graded $L_{\Sigma}\overline{L}_{\Sigma}$-comodule, whose coaction agrees with the coproduct on $L_{\Sigma}\overline{L}_{\Sigma}$,
\beq
\Delta: L_{\Sigma}^{SV} \to L_{\Sigma}^{SV}\otimes L_{\Sigma}\overline{L}_{\Sigma}\,.
\eeq
In Appendix~\ref{app:hopf} we show that the coaction is given by
\beq\label{eq:SVMPL_coaction}
\Delta\bfs(I(a_0;\vec a;z)) = \sum_{\emptyset\subseteq \vec c\subseteq\vec b\subseteq \vec a}\bfs(I_{\vec c}(a_0;\vec b;z))\otimes\left[\tilde{S}(I(a_0;\vec c;z))\,I_{\vec b}(a_0;\vec a;z)\right]\,.
\eeq

Let us now show that $\cG(\vec a; z)=\bfs(G(\vec a;z))$ is single-valued. Following Section~\ref{eq:SV_hyperlogs_KZ} we denote by $\cM_{\sigma}\cG(\vec a; z)$ the result of analytically continuing $\cG(\vec a; z)$ along a small loop (oriented counterclockwise) encircling the singularity $\sigma\in\Sigma$ (and no other singularity). In order to show that $\cG(\vec a; z)$ is single-valued, we need show that
\beq
\cM_{\sigma}\cG(\vec a; z) = \cG(\vec a; z)\,,\quad \forall \sigma\in\Sigma\,,
\eeq
or equivalently 
\beq
\textrm{Disc}_{\sigma}\cG(\vec a; z) = 0\,,\quad \forall \sigma\in\Sigma\,,
\eeq
where the discontinuity operator is $\textrm{Disc}_{\sigma} = \cM_{\sigma} - \textrm{id}$. The proof that $\cG(\vec a; z)$ is single-valued proceeds by induction in the weight. If $|\vec a|=1$, we have
\beq
\cG(a;z) = G(a;z) + \tilde{S}(G(a;z)) = \log\left|1-\frac{z}{a}\right|^2\,,
\eeq
and this function is manifestly single-valued. Let us now assume that all functions $\cG$ are single-valued up to a certain weight $n$, and let us show that a function $\cG(\vec a;z)$ of weight $n+1$ is still single-valued. Since the discontinuity operator only acts in the first factor of the coproduct, $\Delta\textrm{Disc}_{\sigma} = (\textrm{Disc}_{\sigma}\otimes \textrm{id})\Delta$, the graded comodule structure of $L_{\Sigma}^{SV}$ implies that
\beq
\Delta\textrm{Disc}_{\sigma}(\cG(\vec a;z)) = (\textrm{Disc}_{\sigma}\otimes \textrm{id})\Delta(\cG(a;z)) =  \textrm{Disc}_{\sigma}\cG(\vec a;z)\otimes 1\,.
\eeq
From eq.~\eqref{eq:antipode_def} we obtain
\beq
0 = \mu(\textrm{id}\otimes S)\Delta\textrm{Disc}_{\sigma}(\cG(\vec a;z)) = \textrm{Disc}_{\sigma}(\cG(\vec a;z))\,\cdot S(1) = \textrm{Disc}_{\sigma}(\cG(\vec a;z))\,,
\eeq
and so $\cG(\vec a;z)$ is single-valued.

So far we have shown that $\bfs$ respects the multiplication and that the resulting functions are single-valued. We now show that the functions $\cG(\vec a;z)$ agree with the single-valued realisation $\rho_{SV}$ of $\cH\cL_{\Sigma}$, see Section~\ref{eq:SV_hyperlogs_KZ}. In order to see this we need to prove that the single-valued map commutes with holomorphic differentiation,
\beq
\partial_z\,\bfs = \bfs\,\partial_z\,,
\eeq
This follows immediately from the fact that derivatives only act in the second factor of the coproduct, $\Delta\partial_z = (\textrm{id}\otimes \partial_z)\Delta$. We obtain,
\beq
\bfs\,\partial_z = \mu(\tilde{S}\otimes \textrm{id})\Delta\partial_z = \mu(\tilde{S}\otimes\partial_z)\Delta = \partial_z\bfs - \mu(\partial_z\tilde{S}\otimes\textrm{id})\Delta\,,
\eeq
where the last step follows from the Leibniz rule, $\partial_z\mu = \mu(\partial_z\otimes\textrm{id} + \textrm{id}\otimes\partial_z)$. The claim then follows upon noting that $\tilde{S}(G(\vec a;z))$ is always anti-holomorphic, and so $\partial_z\tilde{S}=0$.
Hence, we have shown that $\cG(a,\vec b;z)$ and $G(a,\vec b;z)$ behave in the same way under holomorphic differentiation,
\beq
\partial_z\,\cG(a,\vec b;z) = \frac{1}{z-a}\,\cG(\vec b;z)\,.
\eeq
Moreover, it is easy to check that $\cG(a,\vec b;z)$ vanishes as $z\to0$, and so the functions $\cG(\vec a;z)$ coincide with the single-valued realisation of $\cH\cL_{\Sigma}$ defined in Section~\ref{eq:SV_hyperlogs_KZ}. Note, however, that the single-valued map does not commute with anti-holomorphic derivatives, $\bar{\partial}_{{z}}\bfs \neq (\bfs{\partial}_z)^*$.

Single-valued hyperlogarithms naturally have both anti-holomorphic and holomorphic parts. Hence, they carry a natural action of complex conjugation. We can again decompose a complex conjugated single-valued hyperlogarithm into standard single-valued hyperlogarithms,
\beq\label{eq:cmplx_conj}
\cG({\vec a};\zb) = \sum_{\vec b}c_{\vec a,\vec b}\,\cG({\vec b};z)\,.
\eeq
Note that the fact that complex conjugation acts non-trivially on single-valued hyperlogarithms (in the sense that the complex conjugate of an single-valued hyperlogarithm is a linear combination of single-valued hyperlogarithms) is at the origin of why $\bfs$ does not commute with anti-holomorphic derivatives. In Appendix~\ref{app:hopf} we show that the action of complex conjugation on single-valued hyperlogarithms is encoded in the map $\tilde{S}$. If $\bar{\bfs}$ denotes the complex conjugate of $\bfs$, we find
\beq\label{eq:bfscc}
\bar{\bfs} = \bfs\,\tilde{S}\,.
\eeq
%
As an example, we have
\beq
\cG(\bar{a},\bar{b};\zb) = \bar{\bfs}(G({a},{b};z)) = \cG({b,a};z)+\cG(b;a)\,\cG(a;z)-\cG(a;b)\, \cG(b;z)\,.
\eeq
In the same way, we can also easily compute anti-holomorphic derivatives, because we can reduce the anti-holomorphic derivative to a holomorphic one via the map $\tilde{S}$. For example, we find,
\beq
\bar{\partial}_{{z}}\cG(a,b;z) = \frac{1}{\zb-\bar{a}}\,\cG(b;a)+\frac{1}{\zb-\bar{b}}(\cG(a;z)-\cG(a;b))\,.
\eeq

We conclude this section by commenting on functional equations for single-valued hyperlogarithms. We can of course obtain functional equations by expressing single-valued hyperlogarithms in terms of ordinary hyperlogarithms, and then applying functional equations to the latter. 
There is, however, a simpler way  to obtain functional equations for single-valued hyperlogarithms: assume we are given a relation between ordinary hyperlogarithms. We can then act with $\bfs$ on it, and we obtain a relation among single-valued hyperlogarithms. Since the action of $\bfs$ is, essentially, to replace $G$ by $\cG$, we conclude that single-valued hyperlogarithms satisfy the same identities as ordinary hyperlogarithms. Note that eq.~\eqref{eq:sv_zeta} is crucial for this to work. Let us consider an example to see how this works: we start from the following relation among ordinary hyperlogarithms of weight three (valid on some branch for the logarithm),
\beq\bsp
G\left(0,1,1;\frac{1}{z}\right)&\, = 
-G(0,0,0;z)+G(0,0,1;z)+G(0,1,0;z)-G(0,1,1;z)\\
&\,+i \pi \, [G(0,0;z)-G(0,1;z)]+
\frac{\pi^2}{2}\, G(0;z)+\zeta_3-\frac{i \pi ^3}{6}\,.
\esp\eeq
We can act on both sides with $\bfs$, and we obtain,
\beq\bsp
\cG\left(0,1,1;\frac{1}{z}\right)&\, = 
-\cG(0,0,0;z)+\cG(0,0,1;z)+\cG(0,1,0;z)-\cG(0,1,1;z)+2\zeta_3\,.
\esp\eeq
This is indeed a valid identity among single-valued hyperlogarithms. We stress the importance of eq.~\eqref{eq:sv_zeta} in order for this to be true.



\section{MHV amplitudes in MRK}
\label{sec:mhv}
\subsection{An invitation: the six-point MHV amplitude}
In this section we apply the machinery of single-valued iterated integrals on $\mathfrak{M}_{0,N-2}$ of the previous section to the computation of scattering amplitudes in MRK to LLA. We start by reviewing the six-point MHV amplitude in MRK, and we generalise the discussion to more external legs and other helicity configurations in subsequent sections. Most of the techniques introduced in this paper apply also beyond LLA, and we will comment on how to extend the results of this paper beyond LLA in Section~\ref{sec:conclusion}.

Traditionally, scattering amplitudes in MRK are computed by closing the integration contour in the Fourier-Mellin representation of the amplitude, eq.~\eqref{eq:MRK_conjecture}, and taking residues at the poles of the integrand~\cite{Lipatov:2010ad,Bartels:2011ge,Dixon:2012yy,Pennington:2012zj,Drummond:2015jea,Broedel:2015nfp}. In the case of the six-point amplitude, the resulting multiple sums can all be performed in terms of polylogarithms using standard techniques~\cite{Remiddi:1999ew,Vermaseren:1998uu,Moch:2001zr,Weinzierl:2002hv,Moch:2005uc}. For amplitudes with more external legs, performing the multiple sums, however, soon becomes prohibitive.

The goal of this section is to introduce a new way to compute, or rather to circumvent, the Fourier-Mellin transform of eq.~\eqref{eq:MRK_conjecture}. The main idea is to use the convolution theorem~\eqref{eq:conv_thm} and to perform the computation directly in $z$-space, rather than evaluating the Fourier-Mellin transform explicitly. While in itself this idea is not new, performing the convolution integral~\eqref{eq:conv_def} requires the evaluation of some integral over the whole complex plane, which seems a daunting task. We show that the fact that amplitudes in MRK are single-valued functions on $\mathfrak{M}_{0,N-2}$ reduces the computation to a simple application of Stokes' theorem.

In order to illustrate our method, we apply it in this section to the six-point MHV amplitude. While the results of this section are not new (see for example ref.~\cite{Dixon:2012yy,Pennington:2012zj}), we use them to show all the steps that enter the computation.
We start from eq.~\eqref{eq:g_n_def}, and we obtain a recursion for the coefficients to LLA
\beq
{g}_{++}^{(l)}(z) = -\frac{1}{2}\,\cF\left[\chi^+({\nu,n})\,E_{\nu n}^{l}\,\chi^-({\nu,n})\right] = {g}_{++}^{(l-1)}(z)\ast\cF[E_{\nu n}]\,.
\eeq
We see that increasing the number of loops is equivalent to convoluting the lower loop result with the Fourier-Mellin transform of the BFKL eigenvalue.
In order to start the recursion, we need to know ${g}_{++}^{(l)}(z)$ analytically for some value of $l$. This can easily be achieved by performing explicitly the Fourier-Mellin transform for $l=1$ or $l=2$, cf., e.g., ref.~\cite{Dixon:2012yy},
\beq\bsp
\cF\left[\chi^+({\nu,n})\chi^-({\nu,n})\right] &\,= \cG_1(z)-\frac{1}{2}\,\cG_0(z) \,,\\
\cF\left[\chi^+({\nu,n})\,E_{\nu n}\,\chi^-({\nu,n})\right] &\,=  \frac{1}{2}\cG_{0,1}(z) + \frac{1}{2}\cG_{1,0}(z) - \cG_{1,1}(z)\,,
\esp\eeq
where we use the notation $\cG_{a_1,\ldots,a_w}(z)\equiv \cG(a_1,\ldots,a_w;z)$.
We also need the Fourier-Mellin transform of the LO BFKL eigenvalue, which can easily be obtained by noting that the functions $\chi^{\pm}({\nu,n})$ have a very simple interpretation in terms of Fourier-Mellin transforms: they are related to derivatives in $z$-space,
\beq\bsp\label{eq:FM_derivative}
z\,\partial_z\cF\left[\chi^+({\nu,n})\, F(\nu,n)\right] &\,= \cF\left[F(\nu,n)\right]\,.
\esp\eeq
A similar relation holds when replacing $z$ by $\bar{z}$ and $\chi^+$ by $\chi^-$.
The Fourier-Mellin transform of the LO BFKL eigenvalue is then given by
\beq
\cE(z) \equiv \cF\left[E_{\nu n}\right] = z\,\zb\,\partial_z\,\bar{\partial}_{z}\cF\left[\chi^+({\nu,n})\, E_{\nu n}\,\chi^-({\nu,n})\right]  = -\frac{z+\zb}{2\,|1-z|^2}\,.
\eeq

 Next we discuss how we can evaluate the convolution integral.
We assume for now that in the multi-Regge limit we can express the amplitude to all loop orders in terms of single-valued hyperlogarithms (This will be proven later in Section~\ref{sec:MPL_proof}). In ref.~\cite{Schnetz:2013hqa} it was shown that convolution integrals of this type can be computed using residues.
To see how this works, consider a function $f(z)$ that consists of single-valued hyperlogarithms and rational functions with singularities at $z=a_i$ and $z=\infty$. Close to any of these singularities, $f$ can be expanded into a series of the form
\beq\bsp
f(z) &\,= \sum_{k,m,n}\,c^{a_i}_{k,m,n}\,\log^k\left|1-\frac{z}{a_i}\right|^2\,(z-a_i)^m\,(\zb-\bar{a}_i)^n\,, \quad z\to a_i\,,\\
f(z) &\,= \sum_{k,m,n}\,c^{\infty}_{k,m,n}\,\log^k\frac{1}{|z|^2}\,\frac{1}{z^m}\,\frac{1}{\zb^n}\,, \quad z\to \infty\,.
\esp\eeq
The \emph{holomorphic residue} of $f$ at the point $z=a$ is then defined as the coefficient of the simple holomorphic pole without logarithmic singularities,
\beq
\textrm{Res}_{z=a}f(z) \equiv c^{a}_{0,-1,0}\,.
\eeq
In ref.~\cite{Schnetz:2013hqa} it was shown that the integral of $f$ over the whole complex plane, if it exists, can be computed in terms of its holomorphic residues. More precisely, if $F$ is an anti-holomorphic primitive of $f$, $\bar{\partial}_zF=f$, then
\beq
\int \frac{d^2z}{\pi}\,f(z) = \textrm{Res}_{z=\infty}F(z) - \sum_i\textrm{Res}_{z=a_i}F(z)\,.
\eeq
This result is essentially an application of Stokes' theorem to the punctured complex plane.
Note that the anti-holomorphic primitive is only defined up to an arbitrary holomorphic function. It was shown in ref.~\cite{BrownSVMPLs} that every single-valued hyperlogarithm has a single-valued primitive, and the sum of residues is independent on the choice of the primitive~\cite{Schnetz:2013hqa}. It is clear that we can repeat the previous argument by reversing the roles of holomorphic and anti-holomorphic functions.

As a pedagogical example, let us illustrate how this works on the two-loop remainder function in MRK. Using the convolution theorem, we can write
\beq\bsp
\cF\left[\chi^+({\nu,n})\, E_{\nu n}\,\chi^-({\nu,n})\right] &\,= \cF\left[\chi^+({\nu,n})\chi^-({\nu,n})\right]\ast \cE(z)\\
&\,=\int\frac{d^2w}{\pi }\,\underbrace{\left[\frac{1}{2}\,\cG_0(w)-\cG_1(w)\right]\,\frac{\wb z+w\zb}{2\,|w|^2\,|w-z|^2}}_{=f(w)}\,.
\esp\eeq
First, we need to compute the anti-holomorphic primitive. Since
\beq
\cG_0(w) = \cG_0(\wb) {\rm~~and~~} \cG_1(w) = \cG_1(\wb)\,,
\eeq
and single-valued hyperlogarithms satisfy the same (holomorphic) differential equations as their non-single-valued analogues, we obtain
\beq\bsp
F(w) &\,= \int d\wb\, f(w) = \frac{1}{2w\,(w-z)}\,\int d\wb \left[\frac{1}{2}\,\cG_0(\wb)-\cG_1(\wb)\right]\,\frac{\wb z+w\zb}{\wb\,(\wb-\zb)}\\
&\,= \frac{1}{4 (w-z)}\left[2 \cG_{0,z}(w)-4 \cG_{1,z}(w)-\cG_{0,0}(w)+2 \cG_{1,0}(w)-4 \cG_1(w) \cG_0(z)\right.\\
&\,\left.\qquad+4 \cG_1(w) \cG_1(z)+2 \cG_0(z) \cG_z(w)-4 \cG_1(z) \cG_z(w)\right]\\
&\,+\frac{1}{4 w}\left[-\cG_{0,z}(w)+2 \cG_{1,z}(w)+2 \cG_1(w) \cG_0(z)-2 \cG_1(w) \cG_1(z)-\cG_0(z) \cG_z(w)\right.\\
&\,\qquad\left.+2 \cG_1(z) \cG_z(w)\right]
\,.
\esp\eeq
We anticipate, however, that for higher weights the relation between $\cG_{\vec a}(w)$ and $\cG_{\vec a}(\wb)$ will not be as easy, but we have
\beq
\cG_{\vec a}(w) = \sum_{\vec b}c_{\vec a,\vec b}\,\cG_{\vec b}(\wb)\,.
\eeq

We see that $F(w)$ has potential poles at $w=0$, $w=z$ and $w=\infty$. It is easy to check that the residue at $w=0$ vanishes (because single-valued hyperlogarithms either vanish at $w=0$, or they have logarithmic singularities). The residue at $w=z$ is easy to obtain,
\beq\bsp
\textrm{Res}_{w=z}F(w) &\,= -\frac{1}{4} \cG_{0,0}(z)-\cG_{0,1}(z)-\frac{1}{2} \cG_{1,0}(z)+2 \cG_{1,1}(z)-\cG_{1,z}(z)\\
&\, = -\frac{1}{4} \cG_{0,0}(z)-\frac{1}{2} \cG_{1,0}(z)+\cG_{1,1}(z)\,,
\esp\eeq
where the last step follows from the identity
\beq
\cG_{1,z}(z) = \cG_{1,1}(z)-\cG_{0,1}(z)\,.
\eeq
Finally, the residue at infinity is obtained by letting $w=1/u$ (and including the corresponding Jacobian) and expanding the result around $u=0$. Note that we obtain single-valued hyperlogarithms of the form $\cG(\vec a;1/u)$. In order to proceed, we need inversion relations for single-valued hyperlogarithms, which may be obtained from the inversion relations for ordinary hyperlogarithms and then acting with the single-valued map $\bfs$. We find
\beq
\textrm{Res}_{w=\infty}F(w) = \frac{1}{2}  \cG_{0,1}(z)-\frac{1}{4} \cG_{0,0}(z)\,.
\eeq
Hence,
\beq\bsp
\cF\left[\chi^+({\nu,n})\, E_{\nu n}\,\chi^-({\nu,n})\right] &\,=  \textrm{Res}_{w=\infty}F(w)  - \textrm{Res}_{w=z}F(w) \\
&\,=\frac{1}{2} \cG_{0,1}(z)+\frac{1}{2} \cG_{1,0}(z)-\cG_{1,1}(z)\,,
\esp\eeq
which is indeed the correct result.
This construction is of course not restricted to two loops, but we can now start from the two-loop result we have just computed and obtain the three, and even higher, loop results by convoluting the two-loop result with the BFKL eigenvalue $\cE$.


\subsection{Higher-point MHV amplitudes and the factorisation theorem}
\label{sec:mhv_fac}
The six-point example from the previous section shows that we can bypass the evaluation of the Fourier-Mellin integrals and the multiple sums, and we can entirely work with convolutions and Stokes' theorem. This procedure can of course be extended to amplitudes with more external legs in a straightforward way. In particular, we obtain the recursion
\beq\bsp\label{eq:recursion_all_N}
{g}_{+\ldots+}^{(i_1,\ldots,i_k+1,\ldots,i_{N-5})}(z_1,\ldots,z_{N-5}) &\,= \cE(z_k)\ast {g}_{+\ldots+}^{(i_1,\ldots,i_{N-5})}(z_1,\ldots,z_{N-5})  \,.
\esp\eeq
In the previous equation the convolution is carried out only over the variable $z_k$, even though this is not manifest in the notation. In general, it will always be clear which is the variable that enters the convolution integral.
The starting point of the recursion is the two-loop MHV remainder function in MRK, which is known at LLA for an arbitrary number $N$ of external legs~\cite{Bartels:2011ge,Prygarin:2011gd}, cf. eq.~\eqref{eq:2-loop_MHV}. While a direct evaluation of the Fourier-Mellin transform in terms of multiple sums becomes prohibitive because the number of sums increases with the number of external legs, the recursion~\eqref{eq:recursion_all_N} requires the evaluation of a single convolution integral at every loop order, independently of the number of external legs. This is one of the key properties why the convolution integral combined with Stokes' theorem gives rise to an efficient algorithm to compute scattering amplitudes in MRK.

In practice, however, if we try to evaluate the convolution integral in terms of residues as we have done for the six-point MHV amplitude, then we have to face a conundrum: The convolution and the BFKL eigenvalue are naturally written in terms of the Fourier-Mellin coordinate $z_k$. The residues, however, are most easily computed in simplicial coordinates, where the poles in ${g}_{+\ldots+}^{(i_1,\ldots,i_{N-5})}$ manifest themselves simply as points where simplicial coordinates become equal to $0,1,\infty$ or to each other. In general, the change of variables from the Fourier-Mellin coordinates to simplicial coordinates is highly non-linear, and will introduce complicated Jacobians. In addition, it will obscure the simple form of the BFKL eigenvalue. This problem arises for the first time for seven points, because for the six-point amplitude the simplicial and Fourier-Mellin coordinate systems coincide.

In some cases it is possible to identify a set of coordinates which share the good properties of the simplicial and Fourier-Mellin coordinates even at higher points.
We have seen in Section~\ref{sec:coordinates} that there is always a (non unique) system of simplicial coordinates based at $z_k$ with the property that $t_k^{(k)}=z_k$. This system of coordinates has already some of the properties we want: it leaves the BFKL eigenvalue unchanged, because $t_k^{(k)}=z_k$. However, the change of coordinates may introduce a non-trivial Jacobian, because in general $z_{k-2}$, $z_{k-1}$ and $z_k$ will depend on the new integration variable $t_k^{(k)}$. There is, however, a special case where the Jacobian is trivial: If we perform a convolution with respect to $z_1$, and we change variables to simplicial coordinates based at $z_1$, only $z_1$ will depend on $t_1^{(1)}$, and so the Jacobian is 1. A similar argument can be made for $z_{N-5}$, using a slightly different set of simplicial coordinates. Alternatively, we know that we can exchange the roles of $z_1$ and $z_{N-5}$ using target-projectile symmetry, so it is sufficient to consider $z_1$. Hence, if we perform a convolution with respect to the first or last cross ratio $z_1$ or $z_{N-5}$, we can find a set of simplicial coordinates with the right properties: it leaves the BFKL eigenvalue unchanged, it has a unit Jacobian, and at the same time it exposes all the singularities of ${g}_{+\ldots+}^{(i_1,\ldots,i_{N-5})}$ in a very simple form. The algorithm to evaluate the recursion~\eqref{eq:recursion_all_N} for the first or last cross ratio is then clear: in order to perform the convolution over $z_1$, we change coordinates to the simplicial coordinates based at $z_1$, and we evaluate the integral in terms of residues. The change of coordinates requires the use of functional equations among single-valued polylogarithms, which can be obtained using the techniques described in Section~\ref{sec:SV_MPLs}.

While the previous considerations answer the question of how to perform convolutions with respect to the first or last cross ratio, we still need to discuss the remaining cases. In the following, we argue that all amplitudes can be constructed by convoluting over the first or last cross ratio only. We only discuss from now on the case of $z_1$; the case of $z_{N-5}$ is similar by target-projectile symmetry. The proof of this claim relies on a certain factorisation theorem which we present in the following.

In order to state the factorisation theorem, it is useful to introduce the following graphical representation for the perturbative coefficients,
\beq\label{eq:coefficient_notation}
   \begin{fmffile}{coefficient_notation}
 {g}_{h_1\ldots h_{N-4}}^{(i_1,\ldots,i_{N-5})}(\rho_1,\ldots,\rho_{N-5}) =
                  \parbox{20mm}{ \begin{fmfgraph*}(150,70)
                        \notation
                    \end{fmfgraph*}}
 \end{fmffile}
\eeq
We work with the simplicial MRK coordinates $\rho_k$ defined in Section~\ref{sec:coordinates}. Every face of the dual graph is associated with a point ${\bf x}_k$ (cf. Fig~\ref{fig:dual_coordinates}), and we work in a coordinate patch where $({\bf x}_1,{\bf x}_2,{\bf x}_{N-2})=(1,0,\infty)$. Every outgoing line is labeled by its helicity $h_k$. In addition, to every face we do not only associate its coordinate $\rho_k$ but also the index $i_k$. In the following we will not show the points $0$, $1$ and $\infty$ explicitly. Using this graphical representation of the perturbative coefficients the factorisation theorem takes the simple form
\begin{eqnarray}\label{eq:fac_thm}
   \begin{fmffile}{theorem}
 \parbox{30mm}{
           \begin{fmfgraph*}(60,90)
              \thmleft
               \end{fmfgraph*}
               }
               \parbox{10mm}{$\quad=$}
        \parbox{20mm}{\begin{fmfgraph*}(60,90)
          \thmright
               \end{fmfgraph*}}
 \end{fmffile}
\end{eqnarray}
In other words, whenever the graph representing a perturbative coefficient contains a face with index $i_b=0$ and the lines adjacent to this face have the same helicity, then this perturbative coefficient is equal to the coefficient where this face has been deleted. We stress that the factorisation theorem holds for arbitrary helicity configurations and is not restricted to MHV amplitudes. In Section~\ref{sec:fac_proof} we will prove eq.~\eqref{eq:fac_thm} in the special case of MHV amplitudes, and we defer the proof in the non-MHV case to Section~\ref{sec:nmhv}.

Before turning to the proof of the factorisation theorem, we discuss its implications for MHV amplitudes.
In that particular case, the factorisation theorem implies that we can drop all the faces labeled by a zero,
\beq\label{eq:MHV_factorisation}
{g}_{+\ldots+}^{(0,\ldots,0,i_{a_1},0,\ldots,0,i_{a_2},0,\ldots,0,i_{a_k},0,\ldots,0)}(\rho_1,\ldots,\rho_{N-5}) = {g}_{+\ldots+}^{(i_{a_1},i_{a_2},\ldots ,i_{a_k})}(\rho_{i_{a_1}},\rho_{i_{a_2}},\ldots,\rho_{i_{a_k}})\,.
\eeq
Let us discuss the implications of this result. First, eq.~\eqref{eq:MHV_factorisation} implies that we can compute all MHV amplitudes by performing convolutions over the left-most variable $z_1$.
 Indeed, assume that we know all MHV amplitude with up to $N$ legs. Then we can write
\beq\bsp
{g}_{+\ldots+}^{(1,i_2,\ldots,i_{N-5})}(\rho_1,\ldots,\rho_{N-5}) &\,= \cE(z_1)\ast {g}_{+\ldots+}^{(0,i_2,\ldots,i_{N-5})}(\rho_1,\ldots,\rho_{N-5})\\
&\, = \cE(z_1)\ast {g}_{+\ldots+}^{(i_2,\ldots,i_{N-5})}(\rho_2,\ldots,\rho_{N-5})\,,\\
{g}_{+\ldots+}^{(2,i_2,\ldots,i_{N-5})}(\rho_1,\ldots,\rho_{N-5}) &\,= \cE(z_1)\ast {g}_{+\ldots+}^{(1,i_2,\ldots,i_{N-5})}(\rho_1,\ldots,\rho_{N-5})\\
&\, = \cE(z_1)\ast \cE(z_1)\ast {g}_{+\ldots+}^{(i_2,\ldots,i_{N-5})}(\rho_2,\ldots,\rho_{N-5})\,,
\esp\eeq
and so on. The amplitude in the left-hand side is a known lower-point amplitude. At the beginning of this section we have argued that we can always easily perform convolutions over $z_1$ by going to simplicial coordinates based at $z_1$, because the change of variable has unit Jacobian and leaves the BFKL eigenvalue unchanged. Hence, we conclude that every MHV amplitude can be recursively constructed in this way, and we have thus obtained an efficient algorithm to compute scattering amplitudes in MRK.

Next, let us discuss the implications of the factorisation theorem for the structure of MHV amplitudes. Indeed, since the sum of all indices is related to the loop number, we see that for a fixed number of loops there is a maximal number of non-zero indices, and so there is only a finite number of different perturbative coefficients at every loop order.
This generalises the factorisation observed for the two-loop MHV amplitude in MRK to LLA~\cite{Bartels:2011ge,Prygarin:2011gd,Bargheer:2015djt}. Indeed, if all indices are zero except for one, say $i_a$, then eq.~\eqref{eq:MHV_factorisation} reduces to
\beq
{g}_{+\ldots+}^{(0,\ldots,0,i_a,0,\ldots,0)}(\rho_1,\ldots,\rho_{N-5}) = {g}_{++}^{(i_a)}(\rho_{a})\,,
\eeq
and so at two loops the amplitude completely factorises, in agreement with ref.~\cite{Bartels:2011ge,Prygarin:2011gd,Bargheer:2015djt},
\beq
\cR_{+\ldots+}^{(2)} = \sum_{1\le i \le N-5}\log\tau_i\,{g}_{++}^{(1)}(\rho_i)\,.
\eeq
As anticipated in ref.~\cite{Bartels:2011ge}, the amplitude does no longer factorise completely beyond two loops. However,  we find that at every loop order only a finite number of different functions appear. For example, at three-loop order at most two indices are non-zero, and so we have
\beq\label{eq:factorisation_3loop}
\cR_{+\ldots+}^{(3)} = \frac{1}{2}\sum_{1\le i \le N-5}\log^2\tau_i\,{g}_{++}^{(2)}(\rho_i) + \sum_{1\le i<j\le N-5}\log\tau_i\,\log\tau_j\,{g}^{(1,1)}_{+++}(\rho_i,\rho_j)\,.
\eeq
The only new function that appears at three loops that is not determined by the six-point amplitude is $g^{(1,1)}_{++}$, which is determined by the three-loop seven-point MHV amplitude. At four loops we have
\beq\bsp
\cR_{+\ldots+}^{(4)} &\,= \frac{1}{6}\sum_{1\le i \le N-5}\log^3\tau_i\,{g}_{++}^{(3)}(\rho_i)\\
&\, + \frac{1}{2}\sum_{1\le i<j\le N-5}\left[\log^2\tau_i\,\log\tau_j\,{g}^{(2,1)}_{+++}(\rho_i,\rho_j) + \log\tau_i\,\log^2\tau_j\,{g}^{(1,2)}_{+++}(\rho_i,\rho_j)\right]\\
&\,+\sum_{1\le i<j<k\le N-5}\log\tau_i\,\log\tau_j\,\log\tau_k\,{g}_{++++}^{(1,1,1)}(\rho_i,\rho_j,\rho_k)\,.
\esp\eeq
The four-loop answer is determined for any number of external legs by the six, seven and eight-point amplitudes through four loops. Similar equations can be obtained for higher-loop amplitudes. In general, at $L$ loops $\cR_{+\ldots+}^{(L)}$ is determined for any number of legs by the MHV amplitudes involving up to $(L+4)$ external legs. In particular, the five-loop MHV amplitude is given by
\beq\bsp
\cR_{+\ldots+}^{(5)} &\,= \frac{1}{24}\sum_{1\le i \le N-5}\log^4\tau_i\,{g}_{++}^{(4)}(\rho_i)\\
&\, +\sum_{1\le i<j\le N-5}\Big[ \frac{1}{6}\log^3\tau_i\,\log\tau_j\,{g}^{(3,1)}_{+++}(\rho_i,\rho_j) + \frac{1}{6} \log\tau_i\,\log^3\tau_j\,{g}^{(1,3)}_{++}(\rho_i,\rho_j)\\
&\,\qquad+ \frac{1}{4}  \log^2\tau_i\,\log^2\tau_j\,{g}^{(2,2)}_{++}(\rho_i,\rho_j)\Big]\\
&\,+\frac{1}{2}\sum_{1\le i<j<k\le N-5}\Big[\log^2\tau_i\,\log\tau_j\,\log\tau_k\,{g}_{++++}^{(2,1,1)}(\rho_i,\rho_j,\rho_k)\\
&\,\qquad+\log\tau_i\,\log^2\tau_j\,\log\tau_k\,{g}_{++++}^{(1,2,1)}(\rho_i,\rho_j,\rho_k)\\
&\,\qquad+\log\tau_i\,\log\tau_j\,\log^2\tau_k\,{g}_{++++}^{(1,1,2)}(\rho_i,\rho_j,\rho_k)\Big]\\
&\,+\sum_{1\le i<j<k<l\le N-5}\log\tau_i\,\log\tau_j\,\log\tau_k\,\log\tau_l\,{g}_{+++++}^{(1,1,1,1)}(\rho_i,\rho_j,\rho_k,\rho_l)\,.
\esp\eeq

In Appendix~\ref{app:results} we present the complete set of perturbative coefficients sufficient to compute MHV amplitudes up to three loops for an arbitrary number of external legs. Results up to five loops are included in computer-readable form as ancillary material with the arXiv submission. Up to four loops and eight external legs, we have computed all the perturbative coefficients explicitly, including those that can be reduced to amplitudes with fewer points via eq.~\eqref{eq:MHV_factorisation}. In this way, we have explicitly checked that the factorisation~\eqref{eq:MHV_factorisation} holds. In addition, we have checked that our results have the correct soft limits and are consistent with complex conjugation and target-projectile symmetry.

\subsection{Proof of the factorisation theorem for MHV amplitudes}
\label{sec:fac_proof}
In this section we prove the factorisation theorem in eq.~\eqref{eq:fac_thm} for MHV amplitudes. The extension of the proof to non-MHV amplitudes will be given in Section~\ref{sec:nmhv}. Since the factorisation theorem is equivalent to eq.~\eqref{eq:MHV_factorisation} in the MHV case, we will prove that we can always drop all faces with a zero index in a MHV coefficient.
The proof will rely on two claims, which we will prove separately.

\begin{claim}\label{claim1}
We can always drop sequences of 0's at either end of the list of indices, i.e.,
\beq
{g}_{+\ldots+}^{(0,\ldots,0,i_k,\ldots,i_{N-5})}(\rho_1,\ldots,\rho_{N-5}) =
{g}_{+\ldots+}^{(i_k,\ldots,i_{N-5})}(\rho_k,\ldots,\rho_{N-5})\,,
\eeq
and a similar relation holds if the sequence of indices ends in a 0.
\end{claim}

{\bf Proof of Claim~\ref{claim1}.}
Target-projectile symmetry implies that it is sufficient to prove Claim~\ref{claim1} in the case where the sequence of indices starts with a 0. The proof relies on an analysis of the Fourier-Mellin integral, and the argument is in fact identical to the argument in Appendix~C of ref.~\cite{Bartels:2011ge}, where the particular case of the two-loop seven-point amplitude was obtained.
Let us start from the Fourier-Mellin integral for ${g}_{+\ldots+}^{(0,\ldots,0,i_k,\ldots,i_{N-5})}$, and let us concentrate on the terms that depend on $(\nu_1,n_1)$ and $(\nu_2,n_2)$,
\begin{align}\label{eq:proof_claim1}
&{g}_{+\ldots+}^{(0,\ldots,0,i_k,\ldots,i_{N-5})} \\
\nonumber& \,\,\,\,\,= \ldots\!\!\int\frac{d\nu_1}{2\pi}\sum_{n_1=-\infty}^{+\infty}z_1^{i\nu_1+n_1/2}\,\bar{z}_1^{i\nu_1+n_1/2}\,z_2^{i\nu_2+n_2/2}\,\bar{z}_2^{i\nu_2+n_2/2}\,\chi^+(\nu_1,n_1)\,C^+(\nu_1,n_1,\nu_2,n_2)\ldots\,,
\end{align}
where the dots indicate terms that are independent of $\nu_1$ and $n_1$. Our first goal is to show that the value of the integral is independent of $\rho_1$. From eq.~\eqref{eq:z_to_rho} we see that only $z_1$ and $z_2$ depend on $\rho_1$, so we do not need to consider the cross ratios $z_i$ with $i>2$. Due to the symmetry in $z_1\leftrightarrow \bar{z}_1$, it is sufficient to analyse the  holomorphic part, i.e., we let $\bar{z}_1\to0$, with $z_1$ held fixed. This corresponds to taking only the residue at $i\nu_1=n_1/2$ for $n_1>0$. After taking this residue, the sum over $n_1$ becomes trivial, and we are left with
\beq\bsp
{g}_{+\ldots+}^{(0,\ldots,0,i_k,\ldots,i_{N-5})} \to \ldots {(-1)}\,\chi^+(\nu_2,n_2)\,[\underbrace{(1-z_1)z_2}_{=\rho_2}]^{i\nu_2+n_2/2}\,\dots
\esp\eeq
We see that the integral does not depend on $\rho_1$. Note the appearance of an impact factor. If there is a second 0, we can iterate the procedure, with $z_1$ replaced by $\rho_2$. The result is
\beq\bsp
\ldots\,(-1)\,\chi^+(\nu_3,n_3)\,
[\underbrace{(1-\rho_2)z_3}_{=\rho_3}]^{i\nu_3+n_3/2}\,\dots
\esp\eeq
Continuing this process, we see that
\beq
{g}_{+\ldots+}^{(0,\ldots,0,i_k,\ldots,i_{N-5})}(\rho_1,\ldots,\rho_{N-5}) = f(\rho_k,\ldots,\rho_{N-5})\,,
\eeq
for some function $f$. We still need to show that $f$ is the lower-point perturbative coefficient. This follows from the fact that $f$ must have the correct soft limits. Indeed, we must have
\beq\bsp
{g}_{+\ldots+}^{(0,\ldots,0,i_k,\ldots,i_{N-5})}(\rho_2,\ldots,\rho_{N-5}) &\,= \lim_{\rho_1\to 0}{g}_{+\ldots+}^{(0,\ldots,0,i_k,\ldots,i_{N-5})}(\rho_1,\ldots,\rho_{N-5})\\
&\, = f(\rho_k,\ldots,\rho_{N-5})\,.
\esp\eeq
Similarly, we must have
\beq\bsp
{g}_{+\ldots+}^{(0,\ldots,0,i_k,\ldots,i_{N-5})}(\rho_3,\ldots,\rho_{N-5})&\, = \lim_{\rho_2\to 0}{g}_{+\ldots+}^{(0,\ldots,0,i_k,\ldots,i_{N-5})}(\rho_2,\ldots,\rho_{N-5})\\
&\, = f(\rho_k,\ldots,\rho_{N-5})\,.
\esp\eeq
Continuing this way, we arrive at
\beq\bsp
{g}_{+\ldots+}^{(i_k,\ldots,i_{N-5})}(\rho_k,\ldots,\rho_{N-5}) &\,= \lim_{\rho_{k-1}\to 0}{g}_{+\ldots+}^{(0,i_k,\ldots,i_{N-5})}(\rho_{k-1},\rho_k,\ldots,\rho_{N-5})\\
&\, = f(\rho_k,\ldots,\rho_{N-5})\,,
\esp\eeq
which finishes the proof.

\begin{claim}\label{claim2}
If $f(\rho_1, \rho_{i_1},\ldots,\rho_{i_k})$ depends on a subset of simplicial MRK coordinates, then the convolution with some function $g(z_1)$ will depend on the same subset with $\rho_1$ added, i.e., we have
\beq\label{eq:claim2}
g(z_1)\ast f(\rho_1,\rho_{i_1},\ldots,\rho_{i_k}) = F(\rho_1,\rho_{i_1},\ldots,\rho_{i_k})\,,
\eeq
for some function $F$.
\end{claim}
In the right-hand side of eq.~\eqref{eq:claim2}, the convolution acts on $z_1$, and the simplicial MRK coordinates $\rho_i$ should be interpreted as functions of the MRK coordinates $z_i$. The relation between the two sets of coordinates is given by eq.~\eqref{eq:z_to_rho}.

{\bf Proof of Claim~\ref{claim2}.} We proceed by changing variables to simplicial coordinates based at $z_1$. The relation between the two sets of coordinates is $z_1=t_1$ (we write $t_i$ instead of $t_i^{(1)}$) and,
\beq
\rho_1 = \frac{t_1}{t_{N-5}}\,,\quad \rho_2 = \frac{1-t_1}{1-t_{N-5}}\,, \quad \rho_i = \frac{t_{i-1}-t_1}{t_{i-1}-t_{N-5}}\,, \,2<i< N-5\,.
\eeq
We start by proving the claim in the case $i_1\neq 2$. In that case the convolution integral takes the form
\beq\bsp
g(z_1)&\ast f(\rho_1,\rho_{i_1}\ldots,\rho_{i_k})\\
&\, =\frac{1}{\pi} \int\frac{d^2\tau}{|\tau|^2}\,g\left(\frac{t_1}{\tau}\right)\,f\left(\frac{\tau}{t_{N-5}},\frac{t_{i_1-1}-\tau}{t_{i_1-1}-t_{N-5}},\ldots,\frac{t_{i_k-1}-\tau}{t_{i_k-1}-t_{N-5}}\right)\,.
\esp\eeq
Shifting the integration variable $\tau\to t_{N-5}\,\tau$ and writing $x_j = t_j/t_{N-5}$, we see that the integrand only depends on the $x_j$. Hence, there is a function $\widetilde{F}(x_1,x_{i_1},\ldots,x_{i_k})$ such that
\beq\bsp
g(z_1)\ast f(\rho_1,\rho_{i_1}\ldots,\rho_{i_k}) &\,= \frac{1}{\pi}\int\frac{d^2\tau}{|\tau|^2}\,g\left(\frac{x_1}{\tau}\right)\,f\left({\tau},\frac{x_{i_1-1}-\tau}{x_{i_1-1}-1},\ldots,\frac{x_{i_k-1}-\tau}{x_{i_k-1}-1}\right)\\
&\,\equiv \widetilde{F}(x_1,x_{i_1-1},\ldots,x_{i_k-1})\,.
\esp\eeq
Finally, we can change coordinates back from $t$'s to $\rho$'s. We find
\beq
x_1 = \frac{t_1}{t_{N-5}} = \rho_1\,,\quad x_{j} = \frac{t_{j}}{t_{N-5}} = \frac{\rho_1-\rho_{j+1}}{1-\rho_{j+1}}\,, \quad j\ge 2\,.
\eeq we see that no new $\rho$-coordinate is introduced, so the claim follows upon identifying $\widetilde{F}(x_1,x_{i_1-1},\ldots,x_{i_k-1})$ with $F(\rho_1,\rho_{i_1},\ldots,\rho_{i_k})$.

To complete the proof, we still need to investigate what happens when $i_1=2$. $\rho_2$ is not homogeneous in $t_{N-5}$, and so the function $\widetilde{F}$ will now not only depend on the ratios $x_i$, but also explicitly on $t_{N-5}$,
\beq
g(z_1)\ast f(\rho_1,\rho_2,\rho_{i_2}\ldots,\rho_{i_k}) =\widetilde{F}(x_1,t_{N-5},x_{i_2-1},\ldots,x_{i_k-1})\,.
\eeq
We know already that the $x_i$'s do not introduce any new $\rho$'s. Adding $t_{N-5}$ will only add $\rho_2$,
\beq
t_{N-5} = \frac{1-\rho_2}{\rho_1-\rho_2}\,,
\eeq
which was already present in the original function $f$. Hence the claim is proven.

\paragraph{Proof of the factorisation theorem for MHV amplitudes.}
The factorisation theorem for MHV amplitudes, eq.~\eqref{eq:MHV_factorisation}, now follows from Claims~\ref{claim1} and~\ref{claim2}. Assume that eq.~\eqref{eq:MHV_factorisation} holds for all perturbative MHV coefficients up to a certain number $N-1$ of legs, and let us show that it still holds for coefficients with one more leg. We denote the perturbative coefficient with one more leg by ${g}_{+\ldots+}^{(i_1,\ldots,i_{N-5})}(\rho_1,\ldots,\rho_{N-5})$ and we label the non-zero elements in $(i_2,\ldots,i_{N-5})$ by $i_{a_1},\ldots,i_{a_k}$, $2\le a_{j}\le N-5$. If $i_1=0$, then Claim~\ref{claim1} implies that we can drop the first index. The resulting function is an $(N-1)$-point amplitude, where eq.~\eqref{eq:MHV_factorisation} applies. So we have
\beq
{g}_{+\ldots+}^{(0,i_2,\ldots,i_{N_5})}(\rho_1,\ldots,\rho_{N-5}) = {g}_{+\ldots+}^{(i_2,\ldots,i_{N-5})}(\rho_2,\ldots,\rho_{N-5}) = {g}_{+\ldots+}^{(i_{a_1},\ldots,i_{a_k})}(\rho_{a_1},\ldots,\rho_{a_k})\,,
\eeq
in agreement with eq.~\eqref{eq:MHV_factorisation}. If $i_1\neq 0$, we write the amplitude as a convolution using the recursion~\eqref{eq:recursion_all_N}. For the sake of the example, consider the case $i_1=1$. We have,
\beq\bsp\label{eq:proof_eq}
{g}_{+\ldots+}^{(1,i_2,\ldots,i_{N_5})}(\rho_1,\ldots,\rho_{N-5})&\, = \cE(z_1)\ast {g}_{+\ldots+}^{(0,i_2,\ldots,i_{N_5})}(\rho_2,\ldots,\rho_{N-5})\\
&\, = \cE(z_1)\ast {g}_{+\ldots+}^{(i_{a_1},\ldots,i_{a_k})}(\rho_{a_1},\ldots,\rho_{a_k})\,,
 \esp\eeq
 where we have used the fact that we know that eq.~\eqref{eq:MHV_factorisation} holds for $i_1=0$. From Claim~\ref{claim2} we know that the result of the convolution will only depend on $(\rho_1,\rho_{a_1},\ldots,\rho_{a_k})$. The only thing left to show is that the function $F$ in Claim~\ref{claim2} is precisely the perturbative coefficient ${g}_{+\ldots+}^{(1,i_{a_1},\ldots,i_{a_k})}$. This follows immediately upon noting that the convolution integral used to compute ${g}_{+\ldots+}^{(1,i_{a_1},\ldots,i_{a_k})}$ is exactly the same as the one in eq.~\eqref{eq:proof_eq}, up to a relabelling of the variables. Repeating exactly the same argument for $i_1>1$, we see that eq.~\eqref{eq:MHV_factorisation} holds in general.

\section{Non-MHV amplitudes in MRK}
\label{sec:nmhv}

\subsection{Helicity-flip operations}
\label{sec:H_flip}

So far we have only considered MHV amplitudes. In this section we generalise all the results from the previous section to non-MHV amplitudes. In particular, we extend the factorisation theorem~\eqref{eq:fac_thm} to the non-MHV case. We start by introducing an additional concept before we are ready to prove the factorisation theorem for non-MHV amplitudes.

Let us start by analysing what happens if we start from an MHV amplitude and we flip the helicity on an impact factor. In Fourier-Mellin space, this amounts to replacing $\chi^+({\nu,n})$ by ${\chi}^-({\nu,n})$, 
\begin{eqnarray}\nonumber
\cF\left[\chi^+({\nu,n})\, F(\nu,n)\right] &\longrightarrow &\,\cF\left[{\chi}^-({\nu,n})\, F(\nu,n)\right] \\
&=& \cF\left[{\chi}^-({\nu,n})/{\chi}^+({\nu,n}) \right]\ast\cF\left[{\chi}^+({\nu,n})\,F(\nu,n)\right]\\
\nonumber&=&  \cF\left[\frac{i\nu+\frac{n}{2}}{i\nu-\frac{n}{2}}\right]\ast\cF\left[\chi^+({\nu,n})\,F(\nu,n)\right]\,.
\end{eqnarray}
We see that flipping the helicity on an impact factor amounts to convoluting with the universal \emph{helicity-flip kernel}
\beq
\cH(z) \equiv \cF\left[\frac{i \nu+\frac{n}{2}}{i \nu-\frac{n}{2}}\right]\,.
\eeq
The functional form of $\cH(z)$ can easily be obtained by performing explicitly the Fourier-Mellin transform. The integrand has only a simple pole at $i\nu=n/2$, and so we find
\beq
\cH(z) = \cH(1/z) = -\frac{z}{(1-z)^2}\,.
\eeq
Note that helicity-flip kernel is an involution, i.e., flipping the helicity twice on the same impact factor returns the original helicity configuration, and so
\beq\label{eq:H_involution}
\cH(z)\ast \cH(\bar{z}) =  \cF[1] = \pi \,\delta^{(2)}(1-z)\,.
\eeq

Similarly, if we flip the helicity on one of the central emission blocks and use eq.~\eqref{eq:helflip_C}, we obtain
\beq\bsp
\cF\Big[C^+(\nu,n,\mu,m)&F(\nu,n,\mu,m)\Big] \longrightarrow \cF\left[C^-(\nu,n,\mu,m) F(\nu,n,\mu,m)\right] \\
&\,=  \cF\left[\frac{C^-(\nu,n,\mu,m)}{C^+(\nu,n,\mu,m)}\right]\ast\cF\left[C^+(\nu,n,\mu,m)F(\nu,n,\mu,m)\right]\\
&\,=  \cF\left[\frac{\chi^+(\nu,n)\,\chi^-(\mu,m)}{\chi^-(\nu,n)\,\chi^+(\mu,m)}\right]\ast\cF\left[C^+(\nu,n,\mu,m)F(\nu,n,\mu,m)\right]\\
&\,=  \cH(\bar{z}_1)\ast\cH(z_2)\ast\cF\left[C^+(\nu,n,\mu,m)F(\nu,n,\mu,m)\right]\,.
\esp\eeq
We see that the flipping of the helicity on a central emission block is controlled by the same kernels as for the impact factor. As a consistency check, the helicity flip kernels allow us to show that MHV and $\overline{\textrm{MHV}}$ amplitudes are identical,
\beq\bsp
\cR_{-\ldots-}(z_1,\ldots,z_{N-5}) &\,= \cH(z_1)\ast \cR_{+-\ldots-}(z_1,\ldots,z_{N-5})\\
&\,= \cH(z_1)\ast \cH(\bar{z}_1)\ast \cH(z_2)\ast \cR_{++-\ldots-}(z_1,\ldots,z_{N-5})\\
&\,=  \cH(z_2)\ast \cR_{++-\ldots-}(z_1,\ldots,z_{N-5})\\
&\,=\ldots\\
&\,=  \cH(z_{N-5})\ast \cR_{+\ldots+-}(z_1,\ldots,z_{N-5})\\
&\,=  \cH(z_{N-5})\ast \cH(\bar{z}_{N-5})\ast \cR_{+\ldots+}(z_1,\ldots,z_{N-5})\\
&\,= \cR_{+\ldots+}(z_1,\ldots,z_{N-5})\,.
\esp\eeq

Let us conclude this section by making a comment about some classes of non-MHV amplitudes with a special property.
In ref.~\cite{Bartels:2011ge} it was argued that flipping the helicity on an impact factor to produce an NMHV amplitude from an MHV amplitude is equivalent to differentiating in the holomorphic variable and integrating in the anti-holomorphic one. Let us see how this arises from the helicity-flip kernel. We have
\beq\bsp\label{eq:NMHV_flip}
\cR_{-+\ldots+}(z_1,\ldots,z_{N-5}) &\,= \cH(z_1)\ast \cR_{+\ldots+}(z_1,\ldots,z_{N-5})\\
&\,=-\int\frac{d^2w}{\pi}\,\frac{z_1}{\bar{w}(w-z_1)^2}\,\cR_{+\ldots+}(w,z_2\ldots,z_{N-5})
\,.
\esp\eeq
We can evaluate eq.~\eqref{eq:NMHV_flip} in terms of residues. Let us denote by $F$ the anti-holomorphic primitive,
\beq\label{eq:NMHV_prim}
F(w,z_2\ldots,z_{N-5}) \equiv \int\frac{d\bar{w}}{\bar{w}}\,\cR_{++\ldots+}(w,z_2\ldots,z_{N-5})\,.
\eeq
Then $\cR_{-+\ldots+}$ is obtained by summing over all the holomorphic residues of $F$. As MHV amplitudes are a pure functions, they have no poles, and so $F$ has no poles either. Furthermore, it is easy to check that there is no pole at infinity, and so the only residue we need to take into account comes from the double pole at $w=z_1$ in eq.~\eqref{eq:NMHV_flip},
\beq\bsp\label{eq:NMHV_d_I}
\cR_{-+\ldots+}(z_1,\ldots,z_{N-5}) &\,= \textrm{Res}_{w=z_1}\frac{z_1\,F(w,z_2\ldots,z_{N-5}) }{(w-z_1)^2} \\
&\,= z_1\,\partial_{z_1}F(z_1,z_2\ldots,z_{N-5}) \\
&\,= z_1\partial_{z_1}\int\frac{d\bar{w}}{\bar{w}}\cR_{++\ldots+}(w,z_2\ldots,z_{N-5})\,.
\esp\eeq
We see that we recover the rule of ref.~\cite{Bartels:2011ge}, but with the differentiation and integration given in the reversed order. While this may look like a minor difference, it is crucial in order to get the complete result. In principle, we need to include a boundary condition when computing the anti-holomorphic primitive. However, if the operations of differentiation and integration are performed in the order shown in eq.~\eqref{eq:NMHV_d_I}, then no boundary condition is needed, because the residue is independent of the choice of the anti-holomorphic primitive. This is, however, not the case if the two operations are performed in the order given in ref.~\cite{Bartels:2011ge}, where one needs to include non-trivial boundary information already for six points.

It is of course tantalising to speculate if this simple rule generalises and all non-MHV amplitudes can be computed by this simple differentiation-integration rule without having to include any boundary information. It turns out that this is not the case, because in general the amplitude in the integrand of the convolution integral~\eqref{eq:NMHV_flip} is not a pure function, but may itself have additional poles whose residues need to be taken into account when performing the convolution with the helicity-flip kernel. An explicit counter-example to the simple differentiation-integration rule without boundary terms can be constructed from an eight-point NNMHV amplitude.

Although the simple rule does not hold in general, there are some special cases where it does apply. 
Besides the case of $\cR_{-+\ldots+}$ discussed above, we have identified the following special case in which we can apply the simple differentiation-integration rule without boundary terms: Consider an amplitude whose  helicity configuration is given by
\beq\label{eq:2_interfaces}
h_i=\left\{\begin{array}{ll}
-1\,,&\textrm{ if }a\le i\le b\,,\\
+1\,,&\textrm{ otherwise}\,.
\end{array}\right.
\eeq
This amplitude can be written as
\beq
\cR_{+\ldots+-\ldots-+\ldots+} = \cH(\bar{z}_{a-1})\ast \cH({z}_{b})\ast\cR_{+\ldots+} \,.
\eeq
Let us first discuss the convolution with $\cH({z}_{b})$. We can repeat exactly the same argument as for $\cR_{-+\ldots+}$ and we conclude that 
\beq
\cH({z}_{b})\ast\cR_{+\ldots+} = z_b\,\partial_{z_b}\int\frac{d\bar{z}_b}{\bar{z}_b}\,\cR_{+\ldots+}\,.
\eeq
Next we want to perform the convolution of this function with $\cH(\bar{z}_{a-1})$. The function $\cH({z}_{b})\ast\cR_{+\ldots+}$ will have poles, but all of them are holomorphic because they arise from computing the holomorphic derivative with respect to $z_b$. Hence, they do not give rise to any additional anti-holomorphic poles, and so we have
\beq\bsp
\cR_{+\ldots+-\ldots-+\ldots+} &\,= \bar{z}_{a-1}\,\bar{\partial}_{{z}_{a-1}}\int\frac{d{z}_{a-1}}{{z}_{a-1}}\,\left[\cH({z}_{b})\ast\cR_{+\ldots+}\right] \\
&\,=\bar{z}_{a-1}\,\bar{\partial}_{{z}_{a-1}}\int\frac{d{z}_{a-1}}{{z}_{a-1}}z_b\,\partial_{z_b}\int\frac{d\bar{z}_b}{\bar{z}_b}\,\cR_{+\ldots+}\,.
\esp\eeq
The previous case covers in particular all NMHV amplitudes. Hence, all six and seven-point amplitudes can be computed in this way.

\subsection{The factorisation theorem for non-MHV amplitudes}
\label{sec:nmhv_fac}
In this section we discuss the factorisation theorem~\eqref{eq:fac_thm} in the non-MHV case. In particular, we extend the  proof of Section~\ref{sec:fac_proof} to non-MHV amplitudes.

We proceed by induction in the number of indices. Let us assume that eq.~\eqref{eq:fac_thm} holds for up to $k$ indices, and let us show that the theorem still holds for $k+1$ indices. If the first two helicities are not the same, we can factor out a helicity flip operator, e.g.,
\beq
g_{-+h_{3}\ldots h_{k+2}}^{(i_1\ldots i_{k+1})}(\rho_{1},\ldots,\rho_{k+1}) = \cH(z_1)\ast g_{++h_{3}\ldots h_{k+2}}^{(i_1\ldots i_{k+1})}(\rho_{1},\ldots,\rho_{k+1})\,.
\eeq
Hence, it is enough to prove the theorem in the case where the first two helicities are the same, and Claim~\ref{claim2} implies that it will remain true in the case where the helicities are different. We therefore assume from now on that the first two helicities are the same.

We can use the recursion~\eqref{eq:recursion_all_N} with $z_k=z_1$ to reduce the value of $i_1$ to zero. If $i_1=0$, we can repeat the proof of Claim~\ref{claim1} and we conclude that we can delete the index $i_1=0$. Indeed, we see from eq.~\eqref{eq:proof_claim1} that the proof of Claim~\ref{claim1} only relies on the structure of the part of the Fourier-Mellin integral that depends on $(\nu_1,n_1)$ and the first two helicities, and so we can repeat the proof of Claim~\ref{claim1} independently of the helicity structure of the rest of the amplitude. After using Claim~\ref{claim1} to delete the index $i_1=0$, the number of indices has decreased by one, and so the factorisation theorem~\eqref{eq:fac_thm} applies by induction hypothesis.
To complete the proof, we need to perform the convolution integrals that were introduced to reduce the value of $i_1$. By Claim~\ref{claim2}, this does not introduce any new variables except for $\rho_1$, and so the factorisation theorem is proven.

Let us make a comment about the difference in the factorisation in the MHV and non-MHV cases.
In the MHV-case the factorisation theorem is equivalent to deleting vanishing indices, see eq.~\eqref{eq:MHV_factorisation}. 
This simple rule is no longer true for non-MHV amplitudes. Consider the seven-point two-loop NMHV amplitude $\cR_{-++}^{(2)}$. We can write
\beq
\cR_{-++}^{(2)} = \cH(z_1)\ast \cR_{+++}^{(2)} = \log\tau_1\,\cH(z_1)\ast\mathfrak{g}_{++}^{(1)}(\rho_1) +  \log\tau_2\,\cH(z_1)\ast\mathfrak{g}_{++}^{(1)}(\rho_2)\,.
\eeq
It is easy to check that the first term behaves as expected,
\beq
\cH(z_1)\ast\mathfrak{g}_{++}^{(1)}(\rho_1) = \mathfrak{g}_{-+}^{(1)}(\rho_1)\,.
\eeq
The second term, however, also depends on $\rho_2$, and so Claim~\ref{claim2} implies that this term should be a function of both $\rho_1$ and $\rho_2$. By explicit computation, one establishes that this is indeed the case, and so we obtain a new non-MHV building block with a vanishing index,
\beq \label{eq:7pt_NMHV_1}
\cR_{-++}^{(2)} = \log\tau_1\,\mathfrak{g}_{-+}^{(1)}(\rho_1) + \log\tau_2\,\mathfrak{g}_{-++}^{(0,1)}(\rho_1,\rho_2)\,.
\eeq
Hence, the simple factorisation observed for MHV amplitudes, eq.~\eqref{eq:MHV_factorisation}, is no longer valid for non-MHV amplitudes.

As a consequence, unlike for MHV amplitudes,
the number of building blocks is no longer finite at each loop order in the non-MHV case.
As  eq.~\eqref{eq:MHV_factorisation} is no longer valid for non-MHV amplitudes, the number of different coefficients is no longer bounded. In particular, unless there is another mechanism at work that waits yet to be uncovered, there should be an infinite tower of different non-MHV coefficients already at two loops, because the factorisation theorem does not allow us to reduce the coefficients corresponding to alternating helicities to simpler functions.

We have computed explicitly all non-MHV amplitudes up to eight external legs and four loops. Analytic results for the independent helicity configurations are shown in Appendix~\ref{app:results} up to three loops for six and seven external legs and up to two loops for eight external legs. Results up to four loops and eight points are included as ancillary material in computer-readable form with the arXiv submission. We have checked that in all cases our results have the correct soft limits and symmetry properties. These results are sufficient to compute all two-loop NMHV amplitudes. If $h_i=-1$ and all other helicities are positive, we obtain
\beq\bsp
\cR_{+\ldots-\ldots+}^{(2)} &\,= \log\tau_{i-1}\,g_{+-+}^{(1,0)}(\rho_{i-1},\rho_{i})+\log\tau_{i}\,g_{+-+}^{(0,1)}(\rho_{i-1},\rho_{i})\\
&\,+\sum_{1\le j<i-1} \log\tau_j\,g_{++-+}^{(1,0,0)}(\rho_{j},\rho_{i-1},\rho_{i}) \\
&\,+\sum_{i<j\le N-5} \log\tau_j\,g_{+-++}^{(0,0,1)}(\rho_{i-1},\rho_{i},\rho_{j})\,.
\esp\eeq
The previous formula is not valid for $i\in\{1,2,N-5,N-6\}$, in which case we have
\beq\bsp
\cR_{-+\ldots}^{(2)} &\,= \log\tau_1\,g_{-+}^{(1)}(\rho_{1})
+\sum_{j=2}^{N-5} \log\tau_j\,g_{-++}^{(0,1)}(\rho_{1},\rho_j) \,,\\
\cR_{+-+\ldots}^{(2)} &\,=\log\tau_1\, g_{+-+}^{(1,0)}(\rho_{1},\rho_2)
+\log\tau_2\,g_{+-+}^{(0,1)}(\rho_{1},\rho_2)
+\sum_{j=3}^{N-5}\log\tau_j\, g_{+-++}^{(0,0,1)}(\rho_{1},\rho_2,\rho_j) \,,
\esp\eeq
and the remaining cases can be obtained from target-projectile symmetry.


\subsection{Leading singularities of scattering amplitudes in MRK}
\label{sec:LS}

In the previous section we have shown how we can compute non-MHV amplitudes via convolution with the universal helicity flip kernel $\cH$.
Due to the double pole in the helicity flip kernel, non-MHV amplitudes are no longer pure, but the transcendental functions are multiplied by rational prefactors. 
This is in agreement with the expectation for the structure of scattering amplitudes in full kinematics, where these coefficients are identified with the leading singularities of the amplitudes~\cite{Cachazo:2008vp}.
In this section we present a way to determine the set of all rational prefactors that can appear in a given non-MHV amplitude in MRK at LLA.

Let us start by defining some concepts that are useful to state the main result. 
We define \emph{interfaces} of the perturbative coefficients $g_{h_1\dots h_{N-4}}^{(i_1,\dots,i_{N-5})} ( \rho_1, \dots, \rho_{N-5})$ 
as the faces of its graph (see eq.~\eqref{eq:coefficient_notation}) that are bounded by two external lines with opposite helicities. In the following we refer to a face of the graph simply by the index of the corresponding dual coordinate (cf. Fig.~\ref{fig:dual_coordinates}).
We call an interface \emph{holomorphic} if the helicity changes from $-1$ to $+1$ in the natural order induced by the color ordering, and  \emph{anti-holomorphic} otherwise.
We denote by $I = \lbrace a_1,\dots,a_{\kappa} \rbrace$ the set of all interfaces of the graph (equipped with the natural order induced by the color ordering) and we let $a_0 = \mathbf{x}_2$ and $a_{\kappa+1} = \mathbf{x}_{N-2}$. For $1\le k \le \kappa$ we define the sets 
\begin{align}
&E^{a_k}_\uparrow = \lbrace b\, |\, a_{k-1} \leq b < a_k \rbrace {\rm~~and~~}E^{a_k}_\downarrow = \lbrace b\, |\, a_{k} < b\leq a_{k+1} \rbrace\,.
\end{align}
We also define the cross-ratios 
\begin{align}
R_{bac} =\left\{\begin{array}{ll}\displaystyle v_{bac1}\,, & \text{for holomorphic interfaces } a\,,\\   
& \\
\displaystyle\bar v_{bac1}\,, & \text{for anti-holomorphic interfaces } a \,,
\end{array}\right.
\end{align}
with 
\beq
v_{bac1} = \frac{(\bx_b - \bx_a)(\bx_c -  \bx_1)}{(\bx_b - \bx_c)(\bx_a -  \bx_1)}\,.
\eeq

We are now ready to state the main result of this section. We claim that it is possible to write the perturbative coefficients in such a way  that all rational prefactors multiplying pure functions take the form
\begin{equation}\label{eq:LS_thm}
\prod_{a\in S} R_{bac}\,,\quad b\in E^{a}_\uparrow\,,\quad c\in \widetilde{E}^{a}_\downarrow\,,
\eeq
where $S\subseteq I$ is a (possibly empty) subset of interfaces and we have introduced the definition
\beq
\widetilde{E}^{a}_\downarrow  = \lbrace b\, |\, a < b\rbrace\,.
\eeq
This implies in particular that the building blocks of all rational prefactors in MRK at LLA are contained in the set
\begin{equation}\label{eqn:buildingBlocks}
\cL = \lbrace R_{bac} | a \in I\,, b \in E^a_\uparrow\,, c \in \widetilde{E}^a_\downarrow \rbrace\,.
\end{equation}
The cross ratios in this set are at the same time the building blocks for all leading singularities in MRK at LLA. We emphasise that this set is an upper bound for the rational prefactors that can appear for a given helicity configuration. In particular, one may wonder whether the asymmetry in eq.~\eqref{eq:LS_thm} and eq.~\eqref{eqn:buildingBlocks} between $E^{a}_\uparrow$ and $\widetilde{E}^{a}_\downarrow$ could not be lifted, and we could restrict the building blocks to the more symmetric set
\begin{equation}\label{eq:sym_set}
\cL_{\textrm{sym}}=\lbrace R_{bac} | a \in I\,, b \in E^a_\uparrow\,, c \in {E}^a_\downarrow \rbrace\,.
\end{equation}
Unfortunately, this is incorrect, because the cross ratios $R_{bac}$ are not independent, but they satisfy intricate non-linear relations, e.g., 
\beq\label{eq:relation_cross_ratio}
R_{23c} + R_{234}\,R_{4ac} = R_{23c}\,R_{4ac} + R_{234}\,R_{2ac}\,,\quad a<c\,,\quad a\in I\,.
\eeq
The apparent asymmetry in the set of building blocks in eq.~\eqref{eqn:buildingBlocks} can then be lifted through such relations. 
It would be interesting to have a classification of all the relations among the building blocks $R_{bac}$.
Our building blocks are, however, \emph{linearly} independent, and so we can restrict to the more symmetric set $\cL_{\textrm{sym}}$ in situations where there is at most one interface of a given type (holomorphic or anti-holomorphic). Helicity configurations involving products of building blocks of the same type require at least three interfaces, and the simplest such amplitude is $\cR_{-+-+}$. We observe by explicit computation that in this case the restricted set $\cL_{\textrm{sym}}$ is indeed insufficient and a new building block $R_{236}\notin \cL_{\textrm{sym}}$ appears (see Appendix~\ref{app:results}).

Before we prove our result, let us
discuss some of its implications. First, it is evident from eq.~\eqref{eq:LS_thm} that every interface contributes at most one factor to the product in eq.~\eqref{eq:LS_thm}, i.e., we never encounter higher powers of $R_{bac}$. 

Second, we see that for a given helicity configuration there is always a finite number of different rational prefactors, independently of the number of loops. The complete set of rational prefactors for a given helicity configuration shows up when all indices are non-zero. In particular, we will see that eq.~\eqref{eq:LS_thm} is consistent with the factorisation theorem~\eqref{eq:fac_thm} in the sense that we never need to consider faces $b$ and $c$ bounded by external lines with equal helicities and vanishing index.

Finally, we note that the ratios $R_{bac}$ transform non-trivially under target-projectile symmetry. Target-projectile symmetry obviously maps interfaces to interfaces, and we have
\beq
R_{bac}\mapsto R_{N-b,N-a,N-c} = 1 - R_{N-c,N-a,N-b}\,.
\eeq

Let us now illustrate the content of eq.~\eqref{eq:LS_thm} on some simple examples. 
MHV and $\overline{\textrm{MHV}}$ amplitudes do not have any interfaces, so these amplitudes should not contain any non-trivial rational prefactors, in agreement with known results. The simplest amplitudes having a single interface are NMHV amplitudes of the form $\cR_{-+\dots +}$. Since these amplitudes have a single interface, we have $\cL=\cL_{\textrm{sym}}$. The amplitude must then take the form 
\begin{equation}\label{eq:LS-+++++}
\cR_{-+\dots +} = \mathfrak{a} + \sum_{c=4}^{N-2} R_{23c}\, \mathfrak{b}_c\,,
\end{equation}
where $\mathfrak{a}$ and $\mathfrak{b}_c$ are pure functions to all loop orders.
In the special case $N = 6$ eq.~\eqref{eq:LS-+++++} reduces to the known structure of the six-point NMHV amplitude in MRK~\cite{Lipatov:2012gk},
\begin{equation}
\cR_{-+} = \mathfrak{a} + R_{234}\, \mathfrak{b} = \mathfrak{a} + \frac{\rho_1}{\rho_1-1} \mathfrak{b}\,.
\end{equation}
Equation~\eqref{eq:LS_thm} implies that this structure generalises to an infinite class of N$^k$MHV amplitudes, $k\ge 1$, with a single holomorphic interface,
\begin{equation}
\cR_{-\dots -+\dots +} = \mathfrak{a} + \sum_{b=2}^{a-1}\sum_{c=a+1}^{N-2} R_{bac}\, \mathfrak{b}_{bc}\,,
\end{equation}
where $a$ is the holomorphic interface and $\mathfrak{a}$ and $\mathfrak{b}_{bc}$ are pure functions.
Products of rational prefactors contribute for the first time for amplitudes with two distinct interfaces, which appear  precisely for the helicity configurations considered in eq.~\eqref{eq:2_interfaces}. The interfaces are located at $(a_1,a_2)=(a+1,b+2)$. One of them is holomorphic and the other one anti-holomorphic, so we can work with the symmetric set $\cL_{\textrm{sym}}$. We find
\beq\bsp\label{eq:two_interfaces}
\cR_{+\dots +-\dots -+\dots +} = \mathfrak{a} &+ \sum_{c=2}^{a_1-1}\sum_{d=a_1+1}^{a_2} \overline{R}_{ca_1d}\, \mathfrak{b}^1_{cd} + \sum_{c=a_1}^{a_2-1}\sum_{d=a_2+1}^{N-2} R_{ca_2d}\, \mathfrak{b}^2_{cd} \\
&+ \sum_{c_1=2}^{a_1-1}\sum_{d_1=a_1+1}^{a_2}\sum_{c_2=a_1}^{a_2-1}\sum_{d_2=a_2+1}^{N-2} \overline{R}_{c_1a_1d_1} \,R_{c_2a_2d_2} \,\mathfrak{c}^{12}_{c_1d_1c_2d_2} \,,
\esp\eeq
where we have indicated the anti-holomorphic rational functions by $\overline{R}_{bac}$ for clarity and $\mathfrak{a}$, $\mathfrak{b}_{cd}^i$ nd $\mathfrak{c}_{c_1d_1c_2d_2}^{ij}$ are pure functions

Let us conclude this section by discussing the soft limits of the rational prefactors. First, we can see that $R_{bac}$ has simple poles for $\bx_b=\bx_c$ and $\bx_a=\bx_1$. None of these singularities corresponds to a soft limit. This implies in particular that the weight of the pure functions does not drop when taking a soft limit. Next, we see that
\beq\label{eq:lim_R}
\lim_{\bx_b\to\bx_a}R_{bac} = 0 {\rm~~and~~} \lim_{\bx_c\to\bx_a}R_{bac}=1\,.
\eeq
In order to understand the implication of these relations, let us consider a NMHV amplitude, which can be written in the form of eq.~\eqref{eq:two_interfaces} with $a\equiv a_1=a_2-1$,
\beq\bsp
\cR_{+\dots +-+\dots +} = \mathfrak{a} &+ \sum_{c=2}^{a-1} \overline{R}_{ca(a+1)}\, \mathfrak{b}^1_{ca+1} + \sum_{d=a+2}^{N-3} R_{a(a+1)d}\, \mathfrak{b}^2_{cd} \\
&+ \sum_{c=2}^{a-1}\sum_{d=a+2}^{N-3} \overline{R}_{ca(a+1)} \,R_{a(a+1)d} \,\mathfrak{c}^{12}_{c_1(a+1)ad_2} \,.
\esp\eeq
In the limit where the gluon with negative helicity becomes soft, $\bx_a\to \bx_{a+1}$, the NMHV amplitude reduces to an MHV amplitude, which is a pure function. Equation~\eqref{eq:lim_R} guarantees that this is indeed the case, and we find,
\beq
\lim_{\bx_a\to\bx_{a+1}}\cR_{+\dots +-+\dots +} = \mathfrak{a} + \sum_{c=2}^{a-1} \mathfrak{b}^1_{c(a+1)} \,.
\eeq

\subsubsection{Proof of the structure of leading singularities in MRK}
\label{sec:LS_proof}

Let us now prove our claim about the structure of the rational prefactors in MRK to LLA. Before turning to the proof itself, we make the following observation: every perturbative coefficient can be built up by a finite sequence of the following three operations:
\begin{enumerate}
\item Flipping the leftmost helicity by convolution with $\cH(z_1)$ or $\cH(\bar{z}_1)$ respectively.
\item Increasing the first index by convolution with $\cE(z_1)$.
\item Adding more particles to the left with index zero and equal helicities.
\end{enumerate}
In particular this implies that every non-MHV amplitude can be constructed from a NMHV helicity configuration of the from $-+\ldots+$ by successive application of these three elementary operations (we can always assume the rightmost helicity to be $h_{N-4}=+1$). It is evident that we can reach a similar conclusion by adding more particles to the right and convoluting with $\cE(z_{N-5})$ and $\cH(z_{N-5})$.

Let us illustrate this procedure on a short example. Note that in the following we consider all convolutions to be over $z_1$ and we see all the simplicial MRK coordinates $\rho_i$ as functions of the $z_i$. The perturbative coefficient $g_{++---++}^{(1,0,2,0,1,2)}(\rho_1,\dots,\rho_6)$ can be constructed in the following way: We start with the perturbative coefficient $g_{+++}^{(1,2)}(\rho_1,\rho_2)$ and flip its leftmost helicity ,
\begin{equation}
g_{+++}^{(1,2)}(\rho_1,\rho_2) \rightarrow g_{-++}^{(1,2)}(\rho_1,\rho_2) = \cH(z_1)*g_{+++}^{(1,2)}(\rho_1,\rho_2)\,.
\end{equation}	
Next, we add additional particles with index zero and negative helicity to the left and we use the factorization theorem to remove zero indices,
\begin{equation}
g_{-++}^{(1,2)}(\rho_1,\rho_2)  \rightarrow g_{---++}^{(0,0,1,2)}(\rho_1,\dots,\rho_4) = g_{-++}^{(1,2)}(\rho_3,\rho_4)\,.
\end{equation}
Note that this operation is equivalent to simply shifting the indices of all the simplicial MRK coordinates. Next, we
increase the first index by convolution with $\cE(z_1)$ and perform another shift in simplicial MRK coordinates to add one more external particle,
\beq\begin{split}
g_{---++}^{(0,0,1,2)}(\rho_1,\dots,\rho_4)\rightarrow &\, g_{---++}^{(2,0,1,2)}(\rho_1,\dots,\rho_4)= \cE(z_1) * \cE(z_1) * g_{---++}^{(0,0,1,2)}(\rho_1,\dots,\rho_4) \\
\rightarrow &\, g_{----++}^{(0,2,0,1,2)}(\rho_1,\dots,\rho_5) = g_{---++}^{(2,0,1,2)}(\rho_2,\dots,\rho_5)\,.
\end{split}\eeq
Then we flip the first helicity and perform one more shift in simplicial MRK coordinates,
\beq\begin{split}
g_{----++}^{(0,2,0,1,2)}(\rho_1,\dots,\rho_5) \rightarrow &\, g_{+---++}^{(0,2,0,1,2)}(\rho_1,\dots,\rho_5) = \cH(\bar{z}_1)*g_{----++}^{(0,2,0,1,2)}(\rho_1,\dots,\rho_5)  \\
\rightarrow &\,g_{++---++}^{(0,0,2,0,1,2)}(\rho_1,\dots,\rho_6) = g_{+---++}^{(0,2,0,1,2)}(\rho_2,\dots,\rho_6)\,.
\end{split}\eeq
Finally, we arrive at the desired perturbative coefficient by increasing the first index by one unit,
\begin{equation}
g_{++---++}^{(0,0,2,0,1,2)}(\rho_1,\dots,\rho_6)\rightarrow g_{++---++}^{(1,0,2,0,1,2)}(\rho_1,\dots,\rho_6) = \cE(z_1)*g_{++---++}^{(0,0,2,0,1,2)}(\rho_1,\dots,\rho_6) \,.
\end{equation}

Let us now turn to the proof of the structure of the leading singularities in MRK. More precisely, 
we will proof that the building blocks in eq.~\eqref{eqn:buildingBlocks} exhaust all rational prefactors of non-MHV amplitudes in MRK for any number of particles and any loop order. 

As a warm-up, let us consider the case of a NMHV amplitude where the first gluon emitted along the ladder has a negative helicity, and all other gluons have a positive helicity. We can construct this amplitude by starting from an MHV amplitude and then we flip the first helicity, cf. eq.~\eqref{eq:NMHV_d_I}. As MHV amplitudes are pure functions, the anti-holomorphic primitive will itself be pure. If we work in simplicial coordinates based at $z_1$, where $t_1=z_1$, we can write eq.~\eqref{eq:NMHV_d_I} in the form
\begin{equation}
g_{-+\dots+}^{(i_1,\dots,i_n)}(\rho_1,\dots,\rho_n) = t_1\partial_{t_1} \int d \bar{t}_1 \frac{1}{\bar{t}_1} g_{+\dots+}^{(i_1,\dots,i_n)}(\rho_1,\dots,\rho_n)\,,
\end{equation}
where we interpret the simplicial MRK coordinates $\rho_i$ as functions of the simplicial coordinates based at $z_1$. The rational prefactors are entirely generated by the action of the holomorphic derivative. Since the anti-holomorphic primitive is pure, all rational prefactors take the following form in simplicial coordinates based at $z_1$,
\begin{equation}
\frac{t_1}{t_1-\tau_i} = \frac{(\bx_3-\bx_2)(\bx_{i+3}-\bx_1)}{(\bx_{i+3}-\bx_2)(\bx_3-\bx_1)} = R_{23(i+3)}\,, \quad 0 \leq i \leq n,
\end{equation}
with 
\beq
\tau_i = \left\{\begin{array}{ll}
0\,,& \textrm{ if } i=0\,,\\
1\,,& \textrm{ if } i=1\,,\\
t_i\,,& \textrm{ otherwise}\,,
\end{array}\right.
\eeq
in agreement with eq.~\eqref{eq:LS-+++++}. We thus see that the claim is true for all NMHV amplitudes of this type.
%

At the beginning of the section we have argued that every non-MHV configuration can be obtained from the helicity configuration $(-,+\ldots+)$ through a successive application of the three elementary operators (convolution with $\cE$ and $\cH$ and adding particles). Since the claim holds for this particular helicity configuration, it is then sufficient to show that the elementary operations preserve the structure of the rational prefactors.
As all convolution integrals will be performed in simplicial coordinates based at $z_1$, it is instructive to see what the $R_{bac}$ look like in these coordinates. We find,
\beq\bsp\label{eq:LS_simplicial_at_z1}
R_{2ac} & = \frac{t_1- \tau_{a-3}}{t_1-\tau_{c-3}} \,,\\
R_{bac}  & = \frac{\tau_{b-3}-\tau_{a-3}}{\tau_{c-3}-\tau_{a-3}}\,,\quad b\neq2\,.
\esp\eeq
We see that $R_{bac}$ depends on $t_1$ if and only if $b=2$. This has important implications for the analytic structure of the perturbative coefficients. Indeed, assume that our claim~\eqref{eq:LS_thm} is true for a certain amplitude. Then every term can contain at most one factor of the form $R_{2ac}$, because $a$ must be the first interface (in the natural order on the interfaces). From eq.~\eqref{eq:LS_simplicial_at_z1} we see that all poles in $t_1$ must be simple, and moreover they are all either holomorphic or anti-holomorphic, depending on whether the first interface $a$ is holomorphic or anti-holomorphic. As a consequence, repeating the discussion at the end of Section~\ref{sec:H_flip}, we see that in those cases where our claim is correct, we can always compute the convolution with $\cH(z_1)$ by differentiating and integrating.

Let us now show that if the claim is correct for a given amplitude, then it remains true if we perform any of the three elementary operations on it. In the following we always assume that the poles in $t_1$ are holomorphic. The extension to the anti-holomorphic case is trivial. Let us analyse the effect of each of the elementary operations in turn.

\begin{itemize}
\item \textbf{Adding new particles:} 
Adding $\ell$ new particles with the same helicity and index zero as the first one is equivalent to a shift in the simplicial MRK coordinates, $\rho_i \rightarrow \rho_{i+\ell}$. This shift has a very simple effect on the cross ratios $R_{bac}$. Indeed, if $b>2$, we have
\beq\bsp
R_{bac}&\,= \frac{(\rho_{a-2}-\rho_{b-2})(\rho_{c-2}-1)}{(\rho_{c-2}-\rho_{b-2})(\rho_{a-2}-1)}\\
&\,\rightarrow \frac{(\rho_{a-2+\ell}-\rho_{b-2+\ell})(\rho_{c-2+\ell}-1)}{(\rho_{c-2+\ell}-\rho_{b-2+\ell})(\rho_{a-2+\ell}-1)} =  R_{(b+\ell)(a+\ell)(c+\ell)}\,.
\esp\eeq
In the case $b=2$, we find
\beq\bsp
R_{2ac}&\,= \frac{\rho_{a-2}(\rho_{c-2}-1)}{\rho_{c-2}(\rho_{a-2}-1)}\rightarrow \frac{\rho_{a-2+\ell}(\rho_{c-2+\ell}-1)}{\rho_{c-2+\ell}(\rho_{a-2+\ell}-1)} =  R_{2(a+\ell)(c+\ell)}\,.
\esp\eeq
In order to complete the argument, we need to show that $(a+\ell)$ is the first interface in the shifted amplitude. This is automatic in this case, because we only add particles with the same helicities to the left, and so no new interface is introduced.
Note that at the same time we have shown that we can always drop faces with vanishing index, consistently with the factorisation theorem~\eqref{eq:fac_thm}. Indeed, we see that when adding particles we do not add any new rational prefactors to those that were already present, and we only relabel the variables in the shifted amplitude.

\item \textbf{Increasing the first index:} 
	This is equivalent to convoluting with $\cE(z_1)$,
	\beq\bsp\label{eq:conv_inc_LS}
	g&_{h_1\dots h_{N-4}}^{(i_1+1,\dots,i_{N-5})}(\rho_1,\dots,\rho_{N-5}) = \cE(z_1)\ast g_{h_1\dots h_{N-4}}^{(i_1,\dots,i_{N-5})}(\rho_1,\dots,\rho_{N-5})\\
	&\,= -\int \frac{d^2w}{2 \pi} g_{h_1\dots h_{N-4}}^{(i_1,\dots,i_{N-5})}(w,t_2,\dots,t_{N-5}) 	\frac{\bar{w} t_1+ w \bar{t}_1}{|w|^2|w-t_1|^2} \\
	&\,=-\int \frac{d^2w}{2 \pi} g_{h_1\dots h_{N-4}}^{(i_1,\dots,i_{N-5})}(w,t_2,\dots,t_{N-5})\,\frac{1}{w(w-t_1)}\left(\frac{w+t_1}{\bar{w}-\bar{t}_1} - \frac{w}{\bar{w}}\right)
	\,.
\esp\eeq
We evaluate this integral in terms of residues. By hypothesis, the perturbative coefficient has only holomorphic poles in $w$. Since $\bar{w}$ only enters via linear denominators in the integration kernel in eq.~\eqref{eq:conv_inc_LS}, the anti-holomorphic primitive of the integrand is a pure function (seen as a function of $\bar{w}$).

Next, we have to compute the holomorphic residues. The anti-holomorphic primitive may still contain holomorphic poles in $w$, but we know from eq.~\eqref{eq:LS_simplicial_at_z1} that these poles are all simple and can only enter through rational prefactors of the type $R_{2ab}$ (the integrand may contain other rational prefactors proportional to $R_{bac}$ with $b\neq2$, but those will not spoil the argument, because they do not enter the convolution integral). By induction hypothesis, $a$ is the first interface. Our goal is to show that by taking holomorphic residues, we do not introduce any new rational prefactors. In order to see that this is true, let us multiply $R_{2ac}$ by the two rational functions $\frac{w+t_1}{w(w-t_1)}$ and $\frac{1}{w-t_1}$  in the integrand of eq.~\eqref{eq:conv_inc_LS}, and apply a partial fractioning in $w$. We find,
	\begin{align}
	\nonumber
	\frac{w+t_1}{w(w-t_1)}\,\frac{w-\tau_{a-3}}{w-\tau_{c-3}} 
	= &-\frac{1}{w}\,R_{3ac}+\frac{2}{w-t_1}\,R_{2ac} +\frac{1}{w-\tau_{c-3}} \left(1 - 2R_{2ac} + R_{3ac} \right)\,, \\
	\label{eqn:fracE1}
	\frac{1}{w-t_1}\,\frac{w-\tau_{a-3}}{w-\tau_{c-3}}
	=&\,  \frac{1}{w-t_1}\,R_{2ac} + \frac{1}{w-\tau_{c-3}} \left(1 - R_{2ac}\right) \,.
	\end{align}
	The previous expressions are multiplied by pure functions in $w$, and so the holomorphic residues are obtained by evaluating the pure functions at the simple poles at $w=0$, $w=t_1$ and $w=\tau_{c-3}$. Hence, we see that no new rational prefactors are introduced in the process. In order to complete the argument, we need to check that the residue at $w = \infty$ does not spoil this result. Letting $w = \frac{1}{u}$ and multiplying with the respective Jacobian, the denominators in eq.~\eqref{eqn:fracE1} give rise to a pole at $u=0$. 
It is easy to check that taking the residue at this pole does not introduce any new rational prefactor. Finally, the previous argument can easily be extended to the case where the factor $R_{2ac}$ is absent, i.e., where the anti-holomorphic primitive has no pole in $w$.
We thus conclude that increasing an index does not change the building blocks for the leading singularities.
	
%
%
\item \textbf{Flipping the first helicity:} By hypothesis, we assume that the perturbative coefficient before flipping the helicity has only holomorphic poles in $t_1$. This means that the first interface before the helicity flip is holomorphic, and so the next flipping will produce an anti-holomorphic interface, i.e., we need to convolute with an anti-holomorphic helicity flip kernel $\cH(\bar{z}_1)$.

As all the poles in $t_1$ before the helicity flip are holomorphic and simple, we can apply the result from the end of Section~\ref{sec:H_flip} and compute the convolution with $\cH(\bar{z}_1)$ by first computing a holomorphic primitive, followed by an anti-holomorphic derivative,
\beq\label{eq:h_flip_LS}
\cH(\bar{z}_1) \ast g_{+-\ldots+} = \bar{t}_1\partial_{\bar{t}_1}\int\frac{dt_1}{t_1}\,g_{--\ldots+}\,.
\eeq
After partial fractioning in $t_1$, we see that the holomorphic primitive is pure as a function of $t_1$. Hence, all rational prefactors in $t_1$ are produced by anti-holomorphic differentiation of a pure function, and so they all take the form
\beq\label{eq:LS_hel_flip_rat}
\frac{\bar{t}_1}{\bar{t}_1-\bar{\tau}_c} = R_{23c}\,,
\eeq
which completes the proof.

\end{itemize}
%

Let us conclude this section by commenting on the asymmetry of the set $\cL$. Throughout the proof we have shown that the elementary operations do not produce any new rational building blocks that are not in the set $\cL$ defined in eq.~\eqref{eqn:buildingBlocks}. The proof relies on the fact that we can construct any helicity configuration algorithmically by adding particles to the left and by convolution with $\cE(z_1)$ and $\cH(z_1)$. The asymmetry in the construction is most manifest in the helicity flip operation, because the range of $c$ in the rational functions in eq.~\eqref{eq:LS_hel_flip_rat} is naturally given by the set $\widetilde{E}^3_\downarrow$, and it is not restricted in any obvious way to ${E}^3_\downarrow$. 

One may wonder, however, if we could restrict the range by adding particles to the right and convoluting with $\cE(z_{N-5})$ and $\cH(z_{N-5})$. Applying exactly the same reasoning as in the other case leads to a new set of building blocks
\beq
\cL' = \lbrace R_{bac} | a \in I\,, b \in \widetilde{E}^a_\uparrow\,, c \in {E}^a_\downarrow \rbrace\,.
\eeq
As the result for the amplitude should not depend on the way it was obtained, one would be inclined to believe that the set of building blocks could also be restricted to the symmetric set in eq.~\eqref{eq:sym_set},
\beq
\cL_{\textrm{sym}} = \cL\cap\cL'\,.
\eeq
This is incorrect, because non-linear relations between the cross ratios (cf. eq.~\eqref{eq:relation_cross_ratio}) allow one to express building blocks from $\cL'$ which are not in $\cL$ as non-linear polynomials in elements in $\cL$. 
In the case that there is a single holomorphic interface, however, the \emph{linear} independence of the cross ratios $R_{bac}$ allows one to consider only the restricted set $\cL_{\textrm{sym}}$.

\subsection{Explicit two-loop, seven-point NMHV check}
\label{7ptexpcheck}
In this section
we outline an explicit check of our discussion for the leading logarithmic contribution to the two-loop seven-point NMHV amplitudes in MRK. The symbol of this amplitude was obtained in ref.~\cite{CaronHuot:2011kk}\footnote{We thank the authors for providing a file with the relevant expressions.}. More precisely, the quantity discussed in ref.~\cite{CaronHuot:2011kk} the so-called `BDS-subtracted' amplitude, equivalent to the exponentiated remainder function multiplied by the ratio function. It is given in supersymmetric notation as follows
\be
\cA^{\rm NMHV}_{\rm BDS-subtracted} = [\tfrac{3}{7}(12)+\tfrac{1}{7}(13)+\tfrac{2}{7}(14)+\text{cyc}.]X+[(67)V_{67}+(47)V_{47}+\text{cyc}.]\,.
\label{7ptNMHV}
\ee
In the above formula the quantities $X$, $V_{67}$ and $V_{47}$ are pure functions based on the heptagon alphabet arising from the cluster algebra structure on $\textrm{Gr}(4,7)$, as discussed in ref.~\cite{Golden:2013xva}. The quantities $(ij)$ above represent the R-invariants which encode all the possible NMHV configurations of external states by use of Grassmann odd variables $\eta_i$.

We recall that all on-shell states in $\mathcal{N}=4$ SYM theory can be described by the on-shell supermultiplet written in superspace notation as a function of Grassmann parameters $\eta^A$ transforming in the fundamental representation of $\mathfrak{su}(4)$,
\be
\Phi(\eta) = G^+ + \eta^A \Gamma_A + \tfrac{1}{2} \eta^A \eta^B S_{AB} + \tfrac{1}{3!} \eta^A \eta^B \eta^C \epsilon_{ABCD} \overline{\Gamma}^D + \tfrac{1}{4!} \eta^A \eta^B \eta^C \eta^D \epsilon_{ABCD} G^-\,.
\ee
The R-invariants generically depend on five indices $[ijklm]$. In the seven-point case, however, we may simply denote them by the two indices which are absent, e.g.,
\be
(12) = [34567]\,.
\ee
Furthermore, at seven points all R-invariants are of the form $[r \, s-1\, s\, t-1\, t]$ for some $r,s,t$ and in this regard we find it helpful to employ the formula of eq.~\cite{Drummond:2008vq} which we express as follows,
\be\bsp
[r \, s-1\, s\, t-1\, t]=&\, \frac{\delta^8(q)}{\langle 12 \rangle \ldots \langle n 1 \rangle}\\
\times&\,  \frac{\langle s-1 s\rangle \langle t-1 t\rangle \delta^{(4)}(\Xi_{rst})}{x^2_{st}\langle r |x_{rs}x_{st}|t-1\rangle\langle r |x_{rs}x_{st}|t\rangle\langle r |x_{rt}x_{ts}|s-1\rangle\langle r |x_{rt}x_{ts}|s\rangle} \,.
\label{Rinvariant}
\esp\ee
We have included in the above formula the supersymmetric Parke-Taylor-Nair prefactor to exhibit all the relevant $\eta$ dependence. The delta function $\delta^8(q)$ is a consequence of supersymmetry and is explicitly given by
\be
\delta^8(q) = \delta^8\biggl( \sum_i \lambda_i \eta_i\biggr),
\ee
where the $\lambda_i$ are the spinor-helicity variables introduced in eq.~(\ref{momspinors}). The argument of the other Grassmann delta function in the numerator of eq.~(\ref{Rinvariant}) is given by
\be
\Xi_{rst}^A=\sum_{i=r+1}^{t-1}\langle r |x_{rs}x_{st}|i\rangle\eta_i^A+\sum_{i=r+1}^{s-1}\langle r |x_{rt}x_{ts}|i\rangle\eta_i^A\,.
\ee

Next we calculate the limits of the pure functions $X$, $V_{ij}$ in MRK and we evaluate the R-invariants in this limit. To perform the second task it is helpful to formulate the passage to multi-Regge kinematics in term of spinor-helicity variables. It is sufficient for us to parametrise our spinors with different powers of $\epsilon$ in such a way so as to systematise the strong ordering of the kinematics in the MRK limit, similar to ref.~\cite{Prygarin:2011gd}. For example in the $12\to34567$ kinematics, the $\lambda$ spinors are parametrised as
\begin{align}
\lambda_1=\begin{bmatrix}0 \\ \Big(-\sum_{i=3}^7\frac{|p_i|^2}{p_i^+}\epsilon^{5-i}\Big)^{\frac{1}{2}}\end{bmatrix}\ & \lambda_2=\begin{bmatrix} \Big(-\sum_{i=3}^7 p_i^+ \epsilon^{i-5} \Big)^{\frac{1}{2}} \\ 0 \end{bmatrix}\mkern-16mu &
\lambda_j=\begin{bmatrix} \sqrt{p_j^+}\epsilon^{\frac{j-5}{2}} \\ \frac{p_j}{\sqrt{p_j^+}}\epsilon^{\frac{5-j}{2}}\end{bmatrix}&
\end{align}
where $j=3,\ldots,7$ and the $\tilde{\lambda}$ are obtained by conjugation. After calculating the R-invariants using this parametrisation we can recover the MRK value by taking $\epsilon \to 0$.

Projecting out the components of the $\eta$'s corresponding to the desired helicity configuration and taking the MRK limit we find the following non-vanishing R-invariants for the $(-++)$ configuration,
\begin{align}
(12)\to 1\,,  & \quad (23) \to 1-\frac{\rho_1(1-\rho_2)}{\rho_2(1-\rho_1)}\,, \quad  (17)\to \frac{\rho_1}{\rho_1-1}\,, \notag 
\end{align}
\begin{align}
&(27)\to  \frac{\rho_1(1-\rho_2)}{\rho_2(1-\rho_1)}\,, \mkern-16mu &  (13)\to  1- \frac{\rho_1}{\rho_1-1}\,, & \quad (37) \to \frac{\rho_1(1-\rho_2)}{\rho_2(1-\rho_1)}-\frac{\rho_1}{\rho_1-1}\,.
\end{align}
For the $(+-+)$ configuration we find
\begin{align}
(34) \to \frac{\bar{\rho}_1(1-\bar{\rho}_2)}{\bar{\rho}_2(1-\bar{\rho}_1)}\Big[1+\frac{(\rho_1-\rho_2)}{(\rho_2-1)}\Big]\,, & \quad (24) \to \frac{\bar{\rho}_1(1-\bar{\rho}_2)}{\bar{\rho}_2(1-\bar{\rho}_1)}\,, & (15) \to 1+\frac{(\rho_1-\rho_2)}{(\rho_2-1)}\,,\notag
\end{align}
\begin{align}
(16) \to \frac{(\rho_1-\rho_2)}{(\rho_2-1)}\Big[\frac{\bar{\rho}_1(1-\bar{\rho}_2)}{\bar{\rho}_2(1-\bar{\rho}_1)}-1\Big]\,, &\quad  (25)\to1\,, & (36) \to \Big[1+ \frac{(\rho_1-\rho_2)}{(\rho_2-1)}\Big] \Big[1-\frac{\bar{\rho}_1(1-\bar{\rho}_2)}{\bar{\rho}_2(1-\bar{\rho}_1)}\Big]\,,\notag
\end{align}
\begin{align}
(15) \to \frac{(\rho_2-\rho_1)}{(\rho_2-1)}\,, \quad (26) \to 1-\frac{\bar{\rho}_1(1-\bar{\rho}_2)}{\bar{\rho}_2(1-\bar{\rho}_1)}\,, & \quad (14) \to \frac{(\rho_2-\rho_1)}{(\rho_2-1)}\frac{\bar{\rho}_1(1-\bar{\rho}_2)}{\bar{\rho}_2(1-\bar{\rho}_1)}\,.
\end{align}
Combining these formulas, we find that the two NMHV helicity configurations become
\begin{align}
\cR_{-++}=\hat{X}&+\hat{V}_{12}+\hat{V}_{23}+R_{234}\Big(\hat{V}_{73}-\hat{V}_{23}\Big)+R_{235}\Big(\hat{V}_{71}-\hat{V}_{73}\Big)\,,\notag\\ 
\cR_{+-+}=\hat{X}&+\hat{V}_{25}+\hat{V}_{36}+\hat{V}_{62}+\overline{R}_{234}\Big(\hat{V}_{34}-\hat{V}_{36}-\hat{V}_{62}\Big)\notag\\
&+R_{345}\Big(\hat{V}_{51}-\hat{V}_{36}\Big)+ \overline{R}_{234}R_{345}\Big(\hat{V}_{14}-\hat{V}_{34}+\hat{V}_{36}\Big)\,.
\label{7ptnmhvMRKexplicit}
\end{align}
Here the $\hat{V}_{ij}$ are the MRK limits of the pure functions $V_{ij}$ of eq. (\ref{7ptNMHV}). In Appendix \ref{app:results} we give the explicit form of the $\hat{V}_{ij}$ at LLA (of course, since we started from just the symbol, these formulas are valid up to terms proportional to multiple zeta values). Note that individually these functions may have beyond-leading log divergences. These extra powers of divergent logarithms cancel when combined into the combinations outlined in eq. (\ref{7ptnmhvMRKexplicit}).
These explicit limits may then be compared to the general structure outlined in eq. (\ref{NMHV7ptMRKform}) and the predicted pure functions presented in eqs. (\ref{nmhv7pta02+-+}) onwards.

\section{Analytic structure of scattering amplitudes in MRK}
\label{sec:MPL_proof}

It is believed that MHV and NMHV amplitudes are expressible in terms of multiple polylogarithms~\cite{ArkaniHamed:2012nw}, but it is expected that for more complicated helicity configurations more general classes of special functions may appear~\cite{CaronHuot:2012ab,Nandan:2013ip}. Knowing that in some limit scattering amplitudes can always be expressed in terms of multiple polylogarithms independently of the helicity configuration and the number of external legs can thus provide valuable information and constraints on the analytic structure of scattering amplitudes. A proof of such a property previously only existed for the six-point amplitudes when expanded to leading order around the collinear limit \cite{Papathanasiou:2013uoa} and to LLA in MRK~\cite{Pennington:2012zj,Broedel:2015nfp}.

In Section~\ref{sec:LS_proof} we have argued that it is possible to construct all amplitudes in MRK to LLA via a sequence of three elementary operations. In this section we show that this recursive structure of scattering amplitudes in the multi-Regge limit implies that they can always be expressed in terms of single-valued multiple polylogarithms of maximal and uniform weight, independently of the loop number and the helicity configuration.

Let us start by discussing MHV amplitudes. The algorithm of Section~\ref{sec:LS} allows us to construct all MHV amplitudes by adding particles and by convoluting with $\cE(z_1)$. We now show that the perturbative MHV coefficients $g_{+\ldots+}^{(i_1,\ldots,i_k)}$ are pure polylogarithmic functions of uniform weight $\omega=1+i_1+\ldots+ i_k$. Obviously, the factorisation theorem~\eqref{eq:fac_thm} implies that the claim remains true under the elementary operation of adding particles, so it suffices to show that convolution with $\cE(z_1)$ has the same property. The proof proceeds by induction. Assume that $g_{+\ldots+}^{(i_1,\ldots,i_k)}$ is a pure function of uniform weight $\omega=1+i_1+\ldots+ i_k$, and let us show that $g_{+\ldots+}^{(i_1+1,\ldots,i_k)} = \cE(z_1)\ast g_{+\ldots+}^{(i_1,\ldots,i_k)}$ is a pure function of uniform weight $\omega+1$. We have
	\beq\bsp\label{eq:MHV_pure_1}
	 \cE(z_1)&\,\ast g_{+\dots+}^{(i_1,\dots,i_{N-5})}(\rho_1,\dots,\rho_{N-5})\\
	 &\,= -\int \frac{d^2w}{2 \pi} g_{+\dots+}^{(i_1,\dots,i_{N-5})}(w,t_2,\dots,t_{N-5}) 	\frac{\bar{w} t_1+ w \bar{t}_1}{|w|^2|w-t_1|^2} \\
	&\,=-\int \frac{d^2w}{2 \pi} g_{+\dots+}^{(i_1,\dots,i_{N-5})}(w,t_2,\dots,t_{N-5})\,\frac{1}{w(w-t_1)}\left(\frac{w+t_1}{\bar{w}-\bar{t}_1} - \frac{w}{\bar{w}}\right)
	\,.
\esp\eeq
We evaluate the integral in terms of residues. As $g_{+\dots+}^{(i_1,\dots,i_{N-5})}$ is assumed pure by induction hypothesis and all the denominators are linear in $\bar{w}$, the anti-holomorphic primitive is a pure function (seen as a function of $\bar{w}$) of uniform weight $\omega+1$. The convolution in eq.~\eqref{eq:MHV_pure_1} can then be written in the form
\beq\bsp
\cE(z_1)&\,\ast g_{+\dots+}^{(i_1,\dots,i_{N-5})}(\rho_1,\dots,\rho_{N-5})\\
&\, = -\int \frac{dw}{2\pi}\,\left[\frac{1}{w}\,F_1(w,t_2,\ldots,t_{N-5}) + \frac{1}{w-t_1}\,F_2(w,t_2,\ldots,t_{N-5})\right]\,,
\esp\eeq
where $F_1$ and $F_2$ are pure single-valued polylogarithmic functions of weight $\omega+1$. As all the poles are simple, the holomorphic residues can be computed by simply evaluating the pure functions of weight $\omega+1$ at $w=0$, $w=t_1$ and $w=\infty$ (and dropping all logarithmically divergent terms). Hence, $\cE(z_1)\ast g_{+\dots+}^{(i_1,\dots,i_{N-5})}$ is a pure polylogarithmic function of weight $\omega+1$.

While the previous result is not unexpected for MHV amplitudes, we show in the remainder of this section that we can extend the argument to non-MHV amplitudes, independently of the helicity configuration. More precisely, we show that the pure functions multiplying the rational prefactors defined in Section~\ref{sec:LS} are always pure polylogarithmic functions of uniform weight $\omega=1+i_1+\ldots+ i_k$. The proof in the MHV case relies crucially on the fact that the anti-holomorphic primitive was a pure function of weight $\omega+1$ and that all the holomorphic poles were simple. Since non-MHV amplitudes are in general not pure but contain rational prefactors, it is not obvious that the same conclusion holds for arbitrary helicity configurations. In addition, for non-MHV we also need to analyse the effect of the helicity flip operation, which should not change the weight of the function.

We proceed again by induction. Let us start by showing that also in the non-MHV case a convolution with $\cE(z_1)$ will increase the weight by one unit. From Section~\ref{sec:LS} we know that all poles in $z_1$ are simple and either holomorphic or anti-holomorphic. In the following we discuss the anti-holomorphic case and the extension to the holomorphic case is trivial. 
The integrand of the convolution integral in the non-MHV case may have additional poles in $\bar{w}$ at points where $R_{2ac}$ is singular, see eq.~\eqref{eq:LS_simplicial_at_z1}. It is easy to see from eq.~\eqref{eq:LS_simplicial_at_z1} that none of these additional poles is located at $\bar{w}=0$ or $\bar{w}=t_1$, and so all the anti-holomorphic poles entering the convolution integral are simple. 
We can thus repeat the same argument as in the MHV case, and the anti-holomorphic primitive will be a pure polylogarithmic function of weight $\omega+1$. Moreover, there are no additional holomorphic poles in $w$ introduced by the rational prefactors, and so we can compute all the holomorphic residues by evaluating the pure functions of weight $\omega+1$ at $w\in\{0,t_1,\infty\}$. Hence, a convolution with $\cE(z_1)$ produces pure polylogarithmic functions of weight $\omega+1$ also in the non-MHV case.

To complete the argument, we need to show that flipping the leftmost helicity does not change the weight of the functions. In Section~\ref{sec:LS} we have seen that, since all poles in $t_1$ are simple and either holomorphic or anti-holomorphic, we can always compute the effect of the helicity flip by integrating and differentiating, cf. eq.~\eqref{eq:h_flip_LS}. Since all poles are simple, the integration will increase the weight by one unit. This effect is compensated by the differentiation, so that the total weight of the functions remains unchanged. Hence, we conclude that non-MHV amplitudes in MRK to LLA are polylogarithmic functions of uniform weight $\omega=1+i_1+\ldots+ i_k$ independently of their helicity configuration.

\section{Discussion and conclusion}
\label{sec:conclusion}

In this paper we have introduced a new method to compute scattering amplitudes in MRK kinematics through LLA. The cornerstone of the method is the realisation that an $N$-point kinematic configuration in MRK is equivalent to a configuration of $(N-2)$-points in $\mathbb{CP}^1$, i.e., to a point in $\mathfrak{M}_{0,N-2}$. Using this framework, the evaluation of convolution integrals is reduced to a simple application of Stokes' theorem and the residue theorem. We have proved a factorisation theorem which generalises a factorisation observed for the two-loop MHV amplitude and which allows one to represent a given multi-leg amplitude in terms of building blocks with fewer legs. We have applied our new framework to obtain analytic results for all MHV amplitudes up to five loops and for all non-MHV amplitudes up to four loops and eight points. 
The techniques introduced in this paper are generic and apply to scattering amplitudes in MRK at arbitrary logarithmic accuracy. In the following we comment on how to extend the results of our paper beyond LLA.


Beyond LLA scattering amplitudes in MRK can still be computed using the techniques outlined in this paper, because neither the convolutions nor the kinematic considerations about the moduli space of Riemann spheres with marked points relie on LLA. Nonetheless, there will be some differences which we outline in the following. Although we currently do not know the exact form of the Fourier-Mellin representation of a multi-leg amplitude beyond LLA, it is reasonable to believe that this representation will involve a Fourier-Mellin transform very similar to eq.~\eqref{eq:MRK_conjecture}. The main difference will be the appearance of higher-order corrections to the BFKL eigenvalue, the impact factor and the central emission block. Currently, the BFKL eigenvalue and the impact factor are known to arbitrary order loop from integrability~\cite{Fadin:2011we,Bartels:2008sc,Lipatov:2010qg,Dixon:2012yy,Basso:2014pla}, but the central emission block is only known to LO. The higher-order corrections will give rise to new building blocks that enter the convolution integrals. Once these Fourier-Mellin transforms of building blocks are known, we can increase the loop order at fixed logarithmic accuracy by convoluting with the LO BFKL eigenvalue, just like at LLA. Similarly, we can relate non-MHV amplitudes to MHV amplitudes via helicity flip operators. We emphasise that the helicity flip operators receive quantum corrections beyond leading logarithmic accuracy that are known to all orders~\cite{Basso:2014pla,Dixon:2014iba}. 

The factorisation theorem~\eqref{eq:fac_thm} will, however, break down beyond LLA, at least in its simple form quoted in this paper. Indeed, in ref.~\cite{Bargheer:2015djt} it was observed that unlike at LLA amplitudes at NLLA receive non-factorisable contributions that depend on two simplicial MRK coordinates simultaneously already at two loops. The breaking of the factorisation theorem in its present form at NLLA can be traced back to the following: Quantum corrections to the BFKL eigenvalue and the impact factor contribute already at six points, and so they will not violate the factorisation. The NLO correction to the central emission block, however, couples two Fourier-Mellin integrations, which leads to a violation of eq.~\eqref{eq:fac_thm}. Indeed, the proof of Claim~\ref{claim1} relies crucially on the analytic structure of the LO central emission block, see eq.~\eqref{eq:proof_claim1}. The NLO correction will alter eq.~\eqref{eq:proof_claim1}, thereby invalidating the proof of Claim~\ref{claim1} and thus the factorisation theorem. At the same time, we see that the violation is restricted to the appearance of the NLO corrections to the central emission block, leading to the more general factorisation observed in ref.~\cite{Bargheer:2015djt}. It would be interesting to study these effects to higher orders. Along the same lines, it would be interesting to see if the structure of the leading singularities discussed in Section~\ref{sec:LS} will change beyond LLA. Indeed, the analysis of Section~\ref{sec:LS} took into account the BFKL eigenvalues and helicity flip operators only at LO, so they results of that section may change beyond LLA.

Finally, let us comment on the validity of the analysis of Section~\ref{sec:MPL_proof} beyond LLA. The analysis of Section~\ref{sec:MPL_proof} relies on the fact that, loosely speaking, the space of single-valued iterated integrals on $\mathfrak{M}_{0,n}$ is closed under convolutions due to Stokes' theorem. Since at fixed logarithmic accuracy every amplitude can be written as a convolution of a small set of building blocks, the analysis of Section~\ref{sec:MPL_proof} will fail if at some order in perturbation theory the Fourier-Mellin transform of the BFKL eigenvalue, impact factor, central emission block or helicity-flip operator cannot be expressed in terms of single-valued multiple polylogarithms. From integrability we know that (at least empirically to very high orders) the BFKL eigenvalue, impact factor and helicity-flip operator can be written at every order as a polynomial in the following four building blocks~\cite{Dixon:2012yy,Basso:2014pla}
\beq
E_{\nu n}\,,\quad D_{\nu} = -i\frac{\partial}{\partial \nu}\,,\quad N = \chi^+-\chi^-\,,\quad V = -\frac{1}{2}\left(\chi^++\chi^-\right)\,.
\eeq
Fourier-Mellin transforms of functions that are polynomials in these variables lead to single-valued harmonic polylogarithms~\cite{Dixon:2012yy}, and so there is a strong indication that the Fourier-Mellin transforms of the BFKL eigenvalue, impact factor and helicity-flip operator give rise to single-valued polylogarithms to every order in perturbation theory. Any appearance of functions in MRK that are not polylogarithmic should thus be associated to higher-order corrections to the central emission block. An explicit computation of higher-order corrections to the central emission block is currently under investigation.

\section*{Acknowledgements}
We thank Lance Dixon for discussions. VDD and CD are grateful for the hospitality of the KITP, Santa Barbara, where part of this work was carried out. RM and BV acknowledge the hospitality of the CERN TH department. FD is grateful for the hospitality of ETH Zurich. This work was supported by the ERC grant ``MathAm'', the ERC grant ``IQFT", and the US Department of Energy, contract DE-AC02-76SF00515.

\bibliographystyle{JHEP}
\bibliography{refs}

\providecommand{\href}[2]{#2}\begingroup\raggedright\begin{thebibliography}{100}

\bibitem{Drummond:2006rz}
J.~M. Drummond, J.~Henn, V.~A. Smirnov, and E.~Sokatchev, {\it {Magic
  identities for conformal four-point integrals}},  {\em JHEP} {\bf 01} (2007)
  064, [\href{http://xxx.lanl.gov/abs/hep-th/0607160}{{\tt hep-th/0607160}}].

\bibitem{Bern:2006ew}
Z.~Bern, M.~Czakon, L.~J. Dixon, D.~A. Kosower, and V.~A. Smirnov, {\it {The
  Four-Loop Planar Amplitude and Cusp Anomalous Dimension in Maximally
  Supersymmetric Yang-Mills Theory}},  {\em Phys. Rev.} {\bf D75} (2007)
  085010, [\href{http://xxx.lanl.gov/abs/hep-th/0610248}{{\tt
  hep-th/0610248}}].

\bibitem{Bern:2007ct}
Z.~Bern, J.~J.~M. Carrasco, H.~Johansson, and D.~A. Kosower, {\it {Maximally
  supersymmetric planar Yang-Mills amplitudes at five loops}},  {\em Phys.
  Rev.} {\bf D76} (2007) 125020, [\href{http://xxx.lanl.gov/abs/0705.1864}{{\tt
  0705.1864}}].

\bibitem{Alday:2007hr}
L.~F. Alday and J.~M. Maldacena, {\it {Gluon scattering amplitudes at strong
  coupling}},  {\em JHEP} {\bf 06} (2007) 064,
  [\href{http://xxx.lanl.gov/abs/0705.0303}{{\tt 0705.0303}}].

\bibitem{Drummond:2007aua}
J.~M. Drummond, G.~P. Korchemsky, and E.~Sokatchev, {\it {Conformal properties
  of four-gluon planar amplitudes and Wilson loops}},  {\em Nucl. Phys.} {\bf
  B795} (2008) 385--408, [\href{http://xxx.lanl.gov/abs/0707.0243}{{\tt
  0707.0243}}].

\bibitem{Brandhuber:2007yx}
A.~Brandhuber, P.~Heslop, and G.~Travaglini, {\it {MHV amplitudes in N=4 super
  Yang-Mills and Wilson loops}},  {\em Nucl. Phys.} {\bf B794} (2008) 231--243,
  [\href{http://xxx.lanl.gov/abs/0707.1153}{{\tt 0707.1153}}].

\bibitem{Drummond:2009fd}
J.~M. Drummond, J.~M. Henn, and J.~Plefka, {\it {Yangian symmetry of scattering
  amplitudes in N=4 super Yang-Mills theory}},  {\em JHEP} {\bf 05} (2009) 046,
  [\href{http://xxx.lanl.gov/abs/0902.2987}{{\tt 0902.2987}}].

\bibitem{Drummond:2007cf}
J.~M. Drummond, J.~Henn, G.~P. Korchemsky, and E.~Sokatchev, {\it {On planar
  gluon amplitudes/Wilson loops duality}},  {\em Nucl. Phys.} {\bf B795} (2008)
  52--68, [\href{http://xxx.lanl.gov/abs/0709.2368}{{\tt 0709.2368}}].

\bibitem{Drummond:2007au}
J.~M. Drummond, J.~Henn, G.~P. Korchemsky, and E.~Sokatchev, {\it {Conformal
  Ward identities for Wilson loops and a test of the duality with gluon
  amplitudes}},  {\em Nucl. Phys.} {\bf B826} (2010) 337--364,
  [\href{http://xxx.lanl.gov/abs/0712.1223}{{\tt 0712.1223}}].

\bibitem{Bern:2005iz}
Z.~Bern, L.~J. Dixon, and V.~A. Smirnov, {\it {Iteration of planar amplitudes
  in maximally supersymmetric Yang-Mills theory at three loops and beyond}},
  {\em Phys. Rev.} {\bf D72} (2005) 085001,
  [\href{http://xxx.lanl.gov/abs/hep-th/0505205}{{\tt hep-th/0505205}}].

\bibitem{Beisert:2006ez}
N.~Beisert, B.~Eden, and M.~Staudacher, {\it {Transcendentality and Crossing}},
   {\em J. Stat. Mech.} {\bf 0701} (2007) P01021,
  [\href{http://xxx.lanl.gov/abs/hep-th/0610251}{{\tt hep-th/0610251}}].

\bibitem{Bern:2008ap}
Z.~Bern, L.~J. Dixon, D.~A. Kosower, R.~Roiban, M.~Spradlin, C.~Vergu, and
  A.~Volovich, {\it {The Two-Loop Six-Gluon MHV Amplitude in Maximally
  Supersymmetric Yang-Mills Theory}},  {\em Phys. Rev.} {\bf D78} (2008)
  045007, [\href{http://xxx.lanl.gov/abs/0803.1465}{{\tt 0803.1465}}].

\bibitem{Drummond:2008aq}
J.~M. Drummond, J.~Henn, G.~P. Korchemsky, and E.~Sokatchev, {\it {Hexagon
  Wilson loop = six-gluon MHV amplitude}},  {\em Nucl. Phys.} {\bf B815} (2009)
  142--173, [\href{http://xxx.lanl.gov/abs/0803.1466}{{\tt 0803.1466}}].

\bibitem{Alday:2007he}
L.~F. Alday and J.~Maldacena, {\it {Comments on gluon scattering amplitudes via
  AdS/CFT}},  {\em JHEP} {\bf 11} (2007) 068,
  [\href{http://xxx.lanl.gov/abs/0710.1060}{{\tt 0710.1060}}].

\bibitem{Berkovits:2008ic}
N.~Berkovits and J.~Maldacena, {\it {Fermionic T-Duality, Dual Superconformal
  Symmetry, and the Amplitude/Wilson Loop Connection}},  {\em JHEP} {\bf 09}
  (2008) 062, [\href{http://xxx.lanl.gov/abs/0807.3196}{{\tt 0807.3196}}].

\bibitem{CaronHuot:2010ek}
S.~Caron-Huot, {\it {Notes on the scattering amplitude / Wilson loop duality}},
   {\em JHEP} {\bf 07} (2011) 058,
  [\href{http://xxx.lanl.gov/abs/1010.1167}{{\tt 1010.1167}}].

\bibitem{Mason:2010yk}
L.~J. Mason and D.~Skinner, {\it {The Complete Planar S-matrix of N=4 SYM as a
  Wilson Loop in Twistor Space}},  {\em JHEP} {\bf 12} (2010) 018,
  [\href{http://xxx.lanl.gov/abs/1009.2225}{{\tt 1009.2225}}].

\bibitem{Alday:2010ku}
L.~F. Alday, D.~Gaiotto, J.~Maldacena, A.~Sever, and P.~Vieira, {\it {An
  Operator Product Expansion for Polygonal null Wilson Loops}},  {\em JHEP}
  {\bf 04} (2011) 088, [\href{http://xxx.lanl.gov/abs/1006.2788}{{\tt
  1006.2788}}].

\bibitem{Gaiotto:2010fk}
D.~Gaiotto, J.~Maldacena, A.~Sever, and P.~Vieira, {\it {Bootstrapping Null
  Polygon Wilson Loops}},  {\em JHEP} {\bf 03} (2011) 092,
  [\href{http://xxx.lanl.gov/abs/1010.5009}{{\tt 1010.5009}}].

\bibitem{Gaiotto:2011dt}
D.~Gaiotto, J.~Maldacena, A.~Sever, and P.~Vieira, {\it {Pulling the straps of
  polygons}},  {\em JHEP} {\bf 12} (2011) 011,
  [\href{http://xxx.lanl.gov/abs/1102.0062}{{\tt 1102.0062}}].

\bibitem{Sever:2011da}
A.~Sever, P.~Vieira, and T.~Wang, {\it {OPE for Super Loops}},  {\em JHEP} {\bf
  11} (2011) 051, [\href{http://xxx.lanl.gov/abs/1108.1575}{{\tt 1108.1575}}].

\bibitem{Basso:2010in}
B.~Basso, {\it {Exciting the GKP string at any coupling}},  {\em Nucl. Phys.}
  {\bf B857} (2012) 254--334, [\href{http://xxx.lanl.gov/abs/1010.5237}{{\tt
  1010.5237}}].

\bibitem{Basso:2013vsa}
B.~Basso, A.~Sever, and P.~Vieira, {\it {Spacetime and Flux Tube S-Matrices at
  Finite Coupling for N=4 Supersymmetric Yang-Mills Theory}},  {\em Phys. Rev.
  Lett.} {\bf 111} (2013), no.~9 091602,
  [\href{http://xxx.lanl.gov/abs/1303.1396}{{\tt 1303.1396}}].

\bibitem{Basso:2013aha}
B.~Basso, A.~Sever, and P.~Vieira, {\it {Space-time S-matrix and Flux tube
  S-matrix II. Extracting and Matching Data}},  {\em JHEP} {\bf 01} (2014) 008,
  [\href{http://xxx.lanl.gov/abs/1306.2058}{{\tt 1306.2058}}].

\bibitem{Basso:2014koa}
B.~Basso, A.~Sever, and P.~Vieira, {\it {Space-time S-matrix and Flux-tube
  S-matrix III. The two-particle contributions}},  {\em JHEP} {\bf 08} (2014)
  085, [\href{http://xxx.lanl.gov/abs/1402.3307}{{\tt 1402.3307}}].

\bibitem{Basso:2014jfa}
B.~Basso, A.~Sever, and P.~Vieira, {\it {Collinear Limit of Scattering
  Amplitudes at Strong Coupling}},  {\em Phys. Rev. Lett.} {\bf 113} (2014),
  no.~26 261604, [\href{http://xxx.lanl.gov/abs/1405.6350}{{\tt 1405.6350}}].

\bibitem{Basso:2014nra}
B.~Basso, A.~Sever, and P.~Vieira, {\it {Space-time S-matrix and Flux-tube
  S-matrix IV. Gluons and Fusion}},  {\em JHEP} {\bf 09} (2014) 149,
  [\href{http://xxx.lanl.gov/abs/1407.1736}{{\tt 1407.1736}}].

\bibitem{Basso:2014hfa}
B.~Basso, J.~Caetano, L.~Cordova, A.~Sever, and P.~Vieira, {\it {OPE for all
  Helicity Amplitudes}},  {\em JHEP} {\bf 08} (2015) 018,
  [\href{http://xxx.lanl.gov/abs/1412.1132}{{\tt 1412.1132}}].

\bibitem{Basso:2015rta}
B.~Basso, J.~Caetano, L.~Cordova, A.~Sever, and P.~Vieira, {\it {OPE for all
  Helicity Amplitudes II. Form Factors and Data analysis}},  {\em JHEP} {\bf
  12} (2015) 088, [\href{http://xxx.lanl.gov/abs/1508.02987}{{\tt
  1508.02987}}].

\bibitem{Basso:2015uxa}
B.~Basso, A.~Sever, and P.~Vieira, {\it {Hexagonal Wilson Loops in Planar
  $\mathcal{N}=4$ SYM Theory at Finite Coupling}},  {\em {}} (2015)
  [\href{http://xxx.lanl.gov/abs/1508.03045}{{\tt 1508.03045}}].

\bibitem{Hodges:2009hk}
A.~Hodges, {\it {Eliminating spurious poles from gauge-theoretic amplitudes}},
  {\em JHEP} {\bf 05} (2013) 135,
  [\href{http://xxx.lanl.gov/abs/0905.1473}{{\tt 0905.1473}}].

\bibitem{Golden:2013xva}
J.~Golden, A.~B. Goncharov, M.~Spradlin, C.~Vergu, and A.~Volovich, {\it
  {Motivic Amplitudes and Cluster Coordinates}},  {\em JHEP} {\bf 01} (2014)
  091, [\href{http://xxx.lanl.gov/abs/1305.1617}{{\tt 1305.1617}}].

\bibitem{symbolsC}
K.~T. Chen, {\it Iterated path integrals},  {\em Bull.\ Amer.\ Math.\ Soc.}
  {\bf 83} (1977) 831.

\bibitem{Golden:2014xqa}
J.~Golden, M.~F. Paulos, M.~Spradlin, and A.~Volovich, {\it {Cluster
  Polylogarithms for Scattering Amplitudes}},  {\em J. Phys.} {\bf A47} (2014),
  no.~47 474005, [\href{http://xxx.lanl.gov/abs/1401.6446}{{\tt 1401.6446}}].

\bibitem{cluster1}
S.~Fomin and A.~Zelevinsky, {\it Cluster algebras. i: Foundations},  {\em J.
  Am. Math. Soc.} {\bf 15} (2002), no.~2 497--529.

\bibitem{cluster2}
S.~Fomin and A.~Zelevinsky, {\it Cluster algebras. ii: Finite type
  classification},  {\em Invent. Math.} {\bf 154} (2003), no.~1 63--121.

\bibitem{scott}
J.~S. Scott, {\it Grassmannians and cluster algebras},  {\em Adv. in Appl.
  Math.} {\bf 28} (2002), no.~2 119--144.

\bibitem{gekhtman}
M.~Gekhtman, M.~Shapiro, and A.~Vainshtein, {\it Cluster algebras and poisson
  geometry},  {\em Mosc. Math. J.} {\bf 3} (2003), no.~3 899--934.

\bibitem{keller}
B.~Keller, {\it Cluster algebras, quiver representations and triangulated
  categories},  in {\em Triangulated Categories}.
\newblock Cambridge University Press, 2003.

\bibitem{Goncharov:2001}
A.~B. Goncharov, {\it {Multiple polylogarithms and mixed Tate motives}},  {\em
  {}} (2001) [\href{http://xxx.lanl.gov/abs/math/0103059v4}{{\tt
  math/0103059v4}}].

\bibitem{Brown:2009qja}
F.~C. Brown, {\it {Multiple zeta values and periods of moduli spaces
  $\mathcal{M}_{0,n}(\mathbb{R})$}},  {\em Annales Sci.Ecole Norm.Sup.} {\bf
  42} (2009) 371, [\href{http://xxx.lanl.gov/abs/math/0606419}{{\tt
  math/0606419}}].

\bibitem{ArkaniHamed:2012nw}
N.~Arkani-Hamed, J.~L. Bourjaily, F.~Cachazo, A.~B. Goncharov, A.~Postnikov,
  and J.~Trnka, {\em {Scattering Amplitudes and the Positive Grassmannian}}.
\newblock Cambridge University Press, 2012.

\bibitem{DelDuca:2009au}
V.~Del~Duca, C.~Duhr, and V.~A. Smirnov, {\it {An Analytic Result for the
  Two-Loop Hexagon Wilson Loop in N = 4 SYM}},  {\em JHEP} {\bf 03} (2010) 099,
  [\href{http://xxx.lanl.gov/abs/0911.5332}{{\tt 0911.5332}}].

\bibitem{DelDuca:2010zg}
V.~Del~Duca, C.~Duhr, and V.~A. Smirnov, {\it {The Two-Loop Hexagon Wilson Loop
  in N = 4 SYM}},  {\em JHEP} {\bf 05} (2010) 084,
  [\href{http://xxx.lanl.gov/abs/1003.1702}{{\tt 1003.1702}}].

\bibitem{Goncharov:2010jf}
A.~B. Goncharov, M.~Spradlin, C.~Vergu, and A.~Volovich, {\it {Classical
  polylogarithms for amplitudes and Wilson loops}},  {\em Phys.Rev.Lett.} {\bf
  105} (2010) 151605, [\href{http://xxx.lanl.gov/abs/1006.5703}{{\tt
  1006.5703}}].

\bibitem{Dixon:2011pw}
L.~J. Dixon, J.~M. Drummond, and J.~M. Henn, {\it {Bootstrapping the three-loop
  hexagon}},  {\em JHEP} {\bf 11} (2011) 023,
  [\href{http://xxx.lanl.gov/abs/1108.4461}{{\tt 1108.4461}}].

\bibitem{Dixon:2011nj}
L.~J. Dixon, J.~M. Drummond, and J.~M. Henn, {\it {Analytic result for the
  two-loop six-point NMHV amplitude in N=4 super Yang-Mills theory}},  {\em
  JHEP} {\bf 01} (2012) 024, [\href{http://xxx.lanl.gov/abs/1111.1704}{{\tt
  1111.1704}}].

\bibitem{Dixon:2013eka}
L.~J. Dixon, J.~M. Drummond, M.~von Hippel, and J.~Pennington, {\it {Hexagon
  functions and the three-loop remainder function}},  {\em JHEP} {\bf 12}
  (2013) 049, [\href{http://xxx.lanl.gov/abs/1308.2276}{{\tt 1308.2276}}].

\bibitem{Dixon:2014iba}
L.~J. Dixon and M.~von Hippel, {\it {Bootstrapping an NMHV amplitude through
  three loops}},  {\em JHEP} {\bf 10} (2014) 065,
  [\href{http://xxx.lanl.gov/abs/1408.1505}{{\tt 1408.1505}}].

\bibitem{Dixon:2014voa}
L.~J. Dixon, J.~M. Drummond, C.~Duhr, and J.~Pennington, {\it {The four-loop
  remainder function and multi-Regge behavior at NNLLA in planar N = 4
  super-Yang-Mills theory}},  {\em JHEP} {\bf 06} (2014) 116,
  [\href{http://xxx.lanl.gov/abs/1402.3300}{{\tt 1402.3300}}].

\bibitem{Dixon:2015iva}
L.~J. Dixon, M.~von Hippel, and A.~J. McLeod, {\it {The four-loop six-gluon
  NMHV ratio function}},  {\em JHEP} {\bf 01} (2016) 053,
  [\href{http://xxx.lanl.gov/abs/1509.08127}{{\tt 1509.08127}}].

\bibitem{Golden:2014xqf}
J.~Golden and M.~Spradlin, {\it {An analytic result for the two-loop
  seven-point MHV amplitude in $ \mathcal{N} $ = 4 SYM}},  {\em JHEP} {\bf 08}
  (2014) 154, [\href{http://xxx.lanl.gov/abs/1406.2055}{{\tt 1406.2055}}].

\bibitem{DelDuca:2010zp}
V.~Del~Duca, C.~Duhr, and V.~A. Smirnov, {\it {A Two-Loop Octagon Wilson Loop
  in N = 4 SYM}},  {\em JHEP} {\bf 09} (2010) 015,
  [\href{http://xxx.lanl.gov/abs/1006.4127}{{\tt 1006.4127}}].

\bibitem{Heslop:2010kq}
P.~Heslop and V.~V. Khoze, {\it {Analytic Results for MHV Wilson Loops}},  {\em
  JHEP} {\bf 11} (2010) 035, [\href{http://xxx.lanl.gov/abs/1007.1805}{{\tt
  1007.1805}}].

\bibitem{Caron-Huot:2013vda}
S.~Caron-Huot and S.~He, {\it {Three-loop octagons and $n$-gons in maximally
  supersymmetric Yang-Mills theory}},  {\em JHEP} {\bf 08} (2013) 101,
  [\href{http://xxx.lanl.gov/abs/1305.2781}{{\tt 1305.2781}}].

\bibitem{symbols}
A.~B. Goncharov, {\it {A simple construction of Grassmannian polylogarithms}},
  {\em {}} (2009) [\href{http://xxx.lanl.gov/abs/0908.2238}{{\tt 0908.2238}}].

\bibitem{Duhr:2011zq}
C.~Duhr, H.~Gangl, and J.~R. Rhodes, {\it {From polygons and symbols to
  polylogarithmic functions}},  {\em JHEP} {\bf 1210} (2012) 075,
  [\href{http://xxx.lanl.gov/abs/1110.0458}{{\tt 1110.0458}}].

\bibitem{CaronHuot:2011ky}
S.~Caron-Huot, {\it {Superconformal symmetry and two-loop amplitudes in planar
  N=4 super Yang-Mills}},  {\em JHEP} {\bf 12} (2011) 066,
  [\href{http://xxx.lanl.gov/abs/1105.5606}{{\tt 1105.5606}}].

\bibitem{Drummond:2014ffa}
J.~M. Drummond, G.~Papathanasiou, and M.~Spradlin, {\it {A Symbol of
  Uniqueness: The Cluster Bootstrap for the 3-Loop MHV Heptagon}},  {\em JHEP}
  {\bf 03} (2015) 072, [\href{http://xxx.lanl.gov/abs/1412.3763}{{\tt
  1412.3763}}].

\bibitem{CaronHuot:2012ab}
S.~Caron-Huot and K.~J. Larsen, {\it {Uniqueness of two-loop master contours}},
   {\em JHEP} {\bf 10} (2012) 026,
  [\href{http://xxx.lanl.gov/abs/1205.0801}{{\tt 1205.0801}}].

\bibitem{Kuraev:1976ge}
E.~A. Kuraev, L.~N. Lipatov, and V.~S. Fadin, {\it {Multi-Reggeon processes in
  the Yang-Mills theory}},  {\em Sov. Phys. JETP} {\bf 44} (1976) 443.

\bibitem{Kuraev:1977fs}
E.~A. Kuraev, L.~N. Lipatov, and V.~S. Fadin, {\it {The Pomeranchuk singularity
  in nonabelian gauge theories}},  {\em Sov. Phys. JETP} {\bf 45} (1977) 199.

\bibitem{Balitsky:1978ic}
I.~I. Balitsky and L.~N. Lipatov, {\it {The Pomeranchuk singularity in quantum
  chromodynamics}},  {\em Sov. J. Nucl. Phys.} {\bf 28} (1978) 822.

\bibitem{Fadin:1998py}
V.~S. Fadin and L.~N. Lipatov, {\it {BFKL pomeron in the next-to-leading
  approximation}},  {\em Phys. Lett.} {\bf B429} (1998) 127,
  [\href{http://xxx.lanl.gov/abs/hep-ph/9802290}{{\tt hep-ph/9802290}}].

\bibitem{Camici:1997ij}
G.~Camici and M.~Ciafaloni, {\it {Irreducible part of the next-to-leading BFKL
  kernel}},  {\em Phys. Lett.} {\bf B412} (1997) 396,
  [\href{http://xxx.lanl.gov/abs/hep-ph/9707390}{{\tt hep-ph/9707390}}].

\bibitem{Ciafaloni:1998gs}
M.~Ciafaloni and G.~Camici, {\it {Energy scale(s) and next-to-leading BFKL
  equation}},  {\em Phys. Lett.} {\bf B430} (1998) 349,
  [\href{http://xxx.lanl.gov/abs/hep-ph/9803389}{{\tt hep-ph/9803389}}].

\bibitem{Bartels:2008ce}
J.~Bartels, L.~N. Lipatov, and A.~Sabio~Vera, {\it {BFKL Pomeron, Reggeized
  gluons and Bern-Dixon-Smirnov amplitudes}},  {\em Phys. Rev.} {\bf D80}
  (2009) 045002, [\href{http://xxx.lanl.gov/abs/0802.2065}{{\tt 0802.2065}}].

\bibitem{Bartels:2008sc}
J.~Bartels, L.~N. Lipatov, and A.~Sabio~Vera, {\it {N=4 supersymmetric Yang
  Mills scattering amplitudes at high energies: The Regge cut contribution}},
  {\em Eur. Phys. J.} {\bf C65} (2010) 587--605,
  [\href{http://xxx.lanl.gov/abs/0807.0894}{{\tt 0807.0894}}].

\bibitem{Brower:2008nm}
R.~C. Brower, H.~Nastase, H.~J. Schnitzer, and C.-I. Tan, {\it {Implications of
  multi-Regge limits for the Bern-Dixon-Smirnov conjecture}},  {\em Nucl.
  Phys.} {\bf B814} (2009) 293--326,
  [\href{http://xxx.lanl.gov/abs/0801.3891}{{\tt 0801.3891}}].

\bibitem{Brower:2008ia}
R.~C. Brower, H.~Nastase, H.~J. Schnitzer, and C.-I. Tan, {\it {Analyticity for
  Multi-Regge Limits of the Bern-Dixon-Smirnov Amplitudes}},  {\em Nucl. Phys.}
  {\bf B822} (2009) 301--347, [\href{http://xxx.lanl.gov/abs/0809.1632}{{\tt
  0809.1632}}].

\bibitem{DelDuca:2008jg}
V.~Del~Duca, C.~Duhr, and E.~W.~N. Glover, {\it {Iterated amplitudes in the
  high-energy limit}},  {\em JHEP} {\bf 12} (2008) 097,
  [\href{http://xxx.lanl.gov/abs/0809.1822}{{\tt 0809.1822}}].

\bibitem{Basso:2014pla}
B.~Basso, S.~Caron-Huot, and A.~Sever, {\it {Adjoint BFKL at finite coupling: a
  short-cut from the collinear limit}},  {\em JHEP} {\bf 01} (2015) 027,
  [\href{http://xxx.lanl.gov/abs/1407.3766}{{\tt 1407.3766}}].

\bibitem{Drummond:2015jea}
J.~M. Drummond and G.~Papathanasiou, {\it {Hexagon OPE Resummation and
  Multi-Regge Kinematics}},  {\em JHEP} {\bf 02} (2016) 185,
  [\href{http://xxx.lanl.gov/abs/1507.08982}{{\tt 1507.08982}}].

\bibitem{Bartels:2010ej}
J.~Bartels, J.~Kotanski, and V.~Schomerus, {\it {Excited Hexagon Wilson Loops
  for Strongly Coupled N=4 SYM}},  {\em JHEP} {\bf 01} (2011) 096,
  [\href{http://xxx.lanl.gov/abs/1009.3938}{{\tt 1009.3938}}].

\bibitem{Bartels:2013dja}
J.~Bartels, J.~Kotanski, V.~Schomerus, and M.~Sprenger, {\it {The Excited
  Hexagon Reloaded}},  {\em {}} (2013)
  [\href{http://xxx.lanl.gov/abs/1311.1512}{{\tt 1311.1512}}].

\bibitem{Lipatov:2010ad}
L.~N. Lipatov and A.~Prygarin, {\it {BFKL approach and six-particle MHV
  amplitude in N=4 super Yang-Mills}},  {\em Phys. Rev.} {\bf D83} (2011)
  125001, [\href{http://xxx.lanl.gov/abs/1011.2673}{{\tt 1011.2673}}].

\bibitem{Lipatov:2010qg}
L.~N. Lipatov and A.~Prygarin, {\it {Mandelstam cuts and light-like Wilson
  loops in N=4 SUSY}},  {\em Phys. Rev.} {\bf D83} (2011) 045020,
  [\href{http://xxx.lanl.gov/abs/1008.1016}{{\tt 1008.1016}}].

\bibitem{Lipatov:2012gk}
L.~Lipatov, A.~Prygarin, and H.~J. Schnitzer, {\it {The Multi-Regge limit of
  NMHV Amplitudes in N=4 SYM Theory}},  {\em JHEP} {\bf 01} (2013) 068,
  [\href{http://xxx.lanl.gov/abs/1205.0186}{{\tt 1205.0186}}].

\bibitem{Dixon:2012yy}
L.~J. Dixon, C.~Duhr, and J.~Pennington, {\it {Single-valued harmonic
  polylogarithms and the multi-Regge limit}},  {\em JHEP} {\bf 1210} (2012)
  074, [\href{http://xxx.lanl.gov/abs/1207.0186}{{\tt 1207.0186}}].

\bibitem{Pennington:2012zj}
J.~Pennington, {\it {The six-point remainder function to all loop orders in the
  multi-Regge limit}},  {\em JHEP} {\bf 1301} (2013) 059,
  [\href{http://xxx.lanl.gov/abs/1209.5357}{{\tt 1209.5357}}].

\bibitem{Broedel:2015nfp}
J.~Broedel and M.~Sprenger, {\it {Six-point remainder function in
  multi-Regge-kinematics: an efficient approach in momentum space}},  {\em
  JHEP} {\bf 05} (2016) 055, [\href{http://xxx.lanl.gov/abs/1512.04963}{{\tt
  1512.04963}}].

\bibitem{BrownSVHPLs}
F.~C.~S. Brown, {\it Single-valued multiple polylogarithms in one variable},
  {\em C. R. Acad. Sci. Paris, Ser. I} {\bf 338} (2004) 527.

\bibitem{Prygarin:2011gd}
A.~Prygarin, M.~Spradlin, C.~Vergu, and A.~Volovich, {\it {All Two-Loop MHV
  Amplitudes in Multi-Regge Kinematics From Applied Symbology}},  {\em Phys.
  Rev.} {\bf D85} (2012) 085019, [\href{http://xxx.lanl.gov/abs/1112.6365}{{\tt
  1112.6365}}].

\bibitem{Bartels:2011ge}
J.~Bartels, A.~Kormilitzin, L.~N. Lipatov, and A.~Prygarin, {\it {BFKL approach
  and $2 \to 5$ maximally helicity violating amplitude in ${\cal N}=4$
  super-Yang-Mills theory}},  {\em Phys. Rev.} {\bf D86} (2012) 065026,
  [\href{http://xxx.lanl.gov/abs/1112.6366}{{\tt 1112.6366}}].

\bibitem{Bargheer:2015djt}
T.~Bargheer, G.~Papathanasiou, and V.~Schomerus, {\it {The Two-Loop Symbol of
  all Multi-Regge Regions}},  {\em JHEP} {\bf 05} (2016) 012,
  [\href{http://xxx.lanl.gov/abs/1512.07620}{{\tt 1512.07620}}].

\bibitem{Bartels:2012gq}
J.~Bartels, V.~Schomerus, and M.~Sprenger, {\it {Multi-Regge Limit of the
  n-Gluon Bubble Ansatz}},  {\em JHEP} {\bf 11} (2012) 145,
  [\href{http://xxx.lanl.gov/abs/1207.4204}{{\tt 1207.4204}}].

\bibitem{Bartels:2014ppa}
J.~Bartels, V.~Schomerus, and M.~Sprenger, {\it {Heptagon Amplitude in the
  Multi-Regge Regime}},  {\em JHEP} {\bf 10} (2014) 67,
  [\href{http://xxx.lanl.gov/abs/1405.3658}{{\tt 1405.3658}}].

\bibitem{Bartels:2014mka}
J.~Bartels, V.~Schomerus, and M.~Sprenger, {\it {The Bethe roots of Regge cuts
  in strongly coupled $ \mathcal{N}=4 $ SYM theory}},  {\em JHEP} {\bf 07}
  (2015) 098, [\href{http://xxx.lanl.gov/abs/1411.2594}{{\tt 1411.2594}}].

\bibitem{DelDuca:2013lma}
V.~Del~Duca, L.~J. Dixon, C.~Duhr, and J.~Pennington, {\it {The BFKL equation,
  Mueller-Navelet jets and single-valued harmonic polylogarithms}},  {\em JHEP}
  {\bf 02} (2014) 086, [\href{http://xxx.lanl.gov/abs/1309.6647}{{\tt
  1309.6647}}].

\bibitem{Fadin:2011we}
V.~S. Fadin and L.~N. Lipatov, {\it {BFKL equation for the adjoint
  representation of the gauge group in the next-to-leading approximation at N=4
  SUSY}},  {\em Phys. Lett.} {\bf B706} (2012) 470--476,
  [\href{http://xxx.lanl.gov/abs/1111.0782}{{\tt 1111.0782}}].

\bibitem{DelDuca:1995zy}
V.~Del~Duca, {\it {Equivalence of the Parke-Taylor and the Fadin-Kuraev-Lipatov
  amplitudes in the high-energy limit}},  {\em Phys. Rev.} {\bf D52} (1995)
  1527--1534, [\href{http://xxx.lanl.gov/abs/hep-ph/9503340}{{\tt
  hep-ph/9503340}}].

\bibitem{Bartels:2013jna}
J.~Bartels, A.~Kormilitzin, and L.~Lipatov, {\it {Analytic structure of the
  $n=7$ scattering amplitude in $\mathcal{N}=4$ SYM theory in the multi-Regge
  kinematics: Conformal Regge pole contribution}},  {\em Phys. Rev.} {\bf D89}
  (2014), no.~6 065002, [\href{http://xxx.lanl.gov/abs/1311.2061}{{\tt
  1311.2061}}].

\bibitem{Bartels:2014jya}
J.~Bartels, A.~Kormilitzin, and L.~N. Lipatov, {\it {Analytic structure of the
  $n=7$ scattering amplitude in $\mathcal{N}=4$ theory in multi-Regge
  kinematics: Conformal Regge cut contribution}},  {\em Phys. Rev.} {\bf D91}
  (2015), no.~4 045005, [\href{http://xxx.lanl.gov/abs/1411.2294}{{\tt
  1411.2294}}].

\bibitem{Lipatov:1976zz}
L.~N. Lipatov, {\it {Reggeization of the vector meson and the vacuum
  singularity in nonabelian gauge theories}},  {\em Sov. J. Nucl. Phys.} {\bf
  23} (1976) 338.

\bibitem{Fadin:2006bj}
V.~S. Fadin, R.~Fiore, M.~G. Kozlov, and A.~V. Reznichenko, {\it {Proof of the
  multi-Regge form of QCD amplitudes with gluon exchanges in the NLA}},  {\em
  Phys. Lett.} {\bf B639} (2006) 74--81,
  [\href{http://xxx.lanl.gov/abs/hep-ph/0602006}{{\tt hep-ph/0602006}}].

\bibitem{Caron-Huot:2013fea}
S.~Caron-Huot, {\it {When does the gluon reggeize?}},  {\em JHEP} {\bf 05}
  (2015) 093, [\href{http://xxx.lanl.gov/abs/1309.6521}{{\tt 1309.6521}}].

\bibitem{Parker:2015cia}
D.~Parker, A.~Scherlis, M.~Spradlin, and A.~Volovich, {\it {Hedgehog bases for
  A$_{n}$ cluster polylogarithms and an application to six-point amplitudes}},
  {\em JHEP} {\bf 11} (2015) 136,
  [\href{http://xxx.lanl.gov/abs/1507.01950}{{\tt 1507.01950}}].

\bibitem{BrownSVMPLs}
F.~C.~S. Brown, {\it {Single-valued hyperlogarithms and unipotent differential
  equations}},  {\em \verb+http://www.ihes.fr/~brown/RHpaper5.pdf+}.

\bibitem{Brown:2013gia}
F.~Brown, {\it {Single-valued periods and multiple zeta values}},  {\em {}}
  (2013) [\href{http://xxx.lanl.gov/abs/1309.5309}{{\tt 1309.5309}}].

\bibitem{Brown_Notes}
F.~C.~S. Brown, {\it {Notes on motivic periods}},  {\em {}} (2015)
  [\href{http://xxx.lanl.gov/abs/1512.06410}{{\tt 1512.06410}}].

\bibitem{Remiddi:1999ew}
E.~Remiddi and J.~Vermaseren, {\it {Harmonic polylogarithms}},  {\em
  Int.J.Mod.Phys.} {\bf A15} (2000) 725,
  [\href{http://xxx.lanl.gov/abs/hep-ph/9905237}{{\tt hep-ph/9905237}}].

\bibitem{Vermaseren:1998uu}
J.~A.~M. Vermaseren, {\it {Harmonic sums, Mellin transforms and integrals}},
  {\em Int. J. Mod. Phys.} {\bf A14} (1999) 2037--2076,
  [\href{http://xxx.lanl.gov/abs/hep-ph/9806280}{{\tt hep-ph/9806280}}].

\bibitem{Moch:2001zr}
S.~Moch, P.~Uwer, and S.~Weinzierl, {\it {Nested sums, expansion of
  transcendental functions and multiscale multiloop integrals}},  {\em J. Math.
  Phys.} {\bf 43} (2002) 3363--3386,
  [\href{http://xxx.lanl.gov/abs/hep-ph/0110083}{{\tt hep-ph/0110083}}].

\bibitem{Weinzierl:2002hv}
S.~Weinzierl, {\it {Symbolic expansion of transcendental functions}},  {\em
  Comput. Phys. Commun.} {\bf 145} (2002) 357--370,
  [\href{http://xxx.lanl.gov/abs/math-ph/0201011}{{\tt math-ph/0201011}}].

\bibitem{Moch:2005uc}
S.~Moch and P.~Uwer, {\it {XSummer: Transcendental functions and symbolic
  summation in form}},  {\em Comput. Phys. Commun.} {\bf 174} (2006) 759--770,
  [\href{http://xxx.lanl.gov/abs/math-ph/0508008}{{\tt math-ph/0508008}}].

\bibitem{Schnetz:2013hqa}
O.~Schnetz, {\it {Graphical functions and single-valued multiple
  polylogarithms}},  {\em Commun. Num. Theor. Phys.} {\bf 08} (2014) 589--675,
  [\href{http://xxx.lanl.gov/abs/1302.6445}{{\tt 1302.6445}}].

\bibitem{Cachazo:2008vp}
F.~Cachazo, {\it {Sharpening The Leading Singularity}},  {\em {}} (2008)
  [\href{http://xxx.lanl.gov/abs/0803.1988}{{\tt 0803.1988}}].

\bibitem{CaronHuot:2011kk}
S.~Caron-Huot and S.~He, {\it {Jumpstarting the All-Loop S-Matrix of Planar N=4
  Super Yang-Mills}},  {\em JHEP} {\bf 07} (2012) 174,
  [\href{http://xxx.lanl.gov/abs/1112.1060}{{\tt 1112.1060}}].

\bibitem{Drummond:2008vq}
J.~M. Drummond, J.~Henn, G.~P. Korchemsky, and E.~Sokatchev, {\it {Dual
  superconformal symmetry of scattering amplitudes in N=4 super-Yang-Mills
  theory}},  {\em Nucl. Phys.} {\bf B828} (2010) 317--374,
  [\href{http://xxx.lanl.gov/abs/0807.1095}{{\tt 0807.1095}}].

\bibitem{Nandan:2013ip}
D.~Nandan, M.~F. Paulos, M.~Spradlin, and A.~Volovich, {\it {Star Integrals,
  Convolutions and Simplices}},  {\em JHEP} {\bf 05} (2013) 105,
  [\href{http://xxx.lanl.gov/abs/1301.2500}{{\tt 1301.2500}}].

\bibitem{Papathanasiou:2013uoa}
G.~Papathanasiou, {\it {Hexagon Wilson Loop OPE and Harmonic Polylogarithms}},
  {\em JHEP} {\bf 11} (2013) 150,
  [\href{http://xxx.lanl.gov/abs/1310.5735}{{\tt 1310.5735}}].

\end{thebibliography}\endgroup

\appendix


\section{Fourier-Mellin transforms and convolutions}
\label{app:fourier_mellin}
In this appendix we prove the inversion formula and the convolution theorem for the Fourier-Mellin transform of eq.~\eqref{eq:fourier_mellin_def}. We start by proving the formula for the inverse transform, eq.~
\eqref{eq:inv_FM}.
If $F(\nu,n)=\cF^{-1}[f(z)]$, we have,
\beq\bsp
\cF[F(\nu,n)] &\,= \sum_{n=-\infty}^{\infty}\left(\frac{z}{\zb}\right)^{n/2}\,\int_{-\infty}^{+\infty}\frac{d\nu}{2\pi}\,|z|^{2i\nu}\,F(\nu,n)\\
&\,= \int\frac{d^2w}{\pi \,|w|^2}\,f(w)\,\sum_{n=-\infty}^{+\infty}\left(\frac{z\,\wb}{\zb\,w}\right)^{n/2}\,\int_{-\infty}^{+\infty}\frac{d\nu}{2\pi}\,\left|\frac{z}{w}\right|^{2i\nu}\,.
\esp\eeq
In polar coordinates we have $w=r\,e^{i\varphi}$ and $z=r_0\,e^{i\varphi_0}$,
\beq\bsp
\cF[F(\nu,n)] &\,= \int_0^{\infty}\frac{dr}{r}\,\int_0^{2\pi}\frac{d\varphi}{\pi}\,f(r,\varphi)\,\sum_{n=-\infty}^{+\infty}e^{-in(\varphi-\varphi_0)}\,\int_{-\infty}^{+\infty}\frac{d\nu}{2\pi}\,\left(\frac{r_0}{r}\right)^{2i\nu}\,.
\esp\eeq
The sum and integral over $n$ and $\nu$ can now be performed in terms of $\delta$ functions. Indeed, the sum over $n$ is just a Fourier sum,
\beq
\sum_{n=-\infty}^{+\infty}e^{-in(\varphi-\varphi_0)} = 2\pi\,\delta(\varphi-\varphi_0)\,.
\eeq
Similarly, letting $\nu=i\tilde\nu$, we get
\beq
\int_{-\infty}^{+\infty}\frac{d\nu}{2\pi}\,\left(\frac{r_0}{r}\right)^{2i\nu} = \int_{-i\infty}^{+i\infty}\frac{d\tilde\nu}{2\pi i}\,\left(\frac{r}{r_0}\right)^{2\tilde\nu}
=\delta\left(1-\left(\frac{r}{r_0}\right)^{2}\right) = \frac{r_0}{2}\,\delta(r-r_0)\,.
\eeq
Hence,
\beq\bsp
\cF[F(\nu,n)] &\,= \int_0^{\infty}\frac{dr}{r}\,\int_0^{2\pi}\frac{d\varphi}{\pi}\,f(r,\varphi)\,2\pi\,\delta(\varphi-\varphi_0)\,\frac{r_0}{2}\,\delta(r-r_0) = f(z)\,.
\esp\eeq

Next, we prove the convolution theorem~\eqref{eq:conv_thm} for the Fourier-Mellin transform. We have
\begin{align}
\cF&[F\cdot G] = \sum_{n=-\infty}^{\infty}\left(\frac{z}{\zb}\right)^{n/2}\,\int_{-\infty}^{+\infty}\frac{d\nu}{2\pi}\,|z|^{2i\nu}\,F(\nu,n)\,G(\nu,n)\,,\\
\nonumber&\,= \frac{1}{\pi^2}\,\int\frac{d^2w_1\,d^2w_2}{|w_1|^2|w_2|^2}\,f(w_1)\,g(w_2)\,\sum_{n=-\infty}^{\infty}\left(\frac{z\,\wb_1\,\wb_2}{\zb\,w_1\,w_2}\right)^{n/2}\,\int_{-\infty}^{+\infty}\frac{d\nu}{2\pi}\,\left|\frac{z}{w_1w_2}\right|^{2i\nu}\\
\nonumber&\,= \frac{1}{\pi^2}\,\int\frac{d^2w_1\,d^2w_2}{|w_1|^2|w_2|^2}\,f(w_1)\,g(w_2)\,2\pi\,\delta(\varphi_1+\varphi_2-\varphi_0)\,\delta\left(1-\left(\frac{r_0}{r_1r_2}\right)^{2}\right)\\
\nonumber&\,= \frac{2}{\pi}\,\int_0^\infty\frac{dr_1\,dr_2}{r_1\,r_2}\,\int_0^{2\pi}d\varphi_1\,d\varphi_2\,f\left(r_1e^{i\varphi_1}\right)\,g\left(r_2e^{i\varphi_2}\right)\,\delta(\varphi_1+\varphi_2-\varphi_0)\,\frac{r_0}{2r_1}\,\delta\left(r_2-\frac{r_0}{r_1}\right)\\
\nonumber&\,= \frac{1}{\pi}\,\int_0^\infty\frac{dr_1}{r_1}\,\int_0^{2\pi}d\varphi_1\,f\left(r_1e^{i\varphi_1}\right)\,g\left(\frac{r_0}{r_1}e^{i(\varphi_0-\varphi_1)}\right)\\
\nonumber&\,= \frac{1}{\pi}\,\int\frac{d^2w_1}{|w_1|^2}f\left(w_1\right)\,g\left(\frac{z}{w_1}\right)\\
\nonumber&\,=(f\ast g)(z)\,.
\end{align}


\section{Details on the algebraic construction of single-valued functions}
\label{app:hopf}

In this appendix we present some background material to Section~\ref{sec:period}. We show that it is possible to define analogues of the single-valued map $\bfs$ in a purely algebraic way, without any reference to polylogarithms. This shows that the construction of Section~\ref{sec:period} is purely combinatorial and follows directly from the Hopf algebra structure on hyperlogarithms. In addition, we present some proofs that have been omitted in Section~\ref{sec:period}.

Consider two graded and connected Hopf algebras $(\mathcal{H}_1,\mu_1,\Delta_1,S_1)$ and $(\mathcal{H}_2,\mu_2,\Delta_2,S_2)$, each equipped with their own multiplication $\mu_i$, coproduct $\Delta_i$ and antipode $S_i$, and assume that they are isomorphic via some isomorphism $\phi:\mathcal{H}_1 \rightarrow \mathcal{H}_2$. This implies that
\beq
\Delta_2\phi = (\phi\otimes\phi)\Delta_1\,,\quad \phi\mu_1 = \mu_2(\phi\otimes\phi)\,,\quad S_2\phi = \phi S_1\,.
\eeq
The tensor product $\cH_1\otimes\cH_2$ carries a natural Hopf algebra structure with a coproduct and an antipode given by
\beq
\Delta_{12} = (\textrm{id}\otimes\tau\otimes\textrm{id})(\Delta_1\otimes\Delta_2) {\rm~~and~~} S_{12}=S_1\otimes S_2\,,
\eeq
with $\tau(a\otimes b)=b\otimes a$.
In the following we show how we can construct analogues of the single-valued map $\bfs$ in this very general and abstract setting. The special case of hyperlogarithms considered in Section~\ref{sec:period} is then recovered by considering $\cH_1=L_{\Sigma}$, $\cH_2=\overline{L}_{\Sigma}$ and the isomorphism $\phi$ is simply complex conjugation, and all coproducts are given by eq.~\eqref{eq:coproduct_def}.

Let us now define the analogues of the map $\tilde{S}$ from eq.~\eqref{eq:s_tilde_def}. We define linear maps $\tilde{S}_i$ by
\beq\bsp
\tilde{S}_1:&\,\mathcal{H}_1\rightarrow\mathcal{H}_2 ;\, x \mapsto (-1)^{|x|}\phi S_1(x)\,,\\	
\tilde{S}_2:&\,\mathcal{H}_2\rightarrow\mathcal{H}_1 ;\, x \mapsto (-1)^{|x|}\phi^{-1}  S_2(x)\,,
\esp\eeq
where $|x|$ is the weight of $x$. In addition, it is easy to see that these maps satisfy
\beq
\tilde{S}_1\mu_1 = \mu_2\tau(\tilde{S}_1\otimes \tilde{S}_1) {\rm~~and~~} \Delta_2\tilde{S}_1 = (\tilde{S}_1\otimes \tilde{S}_1)\tau\Delta_1\,,
\eeq
and similar properties hold for $\tilde{S}_2$. 
These properties follow directly from the corresponding properties of the antipodes. It is easy to see that
\beq
\tilde{S}_2\phi = \phi^{-1}\tilde{S}_1\,.
\eeq
Moreover, $\tilde{S}_1$ and $\tilde{S}_2$ are inverses of one another, because
\beq
\tilde{S}_2\tilde{S}_1(x) = (-1)^{2|x|}\,\phi^{-1}S_2\phi S_1(x) = \phi^{-1}S_2^2\phi(x) = \phi^{-1}\phi(x)=x\,.
\eeq

Using the maps $\tilde{S}_i$, we can define the analogues of the single-valued map $\bfs$ for hyperlogarithms defined in eq.~\eqref{eq:SV_def}. More precisely, we define two maps $\bfs_i$, $i=1,2$, by
\beq\bsp
\mathbf{s}_1&=(\tilde{S}_1 \otimes \textrm{id})\Delta_1:\mathcal{H}_1\rightarrow\mathcal{H}_2\otimes\mathcal{H}_1\,, \\	
\mathbf{s}_2&=(\tilde{S}_2 \otimes  \textrm{id})\Delta_2:\mathcal{H}_2\rightarrow\mathcal{H}_1\otimes\mathcal{H}_2\,.
\esp\eeq
In the remainder of this section we show that these maps enjoy all the properties of the single-valued map $\bfs$. 

We start by showing that the $\bfs_i$ are algebra morphisms. We only discuss the case of $\bfs_1$. We have
\beq
\bfs_1\mu_1 = \mu_{12}(\bfs_1\otimes\bfs_1)\,.
\eeq
Indeed, writing $\Delta_1(x)=\sum_{(x)}x_1\otimes x_2$, we obtain
\beq\bsp
\bfs_1(x)\cdot\bfs_1(y)	&\,	=(\mu_2 \otimes \mu_1) (\textrm{id} \otimes \tau \otimes\textrm{id})(\tilde{S}_1 \otimes \textrm{id} \otimes \tilde{S}_1  \otimes \textrm{id})(\Delta_1(x) \otimes \Delta_1(y)) \\
&\,= (\mu_2 \otimes \mu_1) \sum_{(x),(y)}(\tilde{S}_1(x_1)\otimes\tilde{S}_1(y_1)\otimes x_2 \otimes y_2)\\
		&\,= (\tilde{S}_1 \otimes \textrm{id})\Delta_1(x\cdot y)\\
		&\,= \bfs_1(x\cdot y)\,.
\esp\eeq

The maps $\bfs_1$ and $\bfs_2$ are not independent, but they are related by the isomorphism $\phi$,
\beq
\bfs_2\phi = (\phi^{-1}\otimes\phi)\bfs_1\,.
\eeq
Indeed, we have
\beq
\bfs_2\phi = (\tilde{S}_2\otimes\textrm{id})\Delta_2\phi= (\tilde{S}_2\phi\otimes\phi)\Delta_1 = (\phi^{-1}\tilde{S}_1\otimes\phi)\Delta_1 =(\phi^{-1}\otimes\phi)\bfs_1\,.
\eeq
This relation generalises the action of complex conjugation, and in particular $(\phi^{-1}\otimes\phi)\bfs_1$ corresponds to the complex conjugated single-valued map $\bar{\bfs}$ of Section~\ref{sec:period}.

Next, let us show that the maps $\bfs_1$ and $\bfs_2$ are related via $\tilde{S}_1$ and $\tilde{S}_2$ in the same way as $\bfs$ and $\bar{\bfs}$ are related via $\tilde{S}$, cf. eq.~\eqref{eq:bfscc}. More precisely, we have
\begin{equation}
	\mathbf{s}_2  \tilde{S}_1 = \tau  \mathbf{s}_1\,,
\end{equation}
and a similar relation holds for $\tilde{S}_2$. 
Starting from the left-hand side, we see that
\begin{equation}\label{LHS}
\mathbf{s}_2  \tilde{S}_1	= (\tilde{S}_2 \otimes \textrm{id})  \Delta_2  \tilde{S}_1 = (\tilde{S}_2\tilde{S}_1 \otimes \tilde{S}_1) \tau \Delta_1 = (\textrm{id} \otimes \tilde{S}_1)\tau  \Delta_1 = \tau(\tilde{S}_1 \otimes\textrm{id})  \Delta_1 = \tau\bfs_1\,.
\end{equation}

Let us define $\cH_i^{SV}$, $i=1,2$, as the image of $\cH_i$ under the map $\bfs_i$. In the following we only discuss the case $i=1$. We show that the coproduct $\Delta_{12}$ turns  $\cH_1^{SV}$ into a $\cH_2\otimes\cH_1$-comodule. The coaction is given by
\beq\label{eq:generic_coaction}
	\Delta_{12}\mathbf{s}_1=(\mathbf{s}_1\otimes \tilde{S}_1 \otimes \textrm{id})  (\tau \Delta_1 \otimes \textrm{id}) \Delta_1\,.
\end{equation}
$\cH_1^{SV}$ inherits the grading from $\cH_2\otimes\cH_1$ in an obvious way. In the following we give the proof of the formula for the coaction. Note that in the case where $\Delta_1$ is the coproduct on hyperlogarithms defined in eq.~\eqref{eq:coproduct_def}, this proves at the same time the formula for the coaction in eq.~\eqref{eq:SVMPL_coaction}.
	Expanding the left-hand side of eq.~\eqref{eq:generic_coaction}, we get, for $x\in\cH_1$,
\begin{equation}\bsp\label{eq:coaction_proof_1}
\Delta_{12}\mathbf{s}_1(x)&\,	= (\textrm{id}\otimes\tau\otimes\textrm{id}) \textrm{id} (\Delta_2 \otimes \Delta_1)  ( \tilde{S}_1 \otimes \textrm{id})  \Delta_1(x)\\
&\,=
	(\textrm{id}\otimes\tau\otimes\textrm{id})  (\Delta_2 \tilde{S}_1 \otimes \Delta_1)  \Delta_1(x)\\
	&\,=(\textrm{id}\otimes\tau\otimes\textrm{id})  ( (\tau (\tilde{S}_1 \otimes \tilde{S}_1)  \Delta_1) \otimes \Delta_1)  \Delta_1(x)\\
	&\,= \sum_{(x)} \tilde{S}_1(x_{1,2}) \otimes x_{2,1} \otimes  \tilde{S}_1(x_{1,1})\otimes x_{2,2}\,.
\esp\end{equation}
We do the same for the right-hand side of eq.~\eqref{eq:generic_coaction} and we obtain
\begin{equation}\bsp
(\mathbf{s}_1&\otimes \tilde{S}_1 \otimes \textrm{id})  (\tau \Delta_1 \otimes \textrm{id}) \Delta_1(x)=
	((\tilde{S}_1 \otimes \textrm{id})\Delta_1\otimes(\tilde{S}_1 \otimes \textrm{id}))  (\tau \Delta_1 \otimes \textrm{id})  \Delta_1 (x) \\
	&\,=\sum_{(x)} \tilde{S}_1(x_{1,2,1}) \otimes x_{1,2,2} \otimes  \tilde{S}_1(x_{1,1})\otimes x_{2}\\
	&\,=(\tilde{S}_1\otimes\textrm{id}\otimes \tilde{S}_1\otimes\textrm{id})(\Delta_1\otimes\tau)\sum_{(x)}x_{1,2}\otimes x_2\otimes x_{1,1}
\\
&\,=(\tilde{S}_1\otimes\textrm{id}\otimes \tilde{S}_1\otimes\textrm{id})(\textrm{id}\otimes\textrm{id}\otimes\tau)(\textrm{id}\otimes\Delta_1\otimes\textrm{id})\sum_{(x)}x_{1,2}\otimes x_2\otimes x_{1,1}\\
&\,=(\tilde{S}_1\otimes\textrm{id}\otimes \tilde{S}_1\otimes\textrm{id})(\textrm{id}\otimes\textrm{id}\otimes\tau)\sum_{(x)}x_{1,2}\otimes x_{2,1}\otimes x_{2,2}\otimes x_{1,1}\\
&\,=\sum_{(x)}\tilde{S}_1(x_{1,2})\otimes x_{2,1}\otimes \tilde{S}_1(x_{1,1})\otimes x_{2,2}\,,
\esp\end{equation}
and the last line agrees with eq.~\eqref{eq:coaction_proof_1}.

\section{Explicit expression for single-valued hyperlogarithms}
\label{app:sv_examples}
In this appendix we present explicit expressions of single-valued hyperlogarithms up to weight three in terms of ordinary hyperlogarithms.
We only give the results for Lyndon words. All other cases can be reconstructed from the fact that single valued hyperlogarithms form a shuffle algebra.

\subsection{Single-valued hyperlogarithms of weight one}

\begin{align}
\cG_0(z) & = G_0(z) + G_0(\bar{z})\,.\\
\cG_a(z) & = G_a(z) + G_{\bar{a}}(\bar{z})\,.
\end{align}

\subsection{Single-valued hyperlogarithms of weight two}

\begin{align}
\cG_{0,a}(z) & = G_{0,a}(z)+G_{\bar{a},0}\left(\bar{z}\right)-G_0(a) G_{\bar{a}}\left(\bar{z}\right)-G_0\left(\bar{a}\right) G_{\bar{a}}\left(\bar{z}\right)+G_0(z) G_{\bar{a}}\left(\bar{z}\right)\,.\\
\cG_{a,b}(z)&  =G_{a,b}(z)+G_{\bar{b},\bar{a}}\left(\bar{z}\right)+G_b(a) G_{\bar{a}}\left(\bar{z}\right)+G_{\bar{b}}\left(\bar{a}\right) G_{\bar{a}}\left(\bar{z}\right)\\
\nonumber&-G_a(b) G_{\bar{b}}\left(\bar{z}\right)+G_a(z) G_{\bar{b}}\left(\bar{z}\right)-G_{\bar{a}}\left(\bar{b}\right) G_{\bar{b}}\left(\bar{z}\right)\,.
\end{align}

\subsection{Single-valued hyperlogarithms of weight three}
\begin{align}
\cG_{0,0,a}(z) &= G_{0,0}(a) G_{\bar{a}}\left(\bar{z}\right)+G_{0,0}\left(\bar{a}\right) G_{\bar{a}}\left(\bar{z}\right)+G_{0,0}(z) G_{\bar{a}}\left(\bar{z}\right)-G_0(a) G_{\bar{a},0}\left(\bar{z}\right)\\
\nonumber&-G_0\left(\bar{a}\right) G_{\bar{a},0}\left(\bar{z}\right)+G_0(z) G_{\bar{a},0}\left(\bar{z}\right)+G_{\bar{a},0,0}\left(\bar{z}\right)+G_0(a) G_0\left(\bar{a}\right) G_{\bar{a}}\left(\bar{z}\right)\\
\nonumber&-G_0(a) G_0(z) G_{\bar{a}}\left(\bar{z}\right)-G_0(z) G_0\left(\bar{a}\right) G_{\bar{a}}\left(\bar{z}\right)+G_{0,0,a}(z)\,.\\
\nonumber\\
\cG_{0,a,a}(z) &= -G_{0,a}(a) G_{\bar{a}}\left(\bar{z}\right)+G_{\bar{a}}\left(\bar{z}\right) G_{0,a}(z)+G_{0,\bar{a}}\left(\bar{a}\right) G_{\bar{a}}\left(\bar{z}\right)-G_0(a) G_{\bar{a},\bar{a}}\left(\bar{z}\right)\\
\nonumber&-G_0\left(\bar{a}\right) G_{\bar{a},\bar{a}}\left(\bar{z}\right)+G_0(z) G_{\bar{a},\bar{a}}\left(\bar{z}\right)+G_{\bar{a},\bar{a},0}\left(\bar{z}\right)+G_{0,a,a}(z)\,.\\
\nonumber\\
\cG_{0,a,b}(z) &= G_b(a) G_{\bar{a},0}\left(\bar{z}\right)+G_{\bar{b}}\left(\bar{z}\right) G_{0,a}(z)+G_{\bar{b}}\left(\bar{z}\right) G_{0,\bar{a}}\left(\bar{b}\right)-G_{\bar{a}}\left(\bar{z}\right) G_{0,b}(a)\\
\nonumber&-G_{\bar{a}}\left(\bar{z}\right) G_{0,\bar{b}}\left(\bar{a}\right)+G_{\bar{b}}\left(\bar{z}\right) G_{a,0}(b)+G_{\bar{b}}\left(\bar{a}\right) G_{\bar{a},0}\left(\bar{z}\right)-G_{\bar{a}}\left(\bar{z}\right) G_{b,0}(a)\\
\nonumber&-G_{\bar{a}}\left(\bar{z}\right) G_{\bar{b},0}\left(\bar{a}\right)-G_a(b) G_{\bar{b},0}\left(\bar{z}\right)-G_{\bar{a}}\left(\bar{b}\right) G_{\bar{b},0}\left(\bar{z}\right)-G_0(a) G_{\bar{b},\bar{a}}\left(\bar{z}\right)\\
\nonumber&-G_0\left(\bar{a}\right) G_{\bar{b},\bar{a}}\left(\bar{z}\right)+G_0(z) G_{\bar{b},\bar{a}}\left(\bar{z}\right)+G_{\bar{b},\bar{a},0}\left(\bar{z}\right)-G_0\left(\bar{a}\right) G_b(a) G_{\bar{a}}\left(\bar{z}\right)\\
\nonumber&+G_0(z) G_b(a) G_{\bar{a}}\left(\bar{z}\right)-G_0(a) G_{\bar{b}}\left(\bar{a}\right) G_{\bar{a}}\left(\bar{z}\right)+G_0(z) G_{\bar{b}}\left(\bar{a}\right) G_{\bar{a}}\left(\bar{z}\right)\\
\nonumber&+G_0\left(\bar{b}\right) G_a(b) G_{\bar{b}}\left(\bar{z}\right)-G_0(z) G_a(b) G_{\bar{b}}\left(\bar{z}\right)+G_0(a) G_{\bar{a}}\left(\bar{b}\right) G_{\bar{b}}\left(\bar{z}\right)\\
\nonumber&+G_0\left(\bar{a}\right) G_{\bar{a}}\left(\bar{b}\right) G_{\bar{b}}\left(\bar{z}\right)-G_0(z) G_{\bar{a}}\left(\bar{b}\right) G_{\bar{b}}\left(\bar{z}\right)+G_{0,a,b}(z)\,.\\
\nonumber\\
\cG_{a,a,b}(z) &= G_b(a) G_{\bar{a},\bar{a}}\left(\bar{z}\right)+G_{\bar{b}}\left(\bar{z}\right) G_{a,a}(b)+G_{\bar{b}}\left(\bar{z}\right) G_{a,a}(z)+G_{\bar{b}}\left(\bar{z}\right) G_{\bar{a},\bar{a}}\left(\bar{b}\right)\\
\nonumber&+G_{\bar{b}}\left(\bar{a}\right) G_{\bar{a},\bar{a}}\left(\bar{z}\right)-G_{\bar{a}}\left(\bar{z}\right) G_{b,a}(a)+G_{\bar{a}}\left(\bar{z}\right) G_{\bar{b},\bar{a}}\left(\bar{a}\right)-G_a(b) G_{\bar{b},\bar{a}}\left(\bar{z}\right)\\
\nonumber&+G_a(z) G_{\bar{b},\bar{a}}\left(\bar{z}\right)-G_{\bar{a}}\left(\bar{b}\right) G_{\bar{b},\bar{a}}\left(\bar{z}\right)+G_{\bar{b},\bar{a},\bar{a}}\left(\bar{z}\right)+G_b(a) G_a(z) G_{\bar{a}}\left(\bar{z}\right)\\
\nonumber&-G_a(b) G_{\bar{b}}\left(\bar{a}\right) G_{\bar{a}}\left(\bar{z}\right)+G_a(z) G_{\bar{b}}\left(\bar{a}\right) G_{\bar{a}}\left(\bar{z}\right)-G_{\bar{a}}\left(\bar{b}\right) G_{\bar{b}}\left(\bar{a}\right) G_{\bar{a}}\left(\bar{z}\right)\\
\nonumber&-G_a(b) G_a(z) G_{\bar{b}}\left(\bar{z}\right)+G_a(b) G_{\bar{a}}\left(\bar{b}\right) G_{\bar{b}}\left(\bar{z}\right)-G_a(z) G_{\bar{a}}\left(\bar{b}\right) G_{\bar{b}}\left(\bar{z}\right)+G_{a,a,b}(z)\,.\\
\nonumber\\
\cG_{a,b,b}(z) &= G_{\bar{b}}\left(\bar{a}\right) G_{\bar{b},\bar{a}}\left(\bar{z}\right)-G_{\bar{b}}\left(\bar{z}\right) G_{a,b}(b)+G_{\bar{b}}\left(\bar{z}\right) G_{a,b}(z)+G_{\bar{b}}\left(\bar{z}\right) G_{\bar{a},\bar{b}}\left(\bar{b}\right)\\
\nonumber&+G_{\bar{a}}\left(\bar{z}\right) G_{b,b}(a)+G_b(a) G_{\bar{b},\bar{a}}\left(\bar{z}\right)+G_{\bar{a}}\left(\bar{z}\right) G_{\bar{b},\bar{b}}\left(\bar{a}\right)-G_a(b) G_{\bar{b},\bar{b}}\left(\bar{z}\right)\\
\nonumber&+G_a(z) G_{\bar{b},\bar{b}}\left(\bar{z}\right)-G_{\bar{a}}\left(\bar{b}\right) G_{\bar{b},\bar{b}}\left(\bar{z}\right)+G_{\bar{b},\bar{b},\bar{a}}\left(\bar{z}\right)+G_b(a) G_{\bar{b}}\left(\bar{a}\right) G_{\bar{a}}\left(\bar{z}\right)\\
\nonumber&-G_{\bar{a}}\left(\bar{b}\right) G_{\bar{b}}\left(\bar{a}\right) G_{\bar{b}}\left(\bar{z}\right)-G_b(a) G_{\bar{a}}\left(\bar{b}\right) G_{\bar{b}}\left(\bar{z}\right)+G_{a,b,b}(z)\,.\\
\nonumber\\
\cG_{a,b,c}(z) &= G_c(b) G_{\bar{b},\bar{a}}\left(\bar{z}\right)+G_{\bar{c}}\left(\bar{z}\right) G_{a,b}(z)-G_{\bar{b}}\left(\bar{z}\right) G_{a,c}(b)+G_{\bar{c}}\left(\bar{z}\right) G_{\bar{a},\bar{b}}\left(\bar{c}\right)\\
\nonumber&-G_{\bar{b}}\left(\bar{z}\right) G_{\bar{a},\bar{c}}\left(\bar{b}\right)+G_{\bar{c}}\left(\bar{z}\right) G_{b,a}(c)+G_{\bar{a}}\left(\bar{z}\right) G_{b,c}(a)+G_{\bar{c}}\left(\bar{b}\right) G_{\bar{b},\bar{a}}\left(\bar{z}\right)\\
\nonumber&-G_{\bar{b}}\left(\bar{z}\right) G_{c,a}(b)-G_{\bar{b}}\left(\bar{z}\right) G_{\bar{c},\bar{a}}\left(\bar{b}\right)+G_b(a) G_{\bar{c},\bar{a}}\left(\bar{z}\right)-G_b(c) G_{\bar{c},\bar{a}}\left(\bar{z}\right)\\
\nonumber&+G_{\bar{b}}\left(\bar{a}\right) G_{\bar{c},\bar{a}}\left(\bar{z}\right)-G_{\bar{b}}\left(\bar{c}\right) G_{\bar{c},\bar{a}}\left(\bar{z}\right)+G_{\bar{a}}\left(\bar{z}\right) G_{\bar{c},\bar{b}}\left(\bar{a}\right)-G_a(b) G_{\bar{c},\bar{b}}\left(\bar{z}\right)\\
\nonumber&+G_a(z) G_{\bar{c},\bar{b}}\left(\bar{z}\right)-G_{\bar{a}}\left(\bar{b}\right) G_{\bar{c},\bar{b}}\left(\bar{z}\right)+G_{\bar{c},\bar{b},\bar{a}}\left(\bar{z}\right)+G_c(b) G_{\bar{b}}\left(\bar{a}\right) G_{\bar{a}}\left(\bar{z}\right)\\
\nonumber&+G_a(z) G_c(b) G_{\bar{b}}\left(\bar{z}\right)-G_c(b) G_{\bar{a}}\left(\bar{b}\right) G_{\bar{b}}\left(\bar{z}\right)+G_b(a) G_{\bar{c}}\left(\bar{a}\right) G_{\bar{a}}\left(\bar{z}\right)\\
\nonumber&-G_b(c) G_{\bar{c}}\left(\bar{a}\right) G_{\bar{a}}\left(\bar{z}\right)-G_{\bar{c}}\left(\bar{a}\right) G_{\bar{a}}\left(\bar{z}\right) G_{\bar{b}}\left(\bar{c}\right)+G_{\bar{b}}\left(\bar{a}\right) G_{\bar{a}}\left(\bar{z}\right) G_{\bar{c}}\left(\bar{b}\right)\\
\nonumber&-G_a(b) G_{\bar{c}}\left(\bar{b}\right) G_{\bar{b}}\left(\bar{z}\right)+G_a(z) G_{\bar{c}}\left(\bar{b}\right) G_{\bar{b}}\left(\bar{z}\right)-G_b(a) G_{\bar{a}}\left(\bar{c}\right) G_{\bar{c}}\left(\bar{z}\right)\\
\nonumber&-G_a(z) G_b(c) G_{\bar{c}}\left(\bar{z}\right)+G_b(c) G_{\bar{a}}\left(\bar{c}\right) G_{\bar{c}}\left(\bar{z}\right)-G_{\bar{b}}\left(\bar{a}\right) G_{\bar{a}}\left(\bar{c}\right) G_{\bar{c}}\left(\bar{z}\right)\\
\nonumber&+G_a(b) G_{\bar{b}}\left(\bar{c}\right) G_{\bar{c}}\left(\bar{z}\right)-G_a(z) G_{\bar{b}}\left(\bar{c}\right) G_{\bar{c}}\left(\bar{z}\right)+G_{\bar{a}}\left(\bar{b}\right) G_{\bar{b}}\left(\bar{c}\right) G_{\bar{c}}\left(\bar{z}\right)+G_{a,b,c}(z)\,.
\end{align}

\section{Explicit results}
\label{app:results}
In this appendix we present explicit analytic results at two and three loops for all amplitudes with up to seven external legs and for all eight point amplitudes attwo loops.
Additional results through four loops, as well as the five-loop nine-point MHV amplitude, are given as ancillary material in computer-readable from to the arXiv submission.

\subsection{Six-point amplitudes}
The results of this section are not new but they have already been obtained in ref.~\cite{Lipatov:2010ad,Lipatov:2010qg,Lipatov:2012gk,Dixon:2011pw,Dixon:2012yy,Pennington:2012zj,Broedel:2015nfp}. We show them here using the same notation and conventions as the higher point amplitudes.
We only show independent helicity configurations where the last helicity index is positive. The remaining configurations can be obtained by complex conjugation.
The six-point MHV amplitudes at two and three loops are given by
\begin{align}
g_{++}^{(1)}(\rho_1) = &-\frac{1}{4}\cG_{0,1}\left(\rho_1\right)-\frac{1}{4}\cG_{1,0}\left(\rho_1\right)+\frac{1}{2}\cG_{1,1}\left(\rho_1\right)\,.\\
g_{++}^{(2)}(\rho_1) = &-\frac{1}{8}\cG_{0,0,1}\left(\rho_1\right)-\frac{1}{4}\cG_{0,1,0}\left(\rho_1\right)+\frac{1}{2}\cG_{0,1,1}\left(\rho_1\right)-\frac{1}{8}\cG_{1,0,0}\left(\rho_1\right)+\frac{1}{2}\cG_{1,0,1}\left(\rho_1\right)\\
&+\frac{1}{2}\cG_{1,1,0}\left(\rho_1\right)-\cG_{1,1,1}\left(\rho_1\right)\,.
\nonumber
\end{align}
The NMHV amplitudes at two and three loops can be written in the form
\beq\bsp
g_{-+}^{(i)}(\rho_1) = &\fa^{(i)}_{-+}(\rho_1)+R_{234}\,\fb^{(i)}_{-+}(\rho_1)\,,
\esp\eeq
where the pure functions $\fa^{(i)}_{-+}$ and $\fb^{(i)}_{-+}$ are given by
\begin{align}
\fa^{(1)}_{-+}(\rho_1) = &-\frac{1}{4}\cG_{1,0}\left(\rho_1\right)\,,\\
\fb^{(1)}_{-+}(\rho_1) = &-\frac{1}{4}\cG_{0,0}\left(\rho_1\right)+\frac{1}{2}\cG_{1,0}\left(\rho_1\right)\,,\\
\fa^{{(2)}}_{-+}(\rho_1) = &-\frac{1}{8}\cG_{0,1,0}\left(\rho_1\right)-\frac{1}{4}\cG_{1,0,0}\left(\rho_1\right)+\frac{1}{2}\cG_{1,1,0}\left(\rho_1\right)\,,\\
\fb^{{(2)}}_{-+}(\rho_1) = &-\frac{1}{8}\cG_{0,0,0}\left(\rho_1\right)+\frac{1}{2}\cG_{0,1,0}\left(\rho_1\right)+\frac{1}{2}\cG_{1,0,0}\left(\rho_1\right)-\cG_{1,1,0}\left(\rho_1\right)+\zeta_3\,.
\end{align}

\subsection{Seven-point amplitudes}

There is one new perturbative MHV coefficient through three loops that cannot be reduced to six-point by virtue of the factorisation theorem~\eqref{eq:fac_thm}. It is given by
\begin{align}
g_{+++}^{(1,1)}&(\rho_1,\rho_2) = -\frac{1}{8}\cG_{0,1,\rho_2}\left(\rho_1\right)-\frac{1}{8}\cG_{0,\rho_2,1}\left(\rho_1\right)+\frac{1}{8}\cG_{1,1,\rho_2}\left(\rho_1\right)-\frac{1}{8}\cG_{1,\rho_2,0}\left(\rho_1\right)\\
\nonumber&-\frac{1}{8}\cG_{\rho_2,1,0}\left(\rho_1\right)+\frac{1}{8}\cG_{\rho_2,1,1}\left(\rho_1\right)+\frac{1}{4}\cG_{1,\rho_2,1}\left(\rho_1\right)-\frac{1}{4}\cG_1\left(\rho_2\right)\cG_{1,\rho_2}\left(\rho_1\right)\\
\nonumber&+\frac{1}{8}\cG_1\left(\rho_1\right)\cG_{0,0}\left(\rho_2\right)-\frac{1}{8}\cG_0\left(\rho_2\right)\cG_{0,1}\left(\rho_1\right)+\frac{1}{8}\cG_1\left(\rho_2\right)\cG_{0,1}\left(\rho_1\right)-\frac{1}{8}\cG_{\rho_2}\left(\rho_1\right)\cG_{0,1}\left(\rho_2\right)\\
\nonumber&+\frac{1}{8}\cG_1\left(\rho_2\right)\cG_{0,\rho_2}\left(\rho_1\right)-\frac{1}{8}\cG_0\left(\rho_2\right)\cG_{1,0}\left(\rho_1\right)+\frac{1}{8}\cG_1\left(\rho_2\right)\cG_{1,0}\left(\rho_1\right)+\frac{1}{8}\cG_0\left(\rho_2\right)\cG_{1,1}\left(\rho_1\right)\\
\nonumber&-\frac{1}{8}\cG_1\left(\rho_2\right)\cG_{1,1}\left(\rho_1\right)-\frac{1}{8}\cG_1\left(\rho_1\right)\cG_{1,1}\left(\rho_2\right)+\frac{1}{8}\cG_{\rho_2}\left(\rho_1\right)\cG_{1,1}\left(\rho_2\right)+\frac{1}{8}\cG_0\left(\rho_2\right)\cG_{1,\rho_2}\left(\rho_1\right)\\
\nonumber&+\frac{1}{8}\cG_1\left(\rho_2\right)\cG_{\rho_2,0}\left(\rho_1\right)-\frac{1}{8}\cG_1\left(\rho_2\right)\cG_{\rho_2,1}\left(\rho_1\right)\,.
\end{align}
It is worth noting that we have performed an independent check of the above formula, or equivalently of the factorisation (\ref{eq:factorisation_3loop}) of the three-loop MHV seven-point amplitude, by computing the multi-Regge limit of its symbol \cite{Drummond:2014ffa}. In fact, to LLA we can uniquely fix an ansatz of single-valued multiple polylogarithms describing all possible beyond-the-symbol terms, by virtue of the double discontinuity (it determines all terms proportional to $\pi$ multiplied by a function of weight two), and the expected behaviour under soft limits, described in Section \ref{sec:soft_limits}. Thus the aforementioned check extends to full function level.

There are three new NMHV coefficient through three loops,
\begin{align}
g_{+-+}^{(i_1,i_2)}(\rho_1,\rho_2) = &\,\fa^{(i_1,i_2)}_{+-+}(\rho_1,\rho_2)+\overline{R}_{234}\,\fb^{(i_1,i_2)}_{1,+-+}(\rho_1,\rho_2)+R_{345}\,\fb^{(i_1,i_2)}_{2,+-+}(\rho_1,\rho_2)\\
&\nonumber+R_{345}\,\overline{R}_{234}\,\fc^{(i_1,i_2)}_{1,+-+}(\rho_1,\rho_2)\,,\\
&\nonumber\\
g_{-++}^{(i_1,i_2)}(\rho_1,\rho_2) = &\,\fa^{(i_1,i_2)}_{-++}(\rho_1,\rho_2)+R_{234}\,\fb^{(i_1,i_2)}_{1,-++}(\rho_1,\rho_2)+R_{235}\,\fb^{(i_1,i_2)}_{2,-++}(\rho_1,\rho_2)\,,\\
&\nonumber\\
g_{--+}^{(i_1,i_2)}(\rho_1,\rho_2) = &\,\fa^{(i_1,i_2)}_{--+}(\rho_1,\rho_2)+R_{245}\,\fb^{(i_1,i_2)}_{1,--+}(\rho_1,\rho_2)+R_{345}\,\fb^{(i_1,i_2)}_{2,--+}(\rho_1,\rho_2)\,.
\label{NMHV7ptMRKform}
\end{align}
As our convention is to choose the independent helicity configurations to end in a positive helicity, we present the result for the $\overline{\textrm{NMHV}}$ configuration $(-,-,+)$. The pure functions multiplying the rational prefactors are listed below.

%
%
\begin{align}
\fa^{\text{(0,1)}}_{+-+}\left(\rho_1,\rho_2\right) = &-\frac{1}{4}\cG_{1,0}\left(\rho_2\right)-\frac{1}{4}\cG_{1,1}\left(\rho_1\right)+\frac{1}{4}\cG_{1,\rho_2}\left(\rho_1\right)+\frac{1}{4}\cG_0\left(\rho_2\right)\cG_1\left(\rho_1\right)
\,. \\  
%
%
\fb_{1,+-+}^{\text{(0,1)}}\left(\rho_1,\rho_2\right) = &\,\frac{1}{4}\cG_{1,0}\left(\rho_1\right)+\frac{1}{4}\cG_{1,1}\left(\rho_2\right)-\frac{1}{4}\cG_{1,\rho_2}\left(\rho_1\right)-\frac{1}{4}\cG_0\left(\rho_2\right)\cG_1\left(\rho_1\right)\\
 \nonumber &-\frac{1}{4}\cG_0\left(\rho_1\right)\cG_1\left(\rho_2\right)+\frac{1}{4}\cG_1\left(\rho_1\right)\cG_1\left(\rho_2\right)
\,.  \\ 
%
%
\fb_{2,+-+}^{\text{(0,1)}}\left(\rho_1,\rho_2\right) = &-\frac{1}{4}\cG_{0,0}\left(\rho_2\right)+\frac{1}{2}\cG_{1,0}\left(\rho_2\right)+\frac{1}{4}\cG_{1,1}\left(\rho_1\right)-\frac{1}{4}\cG_{1,\rho_2}\left(\rho_1\right)\\
 \nonumber &+\frac{1}{4}\cG_{\rho_2,1}\left(\rho_1\right)-\frac{1}{4}\cG_{\rho_2,\rho_2}\left(\rho_1\right)-\frac{1}{4}\cG_0\left(\rho_2\right)\cG_1\left(\rho_1\right)-\frac{1}{4}\cG_0\left(\rho_2\right)\cG_{\rho_2}\left(\rho_1\right)
\,. \\  
%
%
\fc_{1,+-+}^{\text{(0,1)}}\left(\rho_1,\rho_2\right) = &\,\frac{1}{2}\cG_{0,0}\left(\rho_2\right)-\frac{1}{4}\cG_{0,1}\left(\rho_2\right)-\frac{1}{4}\cG_{1,0}\left(\rho_1\right)-\frac{1}{2}\cG_{1,0}\left(\rho_2\right)\\
 \nonumber &+\frac{1}{4}\cG_{1,\rho_2}\left(\rho_1\right)-\frac{1}{4}\cG_{\rho_2,0}\left(\rho_1\right)+\frac{1}{4}\cG_{\rho_2,\rho_2}\left(\rho_1\right)-\frac{1}{4}\cG_0\left(\rho_1\right)\cG_0\left(\rho_2\right)\\
 \nonumber &+\frac{1}{4}\cG_0\left(\rho_2\right)\cG_1\left(\rho_1\right)+\frac{1}{2}\cG_0\left(\rho_1\right)\cG_1\left(\rho_2\right)-\frac{1}{4}\cG_1\left(\rho_1\right)\cG_1\left(\rho_2\right)\\
 \nonumber &+\frac{1}{2}\cG_0\left(\rho_2\right)\cG_{\rho_2}\left(\rho_1\right)-\frac{1}{4}\cG_1\left(\rho_2\right)\cG_{\rho_2}\left(\rho_1\right)
\,. \\ \nonumber \\ 
%
%
%
%
\fa^{\text{(1,0)}}_{+-+}\left(\rho_1,\rho_2\right) = &\,\frac{1}{4}\cG_{1,1}\left(\rho_1\right)-\frac{1}{4}\cG_{1,\rho_2}\left(\rho_1\right)-\frac{1}{4}\cG_0\left(\rho_2\right)\cG_1\left(\rho_1\right)
\,. \\ 
%
%
\fb_{1,+-+}^{\text{(1,0)}}\left(\rho_1,\rho_2\right) = &-\frac{1}{4}\cG_{1,0}\left(\rho_1\right)+\frac{1}{4}\cG_{1,0}\left(\rho_2\right)-\frac{1}{4}\cG_{1,1}\left(\rho_2\right)+\frac{1}{4}\cG_{1,\rho_2}\left(\rho_1\right)\\
 \nonumber &+\frac{1}{4}\cG_0\left(\rho_2\right)\cG_1\left(\rho_1\right)-\frac{1}{4}\cG_0\left(\rho_1\right)\cG_1\left(\rho_2\right)+\frac{1}{4}\cG_1\left(\rho_1\right)\cG_1\left(\rho_2\right)
\,.  \\ 
%
%
\fb_{2,+-+}^{\text{(1,0)}}\left(\rho_1,\rho_2\right) = &-\frac{1}{4}\cG_{0,1}\left(\rho_1\right)+\frac{1}{4}\cG_{0,\rho_2}\left(\rho_1\right)-\frac{1}{4}\cG_{1,1}\left(\rho_1\right)+\frac{1}{4}\cG_{1,\rho_2}\left(\rho_1\right)\\
 \nonumber &+\frac{1}{4}\cG_{\rho_2,1}\left(\rho_1\right)-\frac{1}{4}\cG_{\rho_2,\rho_2}\left(\rho_1\right)+\frac{1}{4}\cG_0\left(\rho_2\right)\cG_1\left(\rho_1\right)
\,.  \\ 
%
%
\fc_{1,+-+}^{\text{(1,0)}}\left(\rho_1,\rho_2\right) = &-\frac{1}{4}\cG_{0,0}\left(\rho_1\right)-\frac{1}{4}\cG_{0,0}\left(\rho_2\right)+\frac{1}{2}\cG_{0,1}\left(\rho_1\right)+\frac{1}{4}\cG_{0,1}\left(\rho_2\right)\\
 \nonumber &-\frac{1}{4}\cG_{0,\rho_2}\left(\rho_1\right)+\frac{1}{4}\cG_{1,0}\left(\rho_1\right)-\frac{1}{4}\cG_{1,\rho_2}\left(\rho_1\right)+\frac{1}{4}\cG_{\rho_2,0}\left(\rho_1\right)-\frac{1}{2}\cG_{\rho_2,1}\left(\rho_1\right)\\
 \nonumber &+\frac{1}{4}\cG_{\rho_2,\rho_2}\left(\rho_1\right)+\frac{1}{4}\cG_0\left(\rho_1\right)\cG_0\left(\rho_2\right)-\frac{1}{4}\cG_0\left(\rho_2\right)\cG_1\left(\rho_1\right)-\frac{1}{4}\cG_1\left(\rho_1\right)\cG_1\left(\rho_2\right)\\
 \nonumber &-\frac{1}{4}\cG_0\left(\rho_2\right)\cG_{\rho_2}\left(\rho_1\right)+\frac{1}{4}\cG_1\left(\rho_2\right)\cG_{\rho_2}\left(\rho_1\right)
\,. \\ \nonumber \\ 
%
%
%
\fa^{\text{(0,2)}}_{+-+}\left(\rho_1,\rho_2\right) = &-\frac{1}{8}\cG_{0,1,0}\left(\rho_2\right)-\frac{1}{4}\cG_{1,0,0}\left(\rho_2\right)+\frac{1}{2}\cG_{1,1,0}\left(\rho_2\right)-\frac{1}{8}\cG_{1,1,1}\left(\rho_1\right)\label{nmhv7pta02+-+}\\
 \nonumber &+\frac{1}{8}\cG_{1,1,\rho_2}\left(\rho_1\right)-\frac{1}{4}\cG_{1,\rho_2,1}\left(\rho_1\right)+\frac{1}{4}\cG_{1,\rho_2,\rho_2}\left(\rho_1\right)-\frac{1}{8}\cG_{\rho_2,1,1}\left(\rho_1\right)\\
 \nonumber &+\frac{1}{8}\cG_{\rho_2,1,\rho_2}\left(\rho_1\right)+\frac{1}{4}\cG_1\left(\rho_1\right)\cG_{0,0}\left(\rho_2\right)-\frac{3}{8}\cG_1\left(\rho_1\right)\cG_{1,0}\left(\rho_2\right)\\
 \nonumber &-\frac{1}{8}\cG_{\rho_2}\left(\rho_1\right)\cG_{1,0}\left(\rho_2\right)+\frac{1}{8}\cG_0\left(\rho_2\right)\cG_{1,1}\left(\rho_1\right)+\frac{1}{4}\cG_0\left(\rho_2\right)\cG_{1,\rho_2}\left(\rho_1\right)\\
 \nonumber &+\frac{1}{8}\cG_0\left(\rho_2\right)\cG_{\rho_2,1}\left(\rho_1\right)
\,.  
\end{align}
\begin{align}
\fb_{1,+-+}^{\text{(0,2)}}\left(\rho_1,\rho_2\right) = &\,\frac{1}{8}\cG_{0,0,1}\left(\rho_2\right)+\frac{1}{8}\cG_{0,1,0}\left(\rho_2\right)-\frac{1}{8}\cG_{0,1,1}\left(\rho_2\right)+\frac{3}{8}\cG_{1,0,0}\left(\rho_2\right)\\
 \nonumber &+\frac{1}{8}\cG_{1,0,1}\left(\rho_2\right)+\frac{1}{8}\cG_{1,1,0}\left(\rho_1\right)-\frac{1}{4}\cG_{1,1,0}\left(\rho_2\right)-\frac{1}{2}\cG_{1,1,1}\left(\rho_2\right)\\
 \nonumber &-\frac{1}{8}\cG_{1,1,\rho_2}\left(\rho_1\right)+\frac{1}{4}\cG_{1,\rho_2,0}\left(\rho_1\right)-\frac{1}{4}\cG_{1,\rho_2,\rho_2}\left(\rho_1\right)+\frac{1}{8}\cG_{\rho_2,1,0}\left(\rho_1\right)\\
 \nonumber &-\frac{1}{2}\cG_1\left(\rho_1\right)\cG_{0,0}\left(\rho_2\right)-\frac{1}{8}\cG_0\left(\rho_1\right)\cG_{0,1}\left(\rho_2\right)+\frac{1}{4}\cG_1\left(\rho_1\right)\cG_{0,1}\left(\rho_2\right)\\
 \nonumber &+\frac{1}{8}\cG_{\rho_2}\left(\rho_1\right)\cG_{0,1}\left(\rho_2\right)+\frac{1}{4}\cG_0\left(\rho_2\right)\cG_{1,0}\left(\rho_1\right)-\frac{3}{8}\cG_1\left(\rho_2\right)\cG_{1,0}\left(\rho_1\right)\\
 \nonumber &-\frac{1}{4}\cG_0\left(\rho_1\right)\cG_{1,0}\left(\rho_2\right)+\frac{5}{8}\cG_1\left(\rho_1\right)\cG_{1,0}\left(\rho_2\right)+\frac{1}{8}\cG_{\rho_2}\left(\rho_1\right)\cG_{1,0}\left(\rho_2\right)\\
 \nonumber &-\frac{1}{8}\cG_0\left(\rho_2\right)\cG_{1,1}\left(\rho_1\right)+\frac{1}{8}\cG_1\left(\rho_2\right)\cG_{1,1}\left(\rho_1\right)+\frac{1}{2}\cG_0\left(\rho_1\right)\cG_{1,1}\left(\rho_2\right)\\
 \nonumber &-\frac{3}{8}\cG_1\left(\rho_1\right)\cG_{1,1}\left(\rho_2\right)-\frac{1}{8}\cG_{\rho_2}\left(\rho_1\right)\cG_{1,1}\left(\rho_2\right)-\frac{1}{2}\cG_0\left(\rho_2\right)\cG_{1,\rho_2}\left(\rho_1\right)\\
 \nonumber &+\frac{1}{4}\cG_1\left(\rho_2\right)\cG_{1,\rho_2}\left(\rho_1\right)-\frac{1}{8}\cG_1\left(\rho_2\right)\cG_{\rho_2,0}\left(\rho_1\right)-\frac{1}{8}\cG_0\left(\rho_2\right)\cG_{\rho_2,1}\left(\rho_1\right)\\
 \nonumber &+\frac{1}{8}\cG_1\left(\rho_2\right)\cG_{\rho_2,1}\left(\rho_1\right)-\frac{1}{8}\cG_{\rho_2,1,\rho_2}\left(\rho_1\right)
\,.  \\ 
%
%
\fb_{2,+-+}^{\text{(0,2)}}\left(\rho_1,\rho_2\right) = &-\frac{1}{8}\cG_{0,0,0}\left(\rho_2\right)+\frac{1}{2}\cG_{0,1,0}\left(\rho_2\right)+\frac{1}{2}\cG_{1,0,0}\left(\rho_2\right)-\cG_{1,1,0}\left(\rho_2\right)\\
 \nonumber &+\frac{1}{8}\cG_{1,1,1}\left(\rho_1\right)-\frac{1}{8}\cG_{1,1,\rho_2}\left(\rho_1\right)+\frac{3}{8}\cG_{1,\rho_2,1}\left(\rho_1\right)-\frac{3}{8}\cG_{1,\rho_2,\rho_2}\left(\rho_1\right)\\
 \nonumber &+\frac{3}{8}\cG_{\rho_2,1,1}\left(\rho_1\right)-\frac{3}{8}\cG_{\rho_2,1,\rho_2}\left(\rho_1\right)+\frac{1}{8}\cG_{\rho_2,\rho_2,1}\left(\rho_1\right)-\frac{1}{8}\cG_{\rho_2,\rho_2,\rho_2}\left(\rho_1\right)\\
 \nonumber &-\frac{3}{8}\cG_1\left(\rho_1\right)\cG_{0,0}\left(\rho_2\right)-\frac{1}{8}\cG_{\rho_2}\left(\rho_1\right)\cG_{0,0}\left(\rho_2\right)+\frac{1}{2}\cG_1\left(\rho_1\right)\cG_{1,0}\left(\rho_2\right)\\
 \nonumber &+\frac{1}{2}\cG_{\rho_2}\left(\rho_1\right)\cG_{1,0}\left(\rho_2\right)-\frac{1}{8}\cG_0\left(\rho_2\right)\cG_{1,1}\left(\rho_1\right)-\frac{3}{8}\cG_0\left(\rho_2\right)\cG_{1,\rho_2}\left(\rho_1\right)\\
 \nonumber &-\frac{3}{8}\cG_0\left(\rho_2\right)\cG_{\rho_2,1}\left(\rho_1\right)-\frac{1}{8}\cG_0\left(\rho_2\right)\cG_{\rho_2,\rho_2}\left(\rho_1\right)+\zeta_3
\,.  \\ 
%
%
\fc_{1,+-+}^{\text{(0,2)}}\left(\rho_1,\rho_2\right) = &\,\frac{3}{8}\cG_{0,0,0}\left(\rho_2\right)-\frac{5}{8}\cG_{0,0,1}\left(\rho_2\right)-\frac{5}{8}\cG_{0,1,0}\left(\rho_2\right)+\frac{1}{2}\cG_{0,1,1}\left(\rho_2\right)\\
 \nonumber &-\cG_{1,0,0}\left(\rho_2\right)+\frac{1}{2}\cG_{1,0,1}\left(\rho_2\right)-\frac{1}{8}\cG_{1,1,0}\left(\rho_1\right)+\cG_{1,1,0}\left(\rho_2\right)\\
 \nonumber &+\frac{1}{8}\cG_{1,1,\rho_2}\left(\rho_1\right)-\frac{3}{8}\cG_{1,\rho_2,0}\left(\rho_1\right)+\frac{3}{8}\cG_{1,\rho_2,\rho_2}\left(\rho_1\right)-\frac{3}{8}\cG_{\rho_2,1,0}\left(\rho_1\right)\\
 \nonumber &+\frac{3}{8}\cG_{\rho_2,1,\rho_2}\left(\rho_1\right)-\frac{1}{8}\cG_{\rho_2,\rho_2,0}\left(\rho_1\right)+\frac{1}{8}\cG_{\rho_2,\rho_2,\rho_2}\left(\rho_1\right)+\frac{\zeta_3}{4}\\
 \nonumber &-\frac{1}{8}\cG_0\left(\rho_1\right)\cG_{0,0}\left(\rho_2\right)+\frac{3}{4}\cG_1\left(\rho_1\right)\cG_{0,0}\left(\rho_2\right)+\frac{3}{8}\cG_{\rho_2}\left(\rho_1\right)\cG_{0,0}\left(\rho_2\right)\\
 \nonumber &+\frac{1}{2}\cG_0\left(\rho_1\right)\cG_{0,1}\left(\rho_2\right)-\frac{3}{8}\cG_1\left(\rho_1\right)\cG_{0,1}\left(\rho_2\right)-\frac{5}{8}\cG_{\rho_2}\left(\rho_1\right)\cG_{0,1}\left(\rho_2\right)\\
 \nonumber &-\frac{3}{8}\cG_0\left(\rho_2\right)\cG_{1,0}\left(\rho_1\right)+\frac{1}{2}\cG_1\left(\rho_2\right)\cG_{1,0}\left(\rho_1\right)+\frac{1}{2}\cG_0\left(\rho_1\right)\cG_{1,0}\left(\rho_2\right)\\
 \nonumber &-\frac{7}{8}\cG_1\left(\rho_1\right)\cG_{1,0}\left(\rho_2\right)-\frac{5}{8}\cG_{\rho_2}\left(\rho_1\right)\cG_{1,0}\left(\rho_2\right)+\frac{1}{8}\cG_0\left(\rho_2\right)\cG_{1,1}\left(\rho_1\right)\\
 \nonumber &-\frac{1}{8}\cG_1\left(\rho_2\right)\cG_{1,1}\left(\rho_1\right)-\cG_0\left(\rho_1\right)\cG_{1,1}\left(\rho_2\right)+\frac{1}{2}\cG_1\left(\rho_1\right)\cG_{1,1}\left(\rho_2\right)\\
 \nonumber &+\frac{1}{2}\cG_{\rho_2}\left(\rho_1\right)\cG_{1,1}\left(\rho_2\right)+\frac{3}{4}\cG_0\left(\rho_2\right)\cG_{1,\rho_2}\left(\rho_1\right)-\frac{3}{8}\cG_1\left(\rho_2\right)\cG_{1,\rho_2}\left(\rho_1\right)
 \end{align}
 \begin{align}
 \phantom{\fc_{1,+-+}^{\text{(0,2)}}\left(\rho_1,\rho_2\right) = }
 \nonumber &-\frac{1}{8}\cG_0\left(\rho_2\right)\cG_{\rho_2,0}\left(\rho_1\right)+\frac{1}{2}\cG_1\left(\rho_2\right)\cG_{\rho_2,0}\left(\rho_1\right)+\frac{3}{8}\cG_0\left(\rho_2\right)\cG_{\rho_2,1}\left(\rho_1\right)\\
 \nonumber &-\frac{3}{8}\cG_1\left(\rho_2\right)\cG_{\rho_2,1}\left(\rho_1\right)+\frac{1}{4}\cG_0\left(\rho_2\right)\cG_{\rho_2,\rho_2}\left(\rho_1\right)-\frac{1}{8}\cG_1\left(\rho_2\right)\cG_{\rho_2,\rho_2}\left(\rho_1\right)
\,. \\ \nonumber \\ 
%
%
%
%
\fa^{\text{(1,1)}}_{+-+}\left(\rho_1,\rho_2\right) = &\,\frac{1}{8}\cG_{0,1,1}\left(\rho_1\right)-\frac{1}{8}\cG_{0,1,\rho_2}\left(\rho_1\right)+\frac{1}{8}\cG_{1,0,1}\left(\rho_1\right)-\frac{1}{8}\cG_{1,0,\rho_2}\left(\rho_1\right)\\
 \nonumber &+\frac{1}{8}\cG_{1,1,1}\left(\rho_1\right)-\frac{1}{8}\cG_{1,1,\rho_2}\left(\rho_1\right)-\frac{1}{8}\cG_1\left(\rho_1\right)\cG_{0,0}\left(\rho_2\right)-\frac{1}{8}\cG_0\left(\rho_2\right)\cG_{0,1}\left(\rho_1\right)\\
 \nonumber&+\frac{1}{4}\cG_1\left(\rho_1\right)\cG_{1,0}\left(\rho_2\right)-\frac{1}{8}\cG_0\left(\rho_2\right)\cG_{1,1}\left(\rho_1\right)-\frac{1}{8}\cG_0\left(\rho_2\right)\cG_{1,\rho_2}\left(\rho_1\right)
\,.  \\ 
%
%
\fb_{1,+-+}^{\text{(1,1)}}\left(\rho_1,\rho_2\right) = &-\frac{1}{8}\cG_{0,1,0}\left(\rho_1\right)+\frac{1}{8}\cG_{0,1,\rho_2}\left(\rho_1\right)+\frac{1}{8}\cG_{1,0,0}\left(\rho_1\right)+\frac{1}{8}\cG_{1,0,0}\left(\rho_2\right)\\
 \nonumber &-\frac{1}{4}\cG_{1,0,1}\left(\rho_1\right)+\frac{1}{8}\cG_{1,0,\rho_2}\left(\rho_1\right)-\frac{1}{8}\cG_{1,1,0}\left(\rho_1\right)-\frac{1}{4}\cG_{1,1,0}\left(\rho_2\right)\\
 \nonumber &+\frac{1}{8}\cG_{1,1,1}\left(\rho_2\right)+\frac{1}{8}\cG_{1,1,\rho_2}\left(\rho_1\right)-\frac{1}{4}\cG_{1,\rho_2,0}\left(\rho_1\right)+\frac{1}{4}\cG_{1,\rho_2,1}\left(\rho_1\right)\\
 \nonumber &-\frac{1}{8}\cG_1\left(\rho_2\right)\cG_{0,0}\left(\rho_1\right)+\frac{3}{8}\cG_1\left(\rho_1\right)\cG_{0,0}\left(\rho_2\right)+\frac{1}{8}\cG_0\left(\rho_2\right)\cG_{0,1}\left(\rho_1\right)\\
 \nonumber &+\frac{1}{8}\cG_1\left(\rho_2\right)\cG_{0,1}\left(\rho_1\right)-\frac{1}{4}\cG_1\left(\rho_1\right)\cG_{0,1}\left(\rho_2\right)-\frac{1}{4}\cG_0\left(\rho_2\right)\cG_{1,0}\left(\rho_1\right)\\
 \nonumber &+\frac{3}{8}\cG_1\left(\rho_2\right)\cG_{1,0}\left(\rho_1\right)-\frac{1}{4}\cG_1\left(\rho_1\right)\cG_{1,0}\left(\rho_2\right)+\frac{1}{8}\cG_0\left(\rho_2\right)\cG_{1,1}\left(\rho_1\right)\\
 \nonumber &-\frac{1}{8}\cG_1\left(\rho_2\right)\cG_{1,1}\left(\rho_1\right)+\frac{1}{8}\cG_0\left(\rho_1\right)\cG_{1,1}\left(\rho_2\right)-\frac{1}{8}\cG_1\left(\rho_1\right)\cG_{1,1}\left(\rho_2\right)\\
 \nonumber &+\frac{3}{8}\cG_0\left(\rho_2\right)\cG_{1,\rho_2}\left(\rho_1\right)-\frac{1}{4}\cG_1\left(\rho_2\right)\cG_{1,\rho_2}\left(\rho_1\right)
\,.  \\ 
%
%
\fb_{2,+-+}^{\text{(1,1)}}\left(\rho_1,\rho_2\right) = &-\frac{1}{8}\cG_{0,1,1}\left(\rho_1\right)+\frac{1}{8}\cG_{0,1,\rho_2}\left(\rho_1\right)-\frac{1}{8}\cG_{0,\rho_2,1}\left(\rho_1\right)+\frac{1}{8}\cG_{0,\rho_2,\rho_2}\left(\rho_1\right)\\
 \nonumber &-\frac{1}{8}\cG_{1,0,1}\left(\rho_1\right)+\frac{1}{8}\cG_{1,0,\rho_2}\left(\rho_1\right)-\frac{1}{8}\cG_{1,1,1}\left(\rho_1\right)+\frac{1}{8}\cG_{1,1,\rho_2}\left(\rho_1\right)+\frac{1}{4}\cG_{\rho_2,\rho_2,1}\left(\rho_1\right)\\
 \nonumber &-\frac{1}{8}\cG_{\rho_2,0,1}\left(\rho_1\right)+\frac{1}{8}\cG_{\rho_2,0,\rho_2}\left(\rho_1\right)-\frac{1}{8}\cG_{\rho_2,1,1}\left(\rho_1\right)+\frac{1}{8}\cG_{\rho_2,1,\rho_2}\left(\rho_1\right)\\
 \nonumber &+\frac{1}{8}\cG_1\left(\rho_1\right)\cG_{0,0}\left(\rho_2\right)+\frac{1}{8}\cG_0\left(\rho_2\right)\cG_{0,1}\left(\rho_1\right)+\frac{1}{8}\cG_0\left(\rho_2\right)\cG_{0,\rho_2}\left(\rho_1\right)\\
 \nonumber &-\frac{1}{4}\cG_1\left(\rho_1\right)\cG_{1,0}\left(\rho_2\right)+\frac{1}{8}\cG_0\left(\rho_2\right)\cG_{1,1}\left(\rho_1\right)+\frac{1}{8}\cG_0\left(\rho_2\right)\cG_{1,\rho_2}\left(\rho_1\right)\\
 \nonumber &+\frac{1}{8}\cG_0\left(\rho_2\right)\cG_{\rho_2,1}\left(\rho_1\right)-\frac{1}{8}\cG_0\left(\rho_2\right)\cG_{\rho_2,\rho_2}\left(\rho_1\right)-\frac{1}{4}\cG_{\rho_2,\rho_2,\rho_2}\left(\rho_1\right)
\,.  \\ 
%
%
\fc_{1,+-+}^{\text{(1,1)}}\left(\rho_1,\rho_2\right) = &-\frac{3}{8}\cG_{0,0,0}\left(\rho_2\right)+\frac{1}{4}\cG_{0,0,1}\left(\rho_2\right)+\frac{1}{8}\cG_{0,1,0}\left(\rho_1\right)+\frac{1}{4}\cG_{0,1,0}\left(\rho_2\right)\\
 \nonumber &-\frac{1}{8}\cG_{0,1,1}\left(\rho_2\right)-\frac{1}{8}\cG_{0,1,\rho_2}\left(\rho_1\right)+\frac{1}{8}\cG_{0,\rho_2,0}\left(\rho_1\right)-\frac{1}{8}\cG_{0,\rho_2,\rho_2}\left(\rho_1\right)\\
 \nonumber &-\frac{1}{8}\cG_{1,0,0}\left(\rho_1\right)+\frac{1}{4}\cG_{1,0,0}\left(\rho_2\right)+\frac{1}{4}\cG_{1,0,1}\left(\rho_1\right)-\frac{1}{4}\cG_{1,0,1}\left(\rho_2\right)\\
 \nonumber &-\frac{1}{8}\cG_{1,0,\rho_2}\left(\rho_1\right)+\frac{1}{8}\cG_{1,1,0}\left(\rho_1\right)-\frac{1}{8}\cG_{1,1,\rho_2}\left(\rho_1\right)+\frac{1}{4}\cG_{1,\rho_2,0}\left(\rho_1\right)\\
 \nonumber &-\frac{1}{4}\cG_{1,\rho_2,1}\left(\rho_1\right)-\frac{1}{8}\cG_{\rho_2,0,0}\left(\rho_1\right)+\frac{1}{4}\cG_{\rho_2,0,1}\left(\rho_1\right)-\frac{1}{8}\cG_{\rho_2,0,\rho_2}\left(\rho_1\right)\\
 \nonumber &+\frac{1}{8}\cG_{\rho_2,1,0}\left(\rho_1\right)-\frac{1}{8}\cG_{\rho_2,1,\rho_2}\left(\rho_1\right)-\frac{1}{4}\cG_{\rho_2,\rho_2,1}\left(\rho_1\right)+\frac{1}{4}\cG_{\rho_2,\rho_2,\rho_2}\left(\rho_1\right)
 \end{align}
 \begin{align}
 \phantom{\fc_{1,+-+}^{\text{(1,1)}}\left(\rho_1,\rho_2\right) = }
 \nonumber &-\frac{1}{8}\cG_0\left(\rho_2\right)\cG_{0,0}\left(\rho_1\right)+\frac{1}{4}\cG_1\left(\rho_2\right)\cG_{0,0}\left(\rho_1\right)+\frac{1}{4}\cG_0\left(\rho_1\right)\cG_{0,0}\left(\rho_2\right)\\
 \nonumber &-\frac{3}{8}\cG_1\left(\rho_1\right)\cG_{0,0}\left(\rho_2\right)-\frac{3}{8}\cG_{\rho_2}\left(\rho_1\right)\cG_{0,0}\left(\rho_2\right)+\frac{1}{8}\cG_0\left(\rho_2\right)\cG_{0,1}\left(\rho_1\right)\\
 \nonumber &-\frac{3}{8}\cG_1\left(\rho_2\right)\cG_{0,1}\left(\rho_1\right)-\frac{1}{8}\cG_0\left(\rho_1\right)\cG_{0,1}\left(\rho_2\right)+\frac{1}{4}\cG_1\left(\rho_1\right)\cG_{0,1}\left(\rho_2\right)\\
 \nonumber &+\frac{1}{4}\cG_{\rho_2}\left(\rho_1\right)\cG_{0,1}\left(\rho_2\right)-\frac{1}{4}\cG_0\left(\rho_2\right)\cG_{0,\rho_2}\left(\rho_1\right)+\frac{1}{8}\cG_1\left(\rho_2\right)\cG_{0,\rho_2}\left(\rho_1\right)\\
 \nonumber &+\frac{1}{4}\cG_0\left(\rho_2\right)\cG_{1,0}\left(\rho_1\right)-\frac{3}{8}\cG_1\left(\rho_2\right)\cG_{1,0}\left(\rho_1\right)-\frac{1}{4}\cG_0\left(\rho_1\right)\cG_{1,0}\left(\rho_2\right)\\
 \nonumber &+\frac{1}{4}\cG_1\left(\rho_1\right)\cG_{1,0}\left(\rho_2\right)+\frac{1}{4}\cG_{\rho_2}\left(\rho_1\right)\cG_{1,0}\left(\rho_2\right)-\frac{1}{8}\cG_0\left(\rho_2\right)\cG_{1,1}\left(\rho_1\right)\\
 \nonumber &+\frac{1}{8}\cG_1\left(\rho_2\right)\cG_{1,1}\left(\rho_1\right)+\frac{1}{8}\cG_1\left(\rho_1\right)\cG_{1,1}\left(\rho_2\right)-\frac{1}{8}\cG_{\rho_2}\left(\rho_1\right)\cG_{1,1}\left(\rho_2\right)\\
 \nonumber &-\frac{3}{8}\cG_0\left(\rho_2\right)\cG_{1,\rho_2}\left(\rho_1\right)+\frac{1}{4}\cG_1\left(\rho_2\right)\cG_{1,\rho_2}\left(\rho_1\right)+\frac{1}{4}\cG_0\left(\rho_2\right)\cG_{\rho_2,0}\left(\rho_1\right)\\
 \nonumber &-\frac{1}{8}\cG_1\left(\rho_2\right)\cG_{\rho_2,0}\left(\rho_1\right)-\frac{3}{8}\cG_0\left(\rho_2\right)\cG_{\rho_2,1}\left(\rho_1\right)+\frac{1}{8}\cG_1\left(\rho_2\right)\cG_{\rho_2,1}\left(\rho_1\right)\\
 \nonumber &+\frac{1}{8}\cG_0\left(\rho_2\right)\cG_{\rho_2,\rho_2}\left(\rho_1\right)-\frac{\zeta_3}{4}
\,. \\ \nonumber \\ 
%
%
%
%
\fa^{\text{(2,0)}}_{+-+}\left(\rho_1,\rho_2\right) = &\,\frac{1}{4}\cG_{0,1,1}\left(\rho_1\right)-\frac{1}{4}\cG_{0,1,\rho_2}\left(\rho_1\right)-\frac{1}{8}\cG_{1,0,1}\left(\rho_1\right)+\frac{1}{8}\cG_{1,0,\rho_2}\left(\rho_1\right)\\
 \nonumber &+\frac{1}{8}\cG_{1,1,0}\left(\rho_1\right)-\frac{1}{2}\cG_{1,1,1}\left(\rho_1\right)+\frac{3}{8}\cG_{1,1,\rho_2}\left(\rho_1\right)-\frac{1}{8}\cG_{1,\rho_2,0}\left(\rho_1\right)+\frac{1}{4}\cG_{1,\rho_2,1}\left(\rho_1\right)\\
 \nonumber &+\frac{1}{8}\cG_1\left(\rho_1\right)\cG_{0,0}\left(\rho_2\right)-\frac{1}{4}\cG_0\left(\rho_2\right)\cG_{0,1}\left(\rho_1\right)-\frac{1}{8}\cG_1\left(\rho_1\right)\cG_{0,1}\left(\rho_2\right)\\
 \nonumber &-\frac{1}{8}\cG_0\left(\rho_2\right)\cG_{1,0}\left(\rho_1\right)+\frac{3}{8}\cG_0\left(\rho_2\right)\cG_{1,1}\left(\rho_1\right)+\frac{1}{8}\cG_1\left(\rho_2\right)\cG_{1,1}\left(\rho_1\right)\\
 \nonumber &+\frac{1}{8}\cG_0\left(\rho_2\right)\cG_{1,\rho_2}\left(\rho_1\right)-\frac{1}{8}\cG_1\left(\rho_2\right)\cG_{1,\rho_2}\left(\rho_1\right)-\frac{1}{8}\cG_{1,\rho_2,\rho_2}\left(\rho_1\right)
\,.  \\ 
%
%
\fb_{1,+-+}^{\text{(2,0)}}\left(\rho_1,\rho_2\right) = &-\frac{3}{8}\cG_{0,1,0}\left(\rho_1\right)+\frac{3}{8}\cG_{0,1,\rho_2}\left(\rho_1\right)-\frac{3}{8}\cG_{1,0,0}\left(\rho_1\right)-\frac{1}{8}\cG_{1,0,0}\left(\rho_2\right)\\
 \nonumber &+\frac{1}{2}\cG_{1,0,1}\left(\rho_1\right)+\frac{1}{8}\cG_{1,0,1}\left(\rho_2\right)-\frac{1}{8}\cG_{1,0,\rho_2}\left(\rho_1\right)+\frac{1}{2}\cG_{1,1,0}\left(\rho_1\right)\\
 \nonumber &+\frac{1}{8}\cG_{1,1,0}\left(\rho_2\right)-\frac{1}{8}\cG_{1,1,1}\left(\rho_2\right)-\frac{1}{2}\cG_{1,1,\rho_2}\left(\rho_1\right)+\frac{3}{8}\cG_{1,\rho_2,0}\left(\rho_1\right)\\
 \nonumber &-\frac{1}{8}\cG_1\left(\rho_2\right)\cG_{0,0}\left(\rho_1\right)-\frac{3}{8}\cG_1\left(\rho_1\right)\cG_{0,0}\left(\rho_2\right)+\frac{3}{8}\cG_0\left(\rho_2\right)\cG_{0,1}\left(\rho_1\right)\\
 \nonumber &+\frac{1}{8}\cG_1\left(\rho_2\right)\cG_{0,1}\left(\rho_1\right)+\frac{3}{8}\cG_1\left(\rho_1\right)\cG_{0,1}\left(\rho_2\right)+\frac{3}{8}\cG_0\left(\rho_2\right)\cG_{1,0}\left(\rho_1\right)\\
 \nonumber &+\frac{1}{8}\cG_1\left(\rho_2\right)\cG_{1,0}\left(\rho_1\right)+\frac{1}{8}\cG_0\left(\rho_1\right)\cG_{1,0}\left(\rho_2\right)-\frac{1}{8}\cG_1\left(\rho_1\right)\cG_{1,0}\left(\rho_2\right)\\
 \nonumber &-\frac{1}{2}\cG_0\left(\rho_2\right)\cG_{1,1}\left(\rho_1\right)-\frac{1}{2}\cG_1\left(\rho_2\right)\cG_{1,1}\left(\rho_1\right)-\frac{1}{8}\cG_0\left(\rho_1\right)\cG_{1,1}\left(\rho_2\right)\\
 \nonumber &+\frac{1}{8}\cG_1\left(\rho_1\right)\cG_{1,1}\left(\rho_2\right)-\frac{3}{8}\cG_0\left(\rho_2\right)\cG_{1,\rho_2}\left(\rho_1\right)+\frac{3}{8}\cG_1\left(\rho_2\right)\cG_{1,\rho_2}\left(\rho_1\right)\\
\nonumber& -\frac{1}{2}\cG_{1,\rho_2,1}\left(\rho_1\right)+\frac{1}{8}\cG_{1,\rho_2,\rho_2}\left(\rho_1\right)
\,. \end{align}
\begin{align}
\fb_{2,+-+}^{\text{(2,0)}}\left(\rho_1,\rho_2\right) = &-\frac{1}{4}\cG_{0,0,1}\left(\rho_1\right)+\frac{1}{4}\cG_{0,0,\rho_2}\left(\rho_1\right)-\frac{1}{8}\cG_{0,1,0}\left(\rho_1\right)+\frac{1}{4}\cG_{0,1,1}\left(\rho_1\right)\\
 \nonumber &-\frac{1}{8}\cG_{0,1,\rho_2}\left(\rho_1\right)+\frac{1}{8}\cG_{0,\rho_2,0}\left(\rho_1\right)-\frac{1}{4}\cG_{0,\rho_2,1}\left(\rho_1\right)+\frac{1}{8}\cG_{0,\rho_2,\rho_2}\left(\rho_1\right)\\
 \nonumber &+\frac{1}{8}\cG_{1,0,1}\left(\rho_1\right)-\frac{1}{8}\cG_{1,0,\rho_2}\left(\rho_1\right)-\frac{1}{8}\cG_{1,1,0}\left(\rho_1\right)+\frac{1}{2}\cG_{1,1,1}\left(\rho_1\right)\\
 \nonumber &-\frac{3}{8}\cG_{1,1,\rho_2}\left(\rho_1\right)+\frac{1}{8}\cG_{1,\rho_2,0}\left(\rho_1\right)-\frac{1}{4}\cG_{1,\rho_2,1}\left(\rho_1\right)+\frac{1}{8}\cG_{1,\rho_2,\rho_2}\left(\rho_1\right)\\
 \nonumber &+\frac{1}{4}\cG_{\rho_2,0,1}\left(\rho_1\right)-\frac{1}{4}\cG_{\rho_2,0,\rho_2}\left(\rho_1\right)+\frac{1}{8}\cG_{\rho_2,1,0}\left(\rho_1\right)-\frac{1}{2}\cG_{\rho_2,1,1}\left(\rho_1\right)\\
 \nonumber &+\frac{3}{8}\cG_{\rho_2,1,\rho_2}\left(\rho_1\right)-\frac{1}{8}\cG_{\rho_2,\rho_2,0}\left(\rho_1\right)+\frac{1}{4}\cG_{\rho_2,\rho_2,1}\left(\rho_1\right)-\frac{1}{8}\cG_{\rho_2,\rho_2,\rho_2}\left(\rho_1\right)\\
 \nonumber &-\frac{1}{8}\cG_1\left(\rho_1\right)\cG_{0,0}\left(\rho_2\right)+\frac{3}{8}\cG_0\left(\rho_2\right)\cG_{0,1}\left(\rho_1\right)-\frac{1}{8}\cG_1\left(\rho_2\right)\cG_{0,1}\left(\rho_1\right)\\
 \nonumber &+\frac{1}{8}\cG_1\left(\rho_1\right)\cG_{0,1}\left(\rho_2\right)-\frac{1}{8}\cG_0\left(\rho_2\right)\cG_{0,\rho_2}\left(\rho_1\right)+\frac{1}{8}\cG_1\left(\rho_2\right)\cG_{0,\rho_2}\left(\rho_1\right)\\
 \nonumber &+\frac{1}{8}\cG_0\left(\rho_2\right)\cG_{1,0}\left(\rho_1\right)-\frac{3}{8}\cG_0\left(\rho_2\right)\cG_{1,1}\left(\rho_1\right)-\frac{1}{8}\cG_1\left(\rho_2\right)\cG_{1,1}\left(\rho_1\right)\\
 \nonumber &-\frac{1}{8}\cG_0\left(\rho_2\right)\cG_{1,\rho_2}\left(\rho_1\right)+\frac{1}{8}\cG_1\left(\rho_2\right)\cG_{1,\rho_2}\left(\rho_1\right)-\frac{1}{8}\cG_0\left(\rho_2\right)\cG_{\rho_2,1}\left(\rho_1\right)\\
 \nonumber &+\frac{1}{8}\cG_1\left(\rho_2\right)\cG_{\rho_2,1}\left(\rho_1\right)+\frac{1}{8}\cG_0\left(\rho_2\right)\cG_{\rho_2,\rho_2}\left(\rho_1\right)-\frac{1}{8}\cG_1\left(\rho_2\right)\cG_{\rho_2,\rho_2}\left(\rho_1\right)
\,.  \\ 
%
%
\fc_{1,+-+}^{\text{(2,0)}}\left(\rho_1,\rho_2\right) = &-\frac{1}{8}\cG_{0,0,0}\left(\rho_1\right)+\frac{1}{8}\cG_{0,0,0}\left(\rho_2\right)+\frac{1}{2}\cG_{0,0,1}\left(\rho_1\right)-\frac{1}{8}\cG_{0,0,1}\left(\rho_2\right)\\
 \nonumber &-\frac{3}{8}\cG_{0,0,\rho_2}\left(\rho_1\right)+\frac{7}{8}\cG_{0,1,0}\left(\rho_1\right)-\frac{1}{8}\cG_{0,1,0}\left(\rho_2\right)-\cG_{0,1,1}\left(\rho_1\right)\\
 \nonumber &+\frac{1}{8}\cG_{0,1,1}\left(\rho_2\right)+\frac{1}{8}\cG_{0,1,\rho_2}\left(\rho_1\right)-\frac{3}{8}\cG_{0,\rho_2,0}\left(\rho_1\right)+\frac{1}{2}\cG_{0,\rho_2,1}\left(\rho_1\right)\\
 \nonumber &-\frac{1}{8}\cG_{0,\rho_2,\rho_2}\left(\rho_1\right)+\frac{3}{8}\cG_{1,0,0}\left(\rho_1\right)-\frac{1}{2}\cG_{1,0,1}\left(\rho_1\right)+\frac{1}{8}\cG_{1,0,\rho_2}\left(\rho_1\right)\\
 \nonumber &-\frac{1}{2}\cG_{1,1,0}\left(\rho_1\right)+\frac{1}{2}\cG_{1,1,\rho_2}\left(\rho_1\right)-\frac{3}{8}\cG_{1,\rho_2,0}\left(\rho_1\right)+\frac{1}{2}\cG_{1,\rho_2,1}\left(\rho_1\right)\\
 \nonumber &-\frac{1}{8}\cG_{1,\rho_2,\rho_2}\left(\rho_1\right)+\frac{1}{8}\cG_{\rho_2,0,0}\left(\rho_1\right)-\frac{1}{2}\cG_{\rho_2,0,1}\left(\rho_1\right)+\frac{3}{8}\cG_{\rho_2,0,\rho_2}\left(\rho_1\right)\\
 \nonumber &-\frac{1}{2}\cG_{\rho_2,1,0}\left(\rho_1\right)+\cG_{\rho_2,1,1}\left(\rho_1\right)-\frac{1}{2}\cG_{\rho_2,1,\rho_2}\left(\rho_1\right)+\frac{3}{8}\cG_{\rho_2,\rho_2,0}\left(\rho_1\right)\\
 \nonumber &+\frac{1}{8}\cG_0\left(\rho_2\right)\cG_{0,0}\left(\rho_1\right)-\frac{1}{8}\cG_0\left(\rho_1\right)\cG_{0,0}\left(\rho_2\right)+\frac{3}{8}\cG_1\left(\rho_1\right)\cG_{0,0}\left(\rho_2\right)\\
 \nonumber &+\frac{1}{8}\cG_{\rho_2}\left(\rho_1\right)\cG_{0,0}\left(\rho_2\right)-\frac{7}{8}\cG_0\left(\rho_2\right)\cG_{0,1}\left(\rho_1\right)+\frac{3}{8}\cG_1\left(\rho_2\right)\cG_{0,1}\left(\rho_1\right)\\
 \nonumber &+\frac{1}{8}\cG_0\left(\rho_1\right)\cG_{0,1}\left(\rho_2\right)-\frac{3}{8}\cG_1\left(\rho_1\right)\cG_{0,1}\left(\rho_2\right)-\frac{1}{8}\cG_{\rho_2}\left(\rho_1\right)\cG_{0,1}\left(\rho_2\right)\\
 \nonumber &+\frac{3}{8}\cG_0\left(\rho_2\right)\cG_{0,\rho_2}\left(\rho_1\right)-\frac{3}{8}\cG_1\left(\rho_2\right)\cG_{0,\rho_2}\left(\rho_1\right)-\frac{3}{8}\cG_0\left(\rho_2\right)\cG_{1,0}\left(\rho_1\right)\\
 \nonumber &-\frac{1}{8}\cG_1\left(\rho_2\right)\cG_{1,0}\left(\rho_1\right)+\frac{1}{8}\cG_1\left(\rho_1\right)\cG_{1,0}\left(\rho_2\right)-\frac{1}{8}\cG_{\rho_2}\left(\rho_1\right)\cG_{1,0}\left(\rho_2\right)\\
 \nonumber &+\frac{1}{2}\cG_0\left(\rho_2\right)\cG_{1,1}\left(\rho_1\right)+\frac{1}{2}\cG_1\left(\rho_2\right)\cG_{1,1}\left(\rho_1\right)-\frac{1}{8}\cG_1\left(\rho_1\right)\cG_{1,1}\left(\rho_2\right)\\
 \nonumber &+\frac{1}{8}\cG_{\rho_2}\left(\rho_1\right)\cG_{1,1}\left(\rho_2\right)+\frac{3}{8}\cG_0\left(\rho_2\right)\cG_{1,\rho_2}\left(\rho_1\right)-\frac{3}{8}\cG_1\left(\rho_2\right)\cG_{1,\rho_2}\left(\rho_1\right)\\
 \nonumber &-\frac{1}{8}\cG_0\left(\rho_2\right)\cG_{\rho_2,0}\left(\rho_1\right)+\frac{1}{8}\cG_1\left(\rho_2\right)\cG_{\rho_2,0}\left(\rho_1\right)+\frac{1}{2}\cG_0\left(\rho_2\right)\cG_{\rho_2,1}\left(\rho_1\right)
 \label{nmhv7ptc20+-+}
 \end{align}
 \begin{align}
 \phantom{\fc_{1,+-+}^{\text{(2,0)}}\left(\rho_1,\rho_2\right) =}
 \nonumber &-\frac{1}{2}\cG_1\left(\rho_2\right)\cG_{\rho_2,1}\left(\rho_1\right)-\frac{3}{8}\cG_0\left(\rho_2\right)\cG_{\rho_2,\rho_2}\left(\rho_1\right)+\frac{3}{8}\cG_1\left(\rho_2\right)\cG_{\rho_2,\rho_2}\left(\rho_1\right)\\
  \nonumber &-\frac{1}{2}\cG_{\rho_2,\rho_2,1}\left(\rho_1\right)+\frac{1}{8}\cG_{\rho_2,\rho_2,\rho_2}\left(\rho_1\right)+\zeta_3
\,. \\ \nonumber \\ 
%
%
%
%
\fa^{\text{(0,1)}}_{-++}\left(\rho_1,\rho_2\right) = &-\frac{1}{4}\cG_{0,1}\left(\rho_2\right)-\frac{1}{4}\cG_{1,0}\left(\rho_2\right)+\frac{1}{2}\cG_{1,1}\left(\rho_2\right)+\frac{1}{4}\cG_{1,\rho_2}\left(\rho_1\right)\\
 \nonumber &+\frac{1}{4}\cG_{\rho_2,1}\left(\rho_1\right)+\frac{1}{4}\cG_0\left(\rho_2\right)\cG_1\left(\rho_1\right)-\frac{1}{4}\cG_1\left(\rho_1\right)\cG_1\left(\rho_2\right)-\frac{1}{4}\cG_1\left(\rho_2\right)\cG_{\rho_2}\left(\rho_1\right)
\,.  \\ 
%
%
\fb_{1,-++}^{\text{(0,1)}}\left(\rho_1,\rho_2\right) = &\,\frac{1}{4}\cG_{0,1}\left(\rho_1\right)+\frac{1}{4}\cG_{0,1}\left(\rho_2\right)-\frac{1}{4}\cG_{1,1}\left(\rho_2\right)-\frac{1}{4}\cG_{\rho_2,1}\left(\rho_1\right)\\
 \nonumber &-\frac{1}{4}\cG_0\left(\rho_1\right)\cG_1\left(\rho_2\right)+\frac{1}{4}\cG_1\left(\rho_2\right)\cG_{\rho_2}\left(\rho_1\right)
\,.  \\ 
%
%
\fb_{2,-++}^{\text{(0,1)}}\left(\rho_1,\rho_2\right) = &\,\frac{1}{4}\cG_{0,0}\left(\rho_2\right)-\frac{1}{4}\cG_{0,1}\left(\rho_1\right)-\frac{1}{4}\cG_{0,1}\left(\rho_2\right)-\frac{1}{4}\cG_{0,\rho_2}\left(\rho_1\right)\\
 \nonumber &-\frac{1}{4}\cG_0\left(\rho_1\right)\cG_0\left(\rho_2\right)+\frac{1}{2}\cG_0\left(\rho_1\right)\cG_1\left(\rho_2\right)
\,. \\ \nonumber \\ 
%
%
%
%
\fa^{\text{(0,2)}}_{-++}\left(\rho_1,\rho_2\right) = &-\frac{1}{8}\cG_{0,0,1}\left(\rho_2\right)-\frac{1}{4}\cG_{0,1,0}\left(\rho_2\right)+\frac{1}{2}\cG_{0,1,1}\left(\rho_2\right)-\frac{1}{8}\cG_{1,0,0}\left(\rho_2\right) \label{nmhv7pta02-++} \\
 \nonumber &+\frac{1}{2}\cG_{1,0,1}\left(\rho_2\right)+\frac{1}{2}\cG_{1,1,0}\left(\rho_2\right)-\cG_{1,1,1}\left(\rho_2\right)+\frac{1}{8}\cG_{1,1,\rho_2}\left(\rho_1\right)\\
 \nonumber &+\frac{1}{4}\cG_{1,\rho_2,1}\left(\rho_1\right)+\frac{1}{8}\cG_{1,\rho_2,\rho_2}\left(\rho_1\right)+\frac{1}{8}\cG_{\rho_2,1,1}\left(\rho_1\right)+\frac{1}{4}\cG_{\rho_2,1,\rho_2}\left(\rho_1\right)\\
 \nonumber &+\frac{1}{8}\cG_1\left(\rho_1\right)\cG_{0,0}\left(\rho_2\right)-\frac{3}{8}\cG_1\left(\rho_1\right)\cG_{0,1}\left(\rho_2\right)-\frac{1}{8}\cG_{\rho_2}\left(\rho_1\right)\cG_{0,1}\left(\rho_2\right)\\
 \nonumber &-\frac{1}{4}\cG_1\left(\rho_1\right)\cG_{1,0}\left(\rho_2\right)-\frac{1}{4}\cG_{\rho_2}\left(\rho_1\right)\cG_{1,0}\left(\rho_2\right)+\frac{1}{8}\cG_0\left(\rho_2\right)\cG_{1,1}\left(\rho_1\right)\\
 \nonumber &-\frac{1}{8}\cG_1\left(\rho_2\right)\cG_{1,1}\left(\rho_1\right)+\frac{1}{2}\cG_1\left(\rho_1\right)\cG_{1,1}\left(\rho_2\right)+\frac{1}{2}\cG_{\rho_2}\left(\rho_1\right)\cG_{1,1}\left(\rho_2\right)\\
 \nonumber &+\frac{1}{8}\cG_0\left(\rho_2\right)\cG_{1,\rho_2}\left(\rho_1\right)-\frac{3}{8}\cG_1\left(\rho_2\right)\cG_{1,\rho_2}\left(\rho_1\right)+\frac{1}{4}\cG_0\left(\rho_2\right)\cG_{\rho_2,1}\left(\rho_1\right)\\
 \nonumber &-\frac{3}{8}\cG_1\left(\rho_2\right)\cG_{\rho_2,1}\left(\rho_1\right)-\frac{1}{8}\cG_1\left(\rho_2\right)\cG_{\rho_2,\rho_2}\left(\rho_1\right)+\frac{1}{8}\cG_{\rho_2,\rho_2,1}\left(\rho_1\right)
\,. 
 \\ 
%
%
\fb_{1,-++}^{\text{(0,2)}}\left(\rho_1,\rho_2\right) = &\,\frac{1}{4}\cG_{0,0,1}\left(\rho_2\right)+\frac{1}{4}\cG_{0,1,0}\left(\rho_2\right)+\frac{1}{8}\cG_{0,1,1}\left(\rho_1\right)-\frac{5}{8}\cG_{0,1,1}\left(\rho_2\right)\\
 \nonumber &+\frac{1}{4}\cG_{0,1,\rho_2}\left(\rho_1\right)+\frac{1}{8}\cG_{0,\rho_2,1}\left(\rho_1\right)+\frac{1}{4}\cG_{1,0,0}\left(\rho_2\right)-\frac{3}{8}\cG_{1,0,1}\left(\rho_2\right)\\
 \nonumber &-\frac{1}{4}\cG_{1,1,0}\left(\rho_2\right)+\frac{1}{2}\cG_{1,1,1}\left(\rho_2\right)-\frac{1}{8}\cG_{\rho_2,1,1}\left(\rho_1\right)-\frac{1}{4}\cG_{\rho_2,1,\rho_2}\left(\rho_1\right)\\
 \nonumber &+\frac{1}{4}\cG_0\left(\rho_2\right)\cG_{0,1}\left(\rho_1\right)-\frac{3}{8}\cG_1\left(\rho_2\right)\cG_{0,1}\left(\rho_1\right)-\frac{1}{8}\cG_0\left(\rho_1\right)\cG_{0,1}\left(\rho_2\right)\\
 \nonumber &+\frac{1}{8}\cG_{\rho_2}\left(\rho_1\right)\cG_{0,1}\left(\rho_2\right)-\frac{1}{8}\cG_1\left(\rho_2\right)\cG_{0,\rho_2}\left(\rho_1\right)-\frac{1}{4}\cG_0\left(\rho_1\right)\cG_{1,0}\left(\rho_2\right)\\
 \nonumber &+\frac{1}{4}\cG_{\rho_2}\left(\rho_1\right)\cG_{1,0}\left(\rho_2\right)+\frac{1}{2}\cG_0\left(\rho_1\right)\cG_{1,1}\left(\rho_2\right)-\frac{1}{2}\cG_{\rho_2}\left(\rho_1\right)\cG_{1,1}\left(\rho_2\right)\\
 \nonumber &-\frac{1}{4}\cG_0\left(\rho_2\right)\cG_{\rho_2,1}\left(\rho_1\right)+\frac{3}{8}\cG_1\left(\rho_2\right)\cG_{\rho_2,1}\left(\rho_1\right)+\frac{1}{8}\cG_1\left(\rho_2\right)\cG_{\rho_2,\rho_2}\left(\rho_1\right)\\
 \nonumber&-\frac{1}{8}\cG_{\rho_2,\rho_2,1}\left(\rho_1\right)
\,.  \end{align}
\begin{align}
\fb_{2,-++}^{\text{(0,2)}}\left(\rho_1,\rho_2\right) = &\,\frac{1}{4}\cG_{0,0,0}\left(\rho_2\right)-\frac{5}{8}\cG_{0,0,1}\left(\rho_2\right)-\frac{1}{8}\cG_{0,1,0}\left(\rho_2\right)-\frac{1}{8}\cG_{0,1,1}\left(\rho_1\right)\\
 \nonumber &+\frac{1}{2}\cG_{0,1,1}\left(\rho_2\right)-\frac{3}{8}\cG_{0,1,\rho_2}\left(\rho_1\right)-\frac{3}{8}\cG_{0,\rho_2,1}\left(\rho_1\right)-\frac{1}{8}\cG_{0,\rho_2,\rho_2}\left(\rho_1\right)+\frac{5\zeta_3}{4}\\
 \nonumber &-\frac{1}{8}\cG_0\left(\rho_1\right)\cG_{0,0}\left(\rho_2\right)-\frac{3}{8}\cG_0\left(\rho_2\right)\cG_{0,1}\left(\rho_1\right)+\frac{1}{2}\cG_1\left(\rho_2\right)\cG_{0,1}\left(\rho_1\right)\\
 \nonumber &+\frac{1}{2}\cG_0\left(\rho_1\right)\cG_{0,1}\left(\rho_2\right)-\frac{1}{8}\cG_0\left(\rho_2\right)\cG_{0,\rho_2}\left(\rho_1\right)+\frac{1}{2}\cG_1\left(\rho_2\right)\cG_{0,\rho_2}\left(\rho_1\right)\\
 \nonumber &+\frac{1}{2}\cG_0\left(\rho_1\right)\cG_{1,0}\left(\rho_2\right)-\cG_0\left(\rho_1\right)\cG_{1,1}\left(\rho_2\right)
 -\frac{1}{2}\cG_{1,0,0}\left(\rho_2\right)+\frac{1}{2}\cG_{1,0,1}\left(\rho_2\right)
\,. \\ \nonumber \\ 
%
%
%
%
\fa^{\text{(1,1)}}_{-++}\left(\rho_1,\rho_2\right) = &-\frac{1}{8}\cG_{1,0,\rho_2}\left(\rho_1\right)-\frac{1}{8}\cG_{1,1,\rho_2}\left(\rho_1\right)-\frac{1}{8}\cG_{1,\rho_2,0}\left(\rho_1\right)-\frac{1}{8}\cG_{1,\rho_2,1}\left(\rho_1\right)\\
 \nonumber &+\frac{1}{8}\cG_{1,\rho_2,\rho_2}\left(\rho_1\right)-\frac{1}{8}\cG_{\rho_2,0,1}\left(\rho_1\right)-\frac{1}{8}\cG_{\rho_2,1,0}\left(\rho_1\right)+\frac{1}{8}\cG_{\rho_2,\rho_2,1}\left(\rho_1\right)\\
 \nonumber &+\frac{1}{8}\cG_1\left(\rho_1\right)\cG_{0,0}\left(\rho_2\right)-\frac{1}{8}\cG_{\rho_2}\left(\rho_1\right)\cG_{0,1}\left(\rho_2\right)-\frac{1}{8}\cG_0\left(\rho_2\right)\cG_{1,0}\left(\rho_1\right)\\
 \nonumber &+\frac{1}{8}\cG_1\left(\rho_2\right)\cG_{1,0}\left(\rho_1\right)-\frac{1}{8}\cG_0\left(\rho_2\right)\cG_{1,1}\left(\rho_1\right)+\frac{1}{8}\cG_1\left(\rho_2\right)\cG_{1,1}\left(\rho_1\right)\\
 \nonumber &-\frac{1}{8}\cG_1\left(\rho_1\right)\cG_{1,1}\left(\rho_2\right)+\frac{1}{8}\cG_{\rho_2}\left(\rho_1\right)\cG_{1,1}\left(\rho_2\right)+\frac{1}{8}\cG_0\left(\rho_2\right)\cG_{1,\rho_2}\left(\rho_1\right)\\
 \nonumber &+\frac{1}{8}\cG_1\left(\rho_2\right)\cG_{\rho_2,0}\left(\rho_1\right)-\frac{1}{8}\cG_1\left(\rho_2\right)\cG_{\rho_2,\rho_2}\left(\rho_1\right)
\,.  \\ 
%
%
\fb_{1,-++}^{\text{(1,1)}}\left(\rho_1,\rho_2\right) = &\,\frac{1}{8}\cG_{0,0,1}\left(\rho_1\right)-\frac{1}{8}\cG_{0,0,1}\left(\rho_2\right)-\frac{1}{8}\cG_{0,1,0}\left(\rho_1\right)+\frac{1}{8}\cG_{0,1,1}\left(\rho_2\right)\\
 \nonumber &-\frac{1}{8}\cG_{0,\rho_2,1}\left(\rho_1\right)-\frac{1}{4}\cG_{1,0,1}\left(\rho_1\right)+\frac{1}{8}\cG_{1,0,1}\left(\rho_2\right)-\frac{1}{8}\cG_{1,1,1}\left(\rho_2\right)\\
 \nonumber &+\frac{1}{4}\cG_{1,\rho_2,1}\left(\rho_1\right)+\frac{1}{8}\cG_{\rho_2,0,1}\left(\rho_1\right)+\frac{1}{8}\cG_{\rho_2,1,0}\left(\rho_1\right)-\frac{1}{8}\cG_{\rho_2,\rho_2,1}\left(\rho_1\right)\\
 \nonumber &-\frac{1}{8}\cG_1\left(\rho_2\right)\cG_{0,0}\left(\rho_1\right)+\frac{1}{8}\cG_0\left(\rho_1\right)\cG_{0,1}\left(\rho_2\right)-\frac{1}{4}\cG_1\left(\rho_1\right)\cG_{0,1}\left(\rho_2\right)\\
 \nonumber &+\frac{1}{8}\cG_{\rho_2}\left(\rho_1\right)\cG_{0,1}\left(\rho_2\right)+\frac{1}{8}\cG_1\left(\rho_2\right)\cG_{0,\rho_2}\left(\rho_1\right)+\frac{1}{4}\cG_1\left(\rho_2\right)\cG_{1,0}\left(\rho_1\right)\\
 \nonumber &-\frac{1}{8}\cG_0\left(\rho_1\right)\cG_{1,1}\left(\rho_2\right)+\frac{1}{4}\cG_1\left(\rho_1\right)\cG_{1,1}\left(\rho_2\right)-\frac{1}{8}\cG_{\rho_2}\left(\rho_1\right)\cG_{1,1}\left(\rho_2\right)\\
 \nonumber &-\frac{1}{4}\cG_1\left(\rho_2\right)\cG_{1,\rho_2}\left(\rho_1\right)-\frac{1}{8}\cG_1\left(\rho_2\right)\cG_{\rho_2,0}\left(\rho_1\right)+\frac{1}{8}\cG_1\left(\rho_2\right)\cG_{\rho_2,\rho_2}\left(\rho_1\right)
\,.  \\ 
%
%
\fb_{2,-++}^{\text{(1,1)}}\left(\rho_1,\rho_2\right) = &-\frac{1}{8}\cG_{0,0,0}\left(\rho_2\right)-\frac{1}{8}\cG_{0,0,1}\left(\rho_1\right)+\frac{1}{8}\cG_{0,0,1}\left(\rho_2\right)-\frac{1}{8}\cG_{0,0,\rho_2}\left(\rho_1\right)\\
 \nonumber &+\frac{1}{8}\cG_{0,1,0}\left(\rho_1\right)+\frac{1}{8}\cG_{0,1,0}\left(\rho_2\right)-\frac{1}{8}\cG_{0,1,1}\left(\rho_2\right)+\frac{1}{8}\cG_{0,1,\rho_2}\left(\rho_1\right)\\
 \nonumber &+\frac{1}{8}\cG_{0,\rho_2,0}\left(\rho_1\right)-\frac{1}{8}\cG_{0,\rho_2,\rho_2}\left(\rho_1\right)+\frac{1}{4}\cG_{1,0,1}\left(\rho_1\right)+\frac{1}{4}\cG_{1,0,\rho_2}\left(\rho_1\right)+\frac{\zeta_3}{2}\\
 \nonumber &-\frac{1}{8}\cG_0\left(\rho_2\right)\cG_{0,0}\left(\rho_1\right)+\frac{1}{4}\cG_1\left(\rho_2\right)\cG_{0,0}\left(\rho_1\right)+\frac{1}{8}\cG_0\left(\rho_1\right)\cG_{0,0}\left(\rho_2\right)\\
 \nonumber &-\frac{1}{4}\cG_1\left(\rho_1\right)\cG_{0,0}\left(\rho_2\right)+\frac{1}{8}\cG_0\left(\rho_2\right)\cG_{0,1}\left(\rho_1\right)-\frac{1}{8}\cG_1\left(\rho_2\right)\cG_{0,1}\left(\rho_1\right)\\
 \nonumber &-\frac{1}{8}\cG_0\left(\rho_1\right)\cG_{0,1}\left(\rho_2\right)+\frac{1}{4}\cG_1\left(\rho_1\right)\cG_{0,1}\left(\rho_2\right)-\frac{1}{8}\cG_0\left(\rho_2\right)\cG_{0,\rho_2}\left(\rho_1\right)\\
 \nonumber &+\frac{1}{8}\cG_1\left(\rho_2\right)\cG_{0,\rho_2}\left(\rho_1\right)+\frac{1}{4}\cG_0\left(\rho_2\right)\cG_{1,0}\left(\rho_1\right)-\frac{1}{2}\cG_1\left(\rho_2\right)\cG_{1,0}\left(\rho_1\right)\,.
\end{align}
%
%

%
%
\begin{align}
\fa^{\text{(1,0)}}_{--+}\left(\rho_1,\rho_2\right) = &\,\frac{1}{4}\cG_{1,1}\left(\rho_1\right)-\frac{1}{4}\cG_{1,\rho_2}\left(\rho_1\right)-\frac{1}{4}\cG_0\left(\rho_2\right)\cG_1\left(\rho_1\right)
\,.  \\ 
%
%
\fb_{1,--+}^{\text{(1,0)}}\left(\rho_1,\rho_2\right) = &\,\frac{1}{4}\cG_{0,0}\left(\rho_2\right)-\frac{1}{4}\cG_{0,1}\left(\rho_1\right)-\frac{1}{4}\cG_{0,1}\left(\rho_2\right)+\frac{1}{4}\cG_{0,\rho_2}\left(\rho_1\right)\\
 \nonumber &-\frac{1}{4}\cG_0\left(\rho_1\right)\cG_0\left(\rho_2\right)+\frac{1}{2}\cG_0\left(\rho_2\right)\cG_1\left(\rho_1\right)
\,.  \\ 
%
%
\fb_{2,--+}^{\text{(1,0)}}\left(\rho_1,\rho_2\right) = &-\frac{1}{4}\cG_{0,0}\left(\rho_2\right)+\frac{1}{4}\cG_{0,1}\left(\rho_2\right)-\frac{1}{4}\cG_{1,0}\left(\rho_1\right)+\frac{1}{4}\cG_{1,1}\left(\rho_1\right)\\
 \nonumber &+\frac{1}{4}\cG_{\rho_2,0}\left(\rho_1\right)-\frac{1}{4}\cG_{\rho_2,1}\left(\rho_1\right)+\frac{1}{4}\cG_0\left(\rho_1\right)\cG_0\left(\rho_2\right)-\frac{1}{4}\cG_1\left(\rho_1\right)\cG_1\left(\rho_2\right)\\
 \nonumber&-\frac{1}{4}\cG_0\left(\rho_2\right)\cG_{\rho_2}\left(\rho_1\right)+\frac{1}{4}\cG_1\left(\rho_2\right)\cG_{\rho_2}\left(\rho_1\right)
\,. \\ \nonumber \\ 
%
%
%
%
\fa^{\text{(1,1)}}_{--+}\left(\rho_1,\rho_2\right) = &\,\frac{1}{8}\cG_{0,1,1}\left(\rho_1\right)-\frac{1}{8}\cG_{0,1,\rho_2}\left(\rho_1\right)+\frac{1}{8}\cG_{1,1,0}\left(\rho_1\right)-\frac{1}{8}\cG_{1,1,1}\left(\rho_1\right)\\
 \nonumber &-\frac{1}{8}\cG_{1,\rho_2,0}\left(\rho_1\right)+\frac{1}{4}\cG_{1,\rho_2,1}\left(\rho_1\right)-\frac{1}{8}\cG_{1,\rho_2,\rho_2}\left(\rho_1\right)\\
 \nonumber &-\frac{1}{8}\cG_0\left(\rho_2\right)\cG_{0,1}\left(\rho_1\right)-\frac{1}{8}\cG_1\left(\rho_1\right)\cG_{0,1}\left(\rho_2\right)-\frac{1}{8}\cG_0\left(\rho_2\right)\cG_{1,0}\left(\rho_1\right)\\
 \nonumber &+\frac{1}{4}\cG_1\left(\rho_1\right)\cG_{1,0}\left(\rho_2\right)+\frac{1}{8}\cG_1\left(\rho_2\right)\cG_{1,1}\left(\rho_1\right)-\frac{1}{8}\cG_1\left(\rho_2\right)\cG_{1,\rho_2}\left(\rho_1\right)
\,.  \\ 
%
%
\fb_{1,--+}^{\text{(1,1)}}\left(\rho_1,\rho_2\right) = &\,\frac{1}{4}\cG_{0,0,0}\left(\rho_2\right)-\frac{1}{8}\cG_{0,0,1}\left(\rho_2\right)-\frac{1}{8}\cG_{0,1,0}\left(\rho_2\right)-\frac{1}{8}\cG_{0,1,1}\left(\rho_1\right)\\
 \nonumber &+\frac{1}{8}\cG_{0,1,\rho_2}\left(\rho_1\right)-\frac{1}{8}\cG_{0,\rho_2,1}\left(\rho_1\right)+\frac{1}{8}\cG_{0,\rho_2,\rho_2}\left(\rho_1\right)-\frac{1}{4}\cG_{1,0,0}\left(\rho_2\right)\\
 \nonumber &-\frac{1}{8}\cG_0\left(\rho_1\right)\cG_{0,0}\left(\rho_2\right)+\frac{1}{4}\cG_1\left(\rho_1\right)\cG_{0,0}\left(\rho_2\right)+\frac{1}{8}\cG_0\left(\rho_2\right)\cG_{0,1}\left(\rho_1\right)\\
 \nonumber &+\frac{1}{8}\cG_0\left(\rho_2\right)\cG_{0,\rho_2}\left(\rho_1\right)+\frac{1}{4}\cG_0\left(\rho_1\right)\cG_{1,0}\left(\rho_2\right)-\frac{1}{2}\cG_1\left(\rho_1\right)\cG_{1,0}\left(\rho_2\right)\\
  \nonumber &+\frac{1}{4}\cG_{1,0,1}\left(\rho_2\right)+\frac{3\zeta_3}{4}
\,.  \\ 
%
%
\fb_{2,--+}^{\text{(1,1)}}\left(\rho_1,\rho_2\right) = &-\frac{1}{4}\cG_{0,0,0}\left(\rho_2\right)+\frac{1}{8}\cG_{0,0,1}\left(\rho_2\right)+\frac{1}{8}\cG_{0,1,0}\left(\rho_2\right)+\frac{1}{4}\cG_{1,0,0}\left(\rho_2\right)\\
 \nonumber &-\frac{1}{4}\cG_{1,0,1}\left(\rho_2\right)-\frac{1}{8}\cG_{1,1,0}\left(\rho_1\right)+\frac{1}{8}\cG_{1,1,1}\left(\rho_1\right)+\frac{1}{8}\cG_{1,\rho_2,0}\left(\rho_1\right)\\
 \nonumber &-\frac{1}{8}\cG_{1,\rho_2,1}\left(\rho_1\right)-\frac{1}{8}\cG_{\rho_2,1,0}\left(\rho_1\right)+\frac{1}{8}\cG_{\rho_2,1,1}\left(\rho_1\right)+\frac{1}{8}\cG_{\rho_2,\rho_2,0}\left(\rho_1\right)\\
 \nonumber &+\frac{1}{8}\cG_0\left(\rho_1\right)\cG_{0,0}\left(\rho_2\right)-\frac{1}{8}\cG_1\left(\rho_1\right)\cG_{0,0}\left(\rho_2\right)-\frac{1}{4}\cG_{\rho_2}\left(\rho_1\right)\cG_{0,0}\left(\rho_2\right)\\
 \nonumber &+\frac{1}{8}\cG_1\left(\rho_1\right)\cG_{0,1}\left(\rho_2\right)+\frac{1}{8}\cG_{\rho_2}\left(\rho_1\right)\cG_{0,1}\left(\rho_2\right)+\frac{1}{8}\cG_0\left(\rho_2\right)\cG_{1,0}\left(\rho_1\right)\\
 \nonumber &-\frac{1}{4}\cG_0\left(\rho_1\right)\cG_{1,0}\left(\rho_2\right)+\frac{1}{8}\cG_1\left(\rho_1\right)\cG_{1,0}\left(\rho_2\right)+\frac{1}{8}\cG_{\rho_2}\left(\rho_1\right)\cG_{1,0}\left(\rho_2\right)\\
 \nonumber &-\frac{1}{8}\cG_1\left(\rho_2\right)\cG_{1,1}\left(\rho_1\right)-\frac{1}{8}\cG_0\left(\rho_2\right)\cG_{1,\rho_2}\left(\rho_1\right)+\frac{1}{8}\cG_1\left(\rho_2\right)\cG_{1,\rho_2}\left(\rho_1\right)\\
 \nonumber &+\frac{1}{8}\cG_0\left(\rho_2\right)\cG_{\rho_2,0}\left(\rho_1\right)-\frac{1}{8}\cG_1\left(\rho_2\right)\cG_{\rho_2,1}\left(\rho_1\right)-\frac{1}{8}\cG_0\left(\rho_2\right)\cG_{\rho_2,\rho_2}\left(\rho_1\right)\\
 \nonumber &+\frac{1}{8}\cG_1\left(\rho_2\right)\cG_{\rho_2,\rho_2}\left(\rho_1\right)-\frac{1}{8}\cG_{\rho_2,\rho_2,1}\left(\rho_1\right)-\frac{3\zeta_3}{4}
\,. \end{align}
%
%
%
%
\begin{align}
\fa^{\text{(2,0)}}_{--+}\left(\rho_1,\rho_2\right) = &\,\frac{1}{8}\cG_{0,1,1}\left(\rho_1\right)-\frac{1}{8}\cG_{0,1,\rho_2}\left(\rho_1\right)-\frac{1}{8}\cG_{1,0,1}\left(\rho_1\right)+\frac{1}{8}\cG_{1,0,\rho_2}\left(\rho_1\right)\\
 \nonumber &+\frac{1}{4}\cG_{1,1,0}\left(\rho_1\right)-\frac{1}{2}\cG_{1,1,1}\left(\rho_1\right)+\frac{1}{4}\cG_{1,1,\rho_2}\left(\rho_1\right)-\frac{1}{4}\cG_{1,\rho_2,0}\left(\rho_1\right)\\
 \nonumber &+\frac{1}{4}\cG_1\left(\rho_1\right)\cG_{0,0}\left(\rho_2\right)-\frac{1}{8}\cG_0\left(\rho_2\right)\cG_{0,1}\left(\rho_1\right)-\frac{1}{4}\cG_1\left(\rho_1\right)\cG_{0,1}\left(\rho_2\right)\\
 \nonumber &-\frac{1}{4}\cG_0\left(\rho_2\right)\cG_{1,0}\left(\rho_1\right)+\frac{1}{4}\cG_0\left(\rho_2\right)\cG_{1,1}\left(\rho_1\right)+\frac{1}{4}\cG_1\left(\rho_2\right)\cG_{1,1}\left(\rho_1\right)\\
 \nonumber &+\frac{1}{4}\cG_0\left(\rho_2\right)\cG_{1,\rho_2}\left(\rho_1\right)-\frac{1}{4}\cG_1\left(\rho_2\right)\cG_{1,\rho_2}\left(\rho_1\right)+\frac{3}{8}\cG_{1,\rho_2,1}\left(\rho_1\right)\\
 \nonumber&-\frac{1}{8}\cG_{1,\rho_2,\rho_2}\left(\rho_1\right)
\,.  \\ 
%
%
\fb_{1,--+}^{\text{(2,0)}}\left(\rho_1,\rho_2\right) = &-\frac{1}{8}\cG_{0,0,0}\left(\rho_2\right)-\frac{3}{8}\cG_{0,0,1}\left(\rho_1\right)+\frac{1}{8}\cG_{0,0,1}\left(\rho_2\right)+\frac{3}{8}\cG_{0,0,\rho_2}\left(\rho_1\right)\\
 \nonumber &-\frac{3}{8}\cG_{0,1,0}\left(\rho_1\right)+\frac{1}{8}\cG_{0,1,0}\left(\rho_2\right)+\frac{1}{2}\cG_{0,1,1}\left(\rho_1\right)-\frac{1}{8}\cG_{0,1,1}\left(\rho_2\right)\\
 \nonumber &-\frac{1}{8}\cG_{0,1,\rho_2}\left(\rho_1\right)+\frac{3}{8}\cG_{0,\rho_2,0}\left(\rho_1\right)-\frac{1}{2}\cG_{0,\rho_2,1}\left(\rho_1\right)+\frac{1}{8}\cG_{0,\rho_2,\rho_2}\left(\rho_1\right)\\
 \nonumber &-\frac{1}{8}\cG_0\left(\rho_2\right)\cG_{0,0}\left(\rho_1\right)+\frac{1}{8}\cG_0\left(\rho_1\right)\cG_{0,0}\left(\rho_2\right)-\frac{1}{2}\cG_1\left(\rho_1\right)\cG_{0,0}\left(\rho_2\right)\\
 \nonumber &+\frac{7}{8}\cG_0\left(\rho_2\right)\cG_{0,1}\left(\rho_1\right)-\frac{3}{8}\cG_1\left(\rho_2\right)\cG_{0,1}\left(\rho_1\right)-\frac{1}{8}\cG_0\left(\rho_1\right)\cG_{0,1}\left(\rho_2\right)\\
 \nonumber &+\frac{1}{2}\cG_1\left(\rho_1\right)\cG_{0,1}\left(\rho_2\right)-\frac{3}{8}\cG_0\left(\rho_2\right)\cG_{0,\rho_2}\left(\rho_1\right)+\frac{3}{8}\cG_1\left(\rho_2\right)\cG_{0,\rho_2}\left(\rho_1\right)\\
 \nonumber &+\frac{1}{2}\cG_0\left(\rho_2\right)\cG_{1,0}\left(\rho_1\right)-\cG_0\left(\rho_2\right)\cG_{1,1}\left(\rho_1\right)+\frac{1}{2}\cG_{1,0,1}\left(\rho_1\right)-\frac{1}{2}\cG_{1,0,\rho_2}\left(\rho_1\right)
\,.  \\ 
%
%
\fb_{2,--+}^{\text{(2,0)}}\left(\rho_1,\rho_2\right) = &\,\frac{1}{8}\cG_{0,0,0}\left(\rho_2\right)+\frac{1}{4}\cG_{0,0,1}\left(\rho_1\right)-\frac{1}{8}\cG_{0,0,1}\left(\rho_2\right)-\frac{1}{4}\cG_{0,0,\rho_2}\left(\rho_1\right)\\
 \nonumber &+\frac{1}{8}\cG_{0,1,0}\left(\rho_1\right)-\frac{1}{8}\cG_{0,1,0}\left(\rho_2\right)-\frac{1}{8}\cG_{0,1,1}\left(\rho_1\right)+\frac{1}{8}\cG_{0,1,1}\left(\rho_2\right)\\
 \nonumber &-\frac{1}{8}\cG_{0,\rho_2,0}\left(\rho_1\right)+\frac{1}{8}\cG_{0,\rho_2,1}\left(\rho_1\right)-\frac{1}{8}\cG_{1,0,0}\left(\rho_1\right)+\frac{1}{8}\cG_{1,0,1}\left(\rho_1\right)\\
 \nonumber &+\frac{1}{4}\cG_{1,1,0}\left(\rho_1\right)-\frac{1}{2}\cG_{1,1,1}\left(\rho_1\right)+\frac{1}{4}\cG_{1,1,\rho_2}\left(\rho_1\right)-\frac{1}{8}\cG_{1,\rho_2,0}\left(\rho_1\right)\\
 \nonumber &+\frac{1}{8}\cG_{1,\rho_2,1}\left(\rho_1\right)+\frac{1}{8}\cG_{\rho_2,0,0}\left(\rho_1\right)-\frac{3}{8}\cG_{\rho_2,0,1}\left(\rho_1\right)+\frac{1}{4}\cG_{\rho_2,0,\rho_2}\left(\rho_1\right)\\
 \nonumber &-\frac{1}{4}\cG_{\rho_2,1,0}\left(\rho_1\right)+\frac{1}{2}\cG_{\rho_2,1,1}\left(\rho_1\right)-\frac{1}{4}\cG_{\rho_2,1,\rho_2}\left(\rho_1\right)+\frac{1}{8}\cG_{\rho_2,\rho_2,0}\left(\rho_1\right)\\
 \nonumber &+\frac{1}{8}\cG_0\left(\rho_2\right)\cG_{0,0}\left(\rho_1\right)-\frac{1}{8}\cG_0\left(\rho_1\right)\cG_{0,0}\left(\rho_2\right)+\frac{1}{8}\cG_1\left(\rho_1\right)\cG_{0,0}\left(\rho_2\right)\\
 \nonumber &+\frac{1}{8}\cG_{\rho_2}\left(\rho_1\right)\cG_{0,0}\left(\rho_2\right)-\frac{1}{2}\cG_0\left(\rho_2\right)\cG_{0,1}\left(\rho_1\right)+\frac{1}{8}\cG_1\left(\rho_2\right)\cG_{0,1}\left(\rho_1\right)\\
 \nonumber &+\frac{1}{8}\cG_0\left(\rho_1\right)\cG_{0,1}\left(\rho_2\right)-\frac{1}{8}\cG_1\left(\rho_1\right)\cG_{0,1}\left(\rho_2\right)-\frac{1}{8}\cG_{\rho_2}\left(\rho_1\right)\cG_{0,1}\left(\rho_2\right)\\
 \nonumber &+\frac{1}{8}\cG_0\left(\rho_2\right)\cG_{0,\rho_2}\left(\rho_1\right)-\frac{1}{8}\cG_1\left(\rho_2\right)\cG_{0,\rho_2}\left(\rho_1\right)-\frac{1}{8}\cG_0\left(\rho_2\right)\cG_{1,0}\left(\rho_1\right)\\
 \nonumber &-\frac{1}{8}\cG_1\left(\rho_2\right)\cG_{1,0}\left(\rho_1\right)+\frac{1}{8}\cG_1\left(\rho_1\right)\cG_{1,0}\left(\rho_2\right)-\frac{1}{8}\cG_{\rho_2}\left(\rho_1\right)\cG_{1,0}\left(\rho_2\right)\\
 \nonumber &+\frac{1}{4}\cG_0\left(\rho_2\right)\cG_{1,1}\left(\rho_1\right)+\frac{1}{4}\cG_1\left(\rho_2\right)\cG_{1,1}\left(\rho_1\right)-\frac{1}{8}\cG_1\left(\rho_1\right)\cG_{1,1}\left(\rho_2\right)\\
 \nonumber &+\frac{1}{8}\cG_{\rho_2}\left(\rho_1\right)\cG_{1,1}\left(\rho_2\right)+\frac{1}{8}\cG_0\left(\rho_2\right)\cG_{1,\rho_2}\left(\rho_1\right)-\frac{1}{8}\cG_1\left(\rho_2\right)\cG_{1,\rho_2}\left(\rho_1\right)
 \end{align}
 \begin{align}
 \phantom{\fb_{2,--+}^{\text{(2,0)}}\left(\rho_1,\rho_2\right) =}
 \nonumber &-\frac{1}{8}\cG_0\left(\rho_2\right)\cG_{\rho_2,0}\left(\rho_1\right)+\frac{1}{8}\cG_1\left(\rho_2\right)\cG_{\rho_2,0}\left(\rho_1\right)+\frac{1}{4}\cG_0\left(\rho_2\right)\cG_{\rho_2,1}\left(\rho_1\right)\\
 \nonumber &-\frac{1}{4}\cG_1\left(\rho_2\right)\cG_{\rho_2,1}\left(\rho_1\right)-\frac{1}{8}\cG_0\left(\rho_2\right)\cG_{\rho_2,\rho_2}\left(\rho_1\right)+\frac{1}{8}\cG_1\left(\rho_2\right)\cG_{\rho_2,\rho_2}\left(\rho_1\right)\\
  \nonumber &-\frac{1}{8}\cG_{\rho_2,\rho_2,1}\left(\rho_1\right)
\,.  \end{align}

\begin{allowdisplaybreaks}

Here we present the MRK limits of the pure functions appearing in the seven-point NMHV amplitude as described in Section \ref{7ptexpcheck}.
\begin{align}
\hat{X}=&\frac{1}{48} \log ^3(\tau _1)
+\frac{1}{48} \log ^3(\tau _2)
-\frac{1}{32} \log ^2(\tau _1) \log (\tau _2)
-\frac{1}{32} \log (\tau _1) \log ^2(\tau _2)\\
& +\log ^2(\tau _1)\Big(-\frac{1}{16} \cG_0(\rho_1)-\frac{1}{32} \cG_0(\rho _2)+\frac{3}{32} \cG_1(\rho _1)-\frac{1}{16} \cG_1(\rho _2)-\frac{1}{32} \cG_{\rho _2}(\rho _1)\Big) \notag\\
&+\log ^2(\tau _2)\Big(\frac{3}{32} \cG_0(\rho _1)-\frac{1}{32} \cG_0(\rho _2)-\frac{1}{16} \cG_1(\rho _1)+\frac{3}{32} \cG_1(\rho _2)-\frac{1}{32} \cG_{\rho _2}(\rho _1)\Big) 
\notag\\
& +\log (\tau _1) \log (\tau _2)\Big(\frac{3}{16} \cG_0(\rho _1)-\frac{3}{16} \cG_1(\rho _1)-\frac{3}{16} \cG_1(\rho _2)+\frac{1}{4} \cG_{\rho _2}(\rho _1)\Big)
+\log (\tau _1)\Big(-\frac{1}{8} \cG_{0,0}(\rho _1)\notag\\
&+\frac{3}{16} \cG_{0,0}(\rho _2)-\frac{9}{16} \cG_{0,1}(\rho _1)-\frac{3}{16} \cG_{0,1}(\rho _2)+\frac{7}{16} \cG_{0,\rho _2}(\rho _1)-\frac{13}{16} \cG_{1,0}(\rho _1)+\frac{1}{16} \cG_{1,0}(\rho _2)\notag\\
&+\frac{7}{16} \cG_{1,1}(\rho _1)-\frac{3}{8} \cG_{1,1}(\rho _2)-\frac{3}{8} \cG_{1,\rho _2}(\rho _1)+\frac{7}{16} \cG_{\rho _2,0}(\rho _1)-\frac{1}{8} \cG_{\rho _2,1}(\rho _1)-\frac{5}{16} \cG_{\rho _2,\rho _2}(\rho _1)\notag\\
&-\frac{1}{16} \cG_0(\rho _1) \cG_0(\rho _2)+\frac{3}{8} \cG_1(\rho _1) \cG_0(\rho _2)-\frac{9}{16} \cG_{\rho _2}(\rho _1) \cG_0(\rho _2)+\frac{1}{8} \cG_0(\rho _1) \cG_1(\rho _2)+\frac{1}{16} \cG_1(\rho _1) \cG_1(\rho _2)\notag\\
&+\frac{5}{16} \cG_1(\rho _2) \cG_{\rho _2}(\rho _1)\Big)
+\log (\tau _2)\Big(-\frac{5}{16} \cG_{0,0}(\rho _1)-\frac{17}{16} \cG_{0,0}(\rho _2)+\frac{5}{16} \cG_{0,1}(\rho _1)+\frac{3}{8} \cG_{0,1}(\rho _2)\notag\\
&+\frac{1}{4} \cG_{0,\rho _2}(\rho _1)+\frac{9}{16} \cG_{1,0}(\rho _1)+\frac{1}{8} \cG_{1,0}(\rho _2)-\frac{3}{8} \cG_{1,1}(\rho _1)+\frac{7}{16} \cG_{1,1}(\rho _2)+\frac{5}{16} \cG_{1,\rho _2}(\rho _1)+\frac{9}{16} \cG_{\rho _2,1}(\rho _1)\notag\\
&-\frac{5}{16} \cG_{\rho _2,\rho _2}(\rho _1)+\frac{1}{2} \cG_0(\rho _1) \cG_0(\rho _2)-\frac{3}{16} \cG_1(\rho _1) \cG_0(\rho _2)-\frac{1}{16} \cG_{\rho _2}(\rho _1) \cG_0(\rho _2)-\frac{7}{16} \cG_0(\rho _1) \cG_1(\rho _2)\notag\\
&+\frac{1}{16} \cG_1(\rho _1) \cG_1(\rho _2)-\frac{3}{8} \cG_1(\rho _2) \cG_{\rho _2}(\rho _1)\Big)\notag
\end{align}
\begin{align}
\hat{V}_{12}=&-\frac{1}{48} \log ^3(\tau _1)
-\frac{1}{48} \log ^3(\tau _2)
+\log ^2(\tau _1)\Big(\frac{1}{16} \cG_0(\rho _1)-\frac{1}{8} \cG_1(\rho _1)\Big) 
\\
&+\log ^2(\tau _2)\Big(-\frac{1}{16} \cG_0(\rho _1)+\frac{1}{16} \cG_1(\rho _1)-\frac{1}{16} \cG_1(\rho _2)\Big) 
+\log (\tau _1) \log (\tau _2)\Big(-\frac{1}{8} \cG_0(\rho _1)\notag\\
&+\frac{1}{4} \cG_1(\rho _1)-\frac{1}{4} \cG_{\rho _2}(\rho _1)\Big) 
+\log (\tau _1)\Big(\frac{1}{8} \cG_{0,0}(\rho _1)-\frac{1}{8} \cG_{0,0}(\rho _2)+\frac{3}{8} \cG_{0,1}(\rho _1)+\frac{1}{8} \cG_{0,1}(\rho _2)\notag\\
&-\frac{3}{8} \cG_{0,\rho _2}(\rho _1)+\frac{3}{8} \cG_{1,0}(\rho _1)-\frac{1}{8} \cG_{1,0}(\rho _2)-\frac{1}{2} \cG_{1,1}(\rho _1)+\frac{1}{8} \cG_{1,1}(\rho _2)+\frac{1}{4} \cG_{1,\rho _2}(\rho _1)\notag\\
&-\frac{3}{8} \cG_{\rho _2,0}(\rho _1)+\frac{3}{8} \cG_{\rho _2,\rho _2}(\rho _1)+\frac{1}{8} \cG_0(\rho _1) \cG_0(\rho _2)-\frac{1}{2} \cG_1(\rho _1) \cG_0(\rho _2)+\frac{5}{8} \cG_{\rho _2}(\rho _1) \cG_0(\rho _2)\notag\\
&+\frac{1}{4} \cG_1(\rho _1) \cG_1(\rho _2)-\frac{3}{8} \cG_1(\rho _2) \cG_{\rho _2}(\rho _1)\Big)
+\log (\tau _2)\Big(\frac{1}{4} \cG_{0,0}(\rho _1)+\frac{9}{8} \cG_{0,0}(\rho _2)-\frac{1}{8} \cG_{0,1}(\rho _1)\notag\\
&-\frac{5}{8} \cG_{0,1}(\rho _2)-\frac{1}{4} \cG_{0,\rho _2}(\rho _1)-\frac{3}{8} \cG_{1,0}(\rho _1)-\frac{5}{8} \cG_{1,0}(\rho _2)+\frac{3}{8} \cG_{1,1}(\rho _1)+\frac{1}{4} \cG_{1,1}(\rho _2)\notag\\
&-\frac{1}{2} \cG_{\rho _2,1}(\rho _1)+\frac{3}{8} \cG_{\rho _2,\rho _2}(\rho _1)-\frac{1}{2} \cG_0(\rho _1) \cG_0(\rho _2)+\frac{1}{2} \cG_1(\rho _1) \cG_0(\rho _2)+\frac{1}{8} \cG_{\rho _2}(\rho _1) \cG_0(\rho _2)\notag\\
&+\frac{3}{8} \cG_0(\rho _1) \cG_1(\rho _2)-\frac{3}{8} \cG_1(\rho _1) \cG_1(\rho _2)+\frac{1}{8} \cG_1(\rho _2) \cG_{\rho _2}(\rho _1)\Big) \notag 
\end{align}
\begin{align}
\hat{V}_{23}&=\frac{1}{32} \log ^2(\tau _1) \log (\tau _2)
+\frac{1}{32} \log (\tau _1) \log ^2(\tau _2)
+\log ^2(\tau _1)\Big(\frac{1}{32} \cG_0(\rho _2)+\frac{1}{32} \cG_1(\rho _1)\\
&+\frac{1}{16} \cG_1(\rho _2)+\frac{1}{32} \cG_{\rho _2}(\rho _1)\Big)
+\log ^2(\tau _2)\Big(-\frac{1}{32} \cG_0(\rho _1)+\frac{1}{32} \cG_0(\rho _2)-\frac{1}{32} \cG_1(\rho _2)+\frac{1}{32} \cG_{\rho _2}(\rho _1)\Big) 
\notag\\
&+\log (\tau _1) \log (\tau _2)\Big(-\frac{1}{16} \cG_0(\rho _1)-\frac{1}{16} \cG_1(\rho _1)+\frac{3}{16} \cG_1(\rho _2)\Big) 
+\log (\tau _1)\Big(-\frac{1}{16} \cG_{0,0}(\rho _2)\notag\\
&+\frac{3}{16} \cG_{0,1}(\rho _1)+\frac{1}{16} \cG_{0,1}(\rho _2)-\frac{1}{16} \cG_{0,\rho _2}(\rho _1)+\frac{3}{16} \cG_{1,0}(\rho _1)+\frac{1}{16} \cG_{1,0}(\rho _2)+\frac{1}{16} \cG_{1,1}(\rho _1)\notag\\
&+\frac{1}{4} \cG_{1,1}(\rho _2)+\frac{1}{8} \cG_{1,\rho _2}(\rho _1)-\frac{1}{16} \cG_{\rho _2,0}(\rho _1)+\frac{1}{8} \cG_{\rho _2,1}(\rho _1)-\frac{1}{16} \cG_{\rho _2,\rho _2}(\rho _1)-\frac{1}{16} \cG_0(\rho _1) \cG_0(\rho _2)\notag\\
&+\frac{1}{8} \cG_1(\rho _1) \cG_0(\rho _2)-\frac{1}{16} \cG_{\rho _2}(\rho _1) \cG_0(\rho _2)-\frac{1}{8} \cG_0(\rho _1) \cG_1(\rho _2)-\frac{5}{16} \cG_1(\rho _1) \cG_1(\rho _2)+\frac{1}{16} \cG_1(\rho _2) \cG_{\rho _2}(\rho _1)\Big)
\notag\\
&+\log (\tau _2)\Big(\frac{1}{16} \cG_{0,0}(\rho _1)-\frac{1}{16} \cG_{0,0}(\rho _2)-\frac{3}{16} \cG_{0,1}(\rho _1)-\frac{3}{16} \cG_{1,0}(\rho _1)+\frac{1}{4} \cG_{1,0}(\rho _2)-\frac{3}{16} \cG_{1,1}(\rho _2)\notag\\
&-\frac{1}{16} \cG_{1,\rho _2}(\rho _1)+\frac{3}{16} \cG_{\rho _2,1}(\rho _1)-\frac{1}{16} \cG_{\rho _2,\rho _2}(\rho _1)-\frac{1}{16} \cG_0(\rho _2) \cG_1(\rho _1)+\frac{1}{16} \cG_1(\rho _2) \cG_1(\rho _1)\notag\\
&+\frac{1}{16} \cG_0(\rho _1) \cG_1(\rho _2)-\frac{1}{16} \cG_0(\rho _2) \cG_{\rho _2}(\rho _1)\Big)\notag
\end{align}
\begin{align}
\hat{V}_{71}&=\frac{1}{32} \log ^2(\tau _1) \log (\tau _2)
+\frac{1}{32} \log (\tau _1) \log ^2(\tau _2)
+\log ^2(\tau _1)\Big(\frac{1}{32} \cG_0(\rho _2)+\frac{1}{32} \cG_1(\rho _1)\\
&+\frac{1}{16} \cG_1(\rho _2)+\frac{1}{32} \cG_{\rho _2}(\rho _1)\Big) 
+\log ^2(\tau _2)\Big(-\frac{1}{32} \cG_0(\rho _1)+\frac{1}{32} \cG_0(\rho _2)-\frac{1}{32} \cG_1(\rho _2)\notag\\
&+\frac{1}{32} \cG_{\rho _2}(\rho _1)\Big) 
+\log (\tau _1) \log (\tau _2)\Big(-\frac{1}{16} \cG_0(\rho _1)-\frac{1}{16} \cG_1(\rho _1)+\frac{3}{16} \cG_1(\rho _2)\Big)
\notag\\
&+\log (\tau _1)\Big(-\frac{1}{4} \cG_{0,0}(\rho _1)-\frac{1}{16} \cG_{0,0}(\rho _2)+\frac{3}{16} \cG_{0,1}(\rho _1)+\frac{1}{16} \cG_{0,1}(\rho _2)-\frac{1}{16} \cG_{0,\rho _2}(\rho _1)\notag\\
&+\frac{11}{16} \cG_{1,0}(\rho _1)+\frac{1}{16} \cG_{1,0}(\rho _2)+\frac{1}{16} \cG_{1,1}(\rho _1)+\frac{1}{4} \cG_{1,1}(\rho _2)+\frac{1}{8} \cG_{1,\rho _2}(\rho _1)-\frac{1}{16} \cG_{\rho _2,0}(\rho _1)\notag\\
&+\frac{1}{8} \cG_{\rho _2,1}(\rho _1)-\frac{1}{16} \cG_{\rho _2,\rho _2}(\rho _1)-\frac{1}{16} \cG_0(\rho _1) \cG_0(\rho _2)+\frac{1}{8} \cG_1(\rho _1) \cG_0(\rho _2)-\frac{1}{16} \cG_{\rho _2}(\rho _1) \cG_0(\rho _2)\notag\\
&-\frac{1}{8} \cG_0(\rho _1) \cG_1(\rho _2)-\frac{5}{16} \cG_1(\rho _1) \cG_1(\rho _2)+\frac{1}{16} \cG_1(\rho _2) \cG_{\rho _2}(\rho _1)\Big)
+\log (\tau _2)\Big(\frac{1}{16} \cG_{0,0}(\rho _1)\notag\\
&+\frac{3}{16} \cG_{0,0}(\rho _2)-\frac{3}{16} \cG_{0,1}(\rho _1)-\frac{1}{4} \cG_{0,\rho _2}(\rho _1)-\frac{3}{16} \cG_{1,0}(\rho _1)+\frac{1}{4} \cG_{1,0}(\rho _2)-\frac{7}{16} \cG_{1,1}(\rho _2)\notag\\
&-\frac{1}{16} \cG_{1,\rho _2}(\rho _1)-\frac{1}{16} \cG_{\rho _2,1}(\rho _1)-\frac{1}{16} \cG_{\rho _2,\rho _2}(\rho _1)-\frac{1}{4} \cG_0(\rho _1) \cG_0(\rho _2)-\frac{1}{16} \cG_1(\rho _1) \cG_0(\rho _2)\notag\\
&-\frac{1}{16} \cG_{\rho _2}(\rho _1) \cG_0(\rho _2)+\frac{5}{16} \cG_0(\rho _1) \cG_1(\rho _2)+\frac{1}{16} \cG_1(\rho _1) \cG_1(\rho _2)+\frac{1}{4} \cG_1(\rho _2) \cG_{\rho _2}(\rho _1)\Big)\notag 
\end{align}
\begin{align}
\hat{V}_{73}&=\frac{1}{32} \log ^2(\tau _1) \log (\tau _2)
+\frac{1}{32} \log (\tau _1) \log ^2(\tau _2)
+\log ^2(\tau _1)\Big(\frac{1}{32} \cG_0(\rho _2)+\frac{1}{32} \cG_1(\rho _1)\\
&+\frac{1}{16} \cG_1(\rho _2)+\frac{1}{32} \cG_{\rho _2}(\rho _1)\Big)
+\log ^2(\tau _2)\Big(-\frac{1}{32} \cG_0(\rho _1)+\frac{1}{32} \cG_0(\rho _2)-\frac{1}{32} \cG_1(\rho _2)\notag\\
&+\frac{1}{32} \cG_{\rho _2}(\rho _1)\Big) 
+\log (\tau _1) \log (\tau _2)\Big(-\frac{1}{16} \cG_0(\rho _1)-\frac{1}{16} \cG_1(\rho _1)+\frac{3}{16} \cG_1(\rho _2)\Big) 
\notag\\
&+\log (\tau _1)\Big(-\frac{1}{16} \cG_{0,0}(\rho _2)+\frac{3}{16} \cG_{0,1}(\rho _1)+\frac{1}{16} \cG_{0,1}(\rho _2)-\frac{1}{16} \cG_{0,\rho _2}(\rho _1)+\frac{3}{16} \cG_{1,0}(\rho _1)\notag\\
&+\frac{1}{16} \cG_{1,0}(\rho _2)+\frac{1}{16} \cG_{1,1}(\rho _1)+\frac{1}{4} \cG_{1,1}(\rho _2)+\frac{1}{8} \cG_{1,\rho _2}(\rho _1)-\frac{1}{16} \cG_{\rho _2,0}(\rho _1)+\frac{1}{8} \cG_{\rho _2,1}(\rho _1)\notag\\
&-\frac{1}{16} \cG_{\rho _2,\rho _2}(\rho _1)-\frac{1}{16} \cG_0(\rho _1) \cG_0(\rho _2)+\frac{1}{8} \cG_1(\rho _1) \cG_0(\rho _2)-\frac{1}{16} \cG_{\rho _2}(\rho _1) \cG_0(\rho _2)-\frac{1}{8} \cG_0(\rho _1) \cG_1(\rho _2)\notag\\
&-\frac{5}{16} \cG_1(\rho _1) \cG_1(\rho _2)+\frac{1}{16} \cG_1(\rho _2) \cG_{\rho _2}(\rho _1)\Big)
+\log (\tau _2)\Big(\frac{1}{16} \cG_{0,0}(\rho _1)-\frac{1}{16} \cG_{0,0}(\rho _2)+\frac{1}{16} \cG_{0,1}(\rho _1)\notag\\
&+\frac{1}{4} \cG_{0,1}(\rho _2)-\frac{3}{16} \cG_{1,0}(\rho _1)+\frac{1}{4} \cG_{1,0}(\rho _2)-\frac{7}{16} \cG_{1,1}(\rho _2)-\frac{1}{16} \cG_{1,\rho _2}(\rho _1)-\frac{1}{16} \cG_{\rho _2,1}(\rho _1)\notag\\
&-\frac{1}{16} \cG_{\rho _2,\rho _2}(\rho _1)-\frac{1}{16} \cG_0(\rho _2) \cG_1(\rho _1)+\frac{1}{16} \cG_1(\rho _2) \cG_1(\rho _1)-\frac{3}{16} \cG_0(\rho _1) \cG_1(\rho _2)\notag\\
&-\frac{1}{16} \cG_0(\rho _2) \cG_{\rho _2}(\rho _1)+\frac{1}{4} \cG_1(\rho _2) \cG_{\rho _2}(\rho _1)\Big)\notag
\end{align}
\begin{align}
\hat{V}_{14}=&-\frac{1}{48} \log ^3(\tau _2)
+ \log ^2(\tau _2)\Big(-\frac{1}{16} \cG_0(\rho _1)+\frac{1}{16} \cG_1(\rho _1)-\frac{1}{16} \cG_1(\rho _2)\Big)\\
&+\log (\tau _1) \log (\tau _2)\Big(-\frac{1}{8} \cG_0(\rho _1)+\frac{1}{8} \cG_0(\rho _2)+\frac{1}{4} \cG_1(\rho _1)-\frac{1}{8} \cG_1(\rho _2)-\frac{1}{8} \cG_{\rho _2}(\rho _1)\Big)
\notag\\
&+\log (\tau _1)\Big(-\frac{1}{4} \cG_{0,0}(\rho _1)-\frac{1}{2} \cG_{0,0}(\rho _2)+\frac{5}{8} \cG_{0,1}(\rho _1)+\frac{3}{8} \cG_{0,1}(\rho _2)-\frac{3}{8} \cG_{0,\rho _2}(\rho _1)\notag\\
&+\frac{3}{8} \cG_{1,0}(\rho _1)+\frac{3}{8} \cG_{1,0}(\rho _2)-\frac{1}{2} \cG_{1,1}(\rho _1)-\frac{1}{4} \cG_{1,1}(\rho _2)+\frac{1}{8} \cG_{1,\rho _2}(\rho _1)-\frac{1}{8} \cG_{\rho _2,0}(\rho _1)\notag\\
&-\frac{1}{8} \cG_{\rho _2,1}(\rho _1)+\frac{1}{4} \cG_{\rho _2,\rho _2}(\rho _1)+\frac{3}{8} \cG_0(\rho _1) \cG_0(\rho _2)-\frac{5}{8} \cG_1(\rho _1) \cG_0(\rho _2)+\frac{1}{4} \cG_{\rho _2}(\rho _1) \cG_0(\rho _2)\notag\\
&-\frac{1}{4} \cG_0(\rho _1) \cG_1(\rho _2)+\frac{3}{8} \cG_1(\rho _1) \cG_1(\rho _2)-\frac{1}{8} \cG_1(\rho _2) \cG_{\rho _2}(\rho _1)\Big)
+\log (\tau _2)\Big(\frac{1}{4} \cG_{0,0}(\rho _1)\notag\\
&+\frac{11}{8} \cG_{0,0}(\rho _2)-\frac{1}{8} \cG_{0,1}(\rho _1)-\frac{7}{8} \cG_{0,1}(\rho _2)-\frac{1}{8} \cG_{0,\rho _2}(\rho _1)-\frac{3}{8} \cG_{1,0}(\rho _1)-\frac{7}{8} \cG_{1,0}(\rho _2)\notag\\
&+\frac{3}{8} \cG_{1,1}(\rho _1)+\frac{1}{2} \cG_{1,1}(\rho _2)-\frac{1}{8} \cG_{\rho _2,0}(\rho _1)-\frac{1}{4} \cG_{\rho _2,1}(\rho _1)+\frac{3}{8} \cG_{\rho _2,\rho _2}(\rho _1)-\frac{5}{8} \cG_0(\rho _1) \cG_0(\rho _2)\notag\\
&+\frac{1}{2} \cG_1(\rho _1) \cG_0(\rho _2)+\frac{3}{8} \cG_{\rho _2}(\rho _1) \cG_0(\rho _2)+\frac{1}{2} \cG_0(\rho _1) \cG_1(\rho _2)-\frac{3}{8} \cG_1(\rho _1) \cG_1(\rho _2)-\frac{1}{8} \cG_1(\rho _2) \cG_{\rho _2}(\rho _1)\Big)\notag
\end{align}
\begin{align}
\hat{V}_{25}&=\frac{1}{32} \log ^2(\tau _1) \log (\tau _2)
+\frac{1}{32} \log (\tau _1) \log ^2(\tau _2)
+\log ^2(\tau _1)\Big(\frac{1}{32} \cG_0(\rho _2)-\frac{1}{32} \cG_1(\rho _1)\\
&+\frac{1}{32} \cG_{\rho _2}(\rho _1)\Big) 
+\log ^2(\tau _2)\Big(-\frac{1}{32} \cG_0(\rho _1)+\frac{1}{32} \cG_0(\rho _2)-\frac{1}{32} \cG_1(\rho _2)+\frac{1}{32} \cG_{\rho _2}(\rho _1)\Big) 
\notag\\
&+\log (\tau _1) \log (\tau _2)\Big(-\frac{1}{16} \cG_0(\rho _1)+\frac{1}{16} \cG_1(\rho _1)+\frac{1}{16} \cG_1(\rho _2)\Big) 
+\log (\tau _1)\Big(-\frac{1}{16} \cG_{0,0}(\rho _2)\notag\\
&+\frac{1}{16} \cG_{0,1}(\rho _1)+\frac{1}{16} \cG_{0,1}(\rho _2)-\frac{1}{16} \cG_{0,\rho _2}(\rho _1)+\frac{1}{16} \cG_{1,0}(\rho _1)+\frac{1}{16} \cG_{1,0}(\rho _2)-\frac{3}{16} \cG_{1,1}(\rho _1)\notag\\
&+\frac{1}{8} \cG_{1,\rho _2}(\rho _1)-\frac{1}{16} \cG_{\rho _2,0}(\rho _1)+\frac{1}{8} \cG_{\rho _2,1}(\rho _1)-\frac{1}{16} \cG_{\rho _2,\rho _2}(\rho _1)-\frac{1}{16} \cG_0(\rho _1) \cG_0(\rho _2)+\frac{1}{8} \cG_1(\rho _1) \cG_0(\rho _2)\notag\\
&-\frac{1}{16} \cG_{\rho _2}(\rho _1) \cG_0(\rho _2)-\frac{1}{16} \cG_1(\rho _1) \cG_1(\rho _2)+\frac{1}{16} \cG_1(\rho _2) \cG_{\rho _2}(\rho _1)\Big)
+\log (\tau _2)\Big(\frac{1}{16} \cG_{0,0}(\rho _1)-\frac{1}{16} \cG_{0,0}(\rho _2)\notag\\
&-\frac{1}{16} \cG_{0,1}(\rho _1)+\frac{1}{8} \cG_{0,1}(\rho _2)-\frac{1}{16} \cG_{1,0}(\rho _1)+\frac{1}{8} \cG_{1,0}(\rho _2)-\frac{3}{16} \cG_{1,1}(\rho _2)+\frac{1}{16} \cG_{1,\rho _2}(\rho _1)+\frac{1}{16} \cG_{\rho _2,1}(\rho _1)\notag\\
&-\frac{1}{16} \cG_{\rho _2,\rho _2}(\rho _1)+\frac{1}{16} \cG_0(\rho _2) \cG_1(\rho _1)-\frac{1}{16} \cG_1(\rho _2) \cG_1(\rho _1)-\frac{1}{16} \cG_0(\rho _1) \cG_1(\rho _2)-\frac{1}{16} \cG_0(\rho _2) \cG_{\rho _2}(\rho _1)\notag\\
&+\frac{1}{8} \cG_1(\rho _2) \cG_{\rho _2}(\rho _1)\Big)\notag
\end{align}
\begin{align}
\hat{V}_{34}=&-\frac{1}{48} \log ^3(\tau _1)
-\frac{1}{48} \log ^3(\tau _2)
+\log ^2(\tau _1)\Big(\frac{1}{16} \cG_0(\rho _1)-\frac{1}{16} \cG_1(\rho _1)+\frac{1}{16} \cG_1(\rho _2)\Big)\\
&+\log ^2(\tau _2)\Big(-\frac{1}{16} \cG_0(\rho _1)+\frac{1}{16} \cG_1(\rho _1)-\frac{1}{16} \cG_1(\rho _2)\Big) 
+\log (\tau _1) \log (\tau _2)\Big(-\frac{1}{8} \cG_0(\rho _1)\notag\\
&+\frac{1}{8} \cG_1(\rho _1)+\frac{1}{8} \cG_1(\rho _2)-\frac{1}{4} \cG_{\rho _2}(\rho _1)\Big) 
+\log (\tau _1)\Big(\frac{1}{8} \cG_{0,0}(\rho _1)-\frac{1}{8} \cG_{0,0}(\rho _2)+\frac{1}{2} \cG_{0,1}(\rho _1)\notag\\
&+\frac{1}{8} \cG_{0,1}(\rho _2)-\frac{3}{8} \cG_{0,\rho _2}(\rho _1)+\frac{1}{2} \cG_{1,0}(\rho _1)+\frac{1}{8} \cG_{1,0}(\rho _2)+\frac{1}{8} \cG_{1,1}(\rho _2)+\frac{1}{4} \cG_{1,\rho _2}(\rho _1)\notag\\
&-\frac{3}{8} \cG_{\rho _2,0}(\rho _1)+\frac{3}{8} \cG_{\rho _2,\rho _2}(\rho _1)+\frac{1}{8} \cG_0(\rho _1) \cG_0(\rho _2)-\frac{1}{2} \cG_1(\rho _1) \cG_0(\rho _2)+\frac{5}{8} \cG_{\rho _2}(\rho _1) \cG_0(\rho _2)\notag\\
&-\frac{3}{8} \cG_0(\rho _1) \cG_1(\rho _2)+\frac{1}{4} \cG_1(\rho _1) \cG_1(\rho _2)-\frac{3}{8} \cG_1(\rho _2) \cG_{\rho _2}(\rho _1)\Big)
+\log (\tau _2)\Big(\frac{1}{4} \cG_{0,0}(\rho _1)\notag\\
&+\frac{9}{8} \cG_{0,0}(\rho _2)-\frac{1}{4} \cG_{0,1}(\rho _1)-\frac{1}{2} \cG_{0,1}(\rho _2)-\frac{1}{4} \cG_{0,\rho _2}(\rho _1)-\frac{1}{4} \cG_{1,0}(\rho _1)-\frac{1}{2} \cG_{1,0}(\rho _2)\notag\\
&+\frac{1}{8} \cG_{1,1}(\rho _1)-\frac{3}{8} \cG_{1,\rho _2}(\rho _1)-\frac{5}{8} \cG_{\rho _2,1}(\rho _1)+\frac{3}{8} \cG_{\rho _2,\rho _2}(\rho _1)-\frac{1}{2} \cG_0(\rho _1) \cG_0(\rho _2)+\frac{1}{8} \cG_1(\rho _1) \cG_0(\rho _2)\notag\\
&+\frac{1}{8} \cG_{\rho _2}(\rho _1) \cG_0(\rho _2)+\frac{1}{4} \cG_0(\rho _1) \cG_1(\rho _2)+\frac{1}{4} \cG_1(\rho _1) \cG_1(\rho _2)+\frac{1}{4} \cG_1(\rho _2) \cG_{\rho _2}(\rho _1)\Big)
\notag
\end{align}
\begin{align}
\hat{V}_{36}=&-\frac{1}{48} \log ^3(\tau _1)
+\log ^2(\tau _1)\Big(\frac{1}{16} \cG_0(\rho _1)-\frac{1}{16} \cG_1(\rho _1)+\frac{1}{16} \cG_1(\rho _2)\Big)\\ 
&+\log (\tau _1) \log (\tau _2)\Big(-\frac{1}{8} \cG_0(\rho _2)-\frac{1}{8} \cG_1(\rho _1)+\frac{1}{4} \cG_1(\rho _2)-\frac{1}{8} \cG_{\rho _2}(\rho _1)\Big)
\notag\\
&+\log (\tau _1)\Big(\frac{1}{8} \cG_{0,0}(\rho _1)+\frac{1}{8} \cG_{0,0}(\rho _2)+\frac{3}{8} \cG_{0,1}(\rho _1)-\frac{1}{4} \cG_{0,\rho _2}(\rho _1)+\frac{3}{8} \cG_{1,0}(\rho _1)\notag\\
&-\frac{1}{4} \cG_{1,0}(\rho _2)+\frac{1}{2} \cG_{1,1}(\rho _1)+\frac{3}{8} \cG_{1,1}(\rho _2)-\frac{1}{8} \cG_{1,\rho _2}(\rho _1)-\frac{3}{8} \cG_{\rho _2,1}(\rho _1)+\frac{3}{8} \cG_{\rho _2,\rho _2}(\rho _1)\notag\\
&-\frac{1}{8} \cG_0(\rho _2) \cG_1(\rho _1)-\frac{3}{8} \cG_1(\rho _2) \cG_1(\rho _1)-\frac{1}{8} \cG_0(\rho _1) \cG_1(\rho _2)+\frac{1}{8} \cG_0(\rho _2) \cG_{\rho _2}(\rho _1)\Big)
\notag\\
&+\log (\tau _2)\Big(\frac{1}{4} \cG_{0,0}(\rho _2)-\frac{1}{8} \cG_{0,1}(\rho _1)+\frac{1}{8} \cG_{0,1}(\rho _2)-\frac{1}{8} \cG_{0,\rho _2}(\rho _1)-\frac{1}{8} \cG_{1,0}(\rho _1)-\frac{1}{8} \cG_{1,0}(\rho _2)\notag\\
&-\frac{1}{4} \cG_{1,1}(\rho _1)-\frac{1}{2} \cG_{1,1}(\rho _2)-\frac{1}{8} \cG_{1,\rho _2}(\rho _1)-\frac{1}{8} \cG_{\rho _2,0}(\rho _1)-\frac{3}{8} \cG_{\rho _2,1}(\rho _1)+\frac{1}{4} \cG_{\rho _2,\rho _2}(\rho _1)\notag\\
&-\frac{1}{8} \cG_0(\rho _1) \cG_0(\rho _2)-\frac{1}{8} \cG_1(\rho _1) \cG_0(\rho _2)+\frac{1}{4} \cG_{\rho _2}(\rho _1) \cG_0(\rho _2)+\frac{1}{4} \cG_0(\rho _1) \cG_1(\rho _2)+\frac{3}{8} \cG_1(\rho _1) \cG_1(\rho _2)\notag\\
&+\frac{1}{8} \cG_1(\rho _2) \cG_{\rho _2}(\rho _1)\Big)\notag
\end{align}
\begin{align}
\hat{V}_{51}=&-\frac{1}{48} \log ^3(\tau _1)
+\log ^2(\tau _1)\Big(\frac{1}{16} \cG_0(\rho _1)-\frac{1}{16} \cG_1(\rho _1)+\frac{1}{16} \cG_1(\rho _2)\Big) 
\\
&+\log (\tau _1) \log (\tau _2)\Big(-\frac{1}{8} \cG_0(\rho _2)-\frac{1}{8} \cG_1(\rho _1)+\frac{1}{4} \cG_1(\rho _2)-\frac{1}{8} \cG_{\rho _2}(\rho _1)\Big) 
\notag\\
&+\log (\tau _1)\Big(\frac{1}{8} \cG_{0,0}(\rho _1)+\frac{1}{8} \cG_{0,0}(\rho _2)+\frac{1}{8} \cG_{0,1}(\rho _1)+\frac{3}{8} \cG_{1,0}(\rho _1)-\frac{1}{4} \cG_{1,0}(\rho _2)\notag\\
&+\frac{1}{4} \cG_{1,1}(\rho _1)+\frac{3}{8} \cG_{1,1}(\rho _2)+\frac{1}{8} \cG_{1,\rho _2}(\rho _1)-\frac{1}{8} \cG_{\rho _2,1}(\rho _1)+\frac{1}{8} \cG_{\rho _2,\rho _2}(\rho _1)+\frac{1}{8} \cG_0(\rho _2) \cG_1(\rho _1)\notag\\
&-\frac{3}{8} \cG_1(\rho _2) \cG_1(\rho _1)-\frac{1}{8} \cG_0(\rho _1) \cG_1(\rho _2)+\frac{1}{8} \cG_0(\rho _2) \cG_{\rho _2}(\rho _1)\Big)
+\log (\tau _2)\Big(-\frac{1}{8} \cG_{0,1}(\rho _1)\notag\\
&+\frac{1}{8} \cG_{0,1}(\rho _2)-\frac{1}{8} \cG_{0,\rho _2}(\rho _1)-\frac{1}{8} \cG_{1,0}(\rho _1)+\frac{3}{8} \cG_{1,0}(\rho _2)-\frac{1}{2} \cG_{1,1}(\rho _2)-\frac{3}{8} \cG_{1,\rho _2}(\rho _1)\notag\\
&-\frac{1}{8} \cG_{\rho _2,0}(\rho _1)-\frac{1}{8} \cG_{\rho _2,1}(\rho _1)-\frac{1}{8} \cG_0(\rho _1) \cG_0(\rho _2)-\frac{3}{8} \cG_1(\rho _1) \cG_0(\rho _2)+\frac{1}{4} \cG_0(\rho _1) \cG_1(\rho _2)\notag\\
&+\frac{3}{8} \cG_1(\rho _1) \cG_1(\rho _2)+\frac{1}{8} \cG_1(\rho _2) \cG_{\rho _2}(\rho _1)\Big)\notag
\end{align}
\begin{align}
\hat{V}_{62}=&-\frac{1}{48} \log ^3(\tau _2)
+\log ^2(\tau _2)\Big(-\frac{1}{16} \cG_0(\rho _1)+\frac{1}{16} \cG_1(\rho _1)-\frac{1}{16} \cG_1(\rho _2)\Big) 
\\
&+\log (\tau _1) \log (\tau _2)\Big(-\frac{1}{8} \cG_0(\rho _1)+\frac{1}{8} \cG_0(\rho _2)+\frac{1}{4} \cG_1(\rho _1)-\frac{1}{8} \cG_1(\rho _2)-\frac{1}{8} \cG_{\rho _2}(\rho _1)\Big) 
\notag\\
&+\log (\tau _1)\Big(-\frac{1}{4} \cG_{0,0}(\rho _2)+\frac{1}{8} \cG_{0,1}(\rho _1)+\frac{1}{8} \cG_{0,1}(\rho _2)-\frac{1}{8} \cG_{0,\rho _2}(\rho _1)+\frac{3}{8} \cG_{1,0}(\rho _1)\notag\\
&+\frac{1}{8} \cG_{1,0}(\rho _2)-\frac{1}{2} \cG_{1,1}(\rho _1)+\frac{1}{8} \cG_{1,\rho _2}(\rho _1)-\frac{3}{8} \cG_{\rho _2,0}(\rho _1)+\frac{3}{8} \cG_{\rho _2,1}(\rho _1)+\frac{1}{8} \cG_0(\rho _1) \cG_0(\rho _2)\notag\\
&-\frac{5}{8} \cG_1(\rho _1) \cG_0(\rho _2)+\frac{1}{2} \cG_{\rho _2}(\rho _1) \cG_0(\rho _2)+\frac{3}{8} \cG_1(\rho _1) \cG_1(\rho _2)-\frac{3}{8} \cG_1(\rho _2) \cG_{\rho _2}(\rho _1)\Big)
\notag\\
&+\log (\tau _2)\Big(\frac{1}{4} \cG_{0,0}(\rho _1)+\frac{7}{8} \cG_{0,0}(\rho _2)-\frac{1}{8} \cG_{0,1}(\rho _1)-\frac{5}{8} \cG_{0,1}(\rho _2)-\frac{1}{8} \cG_{0,\rho _2}(\rho _1)-\frac{3}{8} \cG_{1,0}(\rho _1)\notag\\
&-\frac{3}{8} \cG_{1,0}(\rho _2)+\frac{3}{8} \cG_{1,1}(\rho _1)+\frac{1}{4} \cG_{1,1}(\rho _2)+\frac{1}{8} \cG_{\rho _2,0}(\rho _1)-\frac{1}{4} \cG_{\rho _2,1}(\rho _1)+\frac{1}{8} \cG_{\rho _2,\rho _2}(\rho _1)\notag\\
&-\frac{3}{8} \cG_0(\rho _1) \cG_0(\rho _2)+\frac{1}{2} \cG_1(\rho _1) \cG_0(\rho _2)-\frac{1}{8} \cG_{\rho _2}(\rho _1) \cG_0(\rho _2)+\frac{1}{4} \cG_0(\rho _1) \cG_1(\rho _2)-\frac{3}{8} \cG_1(\rho _1) \cG_1(\rho _2)\notag\\
&+\frac{1}{8} \cG_1(\rho _2) \cG_{\rho _2}(\rho _1)\Big)\notag
\end{align}
\end{allowdisplaybreaks}

\subsection{Eight-point amplitudes}
There is no new perturbative MHV coefficient through three loops. There are four NMHV-type perturbative coefficients for eight particles,

\begin{align}
g_{++-+}^{(i_1,i_2,i_3)}&(\rho_1,\rho_2,\rho_3) = \fa^{(i_1,i_2,i_3)}_{++-+}(\rho_1,\rho_2,\rho_3)+\overline{R}_{245}\,\fb^{(i_1,i_2,i_3)}_{1,++-+}(\rho_1,\rho_2,\rho_3)\\
\nonumber&+\overline{R}_{345}\,\fb^{(i_1,i_2,i_3)}_{2,++-+}(\rho_1,\rho_2,\rho_3)+R_{456}\,\fb^{(i_1,i_2,i_3)}_{3,++-+}(\rho_1,\rho_2,\rho_3)\\
\nonumber&+\overline{R}_{245}\,R_{456}\,\fc^{(i_1,i_2,i_3)}_{1,++-+}(\rho_1,\rho_2,\rho_3)
+\overline{R}_{345}\,R_{456}\,\fc^{(i_1,i_2,i_3)}_{2,++-+}(\rho_1,\rho_2,\rho_3)\,,\\
&\nonumber\\
g_{+-++}^{(i_1,i_2,i_3)}&(\rho_1,\rho_2,\rho_3) = \fa^{(i_1,i_2,i_3)}_{+-++}(\rho_1,\rho_2,\rho_3)+\overline{R}_{234}\,\fb^{(i_1,i_2,i_3)}_{1,+-++}(\rho_1,\rho_2,\rho_3)\\
\nonumber&+{R}_{345}\,\fb^{(i_1,i_2,i_3)}_{2,+-++}(\rho_1,\rho_2,\rho_3)+R_{346}\,\fb^{(i_1,i_2,i_3)}_{3,+-++}(\rho_1,\rho_2,\rho_3)\\
\nonumber&+\overline{R}_{234}\,{R}_{345}\,\fc^{(i_1,i_2,i_3)}_{1,+-++}(\rho_1,\rho_2,\rho_3)+\overline{R}_{234}\,R_{346}\,\fc^{(i_1,i_2,i_3)}_{2,+-++}(\rho_1,\rho_2,\rho_3)\,,\\
&\nonumber\\
g_{-+++}^{(i_1,i_2,i_3)}&(\rho_1,\rho_2,\rho_3) = \fa^{(i_1,i_2,i_3)}_{-+++}(\rho_1,\rho_2,\rho_3)+R_{234}\,\fb^{(i_1,i_2,i_3)}_{1,-+++}(\rho_1,\rho_2,\rho_3)\\
\nonumber&+R_{235}\,\fb^{(i_1,i_2,i_3)}_{2,-+++}(\rho_1,\rho_2,\rho_3)+{R}_{236}\,\fb^{(i_1,i_2,i_3)}_{3,-+++}(\rho_1,\rho_2,\rho_3)\,,\\
&\nonumber\\
g_{---+}^{(i_1,i_2,i_3)}&(\rho_1,\rho_2,\rho_3) = \fa^{(i_1,i_2,i_3)}_{---+}(\rho_1,\rho_2,\rho_3)+{R}_{256}\,\fb^{(i_1,i_2,i_3)}_{1,---+}(\rho_1,\rho_2,\rho_3)\\
\nonumber&+R_{356}\,\fb^{(i_1,i_2,i_3)}_{2,---+}(\rho_1,\rho_2,\rho_3)+R_{456}\,\fb^{(i_1,i_2,i_3)}_{3,---+}(\rho_1,\rho_2,\rho_3)\,.
\end{align}
For eight external legs, there are for the first time also independent NNMHV helicity configurations,
\begin{align}
g_{+--+}^{(i_1,i_2,i_3)}&\left(\rho_1,\rho_2,\rho_3\right) = \fa^{(i_1,i_2,i_3)}_{+--+}\left(\rho_1,\rho_2,\rho_3\right)+\overline{R}_{234}\,\fb_{1,+--+}^{(i_1,i_2,i_3)}\left(\rho_1,\rho_2,\rho_3\right)\\
\nonumber&+\overline{R}_{235}\,\fb_{2,+--+}^{(i_1,i_2,i_3)}\left(\rho_1,\rho_2,\rho_3\right)+R_{356}\,\fb_{3,+--+}^{(i_1,i_2,i_3)}\left(\rho_1,\rho_2,\rho_3\right)+R_{456}\,\fb_{4,+--+}^{(i_1,i_2,i_3)}\left(\rho_1,\rho_2,\rho_3\right)\\
\nonumber&+\overline{R}_{234}\,R_{356}\,\fc_{1,+--+}^{(i_1,i_2,i_3)}\left(\rho_1,\rho_2,\rho_3\right)+\overline{R}_{234}\,R_{456}\,\fc_{2,+--+}^{(i_1,i_2,i_3)}\left(\rho_1,\rho_2,\rho_3\right)\\
\nonumber&+\overline{R}_{235}\,R_{356}\,\fc_{3,+--+}^{(i_1,i_2,i_3)}\left(\rho_1,\rho_2,\rho_3\right)+\overline{R}_{235}\,R_{456}\,\fc_{4,+--+}^{(i_1,i_2,i_3)}\left(\rho_1,\rho_2,\rho_3\right)\,,\\
\nonumber\\
g_{-+-+}^{(i_1,i_2,i_3)}&\left(\rho_1,\rho_2,\rho_3\right) = \fa^{(i_1,i_2,i_3)}_{-+-+}\left(\rho_1,\rho_2,\rho_3\right)+R_{234}\,\fb_{1,-+-+}^{(i_1,i_2,i_3)}\left(\rho_1,\rho_2,\rho_3\right)\\
\nonumber&+R_{236}\,\fb_{2,-+-+}^{(i_1,i_2,i_3)}\left(\rho_1,\rho_2,\rho_3\right)+\overline{R}_{345}\,\fb_{3,-+-+}^{(i_1,i_2,i_3)}\left(\rho_1,\rho_2,\rho_3\right)+R_{456}\,\fb_{4,-+-+}^{(i_1,i_2,i_3)}\left(\rho_1,\rho_2,\rho_3\right)\\
\nonumber&+R_{234}\,\overline{R}_{345}\,\fc_{1,-+-+}^{(i_1,i_2,i_3)}\left(\rho_1,\rho_2,\rho_3\right)+R_{236}\,\overline{R}_{345}\,\fc_{2,-+-+}^{(i_1,i_2,i_3)}\left(\rho_1,\rho_2,\rho_3\right)\\
\nonumber&+R_{234}\,R_{456}\,\fc_{3,-+-+}^{(i_1,i_2,i_3)}\left(\rho_1,\rho_2,\rho_3\right)+R_{236}\,R_{456}\,\fc_{4,-+-+}^{(i_1,i_2,i_3)}\left(\rho_1,\rho_2,\rho_3\right)\\
\nonumber&+\overline{R}_{345}\,R_{456}\,\fc_{5,-+-+}^{(i_1,i_2,i_3)}\left(\rho_1,\rho_2,\rho_3\right)+R_{234}\,\overline{R}_{345}\,R_{456}\,\fd_{1,-+-+}^{(i_1,i_2,i_3)}\left(\rho_1,\rho_2,\rho_3\right)\\
\nonumber&+R_{236}\,\overline{R}_{345}\,R_{456}\,\fd_{2,-+-+}^{(i_1,i_2,i_3)}\left(\rho_1,\rho_2,\rho_3\right)\,.
\end{align}

%
%
\begin{align}
\fa^{\text{(1,0,0)}}_{++-+}\left(\rho_1,\rho_2,\rho_3\right) = &\frac{1}{4}\cG_{1,1}\left(\rho_1\right)-\frac{1}{4}\cG_{1,\rho_3}\left(\rho_1\right) -\frac{1}{4}\cG_0\left(\rho_3\right)\cG_1\left(\rho_1\right) \,.\\
\fb_{1,++-+}^{\text{(1,0,0)}}\left(\rho_1,\rho_2,\rho_3\right) = &-\frac{1}{4}\cG_{1,0}\left(\rho_1\right)+\frac{1}{4}\cG_{1,0}\left(\rho_3\right)-\frac{1}{4}\cG_{1,1}\left(\rho_3\right)+\frac{1}{4}\cG_{1,\rho_3}\left(\rho_1\right)\\
 \nonumber &+\frac{1}{4}\cG_0\left(\rho_3\right)\cG_1\left(\rho_1\right)-\frac{1}{4}\cG_0\left(\rho_1\right)\cG_1\left(\rho_3\right)+\frac{1}{4}\cG_1\left(\rho_1\right)\cG_1\left(\rho_3\right) \,.\\
\fb_{2,++-+}^{\text{(1,0,0)}}\left(\rho_1,\rho_2,\rho_3\right) = &-\frac{1}{4}\cG_{0,1}\left(\rho_1\right)-\frac{1}{4}\cG_{1,0}\left(\rho_3\right)+\frac{1}{4}\cG_{1,1}\left(\rho_1\right)+\frac{1}{4}\cG_{1,1}\left(\rho_3\right)\\
 \nonumber &+\frac{1}{4}\cG_0\left(\rho_1\right)\cG_1\left(\rho_3\right)-\frac{1}{4}\cG_1\left(\rho_1\right)\cG_1\left(\rho_3\right)\,. \\
\fb_{3,++-+}^{\text{(1,0,0)}}\left(\rho_1,\rho_2,\rho_3\right) = &-\frac{1}{4}\cG_{\rho_2,1}\left(\rho_1\right)+\frac{1}{4}\cG_{\rho_2,\rho_3}\left(\rho_1\right)-\frac{1}{4}\cG_{\rho_2}\left(\rho_1\right)\cG_{\rho_3}\left(\rho_2\right)\\
 \nonumber &+\frac{1}{4}\cG_0\left(\rho_3\right)\cG_1\left(\rho_1\right)-\frac{1}{4}\cG_1\left(\rho_1\right)\cG_1\left(\rho_2\right)+\frac{1}{4}\cG_1\left(\rho_2\right)\cG_{\rho_2}\left(\rho_1\right)\\
 \nonumber &+\frac{1}{4}\cG_1\left(\rho_1\right)\cG_{\rho_3}\left(\rho_2\right)\,. \\
\fc_{1,++-+}^{\text{(1,0,0)}}\left(\rho_1,\rho_2,\rho_3\right) = &\frac{1}{4}\cG_{0,0}\left(\rho_2\right)-\frac{1}{4}\cG_{0,0}\left(\rho_3\right)+\frac{1}{4}\cG_{0,1}\left(\rho_3\right)-\frac{1}{4}\cG_{0,\rho_3}\left(\rho_2\right)\\
 \nonumber &-\frac{1}{4}\cG_{1,0}\left(\rho_2\right)+\frac{1}{4}\cG_{1,\rho_3}\left(\rho_2\right)+\frac{1}{4}\cG_{\rho_2,0}\left(\rho_1\right)-\frac{1}{4}\cG_{\rho_2,\rho_3}\left(\rho_1\right)\\
 \nonumber &-\frac{1}{4}\cG_0\left(\rho_1\right)\cG_0\left(\rho_2\right)+\frac{1}{4}\cG_0\left(\rho_1\right)\cG_0\left(\rho_3\right)+\frac{1}{2}\cG_0\left(\rho_2\right)\cG_1\left(\rho_1\right) \\
 \nonumber &-\frac{1}{2}\cG_0\left(\rho_3\right)\cG_1\left(\rho_1\right)+\frac{1}{4}\cG_0\left(\rho_3\right)\cG_1\left(\rho_2\right)-\frac{1}{4}\cG_1\left(\rho_2\right)\cG_1\left(\rho_3\right)\\
 \nonumber &-\frac{1}{4}\cG_0\left(\rho_2\right)\cG_{\rho_2}\left(\rho_1\right)+\frac{1}{4}\cG_0\left(\rho_1\right)\cG_{\rho_3}\left(\rho_2\right)-\frac{1}{4}\cG_0\left(\rho_3\right)\cG_{\rho_3}\left(\rho_2\right)\\
 \nonumber &-\frac{1}{2}\cG_1\left(\rho_1\right)\cG_{\rho_3}\left(\rho_2\right)+\frac{1}{4}\cG_1\left(\rho_3\right)\cG_{\rho_3}\left(\rho_2\right)+\frac{1}{4}\cG_{\rho_2}\left(\rho_1\right)\cG_{\rho_3}\left(\rho_2\right)\,. 
\end{align}
\begin{align}
\fc_{2,++-+}^{\text{(1,0,0)}}\left(\rho_1,\rho_2,\rho_3\right) = &-\frac{1}{4}\cG_{0,0}\left(\rho_2\right)+\frac{1}{4}\cG_{0,0}\left(\rho_3\right)-\frac{1}{4}\cG_{0,1}\left(\rho_3\right)+\frac{1}{4}\cG_{0,\rho_2}\left(\rho_1\right)\\
 \nonumber &+\frac{1}{4}\cG_{0,\rho_3}\left(\rho_2\right)+\frac{1}{4}\cG_{1,0}\left(\rho_2\right)-\frac{1}{4}\cG_{1,\rho_2}\left(\rho_1\right)-\frac{1}{4}\cG_{1,\rho_3}\left(\rho_2\right)\\
 \nonumber &+\frac{1}{4}\cG_0\left(\rho_1\right)\cG_0\left(\rho_2\right)-\frac{1}{4}\cG_0\left(\rho_1\right)\cG_0\left(\rho_3\right)-\frac{1}{4}\cG_0\left(\rho_2\right)\cG_1\left(\rho_1\right)\\
 \nonumber &+\frac{1}{4}\cG_0\left(\rho_3\right)\cG_1\left(\rho_1\right)-\frac{1}{4}\cG_0\left(\rho_3\right)\cG_1\left(\rho_2\right)+\frac{1}{4}\cG_1\left(\rho_2\right)\cG_1\left(\rho_3\right)\\
 \nonumber &-\frac{1}{4}\cG_0\left(\rho_1\right)\cG_{\rho_3}\left(\rho_2\right)+\frac{1}{4}\cG_0\left(\rho_3\right)\cG_{\rho_3}\left(\rho_2\right)+\frac{1}{4}\cG_1\left(\rho_1\right)\cG_{\rho_3}\left(\rho_2\right)\\
 \nonumber &-\frac{1}{4}\cG_1\left(\rho_3\right)\cG_{\rho_3}\left(\rho_2\right)\,. \\
 \nonumber \, \\
\fa^{\text{(0,0,1)}}_{+-++}\left(\rho_1,\rho_2,\rho_3\right) = &-\frac{1}{4}\cG_{0,1}\left(\rho_3\right)-\frac{1}{4}\cG_{1,0}\left(\rho_3\right)+\frac{1}{2}\cG_{1,1}\left(\rho_3\right)+\frac{1}{4}\cG_{1,\rho_3}\left(\rho_2\right)\\
 \nonumber &+\frac{1}{4}\cG_{\rho_3,1}\left(\rho_2\right)-\frac{1}{4}\cG_1\left(\rho_2\right)\cG_1\left(\rho_3\right)-\frac{1}{4}\cG_1\left(\rho_3\right)\cG_{\rho_3}\left(\rho_2\right)\\
 \nonumber &+\frac{1}{4}\cG_0\left(\rho_3\right)\cG_1\left(\rho_2\right) \,. \\
\fb_{1,+-++}^{\text{(0,0,1)}}\left(\rho_1,\rho_2,\rho_3\right) = &-\frac{1}{4}\cG_{1,\rho_3}\left(\rho_2\right)-\frac{1}{4}\cG_{\rho_3,1}\left(\rho_2\right)+\frac{1}{4}\cG_1\left(\rho_3\right)\cG_{\rho_3}\left(\rho_2\right)\\
 \nonumber &-\frac{1}{4}\cG_0\left(\rho_3\right)\cG_1\left(\rho_2\right)+\frac{1}{4}\cG_1\left(\rho_2\right)\cG_1\left(\rho_3\right)\,. \\
\fb_{2,+-++}^{\text{(0,0,1)}}\left(\rho_1,\rho_2,\rho_3\right) = &\frac{1}{4}\cG_{0,1}\left(\rho_2\right)+\frac{1}{4}\cG_{0,1}\left(\rho_3\right)-\frac{1}{4}\cG_{1,1}\left(\rho_1\right)-\frac{1}{4}\cG_{1,1}\left(\rho_3\right)\\
 \nonumber &+\frac{1}{4}\cG_{1,\rho_2}\left(\rho_1\right)-\frac{1}{4}\cG_{\rho_3,1}\left(\rho_2\right)+\frac{1}{4}\cG_1\left(\rho_3\right)\cG_{\rho_3}\left(\rho_2\right)\\
 \nonumber &+\frac{1}{4}\cG_0\left(\rho_2\right)\cG_1\left(\rho_1\right)-\frac{1}{4}\cG_1\left(\rho_1\right)\cG_1\left(\rho_2\right)-\frac{1}{4}\cG_0\left(\rho_2\right)\cG_1\left(\rho_3\right)\\
 \nonumber &+\frac{1}{4}\cG_1\left(\rho_1\right)\cG_1\left(\rho_3\right)+\frac{1}{4}\cG_1\left(\rho_2\right)\cG_{\rho_2}\left(\rho_1\right)-\frac{1}{4}\cG_1\left(\rho_3\right)\cG_{\rho_2}\left(\rho_1\right)\,. \\
\fb_{3,+-++}^{\text{(0,0,1)}}\left(\rho_1,\rho_2,\rho_3\right) = &\frac{1}{4}\cG_{0,0}\left(\rho_3\right)-\frac{1}{4}\cG_{0,1}\left(\rho_2\right)-\frac{1}{4}\cG_{0,1}\left(\rho_3\right)-\frac{1}{4}\cG_{0,\rho_3}\left(\rho_2\right)\\
 \nonumber &+\frac{1}{4}\cG_{1,1}\left(\rho_1\right)-\frac{1}{4}\cG_{1,\rho_2}\left(\rho_1\right)+\frac{1}{4}\cG_{\rho_3,1}\left(\rho_1\right)-\frac{1}{4}\cG_{\rho_3,\rho_2}\left(\rho_1\right)\\
 \nonumber &-\frac{1}{4}\cG_0\left(\rho_2\right)\cG_0\left(\rho_3\right)-\frac{1}{4}\cG_0\left(\rho_2\right)\cG_1\left(\rho_1\right)+\frac{1}{4}\cG_1\left(\rho_1\right)\cG_1\left(\rho_2\right)\\
 \nonumber &+\frac{1}{2}\cG_0\left(\rho_2\right)\cG_1\left(\rho_3\right)-\frac{1}{4}\cG_1\left(\rho_1\right)\cG_1\left(\rho_3\right)-\frac{1}{4}\cG_0\left(\rho_3\right)\cG_{\rho_2}\left(\rho_1\right)\\
 \nonumber &-\frac{1}{4}\cG_1\left(\rho_2\right)\cG_{\rho_2}\left(\rho_1\right)+\frac{1}{2}\cG_1\left(\rho_3\right)\cG_{\rho_2}\left(\rho_1\right)-\frac{1}{4}\cG_0\left(\rho_2\right)\cG_{\rho_3}\left(\rho_1\right)\\
 \nonumber &+\frac{1}{4}\cG_0\left(\rho_3\right)\cG_{\rho_3}\left(\rho_1\right)-\frac{1}{4}\cG_1\left(\rho_3\right)\cG_{\rho_3}\left(\rho_1\right)-\frac{1}{4}\cG_{\rho_2}\left(\rho_1\right)\cG_{\rho_3}\left(\rho_2\right)\\
 \nonumber &+\frac{1}{4}\cG_{\rho_3}\left(\rho_1\right)\cG_{\rho_3}\left(\rho_2\right)\,. \\
\fc_{1,+-++}^{\text{(0,0,1)}}\left(\rho_1,\rho_2,\rho_3\right) = &-\frac{1}{4}\cG_{0,1}\left(\rho_2\right)+\frac{1}{4}\cG_{1,0}\left(\rho_1\right)-\frac{1}{4}\cG_{1,\rho_2}\left(\rho_1\right)+\frac{1}{4}\cG_{\rho_3,1}\left(\rho_2\right)\\
 \nonumber &-\frac{1}{4}\cG_0\left(\rho_2\right)\cG_1\left(\rho_1\right)+\frac{1}{4}\cG_1\left(\rho_1\right)\cG_1\left(\rho_2\right)-\frac{1}{4}\cG_0\left(\rho_1\right)\cG_1\left(\rho_3\right)
 \end{align}
 
 \begin{align}
 \nonumber &+\frac{1}{4}\cG_0\left(\rho_2\right)\cG_1\left(\rho_3\right)-\frac{1}{4}\cG_1\left(\rho_2\right)\cG_{\rho_2}\left(\rho_1\right)+\frac{1}{4}\cG_1\left(\rho_3\right)\cG_{\rho_2}\left(\rho_1\right)\\
 \nonumber &-\frac{1}{4}\cG_1\left(\rho_3\right)\cG_{\rho_3}\left(\rho_2\right)\,. \\
\fc_{2,+-++}^{\text{(0,0,1)}}\left(\rho_1,\rho_2,\rho_3\right) = &\frac{1}{4}\cG_{0,1}\left(\rho_2\right)+\frac{1}{4}\cG_{0,\rho_3}\left(\rho_2\right)-\frac{1}{4}\cG_{1,0}\left(\rho_1\right)+\frac{1}{4}\cG_{1,\rho_2}\left(\rho_1\right)\\
 \nonumber &-\frac{1}{4}\cG_{\rho_3,0}\left(\rho_1\right)+\frac{1}{4}\cG_{\rho_3,\rho_2}\left(\rho_1\right)-\frac{1}{4}\cG_{\rho_3}\left(\rho_1\right)\cG_{\rho_3}\left(\rho_2\right)\\
 \nonumber &-\frac{1}{4}\cG_0\left(\rho_1\right)\cG_0\left(\rho_3\right)+\frac{1}{4}\cG_0\left(\rho_2\right)\cG_0\left(\rho_3\right)+\frac{1}{4}\cG_0\left(\rho_2\right)\cG_1\left(\rho_1\right)\\
 \nonumber &-\frac{1}{4}\cG_1\left(\rho_1\right)\cG_1\left(\rho_2\right)+\frac{1}{2}\cG_0\left(\rho_1\right)\cG_1\left(\rho_3\right)-\frac{1}{2}\cG_0\left(\rho_2\right)\cG_1\left(\rho_3\right)\\
 \nonumber &+\frac{1}{4}\cG_0\left(\rho_3\right)\cG_{\rho_2}\left(\rho_1\right)+\frac{1}{4}\cG_1\left(\rho_2\right)\cG_{\rho_2}\left(\rho_1\right)-\frac{1}{2}\cG_1\left(\rho_3\right)\cG_{\rho_2}\left(\rho_1\right)\\
 \nonumber &+\frac{1}{4}\cG_0\left(\rho_2\right)\cG_{\rho_3}\left(\rho_1\right)+\frac{1}{4}\cG_{\rho_2}\left(\rho_1\right)\cG_{\rho_3}\left(\rho_2\right)\,. \\
 \nonumber \, \\
\fa^{\text{(0,1,0)}}_{+--+}\left(\rho_1,\rho_2,\rho_3\right) = &-\frac{1}{4}\cG_{1,1}\left(\rho_1\right)+\frac{1}{4}\cG_{1,1}\left(\rho_2\right)+\frac{1}{4}\cG_{1,\rho_3}\left(\rho_1\right)-\frac{1}{4}\cG_{1,\rho_3}\left(\rho_2\right)\\
 \nonumber &+\frac{1}{4}\cG_0\left(\rho_3\right)\cG_1\left(\rho_1\right)-\frac{1}{4}\cG_0\left(\rho_3\right)\cG_1\left(\rho_2\right) \,. \\
\fb_{1,+--+}^{\text{(0,1,0)}}\left(\rho_1,\rho_2,\rho_3\right) = &\frac{1}{4}\cG_{1,1}\left(\rho_2\right)-\frac{1}{4}\cG_{1,\rho_3}\left(\rho_2\right)-\frac{1}{4}\cG_0\left(\rho_3\right)\cG_1\left(\rho_2\right)\\
\fb_{2,+--+}^{\text{(0,1,0)}}\left(\rho_1,\rho_2,\rho_3\right) = &\frac{1}{4}\cG_{1,0}\left(\rho_1\right)-\frac{1}{2}\cG_{1,0}\left(\rho_2\right)+\frac{1}{4}\cG_{1,0}\left(\rho_3\right)-\frac{1}{4}\cG_{1,1}\left(\rho_3\right)\\
 \nonumber &-\frac{1}{4}\cG_{1,\rho_3}\left(\rho_1\right)+\frac{1}{2}\cG_{1,\rho_3}\left(\rho_2\right)+\frac{1}{4}\cG_1\left(\rho_1\right)\cG_1\left(\rho_3\right)\\
 \nonumber &-\frac{1}{4}\cG_0\left(\rho_3\right)\cG_1\left(\rho_1\right)+\frac{1}{2}\cG_0\left(\rho_3\right)\cG_1\left(\rho_2\right)-\frac{1}{4}\cG_0\left(\rho_1\right)\cG_1\left(\rho_3\right)\\
\fb_{3,+--+}^{\text{(0,1,0)}}\left(\rho_1,\rho_2,\rho_3\right) = &\frac{1}{4}\cG_{0,0}\left(\rho_3\right)-\frac{1}{4}\cG_{0,1}\left(\rho_2\right)-\frac{1}{4}\cG_{0,1}\left(\rho_3\right)+\frac{1}{4}\cG_{0,\rho_3}\left(\rho_2\right)\\
 \nonumber &+\frac{1}{4}\cG_{1,1}\left(\rho_1\right)+\frac{1}{4}\cG_{1,\rho_2}\left(\rho_1\right)-\frac{1}{2}\cG_{1,\rho_3}\left(\rho_1\right)+\frac{1}{2}\cG_{\rho_2,1}\left(\rho_1\right)\\
 \nonumber &-\frac{1}{2}\cG_{\rho_2,\rho_3}\left(\rho_1\right)-\frac{1}{4}\cG_{\rho_3,1}\left(\rho_1\right)-\frac{1}{4}\cG_{\rho_3,\rho_2}\left(\rho_1\right)+\frac{1}{2}\cG_{\rho_3,\rho_3}\left(\rho_1\right)\\
 \nonumber &-\frac{1}{4}\cG_0\left(\rho_2\right)\cG_0\left(\rho_3\right)+\frac{1}{4}\cG_0\left(\rho_2\right)\cG_1\left(\rho_1\right)-\frac{1}{2}\cG_0\left(\rho_3\right)\cG_1\left(\rho_1\right)\\
 \nonumber &+\frac{1}{2}\cG_0\left(\rho_3\right)\cG_1\left(\rho_2\right)-\frac{1}{4}\cG_1\left(\rho_1\right)\cG_1\left(\rho_2\right)+\frac{1}{4}\cG_1\left(\rho_1\right)\cG_1\left(\rho_3\right)\\
 \nonumber &-\frac{1}{4}\cG_0\left(\rho_3\right)\cG_{\rho_2}\left(\rho_1\right)-\frac{1}{4}\cG_1\left(\rho_2\right)\cG_{\rho_2}\left(\rho_1\right)-\frac{1}{4}\cG_0\left(\rho_2\right)\cG_{\rho_3}\left(\rho_1\right)\\
 \nonumber &+\frac{1}{4}\cG_0\left(\rho_3\right)\cG_{\rho_3}\left(\rho_1\right)+\frac{1}{2}\cG_1\left(\rho_2\right)\cG_{\rho_3}\left(\rho_1\right)-\frac{1}{4}\cG_1\left(\rho_3\right)\cG_{\rho_3}\left(\rho_1\right)\\
 \nonumber &+\frac{1}{4}\cG_{\rho_2}\left(\rho_1\right)\cG_{\rho_3}\left(\rho_2\right)-\frac{1}{4}\cG_{\rho_3}\left(\rho_1\right)\cG_{\rho_3}\left(\rho_2\right)\,. \\
\fb_{4,+--+}^{\text{(0,1,0)}}\left(\rho_1,\rho_2,\rho_3\right) = &-\frac{1}{4}\cG_{0,0}\left(\rho_3\right)+\frac{1}{4}\cG_{0,1}\left(\rho_3\right)-\frac{1}{4}\cG_{1,0}\left(\rho_2\right)+\frac{1}{4}\cG_{1,1}\left(\rho_2\right)\\
 \nonumber &-\frac{1}{4}\cG_{\rho_2,1}\left(\rho_1\right)+\frac{1}{4}\cG_{\rho_2,\rho_3}\left(\rho_1\right)+\frac{1}{4}\cG_{\rho_3,0}\left(\rho_2\right)-\frac{1}{4}\cG_{\rho_3,1}\left(\rho_2\right)
  \end{align}
\begin{align}
 \nonumber &+\frac{1}{4}\cG_0\left(\rho_2\right)\cG_0\left(\rho_3\right)-\frac{1}{4}\cG_1\left(\rho_2\right)\cG_1\left(\rho_3\right)+\frac{1}{4}\cG_0\left(\rho_3\right)\cG_{\rho_2}\left(\rho_1\right) \\
 \nonumber &-\frac{1}{4}\cG_0\left(\rho_3\right)\cG_{\rho_3}\left(\rho_2\right)+\frac{1}{4}\cG_1\left(\rho_3\right)\cG_{\rho_3}\left(\rho_2\right)\,. \\
\fc_{1,+--+}^{\text{(0,1,0)}}\left(\rho_1,\rho_2,\rho_3\right) = &-\frac{1}{4}\cG_{0,1}\left(\rho_2\right)+\frac{1}{4}\cG_{0,\rho_3}\left(\rho_2\right)+\frac{1}{4}\cG_{1,0}\left(\rho_1\right)-\frac{1}{4}\cG_{1,\rho_2}\left(\rho_1\right)\\
 \nonumber &-\frac{1}{4}\cG_{\rho_3,0}\left(\rho_1\right)+\frac{1}{4}\cG_{\rho_3,\rho_2}\left(\rho_1\right)-\frac{1}{4}\cG_{\rho_3}\left(\rho_1\right)\cG_{\rho_3}\left(\rho_2\right)\\
 \nonumber &-\frac{1}{4}\cG_0\left(\rho_1\right)\cG_0\left(\rho_3\right)+\frac{1}{4}\cG_0\left(\rho_2\right)\cG_0\left(\rho_3\right)-\frac{1}{4}\cG_0\left(\rho_2\right)\cG_1\left(\rho_1\right)\\
 \nonumber &+\frac{1}{4}\cG_1\left(\rho_1\right)\cG_1\left(\rho_2\right)+\frac{1}{4}\cG_0\left(\rho_3\right)\cG_{\rho_2}\left(\rho_1\right)-\frac{1}{4}\cG_1\left(\rho_2\right)\cG_{\rho_2}\left(\rho_1\right)\\
 \nonumber &+\frac{1}{4}\cG_0\left(\rho_2\right)\cG_{\rho_3}\left(\rho_1\right)+\frac{1}{4}\cG_{\rho_2}\left(\rho_1\right)\cG_{\rho_3}\left(\rho_2\right)\,. \\
\fc_{2,+--+}^{\text{(0,1,0)}}\left(\rho_1,\rho_2,\rho_3\right) = &\frac{1}{4}\cG_{0,1}\left(\rho_2\right)-\frac{1}{4}\cG_{0,\rho_3}\left(\rho_2\right)+\frac{1}{4}\cG_{1,0}\left(\rho_2\right)-\frac{1}{2}\cG_{1,1}\left(\rho_2\right)\\
 \nonumber &+\frac{1}{4}\cG_{1,\rho_3}\left(\rho_2\right)-\frac{1}{4}\cG_{\rho_3,0}\left(\rho_2\right)+\frac{1}{4}\cG_{\rho_3,1}\left(\rho_2\right)-\frac{1}{4}\cG_{\rho_2}\left(\rho_1\right)\cG_{\rho_3}\left(\rho_2\right)\\
 \nonumber &+\frac{1}{4}\cG_0\left(\rho_1\right)\cG_0\left(\rho_3\right)-\frac{1}{4}\cG_0\left(\rho_2\right)\cG_0\left(\rho_3\right)-\frac{1}{4}\cG_0\left(\rho_1\right)\cG_1\left(\rho_2\right)\\
 \nonumber &+\frac{1}{4}\cG_0\left(\rho_3\right)\cG_1\left(\rho_2\right)-\frac{1}{4}\cG_0\left(\rho_3\right)\cG_{\rho_2}\left(\rho_1\right)+\frac{1}{4}\cG_1\left(\rho_2\right)\cG_{\rho_2}\left(\rho_1\right)\\
 \nonumber &+\frac{1}{4}\cG_0\left(\rho_1\right)\cG_{\rho_3}\left(\rho_2\right)\,. \\
\fc_{3,+--+}^{\text{(0,1,0)}}\left(\rho_1,\rho_2,\rho_3\right) = &\frac{1}{2}\cG_{0,0}\left(\rho_2\right)-\frac{1}{2}\cG_{0,0}\left(\rho_3\right)+\frac{1}{2}\cG_{0,1}\left(\rho_3\right)-\frac{1}{2}\cG_{0,\rho_3}\left(\rho_2\right)\\
 \nonumber &-\frac{1}{2}\cG_{1,0}\left(\rho_1\right)+\frac{1}{2}\cG_{1,\rho_3}\left(\rho_1\right)-\frac{1}{2}\cG_{\rho_2,0}\left(\rho_1\right)+\frac{1}{2}\cG_{\rho_2,\rho_3}\left(\rho_1\right)\\
 \nonumber &+\frac{1}{2}\cG_{\rho_3,0}\left(\rho_1\right)-\frac{1}{2}\cG_{\rho_3,\rho_3}\left(\rho_1\right)+\frac{1}{2}\cG_{\rho_3}\left(\rho_1\right)\cG_{\rho_3}\left(\rho_2\right)\\
 \nonumber &-\frac{1}{2}\cG_0\left(\rho_1\right)\cG_0\left(\rho_2\right)+\frac{1}{2}\cG_0\left(\rho_1\right)\cG_0\left(\rho_3\right)+\frac{1}{2}\cG_0\left(\rho_3\right)\cG_1\left(\rho_1\right)\\
 \nonumber &+\frac{1}{2}\cG_0\left(\rho_1\right)\cG_1\left(\rho_2\right)-\frac{1}{2}\cG_0\left(\rho_3\right)\cG_1\left(\rho_2\right)-\frac{1}{2}\cG_1\left(\rho_1\right)\cG_1\left(\rho_3\right)\\
 \nonumber &+\frac{1}{2}\cG_0\left(\rho_2\right)\cG_{\rho_2}\left(\rho_1\right)-\frac{1}{2}\cG_0\left(\rho_3\right)\cG_{\rho_3}\left(\rho_1\right)-\frac{1}{2}\cG_1\left(\rho_2\right)\cG_{\rho_3}\left(\rho_1\right)\\
 \nonumber &+\frac{1}{2}\cG_1\left(\rho_3\right)\cG_{\rho_3}\left(\rho_1\right)-\frac{1}{2}\cG_{\rho_2}\left(\rho_1\right)\cG_{\rho_3}\left(\rho_2\right)\,. \\
\fc_{4,+--+}^{\text{(0,1,0)}}\left(\rho_1,\rho_2,\rho_3\right) = &-\frac{1}{4}\cG_{0,0}\left(\rho_2\right)+\frac{1}{4}\cG_{0,0}\left(\rho_3\right)-\frac{1}{4}\cG_{0,1}\left(\rho_3\right)+\frac{1}{4}\cG_{0,\rho_3}\left(\rho_2\right)\\
 \nonumber &+\frac{1}{4}\cG_{1,0}\left(\rho_2\right)-\frac{1}{4}\cG_{1,\rho_3}\left(\rho_2\right)+\frac{1}{4}\cG_{\rho_2,0}\left(\rho_1\right)-\frac{1}{4}\cG_{\rho_2,\rho_3}\left(\rho_1\right)\\
 \nonumber &+\frac{1}{4}\cG_0\left(\rho_1\right)\cG_0\left(\rho_2\right)-\frac{1}{4}\cG_0\left(\rho_1\right)\cG_0\left(\rho_3\right)-\frac{1}{4}\cG_0\left(\rho_3\right)\cG_1\left(\rho_2\right)\\
 \nonumber &+\frac{1}{4}\cG_1\left(\rho_2\right)\cG_1\left(\rho_3\right)-\frac{1}{4}\cG_0\left(\rho_2\right)\cG_{\rho_2}\left(\rho_1\right)-\frac{1}{4}\cG_0\left(\rho_1\right)\cG_{\rho_3}\left(\rho_2\right)\\
 \nonumber &+\frac{1}{4}\cG_0\left(\rho_3\right)\cG_{\rho_3}\left(\rho_2\right)-\frac{1}{4}\cG_1\left(\rho_3\right)\cG_{\rho_3}\left(\rho_2\right)+\frac{1}{4}\cG_{\rho_2}\left(\rho_1\right)\cG_{\rho_3}\left(\rho_2\right)\,. 
\end{align}
\begin{align}
\fa^{\text{(0,0,1)}}_{-+-+}\left(\rho_1,\rho_2,\rho_3\right) = &-\frac{1}{4}\cG_{1,0}\left(\rho_3\right)-\frac{1}{4}\cG_{1,1}\left(\rho_2\right)+\frac{1}{4}\cG_{1,\rho_3}\left(\rho_2\right)+\frac{1}{4}\cG_0\left(\rho_3\right)\cG_1\left(\rho_2\right)\,. \\
\fb_{1,-+-+}^{\text{(0,0,1)}}\left(\rho_1,\rho_2,\rho_3\right) = &\frac{1}{4}\cG_{1,1}\left(\rho_2\right)-\frac{1}{4}\cG_{1,\rho_3}\left(\rho_2\right)-\frac{1}{4}\cG_0\left(\rho_3\right)\cG_1\left(\rho_2\right)\,. \\
\fb_{2,-+-+}^{\text{(0,0,1)}}\left(\rho_1,\rho_2,\rho_3\right) = &-\frac{1}{4}\cG_{0,0}\left(\rho_3\right)+\frac{1}{2}\cG_{1,0}\left(\rho_3\right)\,. \\
\fb_{3,-+-+}^{\text{(0,0,1)}}\left(\rho_1,\rho_2,\rho_3\right) = &\frac{1}{4}\cG_{1,0}\left(\rho_2\right)+\frac{1}{4}\cG_{1,1}\left(\rho_3\right)+\frac{1}{4}\cG_{1,\rho_3}\left(\rho_1\right)-\frac{1}{4}\cG_{1,\rho_3}\left(\rho_2\right)\\
 \nonumber &+\frac{1}{4}\cG_{\rho_2,1}\left(\rho_1\right)+\frac{1}{4}\cG_1\left(\rho_2\right)\cG_1\left(\rho_3\right)-\frac{1}{4}\cG_1\left(\rho_1\right)\cG_1\left(\rho_3\right)\\
 \nonumber &+\frac{1}{4}\cG_0\left(\rho_3\right)\cG_1\left(\rho_1\right)-\frac{1}{4}\cG_1\left(\rho_3\right)\cG_{\rho_2}\left(\rho_1\right)-\frac{1}{4}\cG_0\left(\rho_3\right)\cG_1\left(\rho_2\right)\\
 \nonumber &-\frac{1}{4}\cG_0\left(\rho_2\right)\cG_1\left(\rho_3\right)\,. \\
\fb_{4,-+-+}^{\text{(0,0,1)}}\left(\rho_1,\rho_2,\rho_3\right) = &-\frac{1}{4}\cG_{0,0}\left(\rho_3\right)+\frac{1}{2}\cG_{1,0}\left(\rho_3\right)+\frac{1}{4}\cG_{1,1}\left(\rho_2\right)-\frac{1}{4}\cG_{1,\rho_3}\left(\rho_2\right)\\
 \nonumber &+\frac{1}{4}\cG_{\rho_3,1}\left(\rho_2\right)-\frac{1}{4}\cG_{\rho_3,\rho_3}\left(\rho_2\right)-\frac{1}{4}\cG_0\left(\rho_3\right)\cG_1\left(\rho_2\right)-\frac{1}{4}\cG_0\left(\rho_3\right)\cG_{\rho_3}\left(\rho_2\right)\,. \\
\fc_{1,-+-+}^{\text{(0,0,1)}}\left(\rho_1,\rho_2,\rho_3\right) = &\frac{1}{4}\cG_{0,1}\left(\rho_1\right)-\frac{1}{4}\cG_{1,0}\left(\rho_2\right)+\frac{1}{4}\cG_{1,\rho_3}\left(\rho_2\right)-\frac{1}{4}\cG_{\rho_2,1}\left(\rho_1\right)\\
 \nonumber &+\frac{1}{4}\cG_0\left(\rho_3\right)\cG_1\left(\rho_2\right)-\frac{1}{4}\cG_0\left(\rho_1\right)\cG_1\left(\rho_3\right)+\frac{1}{4}\cG_0\left(\rho_2\right)\cG_1\left(\rho_3\right)\\
 \nonumber &-\frac{1}{4}\cG_1\left(\rho_2\right)\cG_1\left(\rho_3\right)+\frac{1}{4}\cG_1\left(\rho_3\right)\cG_{\rho_2}\left(\rho_1\right)\,. \\
\fc_{2,-+-+}^{\text{(0,0,1)}}\left(\rho_1,\rho_2,\rho_3\right) = &\frac{1}{2}\cG_{0,0}\left(\rho_3\right)-\frac{1}{4}\cG_{0,1}\left(\rho_1\right)-\frac{1}{4}\cG_{0,1}\left(\rho_3\right)-\frac{1}{4}\cG_{0,\rho_3}\left(\rho_1\right)\\
 \nonumber &-\frac{1}{2}\cG_{1,0}\left(\rho_3\right)-\frac{1}{4}\cG_0\left(\rho_1\right)\cG_0\left(\rho_3\right)+\frac{1}{2}\cG_0\left(\rho_1\right)\cG_1\left(\rho_3\right)\,. \\
\fc_{3,-+-+}^{\text{(0,0,1)}}\left(\rho_1,\rho_2,\rho_3\right) = &-\frac{1}{4}\cG_{1,1}\left(\rho_2\right)+\frac{1}{4}\cG_{1,\rho_3}\left(\rho_2\right)-\frac{1}{4}\cG_{\rho_3,1}\left(\rho_2\right)+\frac{1}{4}\cG_{\rho_3,\rho_3}\left(\rho_2\right)\\
 \nonumber &+\frac{1}{4}\cG_0\left(\rho_3\right)\cG_1\left(\rho_2\right)+\frac{1}{4}\cG_0\left(\rho_3\right)\cG_{\rho_3}\left(\rho_2\right)\,. \\
\fc_{4,-+-+}^{\text{(0,0,1)}}\left(\rho_1,\rho_2,\rho_3\right) = &\frac{1}{4}\cG_{0,0}\left(\rho_3\right)-\frac{1}{2}\cG_{1,0}\left(\rho_3\right)\,. \\
\fc_{5,-+-+}^{\text{(0,0,1)}}\left(\rho_1,\rho_2,\rho_3\right) = &\frac{1}{2}\cG_{0,0}\left(\rho_3\right)-\frac{1}{4}\cG_{0,1}\left(\rho_3\right)-\frac{1}{4}\cG_{1,0}\left(\rho_2\right)-\frac{1}{2}\cG_{1,0}\left(\rho_3\right)\\
 \nonumber &+\frac{1}{4}\cG_{1,\rho_3}\left(\rho_2\right)-\frac{1}{4}\cG_{\rho_2,1}\left(\rho_1\right)-\frac{1}{4}\cG_{\rho_2,\rho_3}\left(\rho_1\right)-\frac{1}{4}\cG_{\rho_3,0}\left(\rho_2\right)\\
 \nonumber &+\frac{1}{4}\cG_{\rho_3,\rho_3}\left(\rho_2\right)+\frac{1}{2}\cG_0\left(\rho_3\right)\cG_{\rho_3}\left(\rho_2\right)-\frac{1}{4}\cG_1\left(\rho_3\right)\cG_{\rho_3}\left(\rho_2\right)\\
 \nonumber &-\frac{1}{4}\cG_0\left(\rho_2\right)\cG_0\left(\rho_3\right)+\frac{1}{4}\cG_0\left(\rho_3\right)\cG_1\left(\rho_2\right)+\frac{1}{2}\cG_0\left(\rho_2\right)\cG_1\left(\rho_3\right)\\
 \nonumber &-\frac{1}{4}\cG_1\left(\rho_2\right)\cG_1\left(\rho_3\right)-\frac{1}{4}\cG_0\left(\rho_3\right)\cG_{\rho_2}\left(\rho_1\right)+\frac{1}{2}\cG_1\left(\rho_3\right)\cG_{\rho_2}\left(\rho_1\right)\,. \\
\fd_{1,-+-+}^{\text{(0,0,1)}}\left(\rho_1,\rho_2,\rho_3\right) = &-\frac{1}{4}\cG_{0,1}\left(\rho_1\right)-\frac{1}{4}\cG_{0,\rho_3}\left(\rho_1\right)+\frac{1}{4}\cG_{1,0}\left(\rho_2\right)-\frac{1}{4}\cG_{1,\rho_3}\left(\rho_2\right)\\
 \nonumber &+\frac{1}{4}\cG_{\rho_2,1}\left(\rho_1\right)+\frac{1}{4}\cG_{\rho_2,\rho_3}\left(\rho_1\right)+\frac{1}{4}\cG_{\rho_3,0}\left(\rho_2\right)-\frac{1}{4}\cG_{\rho_3,\rho_3}\left(\rho_2\right)\\
 \nonumber &-\frac{1}{4}\cG_0\left(\rho_1\right)\cG_0\left(\rho_3\right)+\frac{1}{4}\cG_0\left(\rho_2\right)\cG_0\left(\rho_3\right)-\frac{1}{4}\cG_0\left(\rho_3\right)\cG_1\left(\rho_2\right)
 \end{align}
\begin{align}
 \nonumber &+\frac{1}{2}\cG_0\left(\rho_1\right)\cG_1\left(\rho_3\right)-\frac{1}{2}\cG_0\left(\rho_2\right)\cG_1\left(\rho_3\right)+\frac{1}{4}\cG_1\left(\rho_2\right)\cG_1\left(\rho_3\right)\\
 \nonumber &+\frac{1}{4}\cG_0\left(\rho_3\right)\cG_{\rho_2}\left(\rho_1\right)-\frac{1}{2}\cG_1\left(\rho_3\right)\cG_{\rho_2}\left(\rho_1\right)-\frac{1}{2}\cG_0\left(\rho_3\right)\cG_{\rho_3}\left(\rho_2\right)\\
 \nonumber &+\frac{1}{4}\cG_1\left(\rho_3\right)\cG_{\rho_3}\left(\rho_2\right)\,. \\
\fd_{2,-+-+}^{\text{(0,0,1)}}\left(\rho_1,\rho_2,\rho_3\right) = &-\frac{1}{2}\cG_{0,0}\left(\rho_3\right)+\frac{1}{4}\cG_{0,1}\left(\rho_1\right)+\frac{1}{4}\cG_{0,1}\left(\rho_3\right)+\frac{1}{4}\cG_{0,\rho_3}\left(\rho_1\right)\\
 \nonumber &+\frac{1}{2}\cG_{1,0}\left(\rho_3\right)+\frac{1}{4}\cG_0\left(\rho_1\right)\cG_0\left(\rho_3\right)-\frac{1}{2}\cG_0\left(\rho_1\right)\cG_1\left(\rho_3\right)\,. \\
 \nonumber \, \\
\fa^{\text{(0,1,0)}}_{-+-+}\left(\rho_1,\rho_2,\rho_3\right) = &-\frac{1}{4}\cG_{1,1}\left(\rho_1\right)+\frac{1}{4}\cG_{1,1}\left(\rho_2\right)+\frac{1}{4}\cG_{1,\rho_3}\left(\rho_1\right)-\frac{1}{4}\cG_{1,\rho_3}\left(\rho_2\right)\\
 \nonumber &+\frac{1}{4}\cG_0\left(\rho_3\right)\cG_1\left(\rho_1\right)-\frac{1}{4}\cG_0\left(\rho_3\right)\cG_1\left(\rho_2\right)\,. \\
\fb_{1,-+-+}^{\text{(0,1,0)}}\left(\rho_1,\rho_2,\rho_3\right) = &\frac{1}{4}\cG_{1,1}\left(\rho_2\right)-\frac{1}{4}\cG_{1,\rho_3}\left(\rho_2\right)-\frac{1}{4}\cG_0\left(\rho_3\right)\cG_1\left(\rho_2\right)\,. \\
\fb_{2,-+-+}^{\text{(0,1,0)}}\left(\rho_1,\rho_2,\rho_3\right) = &\frac{1}{4}\cG_{0,0}\left(\rho_3\right)+\frac{1}{4}\cG_{0,1}\left(\rho_1\right)-\frac{1}{2}\cG_{0,1}\left(\rho_2\right)-\frac{1}{4}\cG_{0,1}\left(\rho_3\right)\\
 \nonumber &-\frac{1}{4}\cG_{0,\rho_3}\left(\rho_1\right)+\frac{1}{2}\cG_{0,\rho_3}\left(\rho_2\right)-\frac{1}{4}\cG_0\left(\rho_1\right)\cG_0\left(\rho_3\right)+\frac{1}{2}\cG_0\left(\rho_3\right)\cG_1\left(\rho_2\right)\,. \\
\fb_{3,-+-+}^{\text{(0,1,0)}}\left(\rho_1,\rho_2,\rho_3\right) = &-\frac{1}{4}\cG_{1,0}\left(\rho_2\right)+\frac{1}{4}\cG_{1,0}\left(\rho_3\right)+\frac{1}{4}\cG_{1,1}\left(\rho_1\right)-\frac{1}{4}\cG_{1,1}\left(\rho_3\right)\\
 \nonumber &+\frac{1}{2}\cG_{1,\rho_2}\left(\rho_1\right)-\frac{1}{2}\cG_{1,\rho_3}\left(\rho_1\right)+\frac{1}{4}\cG_{1,\rho_3}\left(\rho_2\right)+\frac{1}{4}\cG_{\rho_2,1}\left(\rho_1\right)\\
 \nonumber &+\frac{1}{2}\cG_0\left(\rho_2\right)\cG_1\left(\rho_1\right)-\frac{1}{2}\cG_0\left(\rho_3\right)\cG_1\left(\rho_1\right)+\frac{1}{4}\cG_0\left(\rho_3\right)\cG_1\left(\rho_2\right)\\
 \nonumber &-\frac{1}{2}\cG_1\left(\rho_1\right)\cG_1\left(\rho_2\right)-\frac{1}{4}\cG_0\left(\rho_2\right)\cG_1\left(\rho_3\right)+\frac{1}{4}\cG_1\left(\rho_1\right)\cG_1\left(\rho_3\right)\\
 \nonumber &+\frac{1}{4}\cG_1\left(\rho_2\right)\cG_1\left(\rho_3\right)-\frac{1}{4}\cG_1\left(\rho_3\right)\cG_{\rho_2}\left(\rho_1\right)\,. \\
\fb_{4,-+-+}^{\text{(0,1,0)}}\left(\rho_1,\rho_2,\rho_3\right) = &-\frac{1}{4}\cG_{0,1}\left(\rho_2\right)+\frac{1}{4}\cG_{0,\rho_3}\left(\rho_2\right)-\frac{1}{4}\cG_{1,1}\left(\rho_2\right)+\frac{1}{4}\cG_{1,\rho_3}\left(\rho_2\right)\\
 \nonumber &+\frac{1}{4}\cG_{\rho_2,1}\left(\rho_1\right)-\frac{1}{4}\cG_{\rho_2,\rho_3}\left(\rho_1\right)+\frac{1}{4}\cG_{\rho_3,1}\left(\rho_2\right)-\frac{1}{4}\cG_{\rho_3,\rho_3}\left(\rho_2\right)\\
 \nonumber &-\frac{1}{4}\cG_0\left(\rho_3\right)\cG_1\left(\rho_1\right)+\frac{1}{4}\cG_0\left(\rho_3\right)\cG_1\left(\rho_2\right)+\frac{1}{4}\cG_1\left(\rho_1\right)\cG_1\left(\rho_2\right)\\
 \nonumber &-\frac{1}{4}\cG_1\left(\rho_2\right)\cG_{\rho_2}\left(\rho_1\right)-\frac{1}{4}\cG_1\left(\rho_1\right)\cG_{\rho_3}\left(\rho_2\right)+\frac{1}{4}\cG_{\rho_2}\left(\rho_1\right)\cG_{\rho_3}\left(\rho_2\right)\,. \\
\fc_{1,-+-+}^{\text{(0,1,0)}}\left(\rho_1,\rho_2,\rho_3\right) = &\frac{1}{4}\cG_{0,1}\left(\rho_1\right)-\frac{1}{4}\cG_{1,0}\left(\rho_2\right)+\frac{1}{4}\cG_{1,\rho_3}\left(\rho_2\right)-\frac{1}{4}\cG_{\rho_2,1}\left(\rho_1\right)\\
 \nonumber &+\frac{1}{4}\cG_0\left(\rho_3\right)\cG_1\left(\rho_2\right)-\frac{1}{4}\cG_0\left(\rho_1\right)\cG_1\left(\rho_3\right)+\frac{1}{4}\cG_0\left(\rho_2\right)\cG_1\left(\rho_3\right)\\
 \nonumber &-\frac{1}{4}\cG_1\left(\rho_2\right)\cG_1\left(\rho_3\right)+\frac{1}{4}\cG_1\left(\rho_3\right)\cG_{\rho_2}\left(\rho_1\right)\,. \\
\fc_{2,-+-+}^{\text{(0,1,0)}}\left(\rho_1,\rho_2,\rho_3\right) = &\frac{1}{2}\cG_{0,0}\left(\rho_2\right)-\frac{1}{2}\cG_{0,0}\left(\rho_3\right)-\frac{1}{2}\cG_{0,1}\left(\rho_1\right)+\frac{1}{2}\cG_{0,1}\left(\rho_3\right)\\
 \nonumber &-\frac{1}{2}\cG_{0,\rho_2}\left(\rho_1\right)+\frac{1}{2}\cG_{0,\rho_3}\left(\rho_1\right)-\frac{1}{2}\cG_{0,\rho_3}\left(\rho_2\right)-\frac{1}{2}\cG_0\left(\rho_3\right)\cG_1\left(\rho_2\right)\\
 \nonumber &-\frac{1}{2}\cG_0\left(\rho_1\right)\cG_0\left(\rho_2\right)+\frac{1}{2}\cG_0\left(\rho_1\right)\cG_0\left(\rho_3\right)+\frac{1}{2}\cG_0\left(\rho_1\right)\cG_1\left(\rho_2\right)\,. 
\end{align}
\begin{align}
\fc_{3,-+-+}^{\text{(0,1,0)}}\left(\rho_1,\rho_2,\rho_3\right) = &\frac{1}{4}\cG_{0,1}\left(\rho_1\right)-\frac{1}{4}\cG_{0,\rho_3}\left(\rho_1\right)-\frac{1}{4}\cG_{\rho_2,1}\left(\rho_1\right)+\frac{1}{4}\cG_{\rho_2,\rho_3}\left(\rho_1\right)\\
 \nonumber &-\frac{1}{4}\cG_{\rho_3,1}\left(\rho_2\right)+\frac{1}{4}\cG_{\rho_3,\rho_3}\left(\rho_2\right)-\frac{1}{4}\cG_{\rho_2}\left(\rho_1\right)\cG_{\rho_3}\left(\rho_2\right)\\
%
 \nonumber &-\frac{1}{4}\cG_0\left(\rho_1\right)\cG_1\left(\rho_2\right)+\frac{1}{2}\cG_0\left(\rho_3\right)\cG_1\left(\rho_2\right)-\frac{1}{4}\cG_1\left(\rho_2\right)\cG_1\left(\rho_3\right)\\
 \nonumber &+\frac{1}{4}\cG_1\left(\rho_2\right)\cG_{\rho_2}\left(\rho_1\right)+\frac{1}{4}\cG_0\left(\rho_1\right)\cG_{\rho_3}\left(\rho_2\right)-\frac{1}{4}\cG_0\left(\rho_3\right)\cG_{\rho_3}\left(\rho_2\right)\\
 \nonumber &+\frac{1}{4}\cG_1\left(\rho_3\right)\cG_{\rho_3}\left(\rho_2\right)\,. \\
\fc_{4,-+-+}^{\text{(0,1,0)}}\left(\rho_1,\rho_2,\rho_3\right) = &-\frac{1}{4}\cG_{0,0}\left(\rho_3\right)-\frac{1}{4}\cG_{0,1}\left(\rho_1\right)+\frac{1}{2}\cG_{0,1}\left(\rho_2\right)+\frac{1}{4}\cG_{0,1}\left(\rho_3\right)\\
 \nonumber &+\frac{1}{4}\cG_{0,\rho_3}\left(\rho_1\right)-\frac{1}{2}\cG_{0,\rho_3}\left(\rho_2\right)+\frac{1}{4}\cG_0\left(\rho_1\right)\cG_0\left(\rho_3\right)-\frac{1}{2}\cG_0\left(\rho_3\right)\cG_1\left(\rho_2\right)\,. \\
\fc_{5,-+-+}^{\text{(0,1,0)}}\left(\rho_1,\rho_2,\rho_3\right) = &-\frac{1}{4}\cG_{0,0}\left(\rho_2\right)-\frac{1}{4}\cG_{0,0}\left(\rho_3\right)+\frac{1}{2}\cG_{0,1}\left(\rho_2\right)+\frac{1}{4}\cG_{0,1}\left(\rho_3\right)\\
 \nonumber &-\frac{1}{4}\cG_{0,\rho_3}\left(\rho_2\right)+\frac{1}{4}\cG_{1,0}\left(\rho_2\right)-\frac{1}{4}\cG_{1,\rho_2}\left(\rho_1\right)-\frac{1}{4}\cG_{1,\rho_3}\left(\rho_2\right)\\
 \nonumber &-\frac{1}{2}\cG_{\rho_2,1}\left(\rho_1\right)-\frac{1}{4}\cG_{\rho_2,\rho_2}\left(\rho_1\right)+\frac{1}{2}\cG_{\rho_2,\rho_3}\left(\rho_1\right)+\frac{1}{4}\cG_{\rho_3,0}\left(\rho_2\right)\\
 \nonumber &-\frac{1}{2}\cG_{\rho_3,1}\left(\rho_2\right)+\frac{1}{4}\cG_{\rho_3,\rho_3}\left(\rho_2\right)-\frac{1}{4}\cG_{\rho_2}\left(\rho_1\right)\cG_{\rho_3}\left(\rho_2\right)\\
 \nonumber &+\frac{1}{4}\cG_0\left(\rho_2\right)\cG_0\left(\rho_3\right)-\frac{1}{4}\cG_0\left(\rho_2\right)\cG_1\left(\rho_1\right)+\frac{1}{4}\cG_0\left(\rho_3\right)\cG_1\left(\rho_1\right)\\
 \nonumber &-\frac{1}{4}\cG_0\left(\rho_3\right)\cG_1\left(\rho_2\right)-\frac{1}{4}\cG_1\left(\rho_2\right)\cG_1\left(\rho_3\right)-\frac{1}{4}\cG_0\left(\rho_2\right)\cG_{\rho_2}\left(\rho_1\right)\\
 \nonumber &+\frac{1}{4}\cG_0\left(\rho_3\right)\cG_{\rho_2}\left(\rho_1\right)+\frac{1}{2}\cG_1\left(\rho_2\right)\cG_{\rho_2}\left(\rho_1\right)-\frac{1}{4}\cG_0\left(\rho_3\right)\cG_{\rho_3}\left(\rho_2\right)\\
 \nonumber &+\frac{1}{4}\cG_1\left(\rho_1\right)\cG_{\rho_3}\left(\rho_2\right)+\frac{1}{4}\cG_1\left(\rho_3\right)\cG_{\rho_3}\left(\rho_2\right)\,. \\
\fd_{1,-+-+}^{\text{(0,1,0)}}\left(\rho_1,\rho_2,\rho_3\right) = &\frac{1}{2}\cG_{0,0}\left(\rho_2\right)-\frac{1}{2}\cG_{0,1}\left(\rho_1\right)-\frac{1}{2}\cG_{0,1}\left(\rho_2\right)-\frac{1}{4}\cG_{0,\rho_2}\left(\rho_1\right)\\
 \nonumber &+\frac{1}{2}\cG_{0,\rho_3}\left(\rho_1\right)+\frac{1}{2}\cG_{\rho_2,1}\left(\rho_1\right)+\frac{1}{4}\cG_{\rho_2,\rho_2}\left(\rho_1\right)-\frac{1}{2}\cG_{\rho_2,\rho_3}\left(\rho_1\right)\\
 \nonumber &-\frac{1}{4}\cG_{\rho_3,0}\left(\rho_2\right)+\frac{1}{2}\cG_{\rho_3,1}\left(\rho_2\right)-\frac{1}{4}\cG_{\rho_3,\rho_3}\left(\rho_2\right)+\frac{1}{4}\cG_{\rho_2}\left(\rho_1\right)\cG_{\rho_3}\left(\rho_2\right)\\
 \nonumber &-\frac{1}{4}\cG_0\left(\rho_1\right)\cG_0\left(\rho_2\right)+\frac{1}{4}\cG_0\left(\rho_1\right)\cG_0\left(\rho_3\right)-\frac{1}{4}\cG_0\left(\rho_2\right)\cG_0\left(\rho_3\right)\\
 \nonumber &+\frac{1}{2}\cG_0\left(\rho_1\right)\cG_1\left(\rho_2\right)-\frac{1}{2}\cG_0\left(\rho_3\right)\cG_1\left(\rho_2\right)+\frac{1}{2}\cG_1\left(\rho_2\right)\cG_1\left(\rho_3\right)\\
 \nonumber &+\frac{1}{4}\cG_0\left(\rho_2\right)\cG_{\rho_2}\left(\rho_1\right)-\frac{1}{4}\cG_0\left(\rho_3\right)\cG_{\rho_2}\left(\rho_1\right)-\frac{1}{2}\cG_1\left(\rho_2\right)\cG_{\rho_2}\left(\rho_1\right)\\
 \nonumber &-\frac{1}{4}\cG_0\left(\rho_1\right)\cG_{\rho_3}\left(\rho_2\right)+\frac{1}{2}\cG_0\left(\rho_3\right)\cG_{\rho_3}\left(\rho_2\right)-\frac{1}{2}\cG_1\left(\rho_3\right)\cG_{\rho_3}\left(\rho_2\right)\,. \\
\fd_{2,-+-+}^{\text{(0,1,0)}}\left(\rho_1,\rho_2,\rho_3\right) = &-\frac{1}{2}\cG_{0,0}\left(\rho_2\right)+\frac{1}{2}\cG_{0,0}\left(\rho_3\right)+\frac{1}{2}\cG_{0,1}\left(\rho_1\right)-\frac{1}{2}\cG_{0,1}\left(\rho_3\right)\\
 \nonumber &+\frac{1}{2}\cG_{0,\rho_2}\left(\rho_1\right)-\frac{1}{2}\cG_{0,\rho_3}\left(\rho_1\right)+\frac{1}{2}\cG_{0,\rho_3}\left(\rho_2\right)+\frac{1}{2}\cG_0\left(\rho_3\right)\cG_1\left(\rho_2\right)\\
 \nonumber &+\frac{1}{2}\cG_0\left(\rho_1\right)\cG_0\left(\rho_2\right)-\frac{1}{2}\cG_0\left(\rho_1\right)\cG_0\left(\rho_3\right)-\frac{1}{2}\cG_0\left(\rho_1\right)\cG_1\left(\rho_2\right)\,. 
\end{align}
%
%
%
%
\begin{align}
\fa^{\text{(1,0,0)}}_{-+-+}\left(\rho_1,\rho_2,\rho_3\right) = &\frac{1}{4}\cG_{1,1}\left(\rho_1\right)-\frac{1}{4}\cG_{1,\rho_3}\left(\rho_1\right)-\frac{1}{4}\cG_0\left(\rho_3\right)\cG_1\left(\rho_1\right)\,. \\
\fb_{1,-+-+}^{\text{(1,0,0)}}\left(\rho_1,\rho_2,\rho_3\right) = &0\,. \\
\fb_{2,-+-+}^{\text{(1,0,0)}}\left(\rho_1,\rho_2,\rho_3\right) = &\frac{1}{4}\cG_{0,0}\left(\rho_3\right)-\frac{1}{4}\cG_{0,1}\left(\rho_1\right)-\frac{1}{4}\cG_{0,1}\left(\rho_3\right)+\frac{1}{4}\cG_{0,\rho_3}\left(\rho_1\right)\\
 \nonumber &-\frac{1}{4}\cG_0\left(\rho_1\right)\cG_0\left(\rho_3\right)+\frac{1}{2}\cG_0\left(\rho_3\right)\cG_1\left(\rho_1\right)\,. \\
\fb_{3,-+-+}^{\text{(1,0,0)}}\left(\rho_1,\rho_2,\rho_3\right) = &-\frac{1}{4}\cG_{1,0}\left(\rho_1\right)-\frac{1}{4}\cG_{1,1}\left(\rho_1\right)+\frac{1}{4}\cG_{1,\rho_3}\left(\rho_1\right)+\frac{1}{4}\cG_0\left(\rho_3\right)\cG_1\left(\rho_1\right)\,. \\
\fb_{4,-+-+}^{\text{(1,0,0)}}\left(\rho_1,\rho_2,\rho_3\right) = &-\frac{1}{4}\cG_{\rho_2,1}\left(\rho_1\right)+\frac{1}{4}\cG_{\rho_2,\rho_3}\left(\rho_1\right)-\frac{1}{4}\cG_{\rho_2}\left(\rho_1\right)\cG_{\rho_3}\left(\rho_2\right)\\
 \nonumber &+\frac{1}{4}\cG_0\left(\rho_3\right)\cG_1\left(\rho_1\right)-\frac{1}{4}\cG_1\left(\rho_1\right)\cG_1\left(\rho_2\right)+\frac{1}{4}\cG_1\left(\rho_2\right)\cG_{\rho_2}\left(\rho_1\right)\\
 \nonumber &+\frac{1}{4}\cG_1\left(\rho_1\right)\cG_{\rho_3}\left(\rho_2\right)\,. \\
\fc_{1,-+-+}^{\text{(1,0,0)}}\left(\rho_1,\rho_2,\rho_3\right) = &0\,. \\
\fc_{2,-+-+}^{\text{(1,0,0)}}\left(\rho_1,\rho_2,\rho_3\right) = &-\frac{1}{4}\cG_{0,0}\left(\rho_1\right)-\frac{1}{4}\cG_{0,0}\left(\rho_3\right)+\frac{1}{4}\cG_{0,1}\left(\rho_1\right)+\frac{1}{4}\cG_{0,1}\left(\rho_3\right)\\
 \nonumber &-\frac{1}{4}\cG_{0,\rho_3}\left(\rho_1\right)+\frac{1}{2}\cG_{1,0}\left(\rho_1\right)+\frac{1}{4}\cG_0\left(\rho_1\right)\cG_0\left(\rho_3\right)-\frac{1}{2}\cG_0\left(\rho_3\right)\cG_1\left(\rho_1\right)\,. \\
\fc_{3,-+-+}^{\text{(1,0,0)}}\left(\rho_1,\rho_2,\rho_3\right) = &-\frac{1}{4}\cG_{0,1}\left(\rho_1\right)+\frac{1}{4}\cG_{0,1}\left(\rho_2\right)+\frac{1}{4}\cG_{0,\rho_3}\left(\rho_1\right)-\frac{1}{4}\cG_{0,\rho_3}\left(\rho_2\right)\\
 \nonumber &-\frac{1}{4}\cG_{1,1}\left(\rho_2\right)+\frac{1}{4}\cG_{1,\rho_3}\left(\rho_2\right)+\frac{1}{4}\cG_{\rho_2,1}\left(\rho_1\right)-\frac{1}{4}\cG_{\rho_2,\rho_3}\left(\rho_1\right)\\
 \nonumber &-\frac{1}{4}\cG_0\left(\rho_1\right)\cG_1\left(\rho_2\right)+\frac{1}{4}\cG_0\left(\rho_3\right)\cG_1\left(\rho_2\right)+\frac{1}{2}\cG_1\left(\rho_1\right)\cG_1\left(\rho_2\right)\\
 \nonumber &-\frac{1}{4}\cG_1\left(\rho_2\right)\cG_1\left(\rho_3\right)-\frac{1}{4}\cG_1\left(\rho_2\right)\cG_{\rho_2}\left(\rho_1\right)+\frac{1}{4}\cG_0\left(\rho_1\right)\cG_{\rho_3}\left(\rho_2\right)\\
 \nonumber &-\frac{1}{4}\cG_0\left(\rho_3\right)\cG_{\rho_3}\left(\rho_2\right)-\frac{1}{2}\cG_1\left(\rho_1\right)\cG_{\rho_3}\left(\rho_2\right)+\frac{1}{4}\cG_1\left(\rho_3\right)\cG_{\rho_3}\left(\rho_2\right)\\
 \nonumber &+\frac{1}{4}\cG_{\rho_2}\left(\rho_1\right)\cG_{\rho_3}\left(\rho_2\right)\,. \\
\fc_{4,-+-+}^{\text{(1,0,0)}}\left(\rho_1,\rho_2,\rho_3\right) = &-\frac{1}{4}\cG_{0,0}\left(\rho_3\right)+\frac{1}{4}\cG_{0,1}\left(\rho_1\right)+\frac{1}{4}\cG_{0,1}\left(\rho_3\right)-\frac{1}{4}\cG_{0,\rho_3}\left(\rho_1\right)\\
 \nonumber &+\frac{1}{4}\cG_0\left(\rho_1\right)\cG_0\left(\rho_3\right)-\frac{1}{2}\cG_0\left(\rho_3\right)\cG_1\left(\rho_1\right)\,. \\
\fc_{5,-+-+}^{\text{(1,0,0)}}\left(\rho_1,\rho_2,\rho_3\right) = &\frac{1}{4}\cG_{1,\rho_2}\left(\rho_1\right)+\frac{1}{4}\cG_{\rho_2,0}\left(\rho_1\right)+\frac{1}{4}\cG_{\rho_2,1}\left(\rho_1\right)-\frac{1}{4}\cG_{\rho_2,\rho_2}\left(\rho_1\right)\\
 \nonumber &-\frac{1}{4}\cG_{\rho_2,\rho_3}\left(\rho_1\right)-\frac{1}{4}\cG_1\left(\rho_1\right)\cG_{\rho_3}\left(\rho_2\right)+\frac{1}{4}\cG_{\rho_2}\left(\rho_1\right)\cG_{\rho_3}\left(\rho_2\right)\\
 \nonumber &+\frac{1}{4}\cG_0\left(\rho_2\right)\cG_1\left(\rho_1\right)-\frac{1}{4}\cG_0\left(\rho_3\right)\cG_1\left(\rho_1\right)-\frac{1}{4}\cG_0\left(\rho_2\right)\cG_{\rho_2}\left(\rho_1\right)\,. \\
\fd_{1,-+-+}^{\text{(1,0,0)}}\left(\rho_1,\rho_2,\rho_3\right) = &-\frac{1}{4}\cG_{0,0}\left(\rho_1\right)-\frac{1}{4}\cG_{0,0}\left(\rho_2\right)+\frac{1}{4}\cG_{0,1}\left(\rho_1\right)+\frac{1}{4}\cG_{0,\rho_2}\left(\rho_1\right)\\
 \nonumber &-\frac{1}{4}\cG_{0,\rho_3}\left(\rho_1\right)+\frac{1}{4}\cG_{0,\rho_3}\left(\rho_2\right)+\frac{1}{2}\cG_{1,0}\left(\rho_1\right)+\frac{1}{4}\cG_{1,0}\left(\rho_2\right)\\
 \nonumber &-\frac{1}{2}\cG_{1,\rho_2}\left(\rho_1\right)-\frac{1}{4}\cG_{1,\rho_3}\left(\rho_2\right)-\frac{1}{4}\cG_{\rho_2,0}\left(\rho_1\right)-\frac{1}{4}\cG_{\rho_2,1}\left(\rho_1\right)
 \end{align}
\begin{align}
 \nonumber &+\frac{1}{4}\cG_{\rho_2,\rho_2}\left(\rho_1\right)+\frac{1}{4}\cG_{\rho_2,\rho_3}\left(\rho_1\right)-\frac{1}{4}\cG_{\rho_2}\left(\rho_1\right)\cG_{\rho_3}\left(\rho_2\right)\\
 \nonumber &+\frac{1}{4}\cG_0\left(\rho_1\right)\cG_0\left(\rho_2\right)-\frac{1}{2}\cG_0\left(\rho_2\right)\cG_1\left(\rho_1\right)-\frac{1}{4}\cG_0\left(\rho_3\right)\cG_1\left(\rho_2\right)\\
 \nonumber &+\frac{1}{4}\cG_1\left(\rho_2\right)\cG_1\left(\rho_3\right)+\frac{1}{4}\cG_0\left(\rho_2\right)\cG_{\rho_2}\left(\rho_1\right)-\frac{1}{4}\cG_0\left(\rho_1\right)\cG_{\rho_3}\left(\rho_2\right)\\
 \nonumber &+\frac{1}{4}\cG_0\left(\rho_3\right)\cG_{\rho_3}\left(\rho_2\right)+\frac{1}{2}\cG_1\left(\rho_1\right)\cG_{\rho_3}\left(\rho_2\right)-\frac{1}{4}\cG_1\left(\rho_3\right)\cG_{\rho_3}\left(\rho_2\right)\,. \\
\fd_{2,-+-+}^{\text{(1,0,0)}}\left(\rho_1,\rho_2,\rho_3\right) = &\frac{1}{4}\cG_{0,0}\left(\rho_1\right)+\frac{1}{4}\cG_{0,0}\left(\rho_3\right)-\frac{1}{4}\cG_{0,1}\left(\rho_1\right)-\frac{1}{4}\cG_{0,1}\left(\rho_3\right)\\
 \nonumber &+\frac{1}{4}\cG_{0,\rho_3}\left(\rho_1\right)-\frac{1}{2}\cG_{1,0}\left(\rho_1\right)-\frac{1}{4}\cG_0\left(\rho_1\right)\cG_0\left(\rho_3\right)+\frac{1}{2}\cG_0\left(\rho_3\right)\cG_1\left(\rho_1\right)\,. 
\end{align}

\end{document}